\documentclass[12pt,a4paper,twoside]{cmhbook}
\makeglossary
\usepackage[dvips]{graphics}
\usepackage{doublespace}
\usepackage{array}
\usepackage{multicol}
\usepackage{multirow}
\usepackage{subfigure}
\usepackage{cite}
\usepackage{fancyhdr}
\usepackage{afterpage}
\usepackage{axodraw}
\usepackage{rotating}
\usepackage{psfrag}
\usepackage[draft]{epsfig}
\usepackage[bf,small]{caption2}
\usepackage{amsmath}
\usepackage{amssymb}
\usepackage{url}

\newcommand{\mylinespace}{1.66} 
\fancypagestyle{plain}{
\fancyhf{}
\fancyfoot[C]{\bf \thepage}
\renewcommand{\headrulewidth}{0pt}
}

\newcommand{\capbox}[2]{\parbox{0.85\textwidth}{\caption[#1]{#2}}}

\newcommand{\secref}[1]{\S\ref{#1}}

\begin{document}
\pagestyle{plain}
\setcounter{tocdepth}{2} 
\setcounter{secnumdepth}{3} 

\frontmatter

\begin{singlespace}

\begin{titlepage}
\begin{center}
\begin{spacing}{2.2}
${}^{}$ \\
\vspace{20mm}
{\LARGE\bf Physics Beyond the Standard Model:} \\
{\LARGE\bf Exotic Leptons and Black Holes} \\
{\LARGE\bf at Future Colliders}\\
\end{spacing}
\vspace{40mm}
\large
Christopher Michael Harris\\
${}^{}$\\
Christ's College \\
\vspace{82mm} \normalsize 
A dissertation submitted to the University
of Cambridge \\ for the degree of Doctor of Philosophy \\ December 2003
\end{center}
\end{titlepage}

$\ $
\newpage
\begin{center}
\begin{spacing}{1.4}
${}^{}$ \\
\vspace{1.5cm}
{\Large\bf Physics Beyond the Standard Model:}\\
{\Large\bf Exotic Leptons and Black Holes at Future Colliders}\\
\end{spacing}
\vspace{0.5cm}
{\large Christopher Michael Harris}\\
\vspace{1.0cm}
{\Large\bf Abstract} \\
\end{center}
\normalsize
\begin{spacing}{1.5} 

The Standard Model of particle physics has been remarkably successful in describing present experimental results.  However, it is assumed to be only a low-energy effective theory which will break down at higher energy scales, theoretically motivated to be around 1~TeV\@.  There are a variety of proposed models of new physics beyond the Standard Model, most notably supersymmetric and extra dimension models. 

New charged and neutral heavy leptons are a feature of a number of theories of new physics, including the `intermediate scale' class of supersymmetric models.  Using a time-of-flight technique to detect the charged leptons at the Large Hadron Collider, the discovery range (in the particular scenario studied in the first part of this thesis) is found to extend up to masses of 950~GeV\@.

Extra dimension models, particularly those with large extra dimensions, allow the possible experimental production of black holes.  The remainder of the thesis describes some theoretical results and computational tools necessary to model the production and decay of these miniature black holes at future particle colliders.  The grey-body factors which describe the Hawking radiation emitted by higher-dimensional black holes are calculated numerically for the first time and then incorporated in a Monte Carlo black hole event generator; this can be used to model black hole production and decay at next-generation colliders.  It is hoped that this generator will allow more detailed examination of black hole signatures and help to devise a method for extracting the number of extra dimensions present in nature.

\end{spacing}

\end{singlespace}

\newpage
\vspace*{65mm}
\centerline{\Huge {\bf Declaration}}
\vspace{10mm}
This thesis is the result of my own work, except where explicit
reference is made to the work of others (see Preface), 
and has not been submitted
for another qualification to this or any other university.

It does not exceed the 60,000 word limit prescribed by the Degree Committee for Physics \& Chemistry.
\\
\\
\\
\vspace{30mm}
\hspace{8cm}
Christopher Michael Harris

\newpage
{\ }
\vspace{10mm}
\begin{center}
{\Large\bf Acknowledgements}
\end{center}
\vspace{5mm}

I would like to thank the Particle Physics and Astronomy Research Council for
providing financial support during this degree (studentship number PPA/S/S/2000/03001). Thanks also to Christ's College for their financial assistance with foreign travel, the benefits provided by a BA scholarship, and the opportunity to teach mathematics to first and second year undergraduates.

I am very grateful to my supervisor, Professor Bryan Webber, for all his help, suggestions and encouragement during the last three years.  Thanks are also due to those with whom I have worked closely, particularly Dr.\ Panagiota Kanti and Dr.\ Peter Richardson, who kindly also proof-read this thesis.

My thanks also go to my ATLAS experimental colleagues in Cambridge, particularly Dr.\ Ali Sabetfakhri (who was one of the first users of the black hole event generator) and Dr.\ Christopher Lester (who made many helpful comments on all aspects of this work).

Finally I would like to thank my family, particularly my parents, for all their support and understanding, and all my friends at Christ's College and at St.\ Andrew the Great Church.

\begin{singlespace}
\begingroup
\setlength{\parskip}{0.29cm}
\tableofcontents
\listoffigures
\listoftables
\endgroup
\end{singlespace}
\mainmatter

\pagenumbering{arabic}

\pagestyle{fancy}
\renewcommand{\chaptermark}[1]{\markboth{\bf \chaptername\ \thechapter:  {#1}}{}}
\renewcommand{\sectionmark}[1]{\markright{\bf \thesection\ #1}}
\fancyhead[LE,RO]{\bf\thepage}
\fancyhead[LO]{\leftmark}
\fancyhead[RE]{\rightmark}
\renewcommand{\headrulewidth}{0.4pt}
\fancyfoot{}
\chapter*{Preface}

This constitutes my research for a PhD degree in theoretical high-energy physics.  Chapter~\ref{introch} contains a review of the relevant theoretical background.  It sets the context in which the work described in the rest of the thesis was carried out, and introduces the theoretical models on which the work is based.  Chapter~\ref{leptons} describes work undertaken during the first year of my PhD on the prospects for detecting exotic heavy leptons at the Large Hadron Collider (LHC).  This was published \cite{Allanach:2001sd} as a study performed by members of the Cambridge Supersymmetry Working Group, but the contents of Chapter~\ref{leptons} are entirely my own work.  

The remainder of the 
thesis concentrates on the study of miniature black holes at the LHC---a topic which is introduced in Chapter~\ref{bhintro}.  Chapter~\ref{greybody} describes new theoretical work on black hole grey-body factors (building on previous work by Kanti and March-Russell); much of this work was published in \cite{Harris:2003eg} and contributions from my collaborator, P.~Kanti, are clearly identified in Chapter~\ref{greybody}.  The final chapter gives details of the {\small CHARYBDIS} event generator written to simulate black hole production and decay.  This was published in \cite{Harris:2003db} and, as detailed in Chapter~\ref{generator}, involved some significant contributions from P.~Richardson (these were mainly involved in interfacing {\small CHARYBDIS} to existing Monte Carlo programs).  The chapter also contains a brief discussion of some preliminary experimental studies using the event generator.  Some of this work will be published in \cite{ali}, together with more detailed analyses by members of the ATLAS collaboration.

\clearpage{\pagestyle{empty}\cleardoublepage}
\chapter{Introduction}
\chaptermark{Introduction}
\label{introch}
\section{The Standard Model}
\label{thesm}

The Standard Model (SM) of particle physics has been remarkably successful.  Many of the parameters are measured to great accuracy, and the Standard Model has made many predictions which have been verified experimentally. 
However there are several reasons why it is widely believed that, whilst working well in the energy regimes which have been investigated to date, the Standard Model does not give the full picture.  These issues, which are both technical and aesthetic, will be discussed further in the following sections.  It is the belief that there is physics `Beyond the Standard Model' (BSM) which motivates the continuation of large scale particle physics experiments, most notably the Large Hadron Collider (LHC) at CERN which will collide protons with a centre-of-mass energy of 14~TeV\@.  Of course, the other great aim of the LHC is to discover the famous Higgs boson, the only missing particle of the Standard Model.  Although the Higgs mechanism is as yet unproved, much of the work in this thesis makes the implicit and sometimes explicit assumption that a Standard Model Higgs boson exists.

\subsection{Particle content}

The particle content of the Standard Model is intimately related to the symmetries of the Lagrangian which describes the physics.  The forces between the fermions are mediated by gauge boson fields which allow the Lagrangian to be invariant under gauge transformations---it is believed that all fundamental interactions are described by some form of gauge theory.  These symmetries imply the existence of conserved currents and charges in the theory.

Firstly the situation before the breaking of electroweak symmetry is reviewed.  At this stage all particles are massless as the Higgs mechanism has not yet been invoked to give masses to the particles.

There are three main symmetries of the Standard Model Lagrangian.  Firstly there is
the U(1) hypercharge symmetry;  this is an Abelian symmetry and the corresponding gauge boson field is usually written as $B^{\mu}$.  The conserved charge, the hypercharge, is most commonly called $Y$.  

There is also an SU(2) weak isospin symmetry which is sometimes referred to as SU(2)$_\text{L}$ since the gauge field couples only to left-handed fermions.  This is a non-Abelian symmetry since the Lie group generators are matrices.  There are three accompanying gauge fields: $W_1^{\mu}$, $W_2^{\mu}$ and $W_3^{\mu}$ and the conserved charge is $I_3$, the third component of the weak isospin $I$.

Finally there is the SU(3) non-Abelian symmetry of Quantum Chromodynamics (QCD) which has a conserved colour charge.  The gauge bosons in this case are gluons and are described by the fields $A_a^{\mu}$ where the colour index $a$ runs from 1 to 8.  The gluons themselves carry colour charge---red, green and blue.

The theory outlined above is clearly incomplete since we know that most observed particles are not massless.  It is the introduction of a Higgs field (a complex scalar field with a non-zero vacuum expectation value) to the theory which both gives most of the particles mass and causes the electroweak symmetry breaking.  The $W_3^{\mu}$ from the SU(2) of isospin mixes with the $B^{\mu}$ from the U(1) of hypercharge to produce a massless photon field and a massive
Z boson.  At the same time the $W_1$ and $W_2$ states which can be identified with the W$^{\pm}$ bosons acquire mass, as do all the fermions with the exception of the neutrinos.\footnote{In the minimal Standard Model neutrinos are massless although recent evidence suggests that this is not the case.}  After this mixing there is a residual U(1)$_Q$ electromagnetic gauge symmetry; the coupling of fermions to the photon is found to be proportional to the electromagnetic charge $Q$, which is related to the hypercharge and weak isospin through $Q=Y+I_3$.\footnote{It is not uncommon for $T$ to be used for the isospin rather than $I$, or for hypercharge to be defined such that this relationship becomes $Q=Y/2+I_3$.}  The coupling to the Z boson is more complicated but is proportional to $I_3-Q\sin^2\theta_\text{W}$ where $\theta_\text{W}$ is the Weinberg angle and $\sin^2\theta_\text{W} \sim 0.23$.  Since $I_3$ is different for the left and right-handed fermions, the coupling to the Z boson is also dependent on the fermion helicity.

In Table~\ref{ewtable} the fermionic particle content of the Standard Model is summarized and the various quantum numbers of the different particles are shown.  From this table, in which L and R refer to left- and right-handed states respectively, it is clear why the isospin symmetry is often referred to as SU(2)$_\text{L}$.

\begin{table}
\def\arraystretch{1.25}
\begin{center}
\begin{tabular}{|c|c|c|c|c|}
\hline
& \multicolumn{2}{c|}{SU(2)$_\text{L}$} & U(1)$_Y$ & U(1)$_Q$ \\
\cline{2-5}
& $I$ & $I_3$ & $Y$ & $Q$\\
\hline
$\text{u}_\text{R}$ & $0$ & $\phantom{-}0$ & $\phantom{-}\frac{2}{3}$ & $\phantom{-}\frac{2}{3}$\\
$\text{d}_\text{R}$ & $0$ & $\phantom{-}0$ & $-\frac{1}{3}$ & $-\frac{1}{3}$\\
\hline
$\biggl(\!\begin{array}{c}
\text{u}\\
\text{d}\\
\end{array}\!\biggr)_{\!\text{L}}$ & $\frac{1}{2}$ &
$\begin{array}{c}
\phantom{-}\frac{1}{2}\\
-\frac{1}{2}\\
\end{array}$ &
$\phantom{-}\frac{1}{6}$ &
$\begin{array}{r}
\phantom{-}\frac{2}{3}\\
-\frac{1}{3}\\
\end{array}$\\
\hline
$\text{e}_\text{R}$ & $0$ & $\phantom{-}0$ & $-1$ & $-1$\\
\hline
$\biggl(\!\begin{array}{c}
\nu_\text{e}\\
\text{e}\\
\end{array}\!\biggr)_{\!\text{L}}$ & $\frac{1}{2}$ &
$\begin{array}{r}
+\frac{1}{2}\\
-\frac{1}{2}\\
\end{array}$ &
$-\frac{1}{2}$ &
$\begin{array}{r}
0\\
-1\\
\end{array}$\\
\hline
\end{tabular}
\capbox{Fermionic particle content of the Standard Model}{Fermionic particle content of the Standard Model showing the relevant quantum numbers.\label{ewtable}} 
\end{center}
\def\arraystretch{1.0}
\end{table} 

Apart from the fermions (with spin $s=1/2$) the SM particle content consists of the gauge bosons ($s=1$) and the Higgs boson ($s=0$).  The Higgs field introduced in the Higgs mechanism has four degrees of freedom, but three of these are `eaten' by the isospin gauge bosons to give mass to the W and Z bosons.  Therefore after the symmetry breaking there is a massive neutral scalar with one degree of freedom---this is the particle known as the Higgs boson.  Although the Higgs boson has not yet been discovered and the significance of hints during the final running of the Large Electron-Positron (LEP) collider has been reduced by more careful analyses \cite{Barate:2003sz}, a low Higgs mass ($\sim$~115 GeV) is still favoured.  This is because of tight constraints coming from precision measurements of electroweak radiative corrections.

In addition to the Standard Model gauge symmetries discussed above, there are two further `accidental symmetries' in that all the terms in the Lagrangian are found to conserve both lepton number $L$ and baryon number $B$.  Lepton number is defined to be $+1$ for leptons, $-1$ for anti-leptons and zero for all other particles; baryon number is defined to be $+1/3$ for quarks, $-1/3$ for anti-quarks and zero for all other particles (this means that baryons have $B=1$).  These symmetries were not imposed on the model, but appear naturally given the requirement that the terms in the Lagrangian should be renormalizable, Lorentz invariant and gauge invariant.  In fact $B$ and $L$ are not exactly conserved because they are violated by non-perturbative electroweak effects \cite{'tHooft:1976fv,'tHooft:1976up}, but they can normally be considered as conserved quantum numbers.  Unfortunately for most theories of physics beyond the Standard Model the most general Lagrangian which can be written down includes terms which violate these symmetries.  Potentially this causes various problems, most notably that it can allow protons to decay with a much shorter lifetime than the present experimental limit.

\section{Motivations for new physics}

It has already been mentioned that there are various reasons for believing there is physics beyond the Standard Model.  The work in this thesis will address some, but by no means all, of these issues which are detailed below.

\subsection{Gravity}

A full model for particle physics would be expected to describe all the fundamental forces between particles---electroweak, strong and gravitational.  However the Standard Model fails to describe any details of the gravitational force, and hence cannot be the full fundamental model.  At energy scales of order the Planck mass, $M_\text{P}$, a theory of quantum gravitation will be required to describe the interactions between particles.  This shows that the Standard Model will need to be replaced by an alternative theory at very high energy scales, and it is reasonable to expect that a more complete model than the Standard Model might be required even at energy scales only moderately higher than those which have already been investigated in detail.

\subsection{Hierarchy problem}

One of the most common arguments for physics beyond the Standard Model is what is known as the hierarchy problem.  This can be expressed in two different ways which, whilst clearly related, are in some ways distinct.  They are referred to here as the `aesthetic hierarchy problem' and the `technical hierarchy problem'.

\subsubsection{Aesthetic hierarchy problem}

As has been discussed above, the Standard Model has nothing to say about gravitational interactions.  It also seems aesthetically unappealing that the mass scale with which particle physicists are most familiar (the electroweak mass scale, $\sim 100$~GeV) is so many orders of magnitude below the Planck scale ($\sim 10^{19}$~GeV).  If there is a fundamental `theory of everything' then it would seem more natural if all the energy scales were of the same order of magnitude.  Another way of expressing this is just to say that it seems unnatural for gravity to be so much weaker than the other gauge forces.  Why are there seemingly two different fundamental scales in nature?

\subsubsection{Technical hierarchy problem}

More technically, the hierarchy problem is the result of considering one-loop corrections to the Higgs mass.  It is relatively easy to calculate the correction to the Higgs mass from diagrams like that in Figure~\ref{hmcf}, but since the diagram is quadratically divergent, a cut-off ($\Lambda$) must be introduced in the loop momentum integral to regulate the divergence.  The correction to the `bare' Higgs mass is
\begin{equation}
\delta M^2_\text{Hf}=\frac{|\lambda_\text{f}|^2}{16\pi^2}\left[-2\Lambda^2+6m_\text{f}^2 \ln(\Lambda/m_\text{f})+\cdots\right].
\end{equation}  
where $\lambda_\text{f}$ is the Higgs-fermion coupling and $m_\text{f}$ the fermion mass.
\begin{figure}
\begin{center}
\begin{picture}(440,80)
\SetOffset(120,-10)
\DashArrowLine(40,50)(80,50){5}
\ArrowArcn(100,50)(20,0,180)
\ArrowArcn(100,50)(20,180,360)
\DashArrowLine(120,50)(160,50){5}
\Text(30,50)[c]{H$^0$}
\Text(170,50)[c]{H$^0$}
\Text(100,17)[c]{$\bar{\text{f}}$}
\Text(100,83)[c]{f}
\end{picture}
\capbox{Fermion one-loop correction to the Higgs mass}{Feynman diagram for the fermion one-loop correction to the Higgs boson mass.\label{hmcf}} 
\end{center}
\end{figure}
In itself this mass correction isn't necessarily a problem since the bare Higgs mass can have any value provided the physical Higgs mass is around the electroweak scale.   The Standard Model is only considered to be a low-energy effective theory and so a natural value for $\Lambda$ would be the Planck scale.  However this means that the `bare Higgs mass' would have a natural value of order the Planck scale and therefore, if there is a Higgs boson with mass of order 100~GeV, there would need to be cancellation at the level of 1 part in $10^{16}$.  This is considered to require enormous `fine-tuning' to the parameters of the bare Lagrangian.

Assuming we believe in the existence of the Standard Model Higgs boson and find the idea of fine-tuning unnatural this leave only one possibility: there must be new physics and/or new particle content at energy scales $\Lambda \sim$~1~TeV\@.  In fact even if there is no SM Higgs boson, new TeV-scale physics is still required in order to unitarize the W-W scattering cross section.

\subsection{Unification of couplings}
\label{couplings}

\begin{figure}[t]
\begin{center}
\epsfig{file=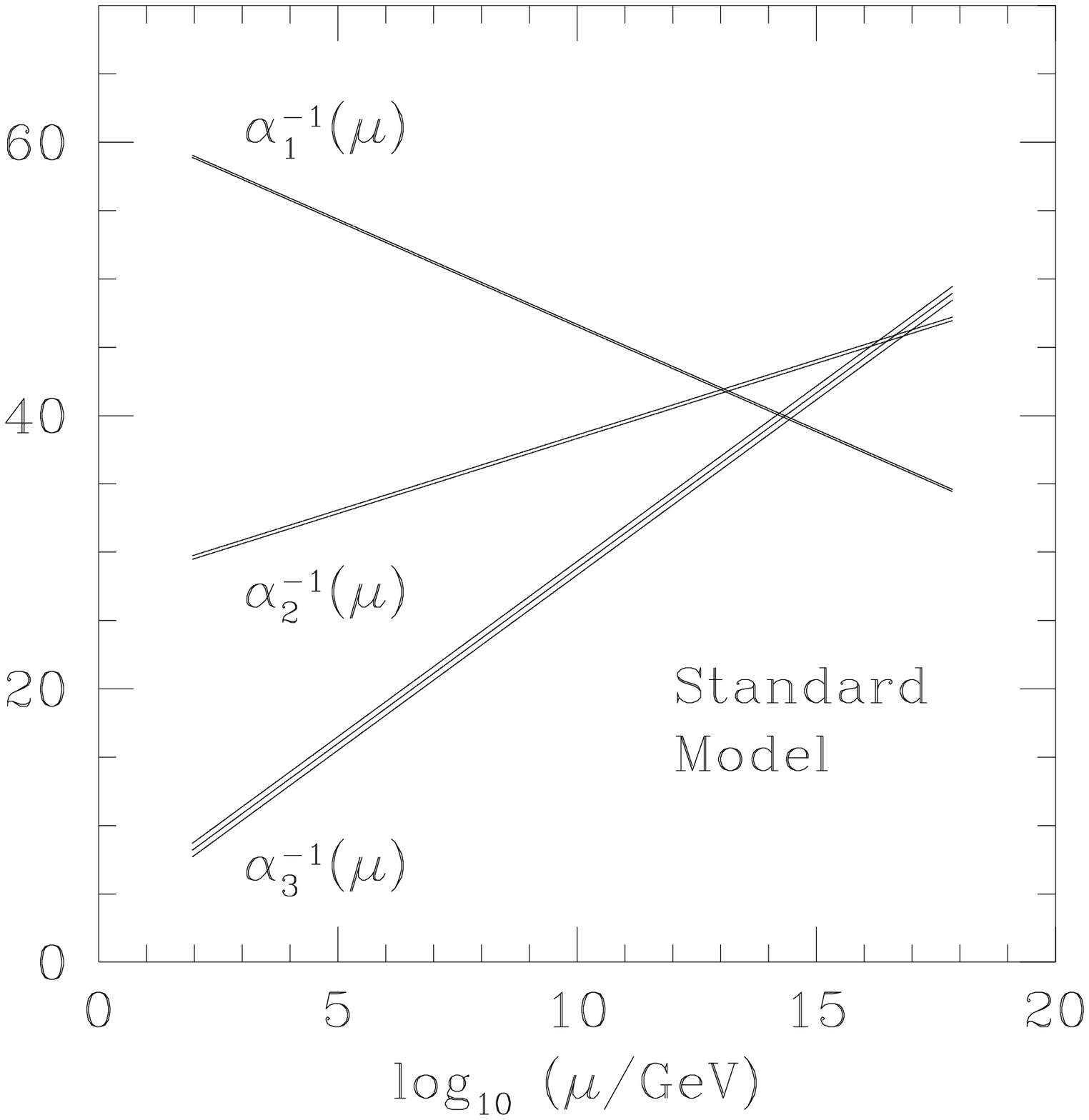, angle=0, width=.55\textwidth}
\capbox{Gauge coupling running with SM particles}{Running of the gauge couplings assuming only Standard Model particle content.  Values of the inverse couplings at energy $\mu$ are determined using the one-loop renormalization group equations.  $\alpha_1$, $\alpha_2$ and $\alpha_3$ are the hypercharge, weak isospin and strong couplings respectively, but the conventional normalization means that $\alpha_1\equiv(5/3)\alpha_Y$. The width of the bands on the plot is due to experimental uncertainties. (Reproduced from \cite{Dienes:1997du}).\label{unifsm}}
\end{center}
\end{figure}

If the strong and electroweak forces are to unify into a single gauge theory at a high energy scale then the gauge couplings must also unify.  The particle content of a model determines how the couplings `run' with energy and although in the Standard Model they almost unify at a scale of about $10^{15}$~GeV, this unification isn't quite exact (as illustrated by Figure~\ref{unifsm}).  This can be seen as a motivation for additional particle content or new physics which will modify the running of the couplings to make the unification more exact.
\section{Supersymmetry and intermediate scale models}
\label{ism}

One type of new physics which is relevant to the work presented in this thesis is supersymmetry (SUSY). This is one of the most popular choices for physics beyond the Standard Model and resolves several of the issues raised in the previous section.  The basic idea is that an additional symmetry is introduced to the theory such that the Lagrangian is invariant under transformations which convert fermions into bosons and vice versa.  This immediately doubles the number of particles and more importantly helps to solve the technical hierarchy problem.  This is because there will be cancellations between Feynman diagrams like that in Figure~\ref{hmcf} and one-loop diagrams with S (the fermion's supersymmetric scalar partner, known as a sfermion) as shown in Figure~\ref{hmcs}.  The Higgs mass correction in this case is
\begin{equation}
\delta M^2_\text{HS}=\frac{\lambda_\text{S}}{16\pi^2}\left[\Lambda^2-2m_\text{S}^2 \ln(\Lambda/m_\text{S})+\cdots\right],
\end{equation}  
with $\lambda_\text{S}$ the Higgs-scalar coupling and $m_\text{S}$ the scalar mass.
\begin{figure}
\begin{center}
\begin{picture}(440,70)
\SetOffset(120,-50)
\DashArrowLine(40,50)(100,50){5}
\DashCArc(100,70)(20,0,180){5}
\DashCArc(100,70)(20,180,360){5}
\DashArrowLine(100,50)(160,50){5}
\Text(30,50)[c]{H$^0$}
\Text(170,50)[c]{H$^0$}
\Text(100,103)[c]{S}
\end{picture}\\[3mm]
\capbox{Scalar one-loop correction to the Higgs mass}{Feynman diagram for the scalar one-loop correction to the Higgs boson mass.\label{hmcs}} 
\end{center}
\end{figure}
Hence provided there are two scalar particles (with coupling $\lambda_\text{S}=|\lambda_\text{f}|^2$) for every fermion, there will be cancellation between the fermion and sfermion loop diagrams.  In fact the cancellation is only between the terms quadratic in $\Lambda$ but the remaining Higgs mass corrections are only logarithmic in $\Lambda$ and hence do not cause the same fine-tuning problems.  Although we know that, if it exists, supersymmetry must be broken (since no superpartners of Standard Model particles have yet been observed) this supersymmetry breaking must occur in such a way that the leading-order cancellation remains.  The requirement for the physical Higgs mass to be at the electroweak scale is now that the superparticles have masses of order 1~TeV or less.  If this is not the case then fine-tuning problems start to resurface and one of the main motivations for SUSY has been lost.  It is also difficult to construct SUSY models where the SM particle masses differ from those of their superpartners by more than about an order of magnitude.

Various different ways to break supersymmetry have been studied.  It has long been known that a viable way to break supersymmetry is in a hidden sector---the SUSY breaking in a particular model is then transmitted to the `outside world' by some means (usually gravitational or gauge interactions).  The supersymmetry breaking means that what is initially a very beautiful theory (and hence very attractive to particle theorists) becomes much less beautiful.  Many new parameters ($\sim$103) are needed to describe a particular supersymmetric model, and only by making simplifying assumptions (e.g.\ regarding various high-scale masses and couplings) can much headway be made in considering the phenomenology of these models.  This makes supersymmetry much less attractive, particularly to SUSY sceptics who argue that a region of parameter space can always be found to ensure that supersymmetry cannot be ruled out by either experimental or cosmological constraints.

\begin{figure}
\begin{center}
\epsfig{file=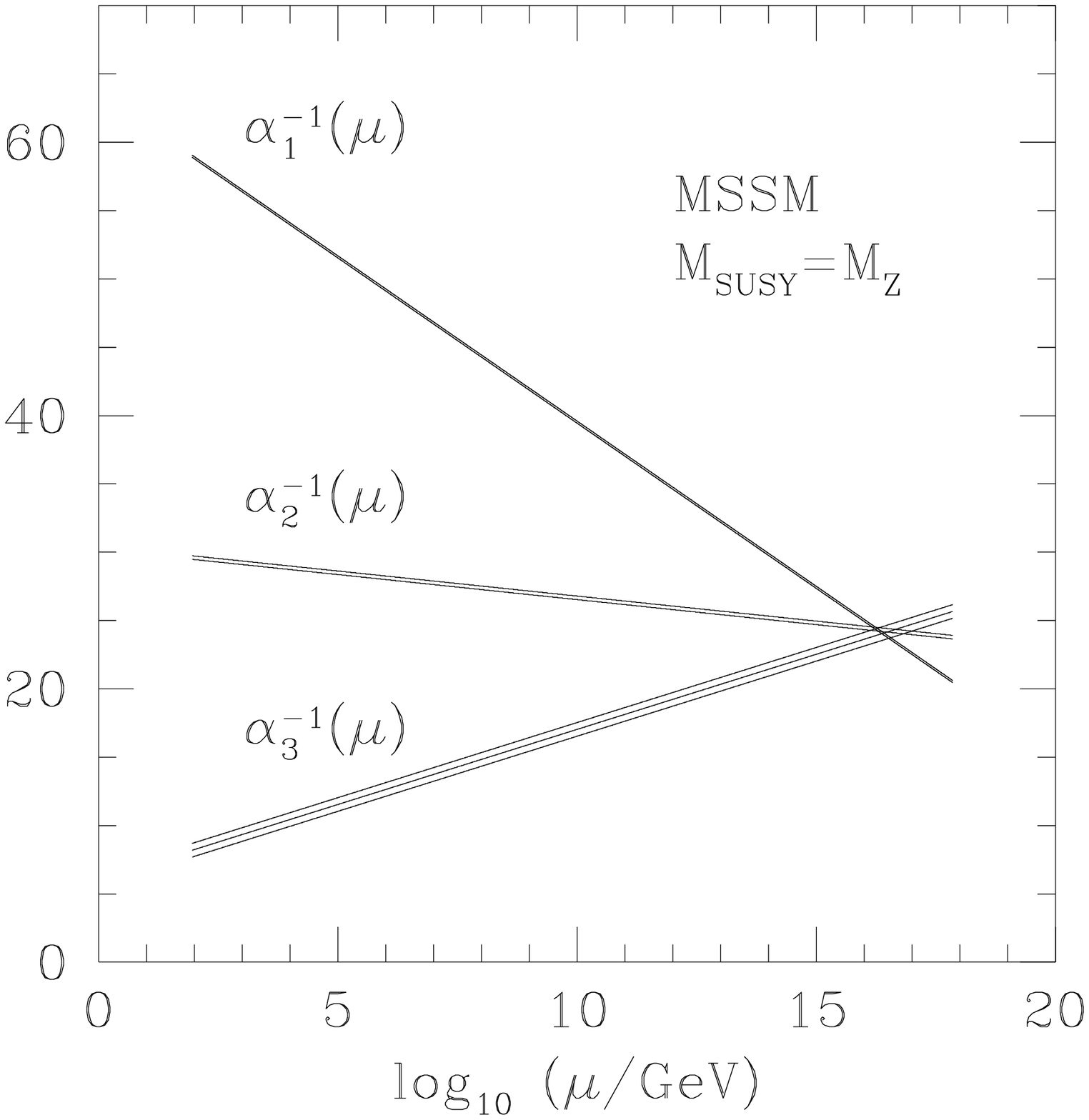, angle=0, width=.55\textwidth}
\capbox{Gauge coupling running with MSSM particles}{Running of the gauge couplings assuming Minimal Supersymmetric Standard Model particle content.  For more details of the quantities plotted, see Figure~\ref{unifsm}. (Reproduced from \cite{Dienes:1997du}).\label{unifsusy}}
\end{center}
\end{figure}

However there are several other arguments in favour of supersymmetry.  It can provide a dark matter candidate in the form of a lightest supersymmetric particle (LSP).  This is the result of the R-parity symmetry which is often imposed to prevent proton decay but which also has the effect of making the LSP stable since it is forbidden from decaying into only SM particles.  One of the other motivations for supersymmetry is that it improves the unification of gauge couplings discussed in \secref{couplings} (although the present value of $\alpha_\text{S}(M_\text{Z})$ seems to be slightly too low to allow perfect unification---see \cite{deBoer:2003xm} for a recent discussion).  Figure~\ref{unifsusy} shows the running of the gauge couplings with energy in the supersymmetric case.
However the couplings do not unify at $\sim 10^{15}$~TeV in all supersymmetric scenarios.  One scenario, motivated by string theory, is that of an intermediate scale \cite{Benakli:1998pw,Burgess:1998px,Aldazabal:1999tw,Aldazabal:2000sk}.  In this type of model, gauge coupling unification is assumed to occur at an `intermediate' energy scale ($\sim 10^{11}$~GeV) which is the geometric mean of the electroweak and Planck energy scales.  This choice is motivated by the scale of supersymmetry breaking in hidden-sector gauge-mediated scenarios and also by the usual solution of the strong Charge Parity (CP) problem.  
\enlargethispage{\baselineskip}

The strong CP problem is that there is an unnaturally small bound ($<10^{-9}$) on the $\Theta$ parameter of the QCD Lagrangian.  The parameter is constrained to be this small because the $\Theta$-term in the Lagrangian is CP-violating and would otherwise lead to a neutron electric dipole moment (on which there are strong experimental limits).  The Peccei-Quinn mechanism \cite{Peccei:1977hh} solves this problem by introducing an additional global chiral symmetry, U(1)$_\text{PQ}$, which is then spontaneously broken generating a Goldstone boson known as the axion field.  The axion acquires a small mass due to non-perturbative effects and and has a lifetime which exceeds the age of the Universe by many orders of magnitude.  The requirement that axion emission does not over-cool stars puts a lower bound bound on the axion decay constant $f_\text{a}$, and an upper bound comes from dark matter constraints.  Therefore $f_\text{a}$ is constrained to be in the approximate energy range $10^9$--$10^{12}$~GeV, as is naturally the case for scalars in intermediate scale string scenarios \cite{Burgess:1998px}.  
\enlargethispage{\baselineskip}

The intermediate scale is also consistent with some neutrino mass mechanisms, and is favoured by certain cosmological inflation models.  High-energy cosmic ray observations and certain non-thermal dark matter candidates are also supporting arguments for an intermediate energy scale.  For more details, see \cite{Benakli:1998pw,Burgess:1998px} and the references contained.

Various intermediate scale models are motivated from different string models
in references \cite{Benakli:1998pw,Burgess:1998px,Aldazabal:1999tw,Aldazabal:2000sk}.  Some phenomenological 
constraints on intermediate scale models are provided by \cite{Abel:2000bj}, and 
constraints from the muon anomalous magnetic moment have also been discussed \cite{Baek:2001nz,Cerdeno:2001aj}.
One of the ways of achieving this intermediate scale unification is to include new particles (with $\sim$~TeV masses) as part of extra supermultiplets added to the Minimal Supersymmetric Standard Model (MSSM).  Extra heavy leptons added in such a scenario are the subject of the work described in Chapter~\ref{leptons}.


\section{Extra dimension models}
\label{edm}

Extra dimension models are another class of model which have recently become quite fashionable and also provide a way of solving the hierarchy problem.\footnote{Many would argue, with good reason, that extra dimensions really just \emph{reformulate} the problem rather than solve it.}  The essence of these models is that the apparent weakness of gravity at macroscopic length scales is due to the presence of extra dimensions.  

All these models involve one or more $(3+1)$-dimensional slices of space-time (known as branes or 3-branes) which are embedded in the full extra-dimensional space-time (the bulk).  In most models the world as we know it and all the Standard Model particles are localized on the brane whilst gravitons can propagate in the bulk.  There are, though, some models where SM particles are not localized on the brane---one will be introduced below.  Another feature of extra dimension models is that the new dimensions are compactified or `curled up'---necessary to explain why they have never been observed in the same way as the infinite dimensions we are familiar with. 

Different models involve extra dimensions of different sizes but it is initially surprising to learn that they can be relatively large ($\sim$1~mm) without being ruled out by experimental measurements of gravity.  Extra dimensions of this size imply that the Newtonian $1/r^2$ force law will fail at these length scales.  However the force law has, even now, only been verified down to distances of $\sim$~100 $\mu$m (see \secref{ssgrav}) and when these extra dimension models were first proposed in 1998, the limit was only about 1~mm.  

As well as providing a solution to the hierarchy problem, extra dimension models can be motivated by string theory.  When combined with supersymmetry, string theory is a candidate for a theory of quantum gravity but it requires the world to be 10- or 11-dimensional.  String theory can also naturally account for some fields (e.g.\ SM particle fields) being fixed on $(3+1)$-dimensional surfaces or branes whilst others, like gravitons, propagate in the extra dimensions.  In both type-I and type-II string theories, matter and the Standard Model interactions are represented by open strings which start and end on a brane, whereas as gravity is described by a closed string which can propagate in all the spatial dimensions.  It is possible that supersymmetry plays an important r\^{o}le in stabilizing the extra-dimensional geometry.  Unfortunately, in common with the majority of proposals for BSM physics, one prediction of most of these theories is a very large value for the vacuum energy density or cosmological constant; this is in contradiction with the very small value calculated by cosmologists.

In fact extra dimension models are not a new idea.  As early as the 1920s a model with one extra compactified dimension was suggested by Kaluza and Klein in an attempt to unify electromagnetism and gravity.  In such a scheme the photon is actually a component of the higher-dimensional graviton.  Although their theory fell out of fashion due to its failure to explain the weak and strong forces or the relative strength of gravity, their ideas seem to have come back into fashion.

The specific geometry of the extra dimensions varies in different models.  The work in this thesis considers phenomenology in large extra dimension models, but two other extra dimension scenarios are briefly discussed in this introductory chapter.

\subsection{Introducing different models and their phenomenology}

Before introducing the different extra dimension models we consider the wave equation for a scalar field in 5 dimensions; this is useful for demonstrating some of the main differences in phenomenology.  We consider the fifth dimension to be compactified on a circle of radius $R$ such that there is a periodic identification $x_5=x_5+2\pi R$.  The scalar field equation for a particle of mass $m$ propagating in the bulk is
\begin{equation}
\left(\frac{\partial ^2}{\partial t^2}-\nabla^2_5+ m^2\right)\,\Phi (x,y,z,x_5,t)=0\,,
\end{equation}
where
\begin{equation}
\nabla^2_5=\frac{\partial ^2}{\partial x^2}+\frac{\partial ^2}{\partial y^2}+\frac{\partial ^2}{\partial z^2}+\frac{\partial ^2}{\partial x_5^2}\,.
\end{equation}
The solution will be of the form
\begin{equation}
\Phi (x,y,z,x_5,t)=\sum_{n \in N} \Phi_n(x,y,z,t)\exp\!\left(\frac{inx_5}{R}\right),
\end{equation}
which gives the following equation of motion:
\begin{equation}
\label{kk}
\sum_{n \in \mathbb{N}} \left(\frac{\partial ^2}{\partial t^2}-\nabla^2_4+ m^2+\frac{n^2}{R^2}\right)\Phi_n(x,y,z,t)\,.
\end{equation}
Viewed from a four-dimensional perspective there is clearly a tower of states of different masses, with mass splitting $1/R^2$;  this is the so-called Kaluza-Klein (KK) tower of states.  It is clear though that the Kaluza-Klein tower is simply a result of viewing a 5-dimensional model in 4-dimensions.  Taking a 5-dimensional view there is only one state of mass $m$ which in the equivalent spin-2 case will be the massless graviton.  

At this point it is worth pointing out another feature of most of these models: there is no requirement for momentum conservation in the extra-dimensions.  Energy-momentum conservation in the four-dimensional world is a consequence of the translational invariance of the four space-time dimensions.  However the 3-brane breaks translational invariance in the extra dimension and hence momentum is these directions need not be conserved in interactions between brane and bulk states.  In effect the brane will be considered as being infinitely heavy, or as being `nailed' in place.\footnote{There are models in which this approximation is not valid, e.g.\ \cite{Murayama:2001av}.}

\subsubsection{`Large' extra dimensions}
\label{led}

The earliest extra dimensions models (at least in recent years) were those of Arkani-Hamed, Dimopoulos and Dvali (known as `large' extra dimension,\footnote{The word `large' may seem to be something of a misnomer, but the extra dimensions are large relative to the fundamental energy scale in the models.} or ADD models \cite{Arkani-Hamed:1998rs,Arkani-Hamed:1998nn,Antoniadis:1998ig}).  These are perhaps the simplest of the extra dimension models and are closest to the ideas of Kaluza and Klein.  

Gravity is the only force which propagates in the full volume of the extra dimensions (the bulk).  Hence the gravitational force in the four-dimensional world (on the brane) is `diluted' and gravity appears weak compared to the other forces which do not propagate in the extra dimensions.  In these models there is only one fundamental energy scale, the electroweak scale, and so there is no longer a fine tuning problem---$\Lambda$ is now of order a TeV\@.
\enlargethispage{-\baselineskip}

For large extra dimensions, equation~(\ref{kk}) shows that the mass splitting will be small, and all graviton modes (up to the mass scale involved in the process) will contribute to the process being considered.  Although each mode has a negligible contribution, the number of gravitons involved can be very large which means care must be taken that the total contribution isn't inconsistent with precision SM measurements.

There are two major classes of experimental signature: direct production of the KK excitations of gravitons (giving jets and missing energy) and virtual exchange of the KK excitations (modifying SM cross~sections and asymmetries).  The missing energy signatures are clearly different to those in other models with invisible particles (like SUSY) since from a four-dimensional point of view the graviton doesn't have a fixed mass---instead it appears to have a continuous mass distribution.

The work in the later chapters of this thesis is almost entirely in the framework of ADD models; as a result there is much further discussion of the details of these models in \secref{add}.

\subsubsection{$\mbox{TeV}^{-1}$ extra dimensions}
\label{tev-1}

The models with extra dimensions of size $\mbox{TeV}^{-1}$ \cite{Antoniadis:1990ew} are somewhat different to other models like those of ADD\@.  These models do not themselves attempt to solve the hierarchy problem; the smaller size of the extra dimensions means that the volume of the bulk is not be large enough to explain the weakness of gravity (although there are models which include a mixture of large and $\mbox{TeV}^{-1}$ extra dimensions).  In fact the extra dimensions are small enough to allow the gauge and Higgs bosons to propagate in the bulk without running into conflict with the constraints from the precision electroweak measurements made at LEP (for example the Z-width).

Although they don't solve the hierarchy problem, there are various advantages of models like this which have the gauge and Higgs bosons propagating in the bulk.  Gauge bosons in the bulk provide a mechanism for early unification of the coupling constants because the logarithmic running is modified by the extra dimensions \cite{Dienes:1998vg}.  If the Higgs field propagates in the bulk, the vacuum expectation value of the Higgs zero mode is able to generate spontaneous symmetry breaking.  Models with two branes with a $\mbox{TeV}^{-1}$ separation can also suppress proton decay if the quarks are localized on one brane and the leptons on the other \cite{Arkani-Hamed:1999za} (alternatively there may be a `thick' brane with quarks and leptons localized at different points but with the Higgs and gauge bosons propagating inside it \cite{Arkani-Hamed:1999dc}).  Such models can also be useful to motivate the three generations of fermions and the hierarchy of fermion masses.

In these models, $R\sim\mbox{TeV}^{-1}$ and so, from equation~(\ref{kk}), the mass splittings $M_\text{c}$ are TeV-scale.  Precision electroweak data requires that $M_\text{c} >$2--5~TeV, but the LHC should be able to probe $M_\text{c}$ up to around 14~TeV by searching for leptonic signatures of KK excitations of the photon and the Z boson. 

A subset of these models which are slightly different are those of universal extra dimensions where all fields are allowed to propagate in the bulk \cite{Appelquist:2000nn}.  These can be constructed to evade current electroweak limits and might provide a dark matter candidate \cite{Cheng:2002ej}.  One particular feature of universal extra dimension models is that KK excitations are pair-produced; this could lead to some distinctive phenomenology at collider experiments.

\subsubsection{`Warped' extra dimensions}

The `warped' extra dimension models of Randall and Sundrum (RS) \cite{Randall:1999ee,Randall:1999vf} solve the hierarchy problem in a different way to the ADD models.  Gravity is not `diluted' by the large volume of the bulk but by the strong curvature of the extra dimensions. RS models involve a second brane; gravity is strong on that brane but a `warp factor' in the metric ensures that gravity on our brane is weak.  These models may provide a way to explain why the cosmological constant seems to be so small.

Equation~(\ref{kk}) cannot strictly be applied in RS models (since the geometry is non-factorizable) but the extra dimensions are again small which corresponds to a large mass splitting.  This gives resonant graviton production where usually only the first graviton resonance is considered since the LHC is unlikely to be of high enough energy to study higher resonances.  Channels like $\text{G}\rightarrow\text{e}^+\text{e}^-$ and $\text{G}\rightarrow\gamma\gamma$ can be studied to find the mass of the first graviton resonance and to confirm its spin-2 nature \cite{Allanach:2000nr}.

\section{Details of ADD models}
\label{add}

\subsection{Gauss' law in extra dimensions}
\label{gauss}

Whilst it is relatively trivial to use Gauss' Law to derive the power dependence of the gravitational force in extra dimensions, slightly more care is required to determine constants of proportionality in the force law and to determine how these relate to the Planck scale.  The difficulties arise both because there are a number of different conventions in use and because it is not trivial to visualize geometries involving more than three spatial dimensions.  In all that follows in this thesis, $n$ toroidally compactified extra dimensions, all of radius $R$, are assumed.\footnote{Much of the extra dimension phenomenology discussed in this and later chapters could be modified by compactification on more complicated manifolds.}
\enlargethispage{-\baselineskip}

In extra dimensions, Gauss' Law can be written as
\begin{equation}
\int_s \mathbf{F} \cdot \mathbf{dS} = \Omega_{n+2} G_{(4+n)} M,
\end{equation}
where $F$ is the force per unit mass, $G_{(4+n)}$ is the $(4+n)$-dimensional Newton constant,\footnote{This is by no means unique as a way of defining the extra-dimensional Newton constant.} $M$ is the mass enclosed by the surface $S$ and $\Omega_p$ (the area of a unit $p$-sphere) is given by
\begin{equation}
\Omega_p=\frac{2\pi^{\frac{p+1}{2}}}{\Gamma(\frac{p+1}{2})}\,.
\label{omegapdef}
\end{equation}
At short distances ($\ll R$) the extra compactified dimensions are effectively flat and so we obtain
\begin{equation}
F_{n+1}(r)=\frac{G_{(4+n)} M}{r^{n+2}}\,,
\end{equation}
whereas at long distances the usual 4-dimensional force law is obtained.

The above suggests that $G_{(4+n)} \sim G_{(4)} R^n$ in order to get a smooth
matching between the two regimes.  However it is worth taking a little more care in deriving the correct relationship between these constants.  We follow the approach of ADD in one of their early papers \cite{Arkani-Hamed:1998nn} and use the analogy of a line of point masses.  First consider an infinite two-dimensional line of masses, each of mass $m$ and separated by a distance $L$.  Provided we are at a large enough distance $r$ from the line, it will appear like a continuous line of mass and the problem has become two dimensional.

Using Gauss's law in three (spatial) dimensions and considering the flux through a cylinder of length $d$ 
we obtain
\begin{equation}
(2 \pi r d) F = 4\pi G_{(4)} m \left(\frac{d}{L}\right) \longrightarrow F(r)=\frac{2}{L} G_{(4)} \frac{m}{r}\,.
\end{equation}
So there is an effective three-dimensional Newton constant (applicable for two spatial dimensions) related to $G_{(4)}$ by
\begin{equation}
G_{(3)}=\frac{2}{L} G_{(4)}\,,
\end{equation}
which is more usefully written as
\begin{equation}
G_{(3)}=\frac{1}{L}\frac{\Omega_2}{\Omega_1} G_{(4)}\,.
\end{equation}

\enlargethispage{-\baselineskip}
The situation with compactified extra dimensions is exactly analogous.  The periodic identification means that the equivalent uncompactified theory has lines of mirror masses $m$ with separation $L=2\pi R$.  So for $r \gg L$ there is an effective 4-dimensional Newton constant related to $G_{(4+n)}$ by
\begin{equation}
G_{(4)}=\frac{1}{L^n}\frac{\Omega_{n+2}}{\Omega_2} G_{(4+n)}=\frac{1}{L^n}\frac{\Omega_{n+2}}{4 \pi} G_{(4+n)}\,.
\label{g4ton}
\end{equation}

\subsection{Planck mass in extra dimensions}

The Planck mass is the scale at which gravity becomes strong, but there are varying definitions even in four dimensions.  Initially we follow the conventions of ADD who use a reduced Planck scale, $\hat{M}_{(4+n)}^{n+2}$ given by
\begin{equation}
\hat{M}_{(4+n)}^{n+2}=\frac{1}{G_{(4+n)}\Omega_{n+2}}\,,
\end{equation}
and hence, using eq.~(\ref{g4ton}),
\begin{equation}
\label{mplconv}
\hat{M}_{(4)}^2=(2 \pi R)^n \hat{M}_{(4+n)}^{n+2}=R^n M_{\text{P}(4+n)}^{n+2}\,,
\end{equation}
where $M_{\text{P}(4+n)}$ is the Planck mass in $4+n$ dimensions.  This is introduced because it is found to be this scale which is most directly constrained by experimental bounds and which is required to be $\sim 1$~TeV in order to solve the hierarchy problem.  However in many ways the reduced Planck mass is more natural since the interactions of the canonically normalized field in the Lagrangian are suppressed by $\sqrt{\hat{M}_{(4+n)}}$\,.  

We can always view interactions in two different ways.  From a 4-dimensional view-point the graviton coupling is very small ($\sim 1/\sqrt{\hat{M}_{(4)}}$) but many Kaluza-Klein modes contribute (for a process with energy scale $E$, there will be contributions from $\sim (ER)^n$ modes).  However from a $(4+n)$-dimensional view-point the graviton coupling is much larger ($\sim 1/\sqrt{\hat{M}_{(4+n)}}$) but there is no tower of graviton states.

As mentioned above there are a number of other ways of defining the Planck scale, and at the same time a number of different definitions of the extra-dimensional Newton constant.  To compare most transparently with other conventions (which in part differ because of alternative definitions of the 4-dimensional Planck mass) equation~(\ref{mplconv}) is more conveniently written as
\begin{equation}
\label{4pG}
M_{\text{P}(4+n)}^{n+2}=\frac{1}{4 \pi G_{(4)}}\frac{1}{R^n}\,.
\end{equation}
For ease of reference in later parts of this thesis, a summary of many of the popular conventions is shown in Table~\ref{consum} (in all cases $n$ toroidally compactified extra dimensions, all of radius $R$, are assumed).

\begin{table}
\def\arraystretch{1.5}
\begin{center}
\begin{tabular}{|l|c|l|}
\hline
&Definition of $M_{\text{P}(4+n)}^{n+2}$ & Comments\\
\hline
a & $\frac{1}{4 \pi G_{(4)}}\frac{1}{R^n}$ & Used by ADD and also GT \cite{Giddings:2001bu}\\
b & $\frac{1}{8 \pi G_{(4)}}\frac{1}{R^n}$ & Used by GRW \cite{Giudice:2001ce}; often known as $M_*$\\
c & $\frac{1}{G_{(4)}}\frac{1}{R^n}$ & Used by EOT-WASH \cite{Adelberger:2002ic} \\
d & $\frac{1}{G_{(4)}}\frac{1}{(2 \pi R)^n}$ & Used by Dimopoulos and Landsberg \cite{Dimopoulos:2001hw}\\
e & $\frac{1}{8 \pi G_{(4)}}\frac{1}{(2\pi R)^n}$ & Used by Han, Kribs and McElrath \cite{Han:2002yy}\\ 
\hline
\end{tabular}
\capbox{Different conventions for the fundamental Planck scale}{Different conventions for the fundamental Planck scale $M_{\text{P}(4+n)}$.  The last column indicates which conventions are used by the authors of some of the papers referenced in this thesis.\label{consum}}
\end{center}
\def\arraystretch{1.0}
\end{table}

\subsection{Constraints on extra dimension models}
\label{edcon}

The differences between the conventions (particularly the first three) are usually relatively small since the conversion factors are close to unity for most allowed values of $n$.  When quoting astrophysical and cosmological limits where there are large uncertainties it is not crucial to be careful about which convention is being used.  However for collider limits the convention should be clearly stated.  For ease of comparison all the limits below are in the $M_*$ convention (convention `b' in Table~\ref{consum}) which is also used by the Particle Data Group.\footnote{These different conventions have unfortunately led to much confusion in the literature, and explain some apparent inconsistencies between the limits mentioned here and those in some other recent reviews, for example \cite{Hewett:2002hv}.}

\subsubsection{Limits from short-scale gravity experiments}
\label{ssgrav}

Short-scale gravity experiments provide limits on the size (i.e.\ the compactification radius) of the extra dimensions and relations like those in Table~\ref{consum} can be used to convert these to limits on the fundamental Planck scale (in a particular compactification scheme).  Seeking to find deviations from Newtonian gravity at short distances is difficult because at these distances gravity is not the dominant force, even between non-magnetic, electrically neutral interacting bodies.  This means it can be necessary to fully understand the Casimir force (taking into account finite conductivity, surface roughness and non-zero temperature corrections) and van der Waals forces which become important at these length scales.  There are also experimental considerations in seeking to isolate the experiments from mechanical, acoustic and thermal vibrations as well as electromagnetic fields.

The best constraints at present come from modern versions of the Cavendish torsion experiment.  These experiments now show no deviation from Newtonian gravity down to about 150~$\mu$m \cite{Adelberger:2002ic} which corresponds to $M_{*(4+2)} > 1.8$~TeV\@.


\subsubsection{Limits from collider experiments and future sensitivity}
\label{collim}

The signature of real graviton production in the process $\text{e}^{+}\text{e}^{-}\rightarrow \text{G} \gamma$ is a photon and missing energy.  Searches in this channel at the LEP collider were able to set limits on the fundamental Planck scale, usually given in the $M_*$ convention.  The best published limits range from $M_{*(4+2)} > 1.1$~TeV through $M_{*(4+4)} > 0.68$~TeV to $M_{*(4+6)} > 0.51$~TeV \cite{Abreu:2000vk}, but some reviews quote values from the later higher energy running at LEP which gives $M_{*(4+2)} > 1.38$~TeV and $M_{*(4+6)} > 0.58$~TeV (DELPHI collaboration, quoted in \cite{Landsberg:2001ma}) or even $M_{*(4+2)} > 1.45$~TeV and $M_{*(4+6)} > 0.61$~TeV (L3 collaboration, in \cite{Hewett:2002hv}).

The Tevatron Run I limits only become competitive with the LEP II limits for the higher values of $n$.  The best limits from the Tevatron so far come from the CDF collaboration; they find, at 95\% confidence level, that $M_{*(4+2)} > 1.00$~TeV but that $M_{*(4+6)} > 0.71$~TeV \cite{Acosta:2003tz}.  These limits are obtained from a study of the `one or two high-energy jets + missing tranverse energy' signature (from processes like $\text{gg}\rightarrow \text{Gg}$, $\text{qg}\rightarrow \text{Gq}$ and $
\text{q}\bar{\text{q}}\rightarrow \text{Gg}$).  This is found to give a better sensitivity than the `photon + missing $E_\text{T}$' signature (from $\text{q}\bar{\text{q}}\rightarrow \text{G} \gamma$)\footnote{The `Z+ missing $E_\text{T}$' signature is less significant because the Z can only be observed via its leptonic decay.} since rates are much lower than in the jet case.  However this signature could be used as confirmation in the event of a discovery in the jet channel.  It is expected that the Tevatron Run II will be able to improve slightly upon these limits.

It should be possible to reach much larger values of $M_*$ by studying the single-jet signature at the LHC (the dominant sub-process is $\text{qg}\rightarrow \text{Gq}$).  With 100~$\mbox{fb}^{-1}$ of integrated luminosity, the ATLAS collaboration expects to be able to probe $M_*$ in the following ranges: $M_{*(4+2)}=$4.0--9.1~TeV, $M_{*(4+3)}=$4.5--7.0~TeV and $M_{*(4+4)}=$5.0--6.0~TeV\cite{Vacavant:2001sd}.  The lower limits reflect the fact that the theory is only an effective one and breaks down at energies $\sim M_*$.  If the fundamental Planck scale is so low that a significant number of events would have $\sqrt{\hat{s}}>M_*$ then it is not possible to accurately extract information.  This also means that nothing can be said for the $n>4$ cases---in effect the LHC would have too high a centre-of-mass energy (this isn't an issue in the pessimistic case in which we are just setting exclusion limits on $M_*$).  The upper limits are due to the signal no longer being observable over the SM background.  In the ranges given, both $M_*$ and $n$ could be extracted at the LHC, but this would require running at two different centre-of-mass energies.

Some limits can also be placed on ADD models by considering virtual graviton emission.  However the sum of KK states involved is found to be divergent; this means a cut-off has to be introduced which makes the process sensitive to ultra-violet effects.  Hence although it is possible to place limits on this cut-off scale (in various different conventions) it is not possible to convert them to direct limits on $M_*$.

\subsubsection{Limits from cosmic rays}
\label{crlim}

Models with large extra dimensions would allow the production of miniature black holes at TeV-scale energies---a topic which will be properly introduced in Chapter~\ref{bhintro}.  Ultra high-energy cosmic rays interact with centre-of-mass energies up to $\sim$~400 TeV and so could produce black holes; their decays could then be observed by experiments like AGASA\@.\footnote{This provides reassurance to those who are unconvinced by the theoretical arguments and worry that black holes produced at the LHC might destroy the world.}  Although there are some uncertainties involved, ref.~\cite{Anchordoqui:2001cg} finds limits on the fundamental scale which, for the larger values of $n$, are more restrictive than present collider limits (e.g.\ $M_{*(4+4)}>$1.3--1.5~TeV and $M_{*(4+7)}>$1.6--1.8~TeV).

\subsubsection{Astrophysical and cosmological constraints}
\label{astcon}

\enlargethispage{\baselineskip}
Some of the strongest constraints on these extra-dimensional models are astrophysical and cosmological.  These put direct lower bounds on $M_{\text{P}(4+n)}$ which cannot absolutely rule out extra dimension models but make them unattractive for two reasons:  firstly the whole motivation for the theory is lost if the fundamental Planck scale is constrained to be significantly larger than the electroweak scale; secondly there would be no obvious phenomenology at the next generation of colliders.

There are however various ways of managing values of $M_{\text{P}(4+n)}$ up to of order 10~TeV\@.  This is because although $\Lambda\sim M_{\text{P}(4+n)}$, the exact form of this relationship depends on the theory being proposed.  For example if the extra dimension model is implemented within a type-I string theory, it can be shown that the scale of new physics ($\Lambda$) is given by the string scale ($M_\text{s}$) but that this can naturally be an order of magnitude smaller than $M_{\text{P}(4+n)}$.  Hence the hierarchy problem is naturally resolved, and new physics is expected at energy scales accessible at the LHC\@.

The first limits we consider here are those from the supernova remnant SN1987A\@.  The rate at which the remnant can lose energy into the extra dimensions by graviton emission must not violate the observed cooling rate which is assumed to be mainly due to neutrinos (the flux of which has been measured by Kamiokande and IMB).  The predominant mechanism for graviton emission is known as `gravi-strahlung' ($\text{N}+\text{N}\rightarrow \text{N}+\text{N}+\text{G}$) and is enhanced over other processes by strong interaction effects.  The graviton emission rate is found to be proportional to $1/(M_{\text{P}(4+n)})^{n+2}$ so this can be used to set a limit on the fundamental Planck scale.  Using the $M_*$ convention once again this gives $M_{*(4+2)} >$25--45~TeV and $M_{*(4+3)} >$2--4~TeV \cite{Cullen:1999hc,Hanhart:2000er} depending on the assumptions made and exactly what value is used for the supernova temperature. 

There are also limits on the fundamental Planck scale from the distortion of the cosmic diffuse background radiation spectrum.  The gravitons produced both by gravi-strahlung and by neutrino annihilation can decay into SM particles and for the low-mass KK modes $\text{G}\rightarrow \gamma\gamma$ is kinematically favoured.
Observations of the cosmic diffuse gamma radiation by EGRET and COMPTEL show that gravi-strahlung must account for less than about 1\% of the total energy emitted by all the supernovae that have exploded during the history of the universe.  This is calculated to set limits of $M_{*(4+2)} > 70$~TeV and $M_{*(4+3)} > 5$~TeV \cite{Hannestad:2001jv} (more recent calculations of the energy emitted by KK states trapped in neutron star haloes suggest $M_{*(4+2)} > 380$~TeV and $M_{*(4+3)} > 24$~TeV \cite{Hannestad:2001xi}).

Other authors have considered the contributions to the gamma radiation background from the gravitons produced by neutrino annihilation and obtained the limits $M_{*(4+2)} > 90$~TeV and $M_{*(4+3)} > 4$~TeV \cite{Hall:1999mk}.  Tighter limits can be obtained by assuming that the universe doesn't enter the radiation epoch until after it has been reheated by the decay of a massive scalar field (or by some other means of entropy production).

Other limits are derived by ensuring that the production of gravitons will not cause the universe to become matter dominated too early---this would lead to a value for the age of the universe which was too low.  The limits are $M_{*(4+2)} > 70$~TeV and $M_{*(4+3)} > 6$~TeV \cite{Fairbairn:2001ct} with the assumption that the universe can be considered as `normal' up to temperatures of 100~MeV\@.

Finally there have been Hubble Space Telescope observations that the surface temperatures of some older neutron stars are higher than conventionally expected.  The explanation for this could be that they are heated by the decay products of gravitons trapped in the neutron star haloes during the supernova collapse.  However the requirement that this heating is not excessive puts limits on the fundamental scale: $M_{*(4+2)} > 1400$~TeV and $M_{*(4+3)} > 50$~TeV \cite{Hannestad:2001xi}.  The equivalent limit for $n=4$ is of order a few TeV\@.  These are clearly the most constraining of the bounds on the fundamental scale although they are subject to various cosmological uncertainties.  Even so, $n=2$ almost certainly seems to be ruled out and $n=3$ is disfavoured.

\clearpage{\pagestyle{empty}\cleardoublepage}
\chapter{Exotic Heavy Leptons at the LHC}
\chaptermark{Heavy Leptons}
\label{leptons}
\section{Introduction}

We have already seen in Chapter~\ref{introch} that supersymmetry is a popular theory for new physics beyond the Standard Model.  It is seen as theoretically attractive, solves the hierarchy problem and improves the unification of the three gauge couplings at high energy scales.

`Intermediate scale' models which can be motivated on various astrophysical and cosmological grounds (see \secref{ism}) are constructed so that the gauge couplings unify at $\sim 10^{11}$~GeV, an intermediate energy scale.  This intermediate scale unification can be achieved by including new leptons as part of extra supermultiplets added to the MSSM\@.  Although 
this choice is not unique, the extra leptons in \cite{Abel:2000bj} are two left-handed SU(2) doublets and three right-handed singlets (all with vector-like copies).
\enlargethispage{-\baselineskip}

By assumption there are no new Yukawa couplings for the model in 
\cite{Abel:2000bj}.  This means that the lightest heavy
lepton is stable and the others will decay into it.  This is possible because
the tree-level mass degeneracy of the charged and neutral leptons will be 
destroyed by electroweak symmetry breaking.
  
As in the MSSM, the renormalization group equations can be used to determine 
the spectrum of supersymmetric particles which will have phenomenological 
implications for the LHC\@.  A study has also been performed \cite{Allanach:2001qe} investigating how to distinguish intermediate scale models from other supersymmetric string scenarios at a future linear collider.  However a characteristic feature of this model is the existence of the new leptons and so this work concentrates on the phenomenology due to the extra leptons themselves.  More specifically it is considered whether it will be possible to detect the charged heavy leptons at the LHC, and for what range of masses.    

Previous limits on the masses of new quasi-stable charged leptons have 
been limited by the energy available for particle production in lepton 
colliders.  The gauge unification arguments in intermediate scale models mean 
that the extra leptons are expected to have masses in the TeV range.  At the 
LHC enough energy should be available to produce and detect these exotic heavy
charged leptons.  Since the charged leptons are expected to be quasi-stable they will decay outside the detector; therefore the experimental procedure is different to that in \cite{Alexa:2001fz,CiezaMontalvo:2001cy} where limits are obtained by considering the decay products of the leptons.

This chapter starts (\secref{litrev}) by reviewing the existing cosmological and experimental limits on heavy particles to ensure that leptons such as those incorporated in this intermediate scale model are not ruled out.  In section~\ref{HERWIG} the theoretical input to the {\small HERWIG} Monte Carlo event generator to take account of new heavy leptons is described.  Then, in \secref{impl}, the use of a `time-of-flight' technique to detect the leptons at the LHC is considered: heavy leptons will arrive at the detector significantly later than relativistic particles, and the prospects for using this time delay as a method of detection are studied, for a range of masses.  There is also a study of how to distinguish such leptons from scalar leptons on the basis of their different angular distributions.  The results and conclusions are presented in \secref{results} and \secref{conchl} respectively.

\section{Cosmological and experimental constraints}
\label{litrev}

The mass degeneracy of the charged and neutral leptons will be broken by 
radiative corrections.  The electromagnetic self-energy correction for the 
charged lepton ensures that its mass will be greater than that of the neutral 
lepton\cite{Sher:1995tc}.  This means that the charged lepton can decay to its 
neutral partner producing either a real or virtual W, depending on the mass 
difference.  The neutral heavy lepton can be assumed to be totally stable 
(i.e.\ it has a lifetime orders of magnitude longer than that of the 
universe) because of the assumption that there are no new Yukawa couplings.

There are a number of cosmological and experimental limits on leptons 
(particularly charged leptons) as a function of both mass and lifetime:

\begin{itemize}

\item
The relic abundance of the leptons (calculated from the self-annihilation 
cross section) must not `over-close' the universe, i.e.\ provide more 
than the critical energy density ($\sim 10^{-5}$ GeV cm$^{-3}$) which is 
presumed to be accounted for by dark matter\cite{Wolfram:1979gp,Griest:1990wd}.

\item
A stable, charged lepton must have a low enough relic abundance for it not to 
have been detected in searches for exotic heavy isotopes in ordinary matter 
(see e.g.\ \cite{Smith:1982qu}).

\item
The massive lepton and any decays it may have must not significantly affect
nucleosynthesis or the synthesized elemental abundances \cite{Ellis:1992nb,Sarkar:1996dd}.

\item
If the lepton decays before the recombination era (at $\sim 10^{12}$~s) it 
must not distort the cosmic microwave background radiation \cite{Ellis:1992nb}.

\item
If the lepton has a longer lifetime it must not contradict the limits from
gamma-ray and neutrino background observations \cite{Kribs:1997ac,Gondolo:1993rn}.

\item
The mass and lifetime of the new leptons must not be such that they would 
have been detected in a previous collider experiment \cite{Abbiendi:2003yd,Achard:2001qw}.

\end{itemize}

Most of these limits are summarized in \cite{Sher:1992yr} and no constraints on charged leptons are found for lifetimes less than $\sim 1$~s, even for 
masses up to the TeV scale.  The mass splitting between the neutral and 
charged leptons is expected to be at least $\sim 10$~MeV \cite{Sher:1995tc} which means that the lifetime of the charged lepton is expected to be shorter than 1~s.

This means the only relevant constraints on the mass of the heavy leptons being studied are the experimental limits.  These are to some extent model-dependent, but the LEP collaborations put a lower limit (95\% confidence) on the mass of a long-lived charged lepton at around 102 GeV \cite{Abbiendi:2003yd,Achard:2001qw} (the Tevatron experiment has not improved upon the LEP limits).

\section{Theoretical input to {\small HERWIG}} 
\label{HERWIG}

\enlargethispage{-\baselineskip}
The general-purpose Monte Carlo event generator {\small HERWIG 6.3} includes 
subroutines for both neutral and charged Drell-Yan processes in a hadron 
collider \cite{Marchesini:1992ch,Corcella:2000bw,Corcella:2001pi}.  As described in \cite{Corcella:2000bw} the 
initial-state parton showers in Drell-Yan processes are matched to the exact 
$\mathcal{O}(\alpha_\text{S})$ matrix element result \cite{Corcella:1999gs}.  However the 
neutral Drell-Yan processes all use the approximation that the two fermions 
produced can be treated as massless.  It is reasonable that this is a valid 
approximation for the SM quarks and leptons but this will not necessarily be 
the case for the proposed new heavy leptons discussed above which may have 
masses of order 1~TeV\@.

For this reason, full Born-level expressions for the Drell-Yan cross sections 
were derived, taking into account the masses of the produced leptons.   The final results (for both neutral- and charged-current cases) were expressed in terms of vector and axial couplings to keep them as general as possible, although in all that follows Standard Model couplings will be assumed.  The details of the derivation are in Appendix~\ref{appa}

\begin{figure}
\begin{center}
\begin{picture} (90,60)
\ArrowLine(30,35)(0,60)
\Text(-3,60)[r]{$\bar{\text{q}}$}
\ArrowLine(0,10)(30,35)
\Text(-3,10)[r]{q}
\Photon(30,35)(90,35){2}{8}
\Text(60,46)[c]{Z$^0$/$\gamma$}
\ArrowLine(120,60)(90,35)
\Text(123,60)[l]{L$^+$}
\ArrowLine(90,35)(120,10)
\Text(123,10)[l]{L$^-$}
\end{picture}
\capbox{Feynman diagram for Drell-Yan production of heavy leptons}{Feynman diagram for Drell-Yan production of heavy leptons.\label{hlfeyn}}
\end{center}
\end{figure}

The derived expressions for both differential and total cross sections for the
neutral-current Drell-Yan processes 
($\text{q}\bar{\text{q}}\rightarrow \text{Z}^0/\gamma\rightarrow \text{L}^-\text{L}^+$, see Figure~\ref{hlfeyn}) are shown below.  The 
notation is influenced by that already used in subroutines within 
{\small HERWIG 6.3}.
\enlargethispage{-\baselineskip}

The differential cross section is given by
\begin{equation}
\frac {d\hat{\sigma}}{d\Omega} (\text{q}\bar{\text{q}}\rightarrow \text{L}^-\text{L}^+) 
= \frac{e^4}{48\pi^2} \frac {1} {\hat{s}^2} \frac {p_3} {E}
\left[ C_1 \left( E_3E_4 + p_3^2\cos^2\theta^*\right) + C_2 m_3 m_4 +
2 C_3 E p_3 \cos\theta^* \right],
\label{dxs}
\end{equation}
where
\begin{gather}
C_1 = \frac {\left[\left(d_\text{V}^\text{f}\right)^2 + \left(d_\text{A}^\text{f}\right)^2 
\right] \left[ \left(d_\text{V}^\text{i}\right)^2 + \left(d_\text{A}^\text{i}\right)^2 \right]\hat{s}^2 } 
{\left(\hat{s}-m_\text{Z}^2\right)^2 + m_\text{Z}^2\Gamma_\text{Z}^2} + \left(q^\text{f}\right)^2
\left(q^\text{i}\right)^2 + \frac {2 q^\text{f} q^\text{i} d_\text{V}^\text{f} d_\text{V}^\text{i}\hat{s}
\left(\hat{s}-m_\text{Z}^2\right)}{\left(\hat{s}-m_\text{Z}^2\right)^2 + m_\text{Z}^2\Gamma_\text{Z}^2}\,,
\\[3mm]
C_2 = \frac {\left[\left(d_\text{V}^\text{f}\right)^2 - \left(d_\text{A}^\text{f}\right)^2 
\right] \left[ \left(d_\text{V}^\text{i}\right)^2 + \left(d_\text{A}^\text{i}\right)^2 \right]\hat{s}^2 } 
{\left(\hat{s}-m_\text{Z}^2\right)^2 + m_\text{Z}^2\Gamma_\text{Z}^2} + \left(q^\text{f}\right)^2
\left(q^\text{i}\right)^2 + \frac {2 q^\text{f} q^\text{i} d_\text{V}^\text{f} d_\text{V}^\text{i}\hat{s} 
\left(\hat{s}-m_\text{Z}^2\right)}{\left(\hat{s}-m_\text{Z}^2\right)^2 + m_\text{Z}^2\Gamma_\text{Z}^2}\,,
\\[3mm]
C_3 = 2 \left( \frac {2 d_\text{V}^\text{f} d_\text{A}^\text{f} d_\text{V}^\text{i} d_\text{A}^\text{i}\hat{s}^2} 
{\left(\hat{s}-m_\text{Z}^2 \right)^2 + m_\text{Z}^2\Gamma_\text{Z}^2} + 
\frac {q^\text{f} q^\text{i} d_\text{A}^\text{f} d_\text{A}^\text{i} \hat{s} \left(\hat{s}-m_\text{Z}^2\right)} 
{\left(\hat{s}-m_\text{Z}^2\right)^2 + m_\text{Z}^2\Gamma_\text{Z}^2} \right).
\end{gather}
In these expressions $q^\text{i}$ and $q^\text{f}$ are the charges (in units of
the electron charge) of the initial- and final-state particles respectively.  
The angle $\theta^*$ is the angle between the out-going lepton (L$^-$) 
direction and the incoming quark direction in the centre-of-mass frame.  
$E_3$ and $p_3$ refer to the centre-of-mass energy and momentum magnitude
of the produced lepton.  Similarly the subscript 4 refers to the produced
anti-lepton. $E$ is the energy of both the colliding quarks in the 
centre-of-mass frame (i.e.\ $\hat{s}=4E^2$).  

The couplings $d_\text{V}$ and $d_\text{A}$ are related to the normal vector and axial 
coupling constants to the Z$^0$ ($c_\text{V}$ and $c_\text{A}$) by relations like
\begin{equation}
d_\text{V} = \frac {c_\text{V} g_\text{Z}} {2e}\,.
\end{equation}
Equation~(\ref{dxs}) is a general expression which includes the Z$^0/\gamma$ interference terms; the massless case is retrieved by
setting $p_3=E_3=E_4=E$ (as well as $m_3=m_4=0$).  The normal charged-current Drell-Yan case ($\text{q}\bar{\text{q}}^\prime \rightarrow \text{W}^\pm\rightarrow 
\text{L}^\pm \text{L}^0$) will not be relevant in this work but can be obtained by setting $c_\text{V}=c_\text{A}=1$ and replacing $g_\text{Z}$,
$m_\text{Z}$, $\Gamma_\text{Z}$ and $q^\text{i/f}$ by $g_\text{W}/\sqrt{2}$, $m_\text{W}$, $\Gamma_\text{W}$ and 0 respectively.

Integration of eq.~(\ref{dxs}) gives the following expression for the parton-level cross section:
\begin{equation}
\hat{\sigma} (\text{q}\bar{\text{q}}\rightarrow \text{L}^-\text{L}^+)= 
\frac {e^4} {12\pi} \frac {1} {\hat{s}^2} \frac {p_3} {E} 
\left[ C_1 \left(E_3E_4 + \frac {p_3^2} {3}\right) + C_2 m_3 m_4 \right].
\label{xs}
\end{equation}

In order to distinguish these heavy leptons from heavy supersymmetric 
partners of SM particles it would be necessary to consider the angular 
distribution of the produced particles.  For the neutral-current production of
left/right-handed scalars the differential cross section is also derived in Appendix~\ref{appa} and is found to be
\begin{equation}
\frac {d\hat{\sigma}}{d\Omega}(\text{q}\bar{\text{q}}\rightarrow\tilde{\text{l}}_\text{L/R}
\tilde{\text{l}}^*_\text{L/R})=\frac {e^4} {96\pi^2} D \frac {1}{\hat{s}^2} \frac {p_3} 
{E} p_3^2 \sin^2\theta^*,
\label{sdxs}
\end{equation}
where
\begin{equation}
D = \frac {h_\text{L/R}^2 \left[\left(d_\text{V}^\text{i}\right)^2 +\left(d_\text{A}^\text{i}\right)^2 \right] 
\hat{s}^2} {\left(\hat{s}-m_\text{Z}^2\right)^2 + m_\text{Z}^2\Gamma_\text{Z}^2} + 
\left(q^\text{f}\right)^2 \left(q^\text{i}\right)^2 +
\frac {2 q^\text{f} q^\text{i} h_\text{L/R} d_\text{V}^\text{i} \hat{s}\left(\hat{s}-m_\text{Z}^2\right)}
{\left(\hat{s}-m_\text{Z}^2\right)^2 + m_\text{Z}^2\Gamma_\text{Z}^2}\,.
 \end{equation}
The couplings $d_\text{V}$ and $d_\text{A}$ are defined as above and $h_\text{L/R}= g_\text{L/R}g_\text{Z}/2e$
where $g_\text{L/R}$ is the coupling to left/right-handed sleptons at the gauge 
boson-slepton-slepton vertex.

The $\sin^2\theta^*$ angular distribution contrasts with the asymmetric 
distribution for the heavy leptons seen in eq.~(\ref{dxs}); this will 
be studied in \secref{angdist}.

\section{Implementation}
\label{impl}

\subsection{Modifications to {\small HERWIG 6.3}}

The above formulae were incorporated into a new subroutine added to 
{\small HERWIG 6.3} together with new particle entries (for the heavy 
leptons) and a new process code (\texttt{IPROC}) for the new process.

Increasing the mass of the new leptons allowed the variation of cross section
with mass to be studied---the mass was varied over a range from of order the 
top quark mass to a maximum of order 1~TeV\@.  The lifetimes of the charged 
leptons produced were set to 1~s so that the number of decays occurring inside
the detector will be negligible.

The angular distribution of the produced leptons was studied over the same 
mass range to investigate the possibility of distinguishing heavy leptons 
from MSSM sleptons.

\subsection{Time-of-flight technique}
\label{tof}

The main aim was to consider the possible detection of charged heavy leptons 
at the LHC and for what mass ranges this might be practical.  This work 
refers specifically to the technical specifications of the ATLAS detector 
\cite{ATLASTDR1,ATLASTDR2,ATLASmTDR}; however it is believed that similar results will
be obtained for the {\small CMS} experiment \cite{CMS}.

The method used was a time-of-flight technique as discussed in 
ref.~\cite{Hinchliffe:1998ys}.  This method utilizes the fact that, when compared to 
relativistic particles, there is a considerable time delay for heavy particles
to reach the muon system.  Heavy charged leptons like those being considered 
in this work will be detected in both the central tracker and the muon 
chambers, and from the measured momentum and time delay it is possible to 
reconstruct the mass.

\begin{figure}
\unitlength1cm
\begin{minipage}[t]{3.05in}
\begin{center}
\includegraphics[width=\linewidth]{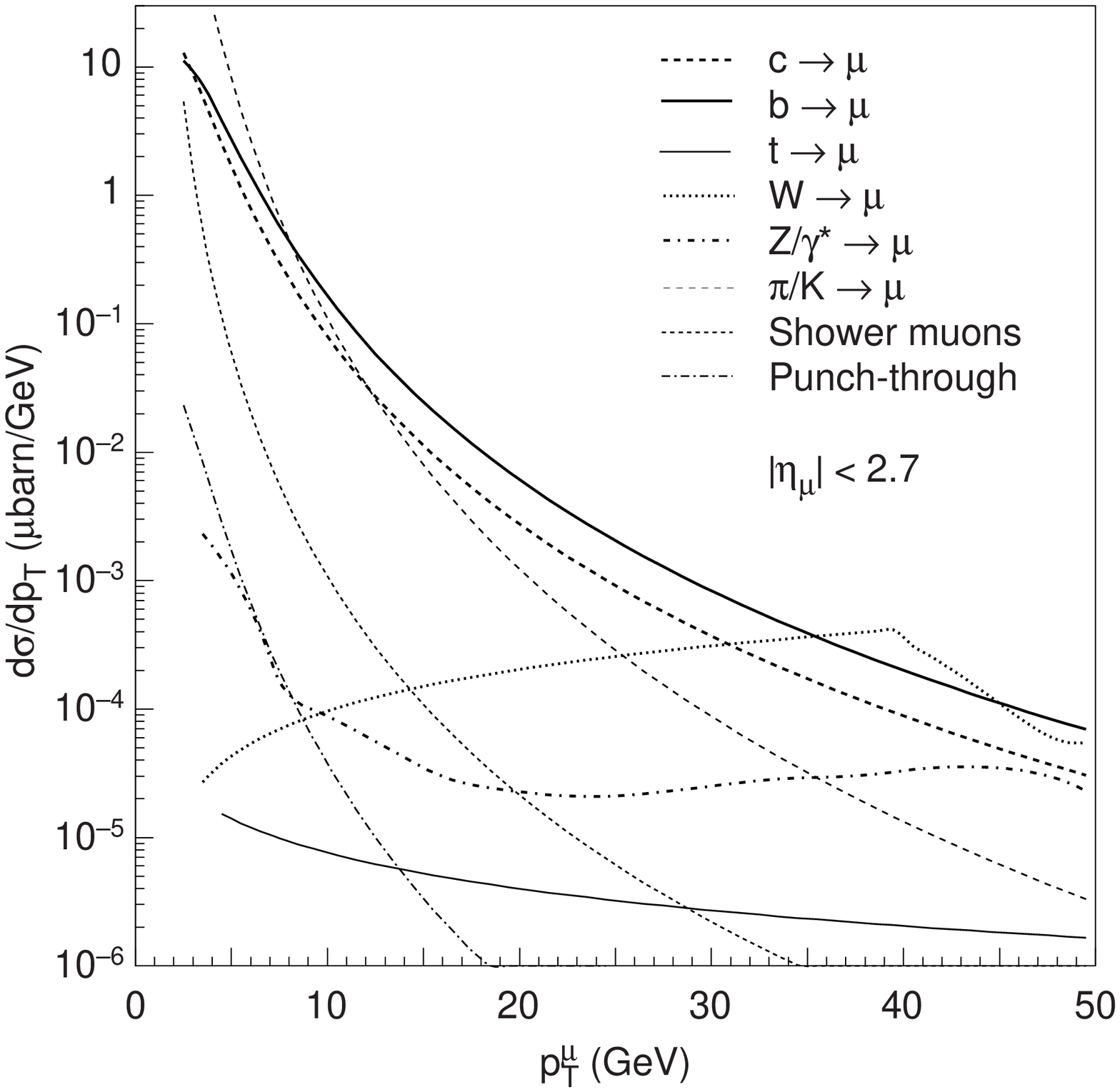}
\capbox{Transverse momentum dependence of the inclusive muon cross section}{Transverse momentum dependence of the inclusive muon cross section 
integrated over $|\eta|< 2.7$.  The horizontal scale is the transverse 
momentum at production. (Reproduced from \cite{ATLASmTDR}).\label{muonpT}}
\end{center}
\end{minipage}
\hfill
\begin{minipage}[t]{3.05in}
\begin{center}
\includegraphics[width=\linewidth]{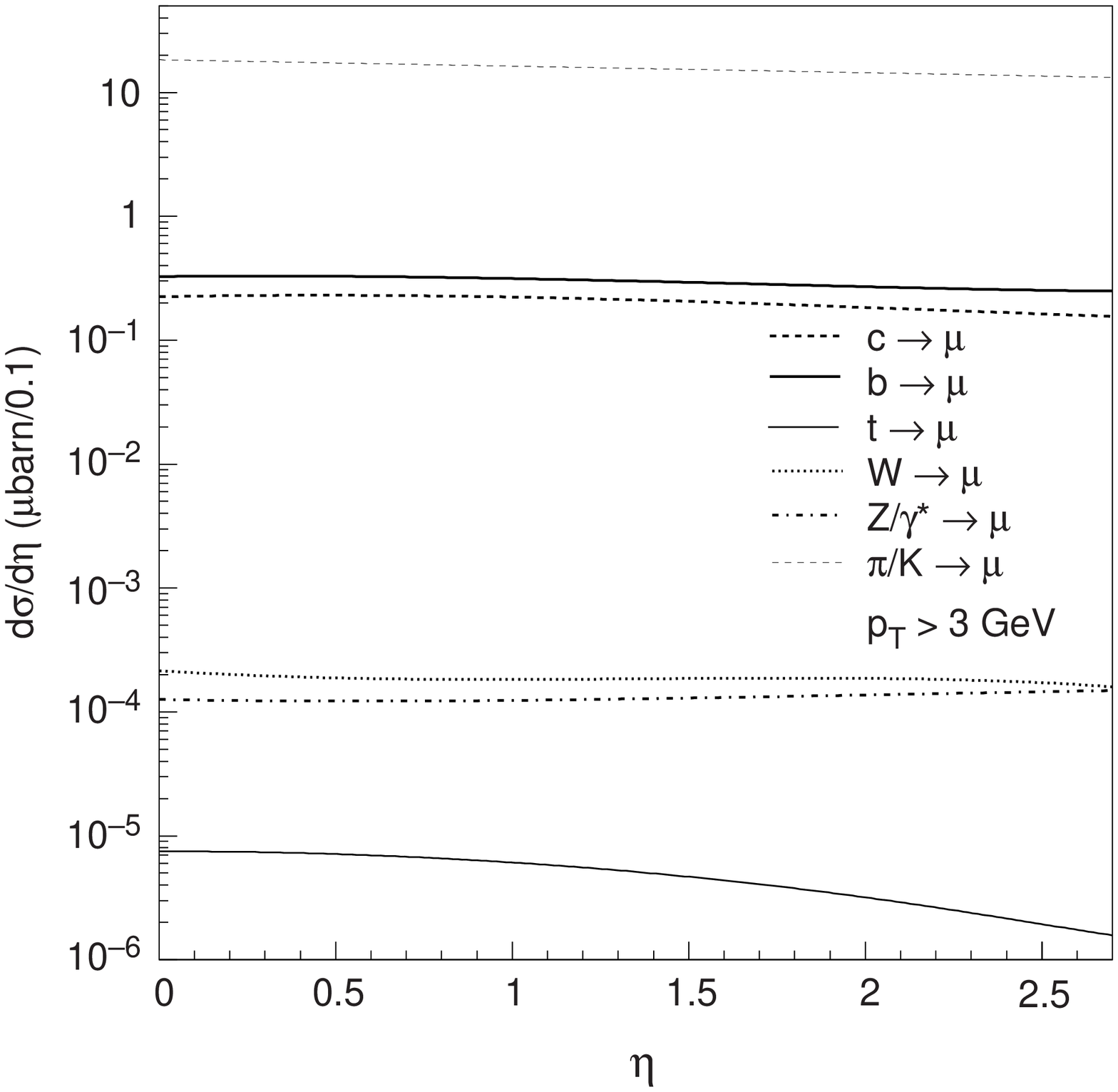}
\capbox{Rapidity-dependence of the inclusive muon cross sections}{Rapidity-dependence of the inclusive muon cross sections, integrated
over \mbox{3 $<p_\text{T}<$ 50~GeV}. (Reproduced from \cite{ATLASmTDR}).
\label{muonbg}}
\end{center}
\end{minipage}
\end{figure}

Imperfections in the time and momentum resolutions will broaden the mass peak.  Uncertainty over which bunch crossing a 
particular detected particle comes from may also provide a background signal 
(from muons produced in Drell-Yan processes or in heavy quark decays).
Plots from \cite{ATLASmTDR} show some differential cross sections for the processes most likely to produce muons which might be mis-identified as heavy leptons.  These are reproduced in Figures~\ref{muonpT} and \ref{muonbg}.

The studies presented here do not attempt to model the background, which it is
not thought will be significant (there is no `physics' background, only 
`detector' background as discussed above), and use a simple cut \cite{Hinchliffe:1998ys} on the range of time delays allowed.

The calculation of a reconstructed mass $m$ from the time delay $\Delta t$ and 
momentum is straightforward.  For a lepton hitting the radial part of the 
muon spectrometer, the time delay with respect to a relativistic 
($\beta=v/c=1$) particle is given by
\begin{equation}
\Delta t = \frac {r} {p_\text{T}} \left(E-p\right), 
\end{equation}
where $r$ is the radius of the outer layer of the muon system, 
$p_\text{T}$ the transverse momentum, and $E$ and $p$ the total energy and momentum 
respectively.  Substituting for $E$ in the energy-momentum invariant 
$E^2-p^2=m^2$ gives the result
\begin{equation}
m^2 = p^2 \left( \frac {1}{\beta^2} - 1 \right)
= \frac {p_\text{T} \Delta t} {r} \left(2p + \frac {p_\text{T} \Delta t} {r} \right).
\label{masssq}
\end{equation}
It is also necessary to check for occasions when the lepton hits the
end-cap of the detector by considering the magnitude of $p_\text{T}/p_z$\@.
The muon system is modelled approximately as a cylinder of radius 10~m and with a half-length of 20~m \cite{ATLASmTDR}.  A pseudo-rapidity cut requiring that 
$|\eta|<2.7$ was applied to take account of the region close to the beam where
particles cannot be detected.

From equation~(\ref{masssq}), the time delay and measured momentum for 
any particle detected in the muon system can be used to calculate its mass.  
As in \cite{Hinchliffe:1998ys} the time delay was smeared with a Gaussian (although a width 
of 0.7~ns  \cite{ATLASmTDR} rather than 1~ns was used) and a cut on 
the smeared time delay was applied such that 
$10 \mbox{ ns} < \Delta t < 50\mbox{ ns}$.  Increasing the lower limit 
on $\Delta t$ would  
reduce the efficiency but improve the mass resolution by removing many of the 
high-$\beta$ leptons for which the time resolution is poor.
The upper limit eliminates very slow particles which lose most of their energy
in the calorimeter.  An alternative upper limit of 25~ns, reflecting practical
concerns, is discussed in \secref{masspeaks}.  

The most important contribution to the momentum uncertainty $\Delta p$ is due
to the measurement error on the sagitta (the deviation from a straight line of
a charged particle in a magnetic field).  $\Delta p_\text{sag}$ was taken into 
account by using 
\begin{equation}
\frac {\Delta p_\text{sag}}{p^2} = 1.1\times10^{-4}, 
\label{sagerror}
\end{equation}
where $p$ is the total momentum in GeV\@.  The constant in 
eq.~(\ref{sagerror}) was obtained from \cite{ATLASTDR1} as an average over the
different parts of the ATLAS detector.  This expression should be valid near 
the discovery limit for heavy leptons when multiple scattering is 
insignificant---the maximum time delay cut is found to remove most of the low-$\beta$ leptons for which multiple scattering would have more effect.  However
to allow investigations over a full mass range a multiple scattering term 
($\Delta p_\text{ms}$) was also incorporated with  
\begin{equation}
\Delta p_\text{ms} = 2\times10^{-2}\sqrt{p^2 + m^2}.
\label{mserror}
\end{equation}
The constant depends on the material distribution in the 
spectrometer---the above value is given in \cite{ATLMUON} for the ATLAS muon 
detector.
\enlargethispage{-\baselineskip}

The two errors in eqs.~(\ref{sagerror}) and (\ref{mserror}) were taken to be 
independent and hence combined in quadrature.  A third possible contribution 
to momentum resolution---due to fluctuation in the energy loss in the 
calorimeter---is neglected as it is always dominated by one of the other 
terms.

Using the momentum and time resolution as described above it was possible to 
satisfactorily reproduce the mass resolution as a function of $\beta$ for 
101~GeV particles as given in \cite{ATLASTDR1}.

The efficiency of the muon detection system is approximated to be 85\%,
independent of the particle momentum \cite{ATLASTDR1}.

\section{Results}
\label{results}

\subsection{Backgrounds}
\label{bg}

Any `detector' background is greatly dependent on the 
details of the muon detection system.  Such a background would come from muons 
produced either in the decays of heavy quarks or directly via the Drell-Yan process (both these processes 
have very high cross sections compared to that for heavy lepton production).  
If timing inadequacies mean that some of these muons are mistakenly thought
to have come from an earlier interaction, i.e.\ they appear to have a 
large time delay, they could obscure the exotic lepton mass peak. For a 
background signal to look like a heavy lepton neutral-current signal, however,
two opposite charge muons would have to be mis-identified at the same time 
which makes it very unlikely that background could be significant. 
\enlargethispage{-2\baselineskip}

\begin{figure}
\begin{center}
\scalebox{0.5}{\rotatebox{-90}{\includegraphics[width=\textwidth]{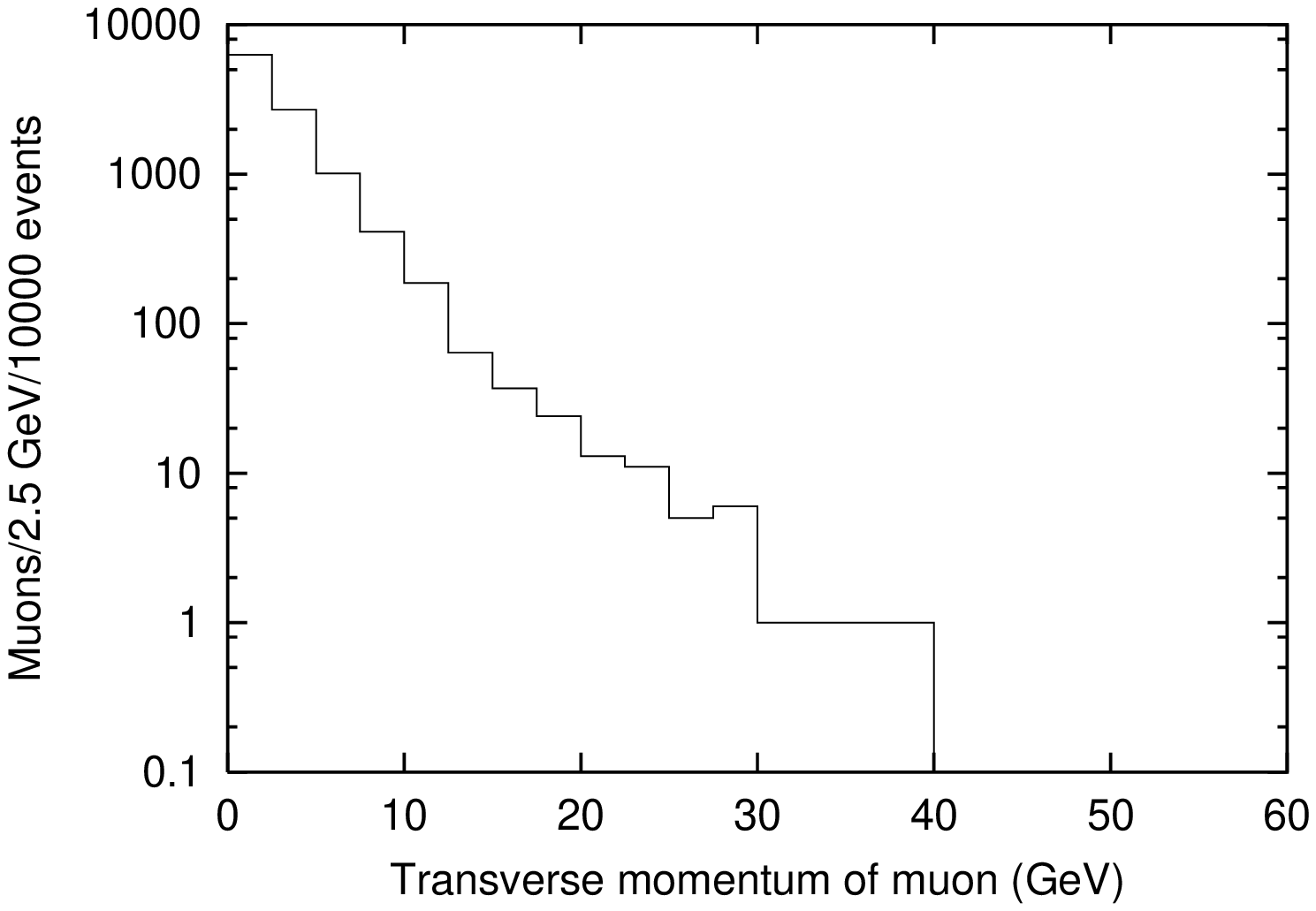}}}
\capbox{$p_\text{T}$ spectrum for muons produced in b decays}{$p_\text{T}$ spectrum for muons produced in b decays.\label{bp}}
\end{center}
\end{figure}

\begin{figure}
\unitlength1cm
\begin{center}
\begin{minipage}[t]{3.05in}
\scalebox{0.6}{\rotatebox{-90}{\includegraphics{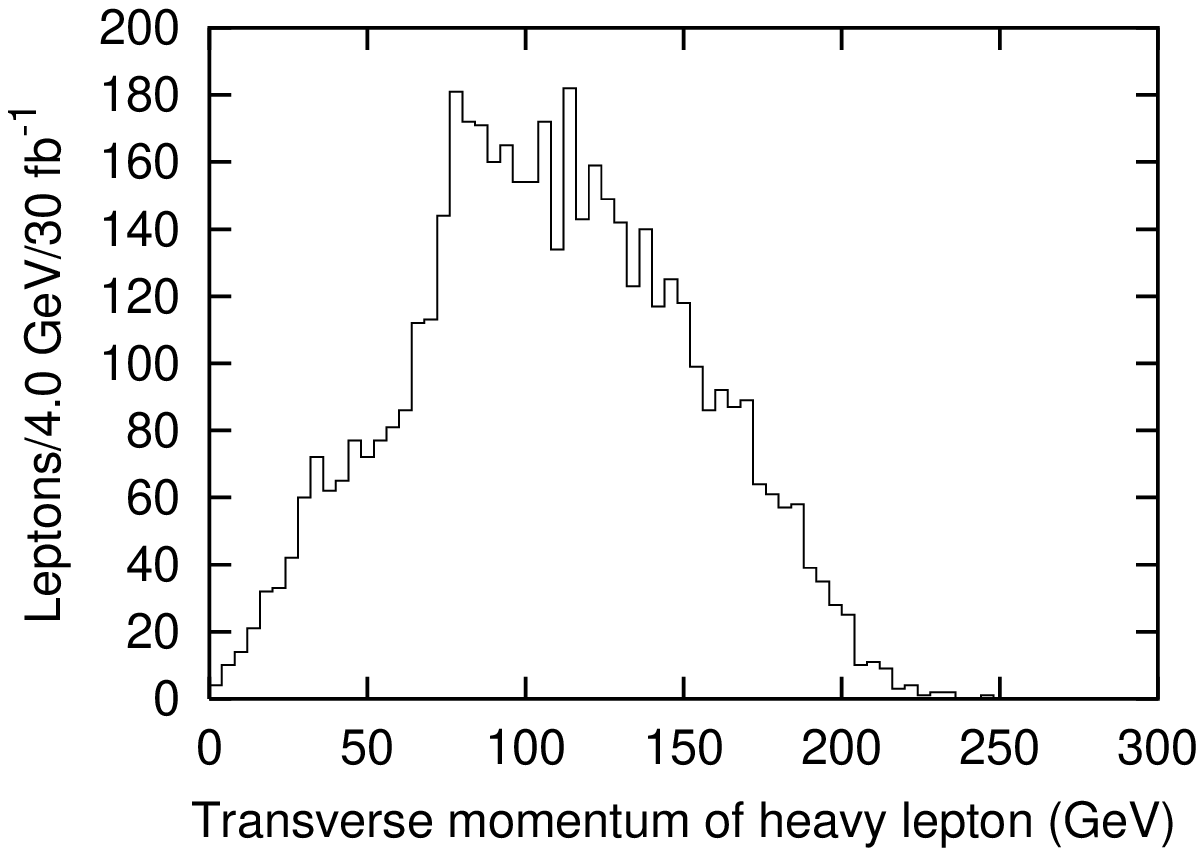}}}
\end{minipage}
\hfill
\begin{minipage}[t]{3.05in}
\scalebox{0.6}{\rotatebox{-90}{\includegraphics{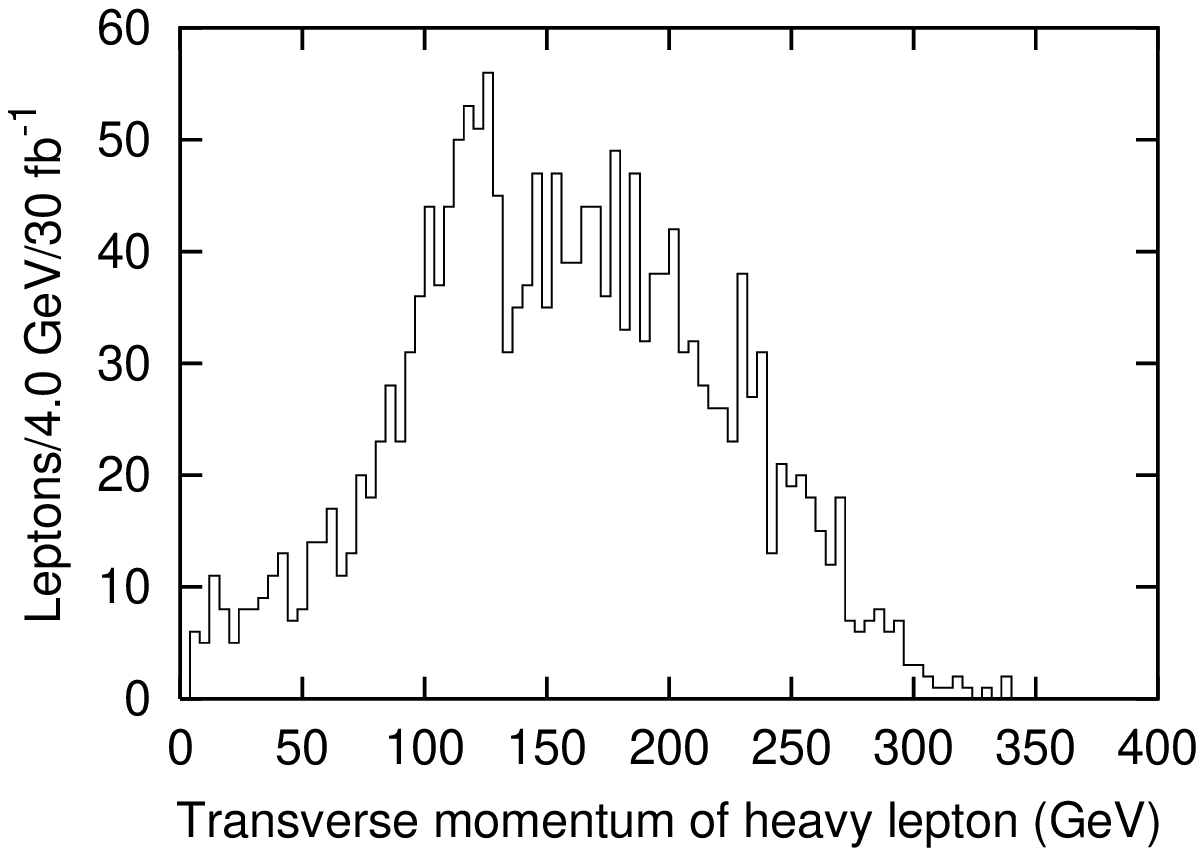}}}
\end{minipage}
\capbox{$p_\text{T}$ spectra for 175 and 250~GeV heavy leptons from Drell-Yan 
processes}{$p_\text{T}$ spectra for 175 and 250~GeV heavy leptons from Drell-Yan 
processes.
\label{p250}}
\end{center}
\end{figure}

To eliminate any possible background from heavy quark decays, a cut requiring 
the transverse momentum to be greater than 50~GeV was made to produce the 
plots in \secref{masspeaks}.  Figures~\ref{bp} and \ref{p250} show that a
$p_\text{T}$ cut at 50~GeV should remove most of the muons from bottom 
quark decay without significantly reducing the exotic heavy lepton signal.  It
can be seen that as the mass of the exotic heavy lepton increases, the peak in 
the $p_\text{T}$ spectrum becomes less well defined and shifts to higher values.  Top
quark decays can give rise to muons at high $p_\text{T}$ but the small top 
cross~section, combined with the improbability of mis-identifying two muons,
should suffice to eliminate this source of background. 

\subsection{Heavy lepton mass peaks}
\label{masspeaks}

The results presented in this section show the reconstructed mass peaks for 
an integrated luminosity of 100~fb$^{-1}$ after the cuts outlined above have 
been applied. 

\begin{figure}
\begin{center}
\scalebox{0.48}{\rotatebox{-90}{\includegraphics[width=\textwidth]
{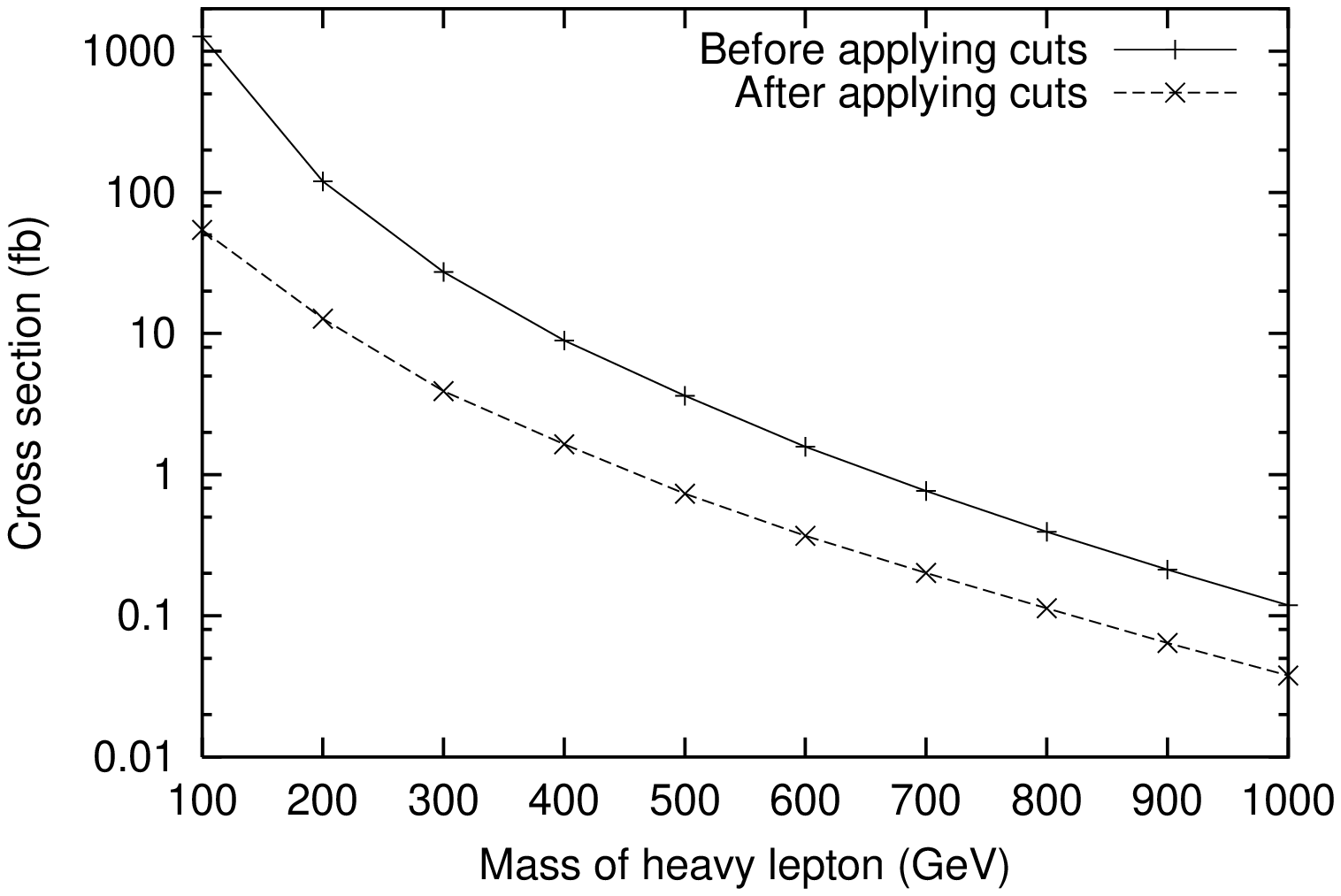}}}
\capbox{Cross~section for Drell-Yan heavy lepton production}{Cross~section for Drell-Yan heavy lepton production as a function of
the lepton mass.\label{cross}}
\end{center}
\end{figure}

The possibility of detection at the LHC is found to be entirely cross~section 
limited---the decrease of cross~section with increasing lepton mass is shown
in Figure~\ref{cross}.  The fraction of leptons passing the cuts actually 
slightly increases with mass due to the longer time delays.

\begin{figure}[t]
\begin{center}
\unitlength1cm
\begin{minipage}[t]{3.05in}
\scalebox{0.6}{\rotatebox{-90}{\includegraphics{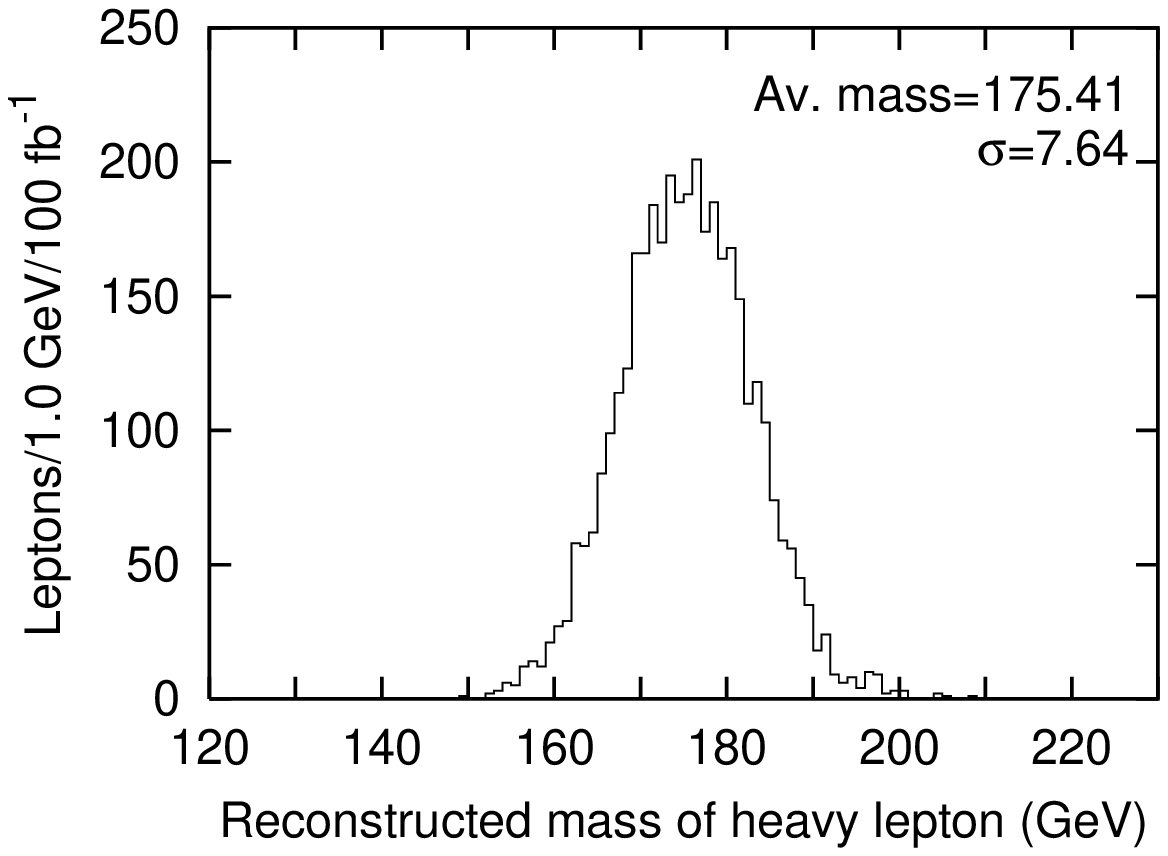}}}
\end{minipage}
\hfill
\begin{minipage}[t]{3.05in}
\scalebox{0.6}{\rotatebox{-90}{\includegraphics{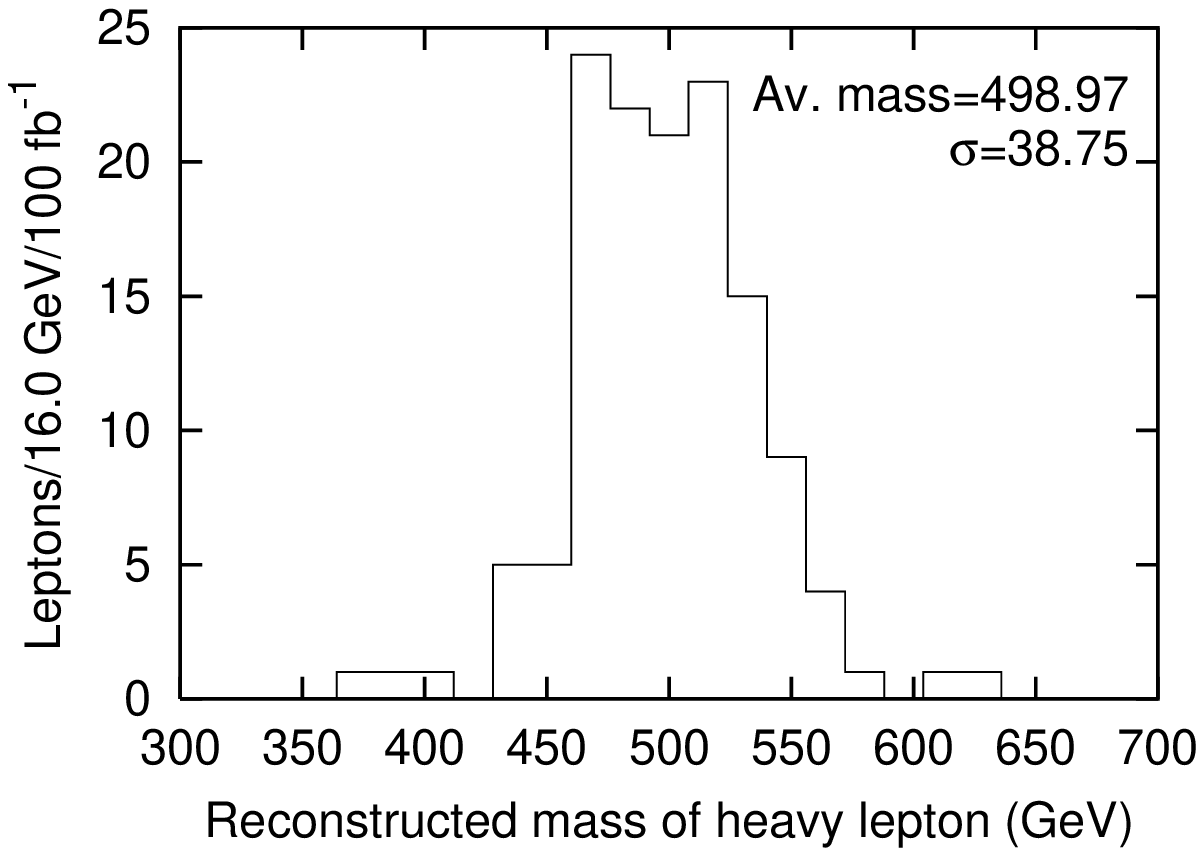}}}
\end{minipage}

\unitlength1cm
\begin{minipage}[t]{3.05in}
\scalebox{0.6}{\rotatebox{-90}{\includegraphics{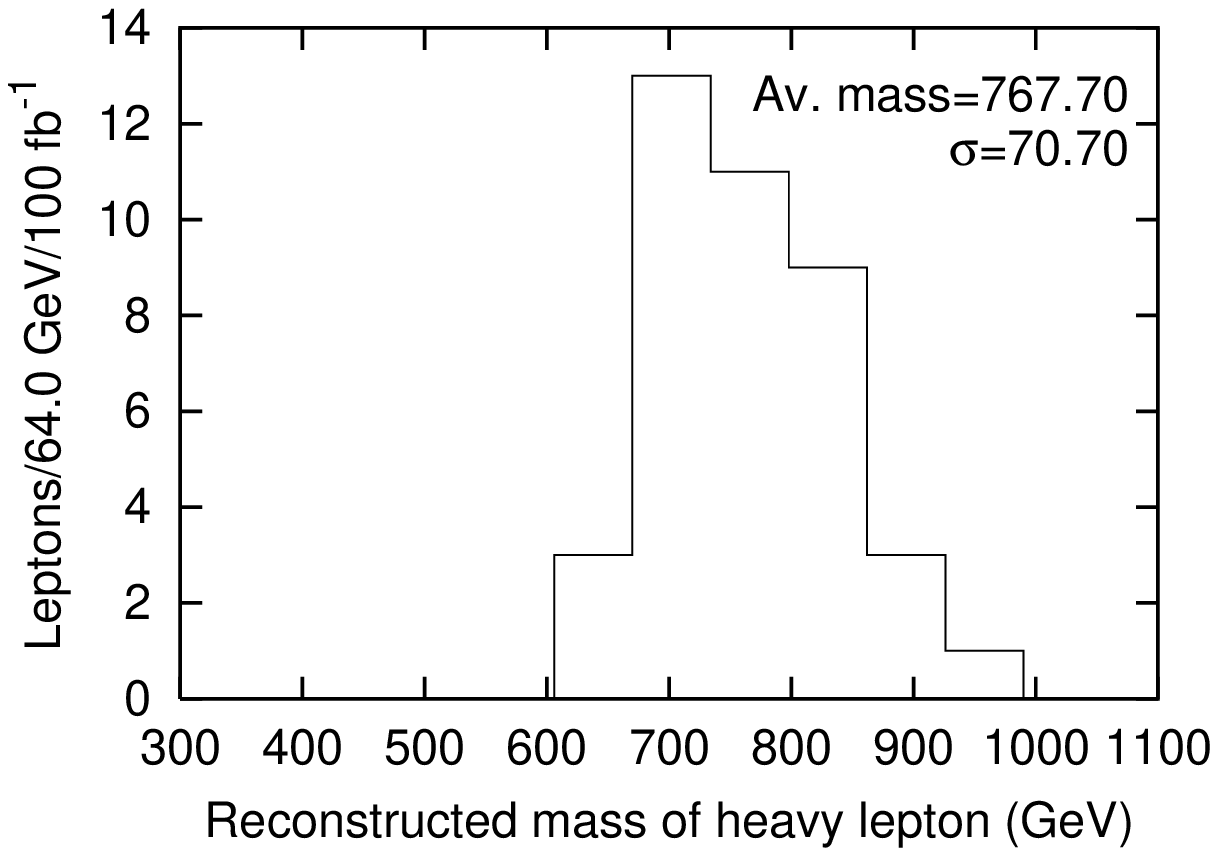}}}
\end{minipage}
\hfill
\begin{minipage}[t]{3.05in}
\scalebox{0.6}{\rotatebox{-90}{\includegraphics{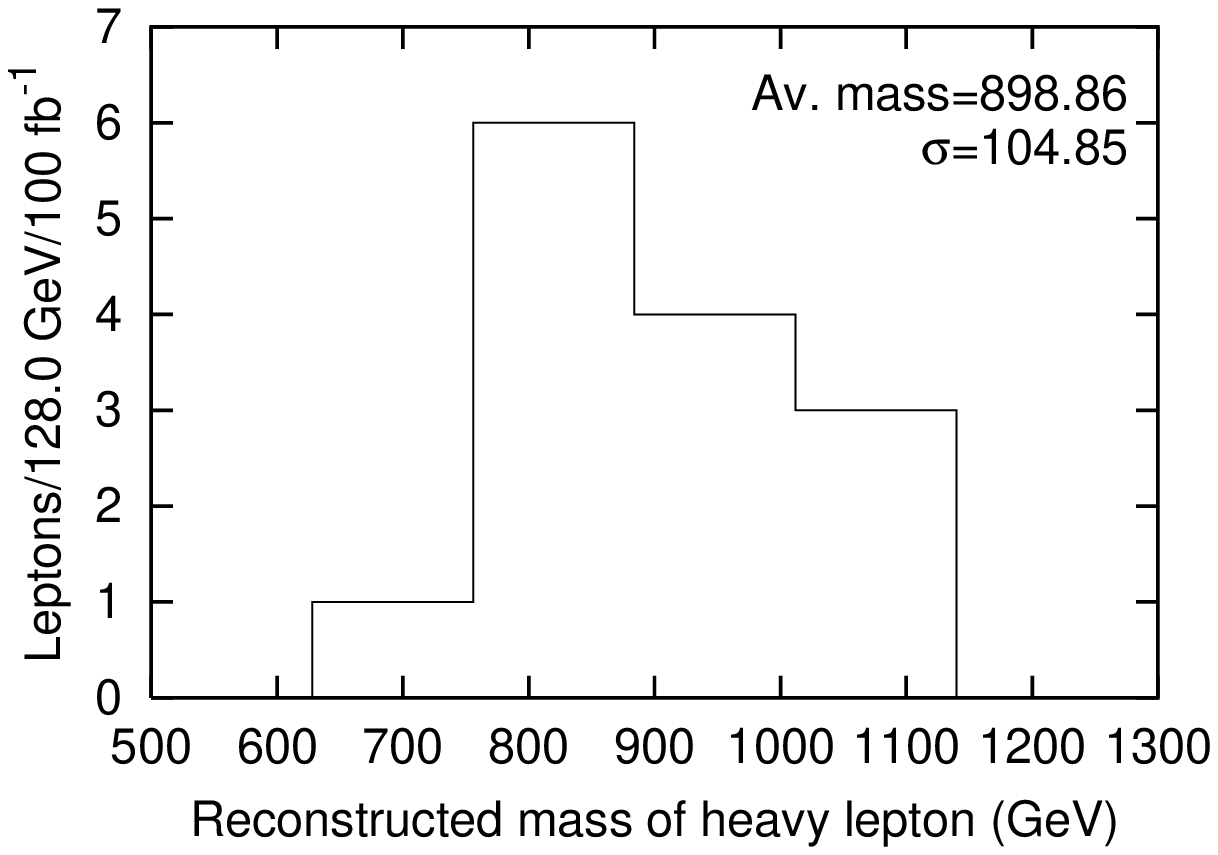}}}
\end{minipage}
\capbox{Reconstructed mass peaks for exotic leptons}{Reconstructed mass peaks for exotic leptons with masses of
175~GeV, 500~GeV, 750~GeV and 900~GeV\@.
\label{1000}}
\end{center}
\end{figure}

Figure~\ref{1000} for the neutral-current Drell-Yan process makes it
clear that for masses up to about 1~TeV it should be possible to detect new
charged heavy leptons at the LHC (particularly if a larger integrated 
luminosity can be obtained).  A discovery criterion of 10 leptons (from 5 
events) in the mass peak is found to give a mass limit of 950~GeV for the 
detection of such leptons using an integrated luminosity of 100~fb$^{-1}$.
\enlargethispage{\baselineskip}

To reduce possible backgrounds the plots are based on events from 
which both leptons produced passed the cuts imposed.

The long drift times in the muon chambers used by ATLAS allow the recording of events with time delays at arrival much greater than the 25~ns beam crossing interval of the LHC \cite{ATLASmTDR}; following \cite{Hinchliffe:1998ys}
a maximum time delay of 50~ns is assumed here. However, to 
correctly associate the delayed track in the muon system with the event 
recorded in the other detector systems would require a specialized trigger
based, for example, on the presence of high-$p_\text{T}$ tracks in the inner
detector (these would identify the true event time). Without such a trigger, the
maximum allowed time delay would be 25~ns, and the discovery reach
would be reduced to a lepton mass of 800~GeV\@. The statistical
sample available for the angular distribution analysis in 
\secref{angdist} would also be reduced by a factor of about 2, depending on the 
lepton mass.

\subsection{Distinguishing leptons from sleptons}
\label{angdist}

As mentioned in \secref{HERWIG} it should be possible to distinguish 
these new heavy leptons from heavy scalar leptons by studying the angular 
distribution with which they are produced.  The two different angular 
distributions in the centre-of-mass frame are shown in eqs.~(\ref{dxs}) and
(\ref{sdxs}).  The application of the cuts described in \secref{tof} as 
well as the effects of the mass and momentum resolution mean that the 
observed angular distributions are somewhat different to these.
\enlargethispage{-\baselineskip}

\begin{figure}
\begin{center}
\scalebox{0.48}{\rotatebox{-90}{\includegraphics[width=\textwidth]
{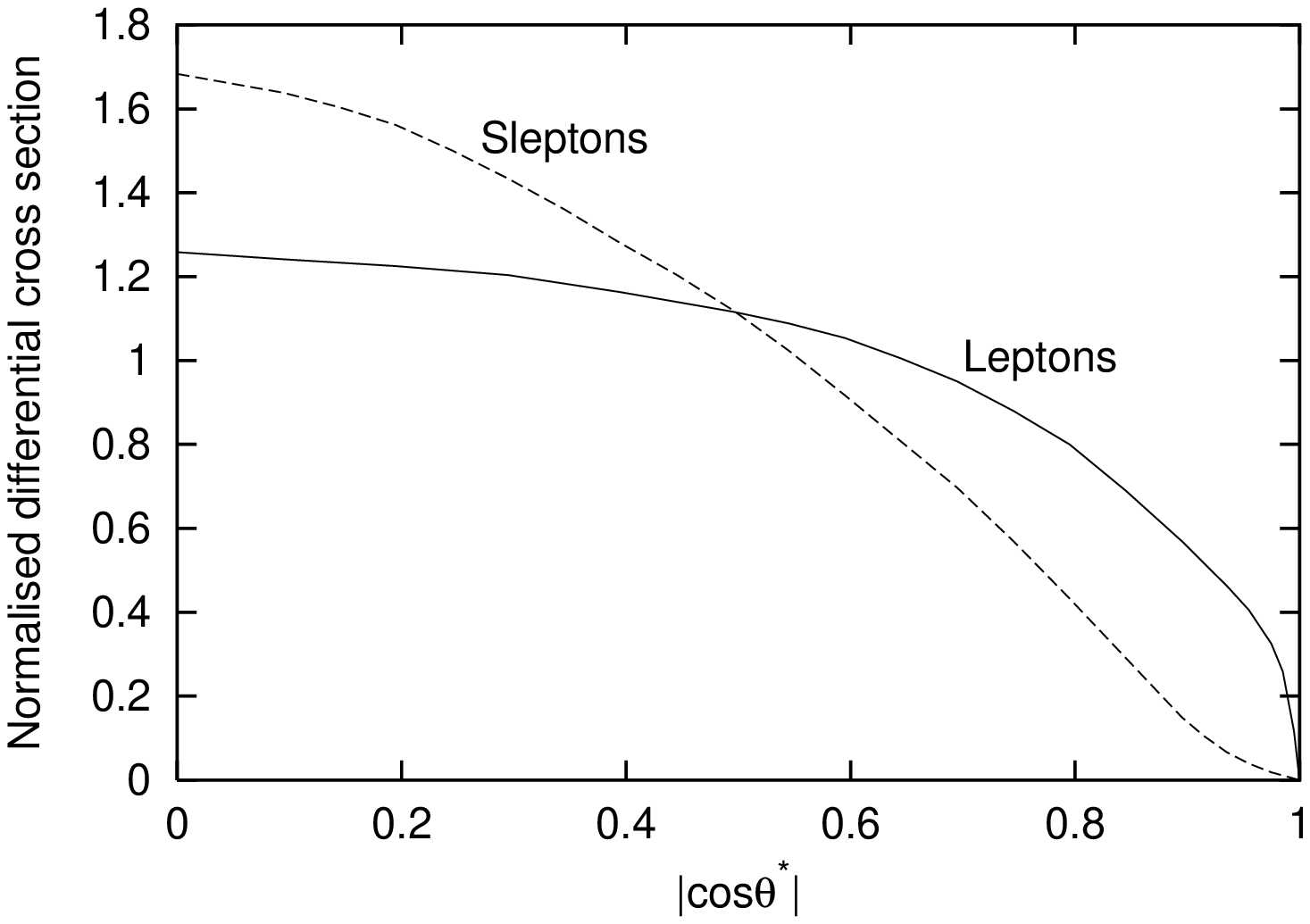}}}
\capbox{Angular distributions of detected pairs of 400 GeV particles}{Angular distributions of detected pairs of 400 GeV particles.
$\theta^*$ is the angle between the out-going particle and the incoming quark in the centre-of-mass frame.\label{a400}}
\end{center}
\end{figure}

The forward-backward asymmetry for the exotic heavy leptons was not initially considered
because of the difficulty in a p-p collider of distinguishing the quark and 
anti-quark in the centre-of-mass frame.  Hence only $|\cos\theta^*|$ was used
when the angular distributions were compared.  Figure~\ref{a400} shows the angular distribution for both exotic leptons and sleptons of mass 400~GeV\@.  In an experimental situation, the exact quark direction is not known and so the angle used in this plot is not exactly $\theta^*$---instead it is the angle between the out-going lepton and the incoming proton.  Examining the event records before and after the initial-state parton shower (in which the quark can acquire some transverse momentum) shows that the difference between the two angular distributions is small.

A chi-squared test was applied to find the probability that a scalar model 
describes the heavy lepton data.  The model was the angular distribution of a large Monte 
Carlo sample of scalars, and the total number of detected scalar pairs was 
rescaled to the expected number of detected pairs of heavy leptons.  An 
`average' data-set was obtained from a rescaled Monte Carlo sample of 
leptons, generated according to their angular distribution.  Both these 
rescaled distributions had negligible theoretical errors compared to the 
(Poisson distributed) statistical errors on the model for 100~fb$^{-1}$ of
integrated luminosity.

\begin{figure}
\begin{center}
\scalebox{0.48}{\rotatebox{-90}{\includegraphics[width=\textwidth]{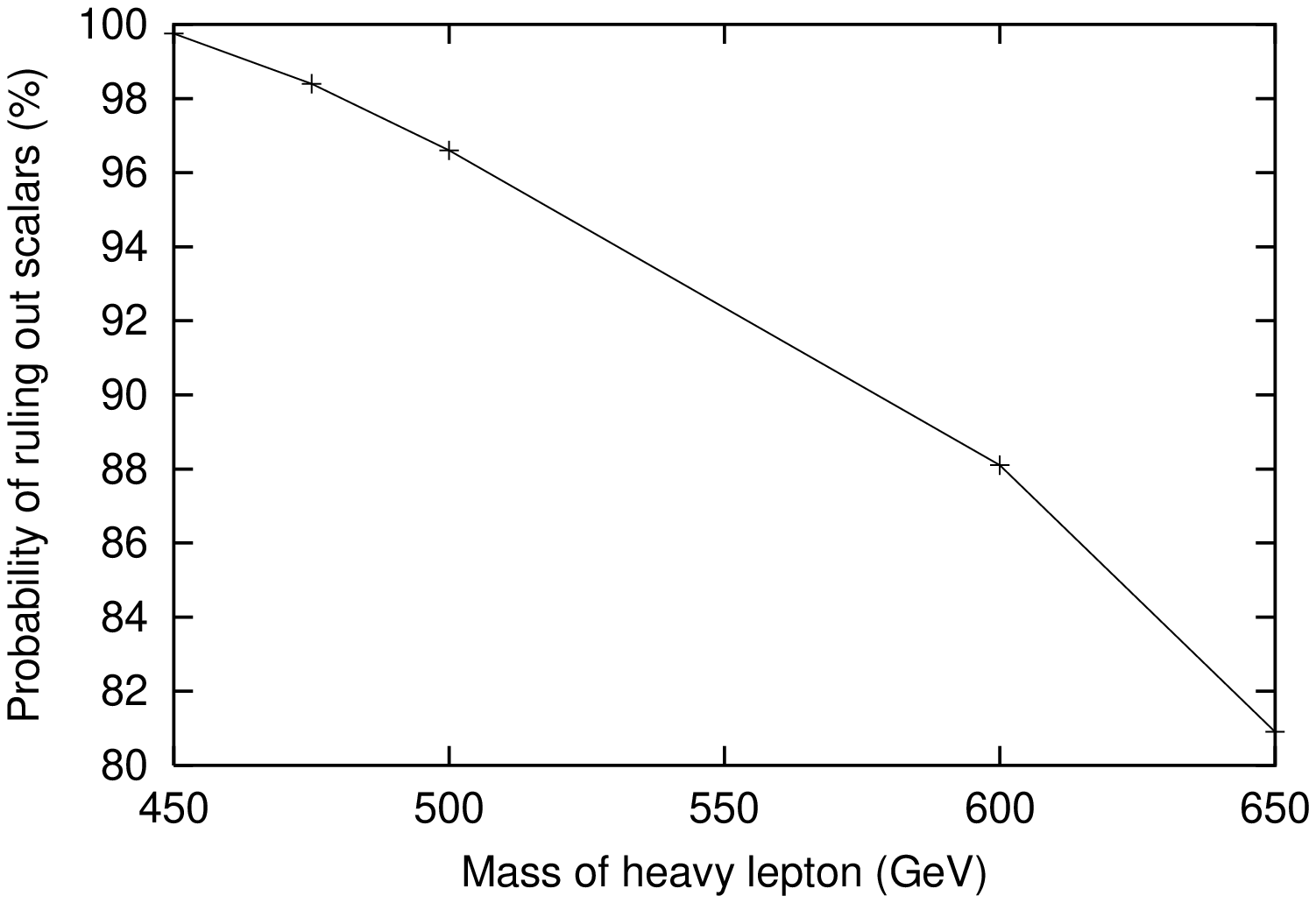}}}
\capbox{Probability of ruling out the scalar hypothesis}{Probability at which it will be possible to rule out the scalar
hypothesis by studying the angular distribution of the detected leptons.
\label{prob}}
\end{center}
\end{figure}

Figure~\ref{prob} shows that it will be possible to rule out scalar leptons 
at a confidence level of better than 90\% up to a mass of 580~GeV\@.  The 
angular distribution does not change significantly over the mass range of this
plot---the decreasing probability of ruling out the scalar hypothesis as the
mass increases is mainly because of the decreasing statistics.

\enlargethispage{-\baselineskip}
An alternative approach to ruling out the scalar hypothesis is to try to measure the forward-backward asymmetry of the lepton pair.  To do this it is necessary to reliably discriminate between the quark and anti-quark directions; this is usually based (e.g.\ in \cite{Dittmar:1997my}) on the assumption that the quark is a valence quark (with relatively high momentum) whereas the anti-quark is a sea quark (with lower momentum).  A method based on the boost of the lepton pair was investigated for 500~GeV leptons and although the discriminator was found to be reliable for large boosts, the original asymmetry ($\sim 0.3$) is not large enough to improve on the distinction discussed above.  Cuts which remove events where the lepton pair has only a small boost (and hence the quark direction is more uncertain) provided only a modest improvement because the increase in measured asymmetry is accompanied by a significant decrease in statistics.

\section{Conclusions}
\label{conchl}

This work was motivated by some of the new models of physics beyond the 
Standard Model in which the gauge couplings unify at an intermediate energy
scale ($\sim 10^{11}$~GeV)\@.  These intermediate scale models can include
additional leptons added to the Minimal Supersymmetric Standard Model and 
which are expected to have TeV scale masses.
\enlargethispage{-\baselineskip}

It is concluded that, assuming Standard Model couplings and a long enough 
lifetime, it should be possible to detect charged heavy leptons up to masses of 950~GeV with 100~fb$^{-1}$ of 
integrated luminosity.  Background from mis-identified muons should be 
negligible and can be reduced by the application of a 
$p_\text{T}^\text{min}$ cut at about 50~GeV\@.  It will be difficult to use 
the angular distribution of the produced particles to distinguish them from 
scalar leptons for masses above 580~GeV because the falling cross section 
limits the available statistics.   

The absence of a specialized trigger in the inner detector could reduce the 
discovery limit to 800~GeV and would reduce the statistics available for the 
angular distribution analysis.

No present cosmological or experimental limits were found to rule out 
additional exotic leptons at the masses considered in this work. 

\clearpage{\pagestyle{empty}\cleardoublepage}
\chapter{Introduction to Black Holes}
\chaptermark{Black Hole Introduction}
\label{bhintro}		
\let\om=\omega	
\let\si=\sigma
\let\Ga=\Gamma	

\section{Black holes in four dimensions}

The work in this thesis relates to extra-dimensional black holes, but it is useful to first summarize some of the properties and features of black holes in four dimensions, as well as the history of the theory describing them.  Black holes (or `frozen stars' as they were originally called) can only be correctly described by Einstein's 1915 theory of general relativity (GR).  Einstein himself was never completely convinced by the idea of black holes but, in 1916, Karl Schwarzschild applied GR to a static non-spinning massive object and obtained his well-known metric showing that an object of mass $M$ has a singularity at the Schwarzschild radius $r_\text{S}=2GM/c^2$.  Therefore an object with radius smaller than $r_\text{S}$ is a black hole with an event horizon at this radius.  The Schwarzschild radius can be calculated for any object but, since most objects do not have the density required to be black holes, it is usually much smaller than the object's physical size---for example, the Schwarzschild radius of the earth is easily calculated to be approximately 1~cm.
\enlargethispage{-\baselineskip}

It is entirely coincidental that the exact horizon radius can be derived using an `escape velocity' argument which would be expected to be only qualitative since it is entirely Newtonian and fails to take into account the relativistic effects.  However both Wheeler (1783) and Laplace (1796) independently used such an argument to postulate the existence of black holes as follows.  An object of mass $m$ propelled vertically upwards (with velocity $v$) from the surface of a mass $M$ with radius $r$ can only escape from the gravitational pull if
\begin{equation}
\frac{1}{2} m v^2 > \frac{GMm}{r}\,.
\end{equation}
If $r$ decreases, $v$ must be larger if the object is to escape.  The radius at which escape is only just possible even if travelling at the speed of light is found by setting $v=c$.  Hence we obtain
\begin{equation}
r=\frac{2GM}{c^2}\,,
\end{equation}
exactly as Schwarzschild derived using the later relativistic theory.

The above argument---that not even light can escape from a black hole---would seem to suggest that black holes can only grow.  However Hawking showed that the combination of general relativity and quantum mechanics predicts that black holes can evaporate by emitting Hawking radiation \cite{Hawking:1975sw}.  The mechanism for this is that a particle anti-particle pair is created at the event horizon; one has positive energy and escapes the black hole's gravitational attractions but the other negative energy one falls back into the black hole leading to a net decrease in its mass. 

Hawking found that the radiation spectrum is almost like that of a black body, and can be described by a characteristic Hawking temperature given by
\begin{equation}
T_\text{H}=\frac{\hbar c}{4\pi k r_\text{S}}\,,
\end{equation}
where $k$ is the Boltzmann constant.  Throughout the rest of this work natural units are used with the following definitions:
\begin{equation}
\hbar=c=k=1\,.
\end{equation}
This makes it possible to use energy units (usually GeV) to express distances ($1\mbox{ GeV}^{-1}=1.97\times 10^{-16}$~m), time intervals ($1\mbox{ GeV}^{-1}=6.58\times 10^{-25}$~s) and temperatures ($1\mbox{ GeV}=1.16\times 10^{13}$~K).

It is important to remember that, despite the wide acceptance of the theory, Hawking radiation has not been experimentally verified.\footnote{Compare this with the comments about the Higgs boson in \secref{thesm}.}  In a four dimensional world we would only be able to observe Hawking radiation by studying astrophysical black holes.  In fact although there are many astrophysical black hole candidates, it is difficult to uniquely identify them as black holes (for a review of some of the difficulties see \cite{Menou:1997qd} and the references therein).  The two main classes of black hole candidate are stellar black holes (with masses of order a few solar masses) in binary star systems, and super-massive black holes ($\sim 10^{12}$ solar masses) at galactic centres which may power quasars and other active galactic nuclei.  
\enlargethispage{-\baselineskip}

The masses of black hole candidates in binary systems can be estimated by studying the Doppler shift of emission lines from the secondary star and, since neutron stars have a maximum mass above which they will collapse to a black hole, this provides indirect evidence for the existence of astrophysical black holes.  Additionally, there seems to be confirmation that the primary stars are black holes from the emission spectra of Doppler-shifted X-rays.  These come from metals (e.g.\ iron) which are stripped off the outer layer of the secondary star and are then ionized due to the high temperatures of the accretion disc around the black hole candidate.

Observation of intense emission from super-massive black hole candidates is consistent with matter accretion by relatively small objects and hence also supports the black hole hypothesis.  The most significant difference between black holes and other massive compact objects is the presence of an event horizon.  Accretion models can provide evidence for both stellar and super-massive black holes by predicting how much energy falling on the central object from the accretion disc is `lost' (as for a black hole event horizon) and how much is re-radiated and observed (as in the case of an ordinary compact object).

Observing astrophysical Hawking radiation would absolutely verify the existence of black holes, but unfortunately the huge masses of these objects make this impossible.  The large masses mean that even the highest black hole temperatures would be $\sim$~nK---lower than the lowest temperatures achieved in the laboratory ($\sim \mu$K) and much lower than the temperature of the cosmic microwave background radiation (2.7~K).  This means that the observation of astronomical Hawking radiation is impossible: it would take $\sim 10^{14}$~years for a single photon from the closest black hole candidate to hit the earth and these photons would be swamped by the X-ray emission from matter being accreted into the disc surrounding the black hole.  This is one of the reasons why the idea of observing Hawking radiation from miniature black holes produced in particle colliders is so exciting.


In spite of the difficulties in studying black holes there is a huge body of work in the mathematical literature considering their properties including the emitted radiation, the singularities, and the effect of charge and angular momentum.  A full review of this literature is not attempted here, but instead a few points particularly relevant for work on extra-dimensional black holes will be highlighted.

\subsection{Grey-body factors}

Black holes are often spoken about as if the Hawking radiation is exactly like that of a black body.  However this is not the case---Hawking's original work showed that the energy spectrum is in fact `grey-body'.  The spectrum is modified from that of a perfect black body by an energy-dependent grey-body factor; these factors are also spin-dependent and the main effect in four dimensions is to suppress the low-energy part of the spectrum (see Figure 1 of \cite{MacGibbon:1990zk}).

Perhaps counter-intuitively the grey-body factors for particle emission from black holes are equal to the absorption cross sections for the same particles incident on the black hole.  However it is this property which ensures that the grey-body factors do not destroy the thermal nature of the black hole; equilibrium with a heat bath at the same temperature is still possible.  Of course this equilibrium is unstable---black holes have a negative heat capacity so if the the temperature of the black hole is slightly below (above) that of the heat bath, there will be a net absorption (emission) of energy so the black hole mass will increase (decrease) with a corresponding further decrease (increase) in Hawking temperature.

A substantial part of the work in this thesis (see Chapter~\ref{greybody}) describes the numerical calculation of extra-dimensional grey-body factors.

\subsection{Super-radiance}
\enlargethispage{-2\baselineskip}

As mentioned above, the grey-body factor for a particular particle can be calculated from the absorption cross section of that particle incident on the black hole.  However in the case of rotating black holes it was found that for incident bosons with low energies it is possible for the absorption cross section to be negative \cite{Zeldovich:1971,Misner:1972kx,Press:1972}.  This is equivalent to the reflection coefficient being greater than unity which means that energy incident on the black hole can be amplified.  This amplification was found to be greatest when the black hole was maximally rotating, and for a particle of spin $s$ was found to occur in the mode with angular momentum quantum numbers $\ell=m=\max(s,1)$.  Perhaps most significant is the increase in magnitude of the super-radiant effect as $s$ increases; peak amplifications quoted in \cite{Press:1974} are 0.3\% for scalars, 4.4\% for gauge bosons but 138\% for gravitons.  In theory, the phenomenon of super-radiance makes it possible to construct a `black hole bomb' \cite{Press:1972} by surrounding a rotating black hole with a spherical reflector; the trapped electromagnetic energy would grow exponentially and eventually the mirror would explode.

\subsection{Back reaction}

The calculation of the Hawking radiation emitted by a black hole relies on the assumption that the back reaction of the black hole metric can be ignored.  Essentially this is a requirement that the black hole mass $M_\text{BH}$ (and equivalently the Hawking temperature $T_\text{H}$) can be considered constant during the emission of each particle.  For this to be a good approximation requires that the energy of the emitted particle satisfies $\omega \ll M_\text{BH}$.\footnote{The spectrum for the emission of higher energy particles (with $\omega \sim M_\text{BH}$) must clearly be modified from Hawking's grey-body spectrum if for no other reason than that $\omega > M_\text{BH}/2$ is kinematically forbidden.}  This will be true for the majority of particle emissions provided that $T_\text{H} \ll M_\text{BH}$ (equivalent to the requirement that $M_\text{BH} \gg M_{P}$ or that $S_\text{BH} \gg 1$ \cite{Preskill:1991tb}, where $M_{P}$ and $S_\text{BH}$ are the Planck mass and black hole entropy respectively).  Regardless of the initial black hole mass, the radiation will no longer be described by Hawking's formula in the later stages of the decay.

\subsection{Black hole charges}

The `no-hair' theorem states that the only meaningful characteristics of black holes are the charges related to continuous local symmetries.  This means that we expect the electric charge of a black hole to be a meaningful quantity, but the same is not true for baryon and lepton number, for example.  Black hole production and decay cannot violate Abelian gauge symmetries so electric charge must be conserved.  Coloured black holes have to be treated slightly differently to electrically charged holes because of the non-Abelian nature of the gauge symmetry (of course colour charges are irrelevant for very massive 4D black holes, but they will be more significant for the extra-dimensional holes to be introduced in \secref{bhined}).  Although coloured black hole solutions can be found \cite{Kanti:1997gs} the charge can never be observed by an observer at infinity, and so is not considered to characterize the black hole.

The Hawking emission process occurs just outside the event horizon where the metric is described only by the charge, mass and angular momentum of the black hole.  Hence the emission from, and decay of, the black hole cannot be dependent on any other properties of the black hole (or the matter from which it was originally made up).  This is equivalent to saying that these properties (e.g.\ baryon number) are meaningless and will be violated in the decay.

\subsection{Information paradox}
\label{inpar}
\enlargethispage{-\baselineskip}

There are many unanswered questions regarding the black hole information loss paradox, and these are intimately related to how the black hole decays, particularly in its final stages.  The essence of the problem is that since a black hole evaporates to a mixture of thermal radiation (according to Hawking) any information which has fallen into the black hole appears to be lost.  Since mass, angular momentum and charge are the only observable characteristics of a black hole, any other information carried by matter which becomes trapped would appear to be lost.

Many of the suggested solutions to this problem relate to what happens once the black hole mass becomes comparable with the Planck mass.  Hawking's semi-classical calculations are no longer valid but without a quantum theory of gravity it is unclear what will happen to the remnant.  Some argue that information is not lost but might be stored in a stable remnant of the black hole.  A more popular idea is that information actually comes out of the black hole as non-thermal correlations within the Hawking radiation, whilst other theories accept and incorporate the loss of quantum information (for reviews of all these issues and possible resolutions, see \cite{Preskill:1992tc,Page:1993up,Stephens:1994an,Strominger:1994tn,Banks:1995ph}).

\section{Black holes in extra dimensions}
\label{bhined}

The extra-dimensional models of Arkani-Hamed, Dimopoulos and Dvali (ADD) \cite{Arkani-Hamed:1998rs,Arkani-Hamed:1998nn,Antoniadis:1998ig} and
Randall and Sundrum (RS) \cite{Randall:1999ee,Randall:1999vf} which were introduced in Chapter~\ref{introch} were motivated by the desire to explain the hierarchy problem---that is, the sixteen orders of magnitude difference between the electroweak energy scale and the Planck scale.


In the standard version of those works the Standard Model fields are
localized on a 3-brane, which plays the role of our 4-dimensional world,
while gravity can propagate both on the brane and in the bulk---the
space-time transverse to the brane. 
In theories with large extra dimensions the traditional reduced Planck scale, 
$\hat{M}_{(4)}\sim 10^{18}$ GeV, is only an effective energy scale derived
from the fundamental higher-dimensional one, $M_{\text{P}(4+n)}$, through the
relation
\begin{equation}
\hat{M}_{(4)}^2 \sim  M_{\text{P}(4+n)}^{n+2}\,R^n.
\label{MPl}
\end{equation}
The above relation involves the volume of the extra dimensions, $V \sim R^n$,
under the assumption that $R$ is the common size of all $n$ extra compact
dimensions. Therefore, if the volume of the internal space is large (i.e.\ if 
$R \gg \ell_\text{P}$, where $\ell_\text{P} =10^{-33}$ cm is the Planck length) 
then $M_{\text{P}(4+n)}$ can be substantially lower than $\hat{M}_{(4)}$.  Only if $M_{\text{P}(4+n)}$ can be as low as $\sim$~TeV does such a model provide a solution to the hierarchy problem (and permit study at the next generation of colliders).

In the regime $r\ll R$, the extra dimensions `open up' and gravity becomes
strong.  Hence, Newton's law for the gravitational interactions in this
regime is modified, with the gravitational potential assuming a $1/r^{n+1}$
dependence on the radial separation between two massive particles.   As a result there are limits on the fundamental scale from short-scale gravity experiments, but the most stringent constraints (at least for $n=2$ and 3) were found in \secref{edcon} to be cosmological, specifically from observations of neutron stars.  Although there are many uncertainties these tend to exclude even the $n=3$ case, while allowing models with $M_{\text{P}(4+n)} \sim 1$~TeV for $n \geq 4$.

If extra dimensions with $R \gg \ell_\text{P}$  exist, then black holes with
a horizon radius $r_\text{h}$ smaller than the size of the extra dimensions $R$
are higher-dimensional objects centred on the brane and extending
along the extra dimensions. It is these black holes which are the subject of the remaining chapters of this thesis.  It has been shown that extra-dimensional black holes have modified properties, e.g.\ they are larger and colder than four-dimensional black holes of the same mass \cite{Argyres:1998qn}.
One striking consequence of the theories with large extra dimensions is that
the lowering of the fundamental gravity scale allows for the production of
such miniature black holes during scattering processes
with centre-of-mass energy $\sqrt{\hat{s}} \gg M_{\text{P}(4+n)}$ 
\cite{Banks:1999gd,Giddings:2001bu,Dimopoulos:2001hw,Voloshin:2001vs,Voloshin:2001fe,Giddings:2001ih,Dimopoulos:2001qe,Giudice:2001ce}. 

There has already been a significant amount of work on production of black holes at future particle colliders \cite{Hossenfelder:2001dn,Kim:2001pg,Cheung:2001ue,Bleicher:2001kh,Casadio:2001wh,Park:2001xc,Landsberg:2001sj,Ahn:2002mj,Rizzo:2002kb,Solodukhin:2002ui,Cardoso:2002ay,Cheung:2002aq,Konoplya:2002zu,Uehara:2002gv,Guedens:2002km,Kotwal:2002wg,Frolov:2002as,Chamblin:2002ad,Han:2002yy,Frolov:2002gf,Anchordoqui:2002cp,Frolov:2002xf,Cardoso:2002jr,Cardoso:2002pa,Frolov:2003en,Chamblin:2003wg,Mocioiu:2003gi,Konoplya:2003ii,Casadio:2003vk,Vasilenko:2003ak,Cavaglia:2003qk,Konoplya:2003dd} and that is also the focus of the work in this thesis.  However, as mentioned in \secref{crlim}, a low fundamental energy scale would also make it possible for miniature black holes to be created in the earth's atmosphere \cite{Goyal:2000ma,Feng:2001ib,Anchordoqui:2001ei,Emparan:2001kf,Anchordoqui:2001cg,Uehara:2001yk,Alvarez-Muniz:2002ga,Ringwald:2001vk,Kowalski:2002gb,Kazanas:2001ep,Jain:2002kf,Ringwald:2002if,Ahn:2003qn,Nicolaidis:2003hi,Anchordoqui:2003jr,Mironov:2003jw}.  For the highest energy cosmic neutrinos ($>10^4$~TeV) the black hole production cross section from neutrino-nucleon interactions would be larger than the Standard Model interaction rate.  There are several different techniques involved in detecting cosmic ray particle showers (for a review see \cite{Anchordoqui:2002hs}) but showers from black hole production and decay are expected to be characterized by particular particle content and multiplicity, and a favoured quasi-horizontal direction.  Before the LHC becomes operational cosmic ray experiments should be able to improve their present limits \cite{Anchordoqui:2001cg} on the Planck scale or, more optimistically, observe black hole production.  Recently it has been suggested \cite{Mironov:2003jw} that black hole production could be one explanation for the hadron-rich `Centauro' cosmic ray events.  The theoretical results derived in Chapter~\ref{greybody} should also be useful for the study of black hole decay in a cosmic ray context.

\subsection{Black hole production}
\label{bhprod}

There has been much discussion in the literature (e.g.\ \cite{Giddings:2001bu,Voloshin:2001vs,Giddings:2001ih,Voloshin:2001fe,Eardley:2002re}) about what the cross section for black hole production is, but the consensus opinion that the geometrical $\sigma\sim\pi r_\text{S}^2$ is valid seems to have been confirmed by the numerical work in \cite{Yoshino:2002tx}.  More formally we can write
\begin{equation}
\label{fndef}
\sigma=F_n\pi r_\text{S}^2\,.
\end{equation}
In order to obtain the coefficient $F_n$ it is necessary to consider the black hole production process in detail.  Various analytic techniques are available but for collisions with non-zero impact parameter in more than four dimensions, the collision must be modelled numerically \cite{Yoshino:2002tx}.  This allows the determination of both $F_n$ and $b_\text{max}$, the maximum impact parameter for which black hole formation will occur.  It is found that $F_n$ is of order 1 (for example $F_0=0.647$ and $F_7=1.883$) and that $b_\text{max}/r_\text{S}$ ranges from 0.804 ($n=0$) to 1.37 ($n=7$).

Although a full numerical relativistic approach was necessary to obtain these values, a simple model of the collision can give good agreement.  In four dimensions Thorne's hoop conjecture \cite{Thorne:1972ji} predicts the creation of a black hole in the case where any two partons from the colliding particles pass within the horizon radius corresponding to their centre-of-mass energy.  In the extra-dimensional case there is probably an equivalent `volume conjecture' \cite{Ida:2002hg,Yoshino:2002br}, but a similar prediction is expected.

Here we follow the approach in \cite{Ida:2002ez} and assume that a black hole will form if $b < 2r_\text{h}(M_\text{BH},J)$, where $r_\text{h}$ is the horizon radius for a black hole with mass $M_{BH}$ and angular momentum $J$.  This differs from that in \cite{Anchordoqui:2001cg} which assumes $b < r_\text{h}(M_\text{BH},J)$, but provides less good agreement with the numerical results.\footnote{Reference \cite{Anchordoqui:2001cg} was published before the numerical results in \cite{Yoshino:2002tx}, whereas \cite{Ida:2002ez} was published afterwards.}   The angular momentum $J$ can be usefully related to a dimensionless rotation parameter $a_*$ by
\begin{equation}
\label{astardef}
a_*=\frac{(n+2)J}{2 r_\text{h} M_\text{BH}}\,.
\end{equation}
There will be more discussion of this in \secref{rotbh}, but $r_\text{h}$ is found to be related to $r_\text{S}$ by
\begin{equation}
r_\text{h}=\frac{r_\text{S}}{(1+a_*^2)^{\frac{1}{n+1}}}\,.
\end{equation}
It is therefore possible to determine $b_\text{max}$ by setting $b=2r_\text{h}$ and assuming that $J=bM_\text{BH}/2$.  This gives
\begin{equation}
b_\text{max}=2\left[1+\left(\frac{n+2}{2}\right)^2\right]^{-\frac{1}{n+1}}r_\text{S}\,,
\end{equation}
which is in excellent agreement with \cite{Yoshino:2002tx} for $n\ge 1$.  It is then possible to estimate $F_n$ (assuming $\sigma=\pi b_\text{max}^2$) and again the agreement is very good.

Above we have assumed that the mass of the black hole will be equal to the centre-of-mass energy of the colliding partons.  It is possible to put a lower bound (dependent on the ratio $b/b_\text{max}$) on the fraction of the energy which is trapped by the black hole.  Lower bounds were derived for head-on ($b=0$) collisions in \cite{penrose,D'Eath:1992hb,D'Eath:1992hd,D'Eath:1992qu,D'Eath:1993gr} (for four dimensions) and in \cite{Eardley:2002re} (for the extra-dimensional case).  However the value when the impact factor is close to its maximum is most important, since this is expected to give the largest contribution to the cross section.  The 4D case was studied in \cite{Eardley:2002re} and then the extra-dimensional case in \cite{Yoshino:2002tx} showing that the bound on the fraction of trapped energy can be as low as $\sim 0.1$ for larger values of $n$.  A significant amount of angular momentum could also be lost at this stage.  More work is required in this area since if this bound is saturated the black hole decay discussed in this work would be significantly affected \cite{Anchordoqui:2003ug} since only if the black hole mass is larger than a few times the fundamental Planck mass can these objects be treated semi-classically.  

At this point it is worth stressing the difference between black holes as a signature for extra dimensions and the graviton emission signatures briefly mentioned in \secref{led} and \secref{collim}.  The graviton emission signatures are necessarily in the `cis-Planckian' or `sub-Planckian' regime ($\sqrt{\hat{s}} \ll M_{\text{P}(4+n)}$) where the low-energy effective theory is valid and graviton emission can be reliably calculated with a perturbative expansion.  This means that if $M_{\text{P}(4+n)}$ is too \emph{low}, the ability of the LHC to extract information from these signatures will be severely limited.  In contrast, black holes would be produced in the `trans-Planckian' or `super-Planckian' regime ($\sqrt{\hat{s}} \gg M_{\text{P}(4+n)}$) where a semi-classical approach is valid.  If $M_{\text{P}(4+n)}$ is too \emph{high} it will not be possible to investigate this regime at the LHC, although it is hard to make a good estimate of what the minimum value of $\sqrt{\hat{s}}/M_{\text{P}(4+n)}$ should be if we are to be confident of examining trans-Planckian physics.  Black holes are not the only feature of such a regime---for colliding partons with larger impact parameter (i.e.\ large compared to the horizon radius corresponding to their centre-of-mass energy) there will instead be gravitational elastic scattering with small momentum transfer.  This can be calculated using the eikonal approximation and would result in jet-jet production close to the beam and with high centre-of-mass energy \cite{Giudice:2001ce}.  There have been some preliminary experimental studies of this by the ATLAS collaboration which suggest that it might be possible to extract information on $n$ and $M_{\text{P}(4+n)}$ from these signatures \cite{azuelos}.

Another possibility in the trans-Planckian regime is the production of `black' $p$-branes ($p$-dimensional, spatially extended solutions of gravitational theories in extra dimensions).  The production and decay of $p$-branes would exhibit some similar features to black holes (which can be considered as 0-branes) but the details of the decay are more model-dependent.  Although the cross section for the production of $p$-branes can dominate that for black holes this is only found to be the case in compactifications with a mixture of small and large extra dimensions \cite{Ahn:2002mj,Ahn:2002zn,Cheung:2002aq,Cheung:2002uq}.  Qualitatively the explanation for this is that symmetric compactifications (with all extra dimensions of a similar size) favour the production of 0-branes (spherically-symmetric black holes) whereas non-spherically-symmetric $p$-branes are more likely to be produced in models with an asymmetric compactification, e.g.\ \cite{Lykken:1999ms}.  There will be no further discussion of $p$-brane production in this work, although it is an area in need of more study since string theory would seem to favour the existence of more than one compactification scale.

The trans-Planckian and cis-Planckian energy regimes in which it might be possible to extract parameters in a relatively model-independent way are separated by the Planckian regime in which quantum gravity effects become important.  A theory of quantum gravity is required to predict cross sections and experimental signals in this regime.  One possibility is string theory; this would introduce another energy scale, the string scale $M_\text{s}$, which could naturally be slightly smaller than $M_{\text{P}(4+n)}$ (see \secref{astcon}).  Therefore, in the Planckian regime it might be possible for the colliding partons to be excited into string modes or even very excited `string balls'.


\subsection{Black hole decay}

Once produced, these miniature black holes are expected to decay almost
instantaneously (typical lifetimes are $\sim 10^{-26}$~s).\footnote{This is only true in the ADD model; in the RS model the black holes can be stable on collider time scales\cite{Casadio:2001wh}.} According to
refs. \cite{Giddings:2001bu,Giddings:2001ih}, the produced black holes will go through a number
of phases before completely evaporating.

\begin{itemize}

\item
{\it Balding phase}\,: The black hole emits mainly gravitational radiation
and sheds the `hair' inherited from the original particles, and the asymmetry
due to the violent production process.\footnote{This phase can also be considered as a production phase since it is predominantly the emission of the `junk energy' which is not trapped by the event horizon as the black hole forms.}

\item
{\it Spin-down phase}\,: The typically non-zero impact parameter
of the colliding partons leads to black holes with some angular momentum
about an axis perpendicular to the plane. During this phase, the black
hole loses its angular momentum through the emission of Hawking radiation
\cite{Hawking:1975sw} and, possibly, through super-radiance.

\item
{\it Schwarzschild phase}\,: A spherically-symmetric black hole loses
energy due to the emission of Hawking radiation. This results in the gradual
decrease of its mass and the increase of its temperature.

\item
{\it Planck phase}\,: The mass and/or the Hawking temperature approach
the Planck scale.  A theory of quantum gravity is necessary to study this
phase in detail but it is suggested that the black hole will decay to a few quanta with Planck-scale energies \cite{Giddings:2001bu}.

\end{itemize}

As in the 4-dimensional case \cite{Page:1976df}, it is reasonable to expect
that the Schwarzschild
phase in the life of a small higher-dimensional black hole will be the
longest one, and will account for the greatest proportion of the mass loss 
through the emission of Hawking radiation.  

\subsection{Experimental signatures}
\label{expsig}

The phases of black hole decay described in the previous section combine to produce distinctive experimental signatures which make it unlikely that black hole events would be mistaken for many other processes.  Some of these features are outlined below:

\begin{itemize}
\item{There is a very large total cross section, particularly at high centre-of-mass energies (because the parton-level cross section grows with energy);}
\item{The ratio of hadronic to leptonic activity is roughly 5:1 and as a result the amount of energy visible in the detector is large (the relative emission into gravitons is also relatively small);}
\item{Most events have a relatively high multiplicity with many hard jets and hard prompt leptons;}
\item{As the total event transverse energy increases the average multiplicity increases and the average energy of each primary emitted parton decreases---a manifestation of the infra-red ultra-violet connection of gravity;}
\item{Events have a high sphericity since most black holes are produced almost at rest;}
\item{Hard perturbative scattering processes are suppressed by the non-perturbative black hole production at high energies.}
\end{itemize}

It has been suggested \cite{Anchordoqui:2002cp} that the high temperature of the black hole ($\gg \Lambda_\text{QCD}$, the scale above which physics becomes perturbative) might cause a quark-gluon `chromosphere' to form around the decaying black hole; the Hawking radiation would then thermalize leading to a suppression of the number of hard hadronic jets.  However this would require $\sim 10$ quarks to be emitted as primary partons \cite{Anchordoqui:2002cp,Anchordoqui:2003ug} which is unlikely to be a problem for many black hole events at the LHC because of the lower entropy of higher-dimensional black holes (see the multiplicity discussion in \secref{totfandp}).  For future higher energy particle colliders, the black hole signatures are more likely to be modified although there would still be hard hadronic jets from W, Z and Higgs bosons (as well as $\tau$ leptons) which would penetrate the chromosphere before decaying \cite{Landsberg:2002sa}.

Experimental cuts proposed to isolate black hole events with negligible SM background include requiring the total energy deposited in the calorimeter to be $\ge 1$~TeV and there to be $\ge 4$ jets with energies above 100~GeV (including a high-energy lepton or photon which is useful for triggering) \cite{Dimopoulos:2001hw}. The theoretical work in Chapter~\ref{greybody} and the event generator described in Chapter~\ref{generator} will eventually make it possible for more quantitative predictions to be made about typical black hole events, and for experimental cuts to be tailored accordingly.

\subsection{Proton decay}
\label{protonbh}

Proton decay is always a potential problem for models which describe physics beyond the Standard Model.  The present limits on proton lifetime require that, at a 90\% confidence level, the partial lifetime for the decay $\text{p}\rightarrow \text{e}^+ \pi^0$ is greater than $1.6 \times 10^{33}$~years \cite{Shiozawa:1998si}.  In general introducing new physics at a scale of $\Lambda$ can allow the decay of the proton (mass $m_\text{p}$) with a width $\Gamma \sim (m_\text{p}^5/\Lambda^4)$ unless the operators in the Lagrangian which would allow this are explicitly forbidden or at least heavily suppressed.

To stop present experimental limits on proton decay being violated in models of new physics it is usually necessary to artificially impose symmetries which forbid the problematic higher dimensions operators (as in the Standard Model this does not completely forbid proton decay since $B$ and $L$ symmetries can be broken by radiative effects).

In extra dimension models there is new physics at the fundamental Planck scale (so $\Lambda\sim M_{\text{P}(4+n)} \sim 1$~TeV) and the possible production of black holes in extra dimension models means that gravity itself could be responsible for proton decay.  This could occur through the production of virtual black holes in processes like $\text{q} + \text{q} \rightarrow \mbox{Black Hole} \rightarrow \bar{\text{q}} + \ell + \ldots$, where the ellipsis represents possible additional particles (e.g.\ gravitons, gluons, photons or neutrinos).  The consequence of the `no hair' theorem is that the only requirement for such processes is the conservation of charge, energy and angular momentum.

The usual method of suppressing proton decay in models of new physics is by imposing symmetries on the high-energy theory---for example, R-parity in supersymmetry.  However in the extra-dimensional case the high-energy theory is an unknown theory of quantum gravity so it is not clear whether it is possible to impose such a symmetry to suppress but not totally forbid proton decay.  The safest approach is to explicitly forbid proton decay by imposing a symmetry on the low-energy effective theory but again this is not trivial.  A global symmetry for baryon number does not seem to be viable since the black hole decay is not guaranteed to conserve quantum numbers associated with global symmetries.   This leaves discrete gauge symmetries involving combinations of $B$ and $L$ as apparently the only possibilities, since it is argued in \cite{Krauss:1989zc} that quantum numbers associated with discrete gauge symmetries as well as continuous gauge symmetries will be conserved in the decay of black holes.  Care must be taken to ensure that any such symmetries introduced are anomaly free, and that they still allow the generation of baryon asymmetry and neutrino masses.
\enlargethispage{-\baselineskip}

Of course the above argument has made the usual assumption that quantum numbers associated with global symmetries can be violated, i.e.\ information is lost in the black hole decay process.  Some would argue \cite{Kobakhidze:2001yk} that the solution of the Hawking information loss paradox discussed in \secref{inpar} might involve the conservation of global charges like baryon number in which case many of the above problems would be avoided.

Some of the $\mbox{TeV}^{-1}$ extra dimension models introduced in \secref{tev-1} provide another way of avoiding proton decay issues.  The models involve physically separating the quarks and leptons \cite{Arkani-Hamed:1999za,Arkani-Hamed:1999dc} which exponentially suppresses their wave-function overlap and hence also the possibility of proton decay.   To obtain normal SM physics the Higgs and gauge bosons must propagate in the bulk (or the `thick' brane) between the lepton and quark fields.  This requires the mass of the lowest KK modes for the bosons to be high enough to avoid the experimental constraints from the precision electroweak measurements, and translates to a lower limit on $M_{\text{P}(4+n)}$ (at least of order 20~TeV \cite{Han:2002yy}).  As a result such models are disfavoured phenomenologically because the high value of $M_{\text{P}(4+n)}$ rules out black hole production at the next generation of colliders. 

\section{Basic formulae and assumptions}
\label{basic}

Although extra dimension models have only recently become popular among particle physicists, Myers and Perry \cite{Myers:1986un} had previously worked on black holes in higher-dimensional space-times.  They considered the form of the gravitational background around an uncharged $(4+n)$-dimensional black hole.  In the non-rotating (Schwarzschild) case the line-element is found to be given by
\begin{equation}
ds^2=- h(r)\,dt^2 + h(r)^{-1}\,dr^2 + r^2\,d \Omega^2_{2+n}\,,
\label{metric-D}
\end{equation}
where 
\begin{equation}
h(r) = 1-\biggl(\frac{r_\text{S}}{r}\biggr)^{n+1},
\label{h-fun}
\end{equation}
and
\begin{eqnarray}
d\Omega_{2+n}^2=d\theta^2_{n+1} + \sin^2\theta_{n+1} \,\biggl(d\theta_n^2 +
\sin^2\theta_n\,\Bigl(\,... + \sin^2\theta_2\,(d\theta_1^2 + \sin^2 \theta_1
\,d\varphi^2)\,...\,\Bigr)\biggr).
\end{eqnarray}
In the above, $0 <\varphi < 2 \pi$ and $0< \theta_i < \pi$, for 
$i=1, ..., n+1$.  As shown in \cite{Myers:1986un}, the extension of the usual 4-dimensional Schwarzschild calculation gives the following horizon radius: 
\begin{equation}
r_\text{h} = r_\text{S} =\frac{1}{\sqrt{\pi}M_{P(4+n)}}\left(\frac{M_\text{BH}}{M_{P(4+n)}}\right)^
{\frac{1}{n+1}}\left(\frac{8\Gamma\left(\frac{n+3}{2}\right)}{n+2}\right)
^{\frac{1}{n+1}},
\end{equation}
where $M_{P(4+n)}$ is the fundamental $(4+n)$-dimensional Planck scale in convention `d' of Table~\ref{consum}. 

The black holes being considered in this work are assumed to have horizon radii satisfying the relation $\ell_\text{P} \ll r_\text{S} \ll R$.
The former inequality guarantees that quantum corrections are not important
in the calculations, while the latter is necessary for the black holes to
be considered as higher-dimensional objects (i.e.\ the curvature of the extra dimensions can be ignored on the scale of the black hole).  The tension of the brane on which the black hole is centred is assumed to be much smaller than the black hole mass which means it can be neglected in this analysis, and a zero bulk cosmological constant is also assumed.  This is essentially equivalent to the assumption of the ADD scenario although much of what follows can also be applied to the RS models provided the bulk cosmological constant is small; this corresponds to a small warp factor and so the space-time can be considered as almost spherically symmetric.  If this is not the case, singularity problems emerge and the black hole solutions lose their spherical symmetry (the horizon becomes flattened on the brane in a pancake shape \cite{Giddings:2000mu}).

A black hole of a particular horizon radius is characterized by a
Hawking temperature related by
\begin{equation}
T_\text{H}=\frac{(n+1)}{4\pi\,r_\text{S}}\,,
\end{equation}
and the black hole entropy is given by
\begin{equation}
S_\text{BH}=\frac{4\pi r_\text{S} M_\text{BH}}{n+2}=\left(\frac{n+1}{n+2}\right)\frac{M_\text{BH}}{T_\text{H}}\,.
\label{bhent}
\end{equation}

The emitted Hawking radiation is {\it almost} like that of a black body at this temperature.  In fact the flux spectrum, i.e.\ the number of particles emitted per unit time, is given by \cite{Hawking:1975sw}
\begin{equation}
\label{3flux}
 \frac{dN^{(s)}(\om)}{dt} = \sum_{\ell} \sigma^{(s)}_{\ell}(\om)\,
\frac{1}{\exp\left(\om/T_\text{H}\right) \mp 1} 
\,\frac{d^{n+3}p}{(2\pi)^{n+3}}\,,
\end{equation}
while the power spectrum, i.e.\ the energy emitted per unit time, is
\begin{equation}
\frac{dE^{(s)}(\om)}{dt} = \sum_{\ell} \sigma^{(s)}_{\ell}(\om)\,
\frac{\om}{\exp\left(\om/T_\text{H}\right) \mp 1}\,\frac{d^{n+3}p}{(2\pi)^{n+3}}\,.
\label{3power}
\end{equation}
In the above, $s$ is the spin of the degree of freedom being considered
and $\ell$ is the angular momentum quantum number. The spin statistics factor
in the denominator is $-1$ for bosons and $+1$ for fermions. For 
massless particles $|p|=\om$ and the phase-space integral reduces to
an integral over $\omega$. The term in
front, $\sigma^{(s)}_{\ell} (\om) $, is the grey-body factor which encodes valuable information about the structure of the surrounding space-time including its dimensionality. Chapter~\ref{greybody} will give more detail on how the grey-body factors are calculated.

As the decay progresses, the black hole mass decreases and the Hawking temperature rises.  The work performed here will allow some comments to be made on whether the usual `quasi-stationary' approach to the decay is valid; if such an approach is correct then the black hole has time to come into equilibrium at each new temperature before the next particle is emitted. 

An important point to stress is that eqs.~(\ref{3flux}) and (\ref{3power}) 
refer to individual degrees of freedom and not to elementary
particles, like electrons or quarks, which contain more than one polarization.
Combining the necessary degrees of freedom and their corresponding flux
or power spectra, the relative numbers of different elementary particles 
produced, and the energy they carry, can be easily computed.  Since the higher-dimensional black holes which might be produced at the LHC have relatively high Hawking temperatures ($\sim$100~GeV) in most cases it is possible for all elementary particles to be produced.  The numbers of degrees of freedom $d_s$ are summarized in Table~\ref{pprobs}.  These take account of the fact that for each massive gauge bosons one of the degrees of freedom comes from the Higgs mechanism.
The numbers of degrees of freedom will be important in calculating total flux and power emission in \secref{totfandp}, and for the construction of the black hole event generator (see Chapter~\ref{generator}).

\begin{table}
\begin{center}
\begin{tabular}{|l|c|c|c|}
\hline
Particle type & $d_0$ & $d_{1/2}$ & $d_1$  \\
\hline
Quarks & 0 & 72 & 0\\
Gluons & 0 & 0 & 16\\
Charged leptons & 0 & 12 & 0\\
Neutrinos\footnotemark & 0 & 6 & 0\\
Photon & 0 & 0 & 2\\
Z$^0$ & 1 & 0 & 2\\
W$^+$ and W$^-$ & 2 & 0 & 4\\
Higgs boson & 1 & 0 & 0\\
\hline
Total & 4 & 90 & 24\\
\hline
\end{tabular}
\capbox{Degrees of freedom for different particle types}{Degrees of freedom for different particle types emitted from a black hole.\label{pprobs}} 
\end{center}
\end{table}

\footnotetext{It is possible that $d_{1/2}=12$ should be used for the neutrinos since right-handed neutrinos, if they exist, will also be emitted by the black hole.}\stepcounter{footnote}
\enlargethispage{-\baselineskip}

In reference \cite{Emparan:2000rs} it was argued that the majority of energy during
the emission of Hawking radiation from a higher-dimensional black hole is
emitted into modes on the brane (i.e.\ Standard Model fermions and gauge bosons,
zero-mode gravitons and scalar fields). This argument was based on their
result that a single brane particle carries as much energy as the
whole Kaluza-Klein tower of massive excitations propagating in the bulk.  Essentially the reasoning is that the KK tower is simply the result of the projection onto four dimensions; from the extra-dimensional view-point there is only one graviton mode.  Since there are many brane modes for the Standard Model particles the majority of energy is expected to be emitted on the brane.

Chapter~\ref{greybody} of this thesis is concerned with the numerical calculation of the extra-dimensional grey-body factors.  These will be a key part of the state-of-the-art black hole event generator described in Chapter \ref{generator}.  The event generator simulates both the production and decay of small black holes at hadronic colliders and, by using the new results for the grey-body factors provides estimates for the spectra and relative numbers of the different types of elementary particles emitted.

\clearpage{\pagestyle{empty}\cleardoublepage}
\chapter{Grey-body Factors in $(4+n)$ Dimensions}
\chaptermark{Grey-body Factors}
\label{greybody}		
\let\om=\omega	
\let\si=\sigma
\let\Ga=\Gamma	

\section{Introduction}
\label{gbfintro}

The idea of grey-body factors, both for four-dimensional and extra-dimensional black holes, was introduced in Chapter~\ref{bhintro}.  Extra-dimensional black holes are of interest because, if scenarios like that of ADD \cite{Arkani-Hamed:1998rs,Arkani-Hamed:1998nn,Antoniadis:1998ig} are realized in nature, they could possibly be produced at high-energy colliders.  To understand the particle emission probabilities and spectra for these black holes it is necessary to calculate the extra-dimensional grey-body factors.

For ease of reference, equations~(\ref{3flux}) and (\ref{3power}) describing the flux and power spectra are reproduced below:
\begin{equation}
\label{4flux}
 \frac{dN^{(s)}(\om)}{dt} = \sum_{\ell} \sigma^{(s)}_{\ell}(\om)\,
\frac{1}{\exp\left(\om/T_\text{H}\right) \mp 1} 
\,\frac{d^{n+3}p}{(2\pi)^{n+3}}\,;
\end{equation}
and
\begin{equation}
\frac{dE^{(s)}(\om)}{dt} = \sum_{\ell} \sigma^{(s)}_{\ell}(\om)\,
\frac{\om}{\exp\left(\om/T_\text{H}\right) \mp 1}\,\frac{d^{n+3}p}{(2\pi)^{n+3}}\,.
\label{4power}
\end{equation}
$\sigma^{(s)}_{\ell} (\om) $ is the grey-body factor\footnote{The quantity $\sigma^{(s)}_{\ell} (\om) $ is alternatively called the absorption cross section. It is also common in the literature to refer to the absorption probability $|{\cal A}^{(s)}_\ell |^2$, related to $\sigma^{(s)}_{\ell} (\om) $ through eq.~(\ref{greydef}), as the grey-body factor.  The notation $\Gamma_\ell^{(s)}$ is sometimes used for $|{\cal A}^{(s)}_\ell|^2$.} which can be determined by solving the 
equation of motion of a particular degree of freedom in the gravitational 
background and computing the corresponding absorption coefficient
${\cal A}^{(s)}_\ell$.  The absorption coefficient is related to the grey-body factor by \cite{Gubser:1997yh} 
\begin{equation}
\si^{(s)}_\ell(\om) = \frac{2^{n}\pi^{(n+1)/2}\,\Ga[(n+1)/2]}{ n!\, \om^{n+2}}\,
\frac{(2\ell+n+1)\,(\ell+n)!}{\ell !}\, |{\cal A}^{(s)}_\ell|^2.
\label{greydef}
\end{equation}
 
As already mentioned, the grey-body factors modify the spectra of emitted particles from that of a perfect thermal black body \emph{even} in four dimensions \cite{Hawking:1975sw}.  For a 4-dimensional Schwarzschild black hole, geometric arguments show that, in the high-energy ($\omega\gg T_\text{H}$) regime, 
$\Sigma_{\ell}\,|{\cal A}_\ell^{(s)}|^2 \propto(\omega r_\text{h})^2$.  Therefore
the grey-body factor at high energies is independent of $\omega$ and the
spectrum is exactly like that of a black body for every particle species
\cite{misner,Sanchez:1978si,Sanchez:1978vz,Page:1976df}. The low-energy behaviour, on the other hand, is
strongly spin-dependent. A common feature for fields with spin $s=0$,
1/2 and 1 is that the grey-body factors reduce
the low-energy emission rate significantly below the geometrical optics value 
\cite{Page:1976df,MacGibbon:1990zk}.  The result is that both the power and flux spectra peak at higher energies than those for a black body at the same temperature.  As $s$ increases, the low-energy suppression increases and so the emission probability decreases.  These particle emission probabilities have been available in the literature for over twenty years \cite{Page:1976df,Sanchez:1978si,Sanchez:1978vz}.

These characteristics of the four-dimensional grey-body factors demonstrate why it is important to calculate the equivalent extra-dimensional factors.  Studying their dependence on the dimensionality of space-time is important if detected Hawking radiation emitted by small black holes is to be used to try to determine the value of $n$.  Both the energy spectra and the relative emissivities of scalars, fermions and gauge bosons will be potentially useful in this task.  The main motivation for this work was to obtain grey-body factors which could be used in the Monte Carlo event generator described in Chapter~\ref{generator}, thereby allowing the possible determination of $n$ to be studied in detail.

The procedure for calculating grey-body factors needs to be generalized to include emission from small higher-dimensional black holes with $r_\text{h} < R$;  there is an added complication in this case because black holes can emit radiation either on the brane or in the bulk.  The emission of particle modes on the brane is the most phenomenologically interesting effect since it involves Standard Model particles.  However gravitons will certainly be emitted into the bulk (since it is the propagation of gravity in the bulk which accounts for its apparent weakness and hence solves the hierarchy problem) and it is possible that a $(4+n)$-dimensional black hole will also emit bulk scalar modes.  These are potentially important because in some extra dimension models there are additional scalars propagating in the bulk; for example, supersymmetric models inspired by string theory involve graviton supermultiplets which can include bulk scalars (see e.g.\ \cite{Scherk:1979aj,Scherk:1979rh}).\footnote{Scalars propagating in the bulk would affect many of the constraints described in \secref{edcon}, although in model-dependent ways since the couplings to SM particles would not be universal (unlike those of the graviton).}  Numerical results for bulk and brane scalar modes are necessary to allow comparison of the total bulk and brane emissivities.  This would allow the usual heuristic argument for brane dominance (because there are many more SM brane degrees of freedom than there are
degrees of freedom in the bulk---see \secref{basic}) to be confirmed or disproved.

There have already been some analytic studies for extra-dimensional black holes \cite{Kanti:2002nr,Kanti:2002ge,Ida:2002ez} which have suggested that the grey-body factors can have a strong dependence on the number of extra dimensions.  However it is known that results from the power series expansions in these papers are only accurate in a very limited low-energy regime (especially in the case of scalar fields) and even the full analytic results are not necessarily reliable above the intermediate-energy regime.

The work described in this chapter focuses on the emission of energy from uncharged $(4+n)$-dimensional black holes; initially only Schwarzschild-like black holes are considered although there are also some results for rotating holes.  Section~\ref{branegbf} gives some details of a master equation describing the motion of scalars, fermions and gauge bosons in the 4-dimensional induced background---the basic steps of Kanti's calculation \cite{Harris:2003eg} are given in Appendix~\ref{appb}.  This is particularly useful in that it helps to resolve ambiguities present in similar equations which have previously appeared in the literature.  The brane grey-body factors and emission rates are also defined in \secref{branegbf}, and analytic and numerical methods for computing these quantities are discussed.  Exact numerical results for emission on the brane (for scalars, fermions and gauge bosons) are presented in \secref{numres}.  This allows a comparison with  the earlier analytic studies of the Schwarzschild phase as well as a discussion of the relative emissivities and their dependence on $n$.  A calculation of the total power and flux emitted (and hence the lifetime and number of emitted particles expected for extra-dimensional black holes) is also possible in this section.  The emission of bulk scalar modes is studied in \secref{embulk}, and the results for the relative bulk-to-brane emissivities are also presented in this section.  The analysis provides, for the first time, exact results for the relative bulk and brane emissivities in all energy regimes and for various values of $n$.  Finally section~\ref{rotbh} includes a discussion of the emission of scalars from rotating black holes; this allows some comments to be made on super-radiance for extra-dimensional black holes.  The conclusions of the work in this chapter are summarized in \secref{gbfconc}. 

\section{Grey-body factors for emission on the brane}
\label{branegbf}

This section and \secref{numres} focus on the emission of brane-localized modes, leaving the study of bulk emission and of relative bulk-to-brane emissivity until \secref{embulk}.

The brane-localized modes propagate in a 4-dimensional black hole background
which is the projection of the higher-dimensional one, given in eq.~(\ref{metric-D}), onto the brane. The induced metric tensor follows by fixing
the values of the extra angular co-ordinates ($\theta_i=\pi/2$ for $i \geq 2$)
and is found to have the form
\begin{equation}
ds^2=-h(r)\,dt^2+h(r)^{-1}dr^2+r^2\,(d\theta^2 + \sin^2\theta\,d \varphi^2)\,.
\label{non-rot}
\end{equation}

The grey-body factors are determined from the amplitudes of in-going and
out-going waves at infinity so the essential requirement is to solve the
equation of motion for a particle propagating in the above background. 
For this purpose, a generalized {\it master equation} was derived by Kanti (see Appendix~\ref{appb}) for a particle with arbitrary spin $s$; this equation is similar to the 4-dimensional one derived by Teukolsky \cite{Teukolsky:1973ha}.
For $s=$1/2 and 1, the equation of motion was
derived using the Newman-Penrose method \cite{Newman:1962qr,chandra}, while for $s=0$
the corresponding equation follows by the evaluation of the double covariant
derivative $g^{\mu\nu} D_\mu D_\nu$ acting on the scalar field. The derived
master equation is separable in each case and, by using the factorization
\begin{equation}
\label{sepsoln}
\Psi_s=e^{-i\omega t}\,e^{im\varphi}\,R_s(r)\,{}_sS^m_{\ell}(\theta)\,,
\end{equation}
we obtain the radial equation
\begin{equation}
\label{radial}
\Delta^{-s} \frac{d}{dr}\left(\Delta^{s+1}\,\frac{d R_s}{dr}\right)+
\left(\frac{\omega^2 r^2}{h}+2i\omega s r-\frac{is\omega r^2 h'}{h}+
s(\Delta''-2)-{}_s\lambda_{\ell} \right)R_s (r)=0\,,
\end{equation}
where $\Delta=hr^2$.  The corresponding angular equation has the form
\begin{equation}
\frac{1}{\sin\theta}\,\frac{d}{d\theta}\left(\sin\theta\,\,
\frac{d\,{}_sS_{\ell}^m}{d\theta}\right)+
\left(-\frac{2ms\cot\theta}{\sin\theta}-\frac{m^2}{\sin^2\theta}+
s-s^2\cot^2\theta+{}_s\lambda_{\ell}\right){}_sS^m_{\ell}(\theta)=0\,,
\label{angular}
\end{equation}
where $e^{im\varphi}\,{}_sS^m_{\ell}(\theta)={}_sY^m_{\ell}(\Omega_2)$ are known as the spin-weighted spherical harmonics and ${}_s\lambda_{\ell}$ is a separation constant which is found to have the value ${}_s\lambda_{\ell}=\ell(\ell+1)-s(s+1)$ \cite{Goldberg:1967uu}.

For $s=0$, equation~(\ref{radial}) reduces as expected to eq. (41) of 
ref. \cite{Kanti:2002nr} which was used for the analytical study of the emission of 
brane-localized scalar modes from a spherically-symmetric higher-dimensional
black hole.  Under the redefinition $R_s=\Delta^{-s} P_s$, eq.~(\ref{radial})
assumes a form similar to eq. (11) of ref. \cite{Kanti:2002ge} which was
used for the study of brane-localized fermion and gauge boson emission.
The two equations differ due to an extra term in the expression of the latter
one, which although vanishing for $s=$1/2 and 1 (leading to the
correct results for fermion and gauge boson fields)
gives a non-vanishing contribution
for all other values of $s$. Therefore, the generalized equation derived by
Cvetic and Larsen \cite{Cvetic:1998ap} cannot be considered as a master equation
valid for all types of fields.\footnote{A similar equation was derived in 
ref. \cite{Ida:2002ez} but due to a typographical error the multiplicative factor $s$ in front of the $\Delta''$-term was missing, leading to an apparently different equation for general $s$.} Hence the derivation of a consistent master equation was imperative, before addressing the question of the grey-body factors in the brane background.  This task was performed by Kanti and led to eqs.~(\ref{radial}) and (\ref{angular}) above.  

For the derivation of grey-body factors associated with the emission of
fields from the projected black hole, we need to know the asymptotic solutions
of eq.~(\ref{radial}) both as $r\rightarrow r_\text{h}$ and as $r\rightarrow \infty$. 
In the former case, the solution is of the form
\begin{equation}
\label{near}
R_s^{(\text{h})}=A_\text{in}^{(\text{h})}\,\Delta^{-s}\,e^{-i\omega r^{*}}+
A_{out }^{(\text{h})}\,e^{i\omega r^{*}},
\end{equation}
where
\begin{equation}
\label{rstarr}
\frac{dr^*}{dr}=\frac{1}{h(r)}\,.
\end{equation}
We impose the boundary condition that there is no out-going solution near the
horizon of the black hole, and therefore set $A_\text{out}^{(\text{h})}=0$.
The solution at infinity is of the form
\begin{equation}
\label{far}
R_s^{(\infty)}=A_\text{in}^{(\infty)}\,\frac{e^{-i\omega r}}{r}+
A_\text{out}^{(\infty)}\,\frac{e^{i\omega r}}{r^{2s+1}}\,,
\end{equation}
and comprises both in-going and out-going modes. 
\enlargethispage{-2\baselineskip}

The grey-body factor $\sigma^{(s)}_{\ell}(\om)$ for the emission of brane-localized
modes is related to the energy absorption coefficient ${\cal A}_\ell$
through the simplified relation
\begin{equation}
\hat \si^{(s)}_\ell(\om) =\frac{\pi}{\om^2}\,(2 \ell +1)\,
|\hat {\cal A}^{(s)}_\ell|^2,
\label{brane-loc}
\end{equation}
where from now on a `hat' will denote quantities associated with
the emission of brane-localized modes. The above relation follows from
eq.~(\ref{greydef}) by setting $n=0$ since the emission of brane-localized
modes is a 4-dimensional process.  The absorption coefficient itself is defined as
\begin{equation}
|\hat {\cal A}^{(s)}_\ell|^2=1-\frac{{\cal F}_\text{out}^{(\infty)}}
{{\cal F}_\text{in}^{(\infty)}}=
\frac{{\cal F}_\text{in}^{(\text{h})}}{{\cal F}_\text{in}^{(\infty)}}\,,
\label{absorption}
\end{equation}
in terms of the out-going and in-going energy fluxes evaluated either at infinity or at the horizon.
The two definitions are related by simple energy conservation and lead to
the same results.  The choice of which is used for the determination of the absorption coefficient may depend on numerical issues.  

The flux and power spectra (equations~(\ref{4flux}) and (\ref{4power}) respectively) of the Hawking radiation emitted on the brane can be computed, for massless particles, from the
4-dimensional expressions: 
\begin{align}
\frac{d \hat N^{(s)}(\om)}{dt}&= \sum_{\ell} (2\ell +1) |\hat {\cal A}^{(s)}_\ell|^2\,
\frac{1}{\exp\left(\om/T_\text{H}\right) \mp 1}\,\frac{d\om}{2\pi}\,;
\label{fdecay-brane}\\
\frac{d \hat E^{(s)}(\om)}{dt}&= \sum_{\ell} (2\ell +1) |\hat {\cal A}^{(s)}_\ell|^2\,
\frac{\om}{\exp\left(\om/T_\text{H}\right) \mp 1}\,\frac{d\om}{2\pi}\,.
\label{pdecay-brane}
\end{align}
Note, however, that the absorption coefficient $\hat {\cal A}_\ell^{(s)}$ still depends on the number of extra dimensions since the projected metric tensor of eq.~(\ref{non-rot}) carries a signature of the dimensionality of the bulk space-time through the expression of the metric function $h(r)$. 

\subsection{Analytic calculation}

The grey-body factors have been determined analytically for the 4-dimensional
case in refs. \cite{Page:1976df,Sanchez:1978si,Sanchez:1978vz} both for a rotating and non-rotating black
hole. In the $(4+n)$-dimensional case, refs. \cite{Kanti:2002nr,Kanti:2002ge} have provided
analytic expressions for grey-body factors for the emission of scalars,
fermions and gauge bosons from a higher-dimensional Schwarzschild-like
black hole.  Both bulk and brane emission were considered for an arbitrary number of extra dimensions $n$. Reference \cite{Ida:2002ez} presented
analytic results for brane-localized emission from a Kerr-like
black hole, but only in the particular case of $n=1$. All the above results were derived
in the low-energy approximation where the procedure used is as follows:

\begin{itemize}
\item{Find an analytic solution in the near-horizon regime and expand it as
in-going and out-going waves so that the $A^{(\text{h})}$ coefficients can be extracted;}
\item{Apply the boundary condition on the horizon so that the wave is purely
out-going;}
\item{Find an analytic solution in the far-field regime and again expand it as
in-going and out-going waves;}
\item{Match the two solutions in the intermediate regime;}
\item{Extract $|{\cal A}_\ell|^2$ and expand it in powers of $\omega r_\text{h}$.}
\end{itemize}

The solution obtained by following the above approximate method is a power
series in $\omega r_\text{h}$, which is only valid for low energies and expected to
significantly deviate from the exact result as the energy of the emitted
particle increases (this was pointed out in \cite{Page:1976df,MacGibbon:1990zk} for the
4-dimensional case). In \cite{Kanti:2002ge}, the full analytic result for the absorption
coefficient (before the final expansion was made) was used for the evaluation
of the emission rates for all particle species.\footnote{This full analytic approach was pursued in \cite{Kanti:2002ge} as a result of some of the numerical results presented in this thesis.} The range of validity of these
results, although improved compared to the power series, was still limited since the assumption of small
$\omega r_\text{h}$ was {\it still} made during the matching of the two solutions
in the intermediate regime. As the equations of motion are too complicated to solve analytically for all values of $\om r_\text{h}$, it becomes clear that only an exact numerical analysis can give the full exact results for grey-body factors and emission rates. 


\subsection{Numerical calculation}
\label{numcal}

There are various numerical issues which arise while trying to do the full
calculation of the absorption coefficient and the complexity of these problems
depends strongly on the spin $s$ of the emitted particle.  The usual numerical 
procedure starts by applying the relevant horizon boundary condition (a 
vanishing out-going wave) to $R_s (r)$.  Then, the solution is integrated out 
to `infinity' and the asymptotic coefficient $A_\text{in}^{(\infty)}$ is extracted 
in terms of which ${\cal A}_\ell^{(s)}$ can be calculated.

By looking at the asymptotic solution at infinity in eq.~(\ref{far}) we can
see that, for the scalar case, the in-going and out-going waves are of
comparable magnitude.  Therefore it is relatively easy to extract the
coefficients at infinity and thus determine the grey-body factor.  However for 
fields with non-vanishing spin, i.e.\ fermions, vector bosons and
gravitons, this is not an easy task. First of all, different components carry
a different
part of the emitted field: the upper component $\Psi_{+s}$ consists mainly of
the in-going wave with the out-going one being greatly suppressed, while for
the lower component $\Psi_{-s}$ the situation is reversed (for a field with
spin $s \neq 0$, only the upper and lower components are radiative). 
Distinguishing between the two parts of the solution (in-going
and out-going) is not an easy task no matter which component is used, and it
becomes more difficult as the magnitude of the spin increases.  In addition,
the choice of either positive or negative $s$ to extract the grey-body factor (i.e.\ using either the upper or the lower component) affects
the numerical issues.

If $s$ is negative, then the horizon boundary condition is easy to apply 
because components of the in-going solution on the horizon will be 
exponentially suppressed, as can be seen from eq.~(\ref{near}).  However eq.~(\ref{far}) shows that negative $s$ also means the out-going solution at infinity is enhanced by $r^{-2s}$ 
with respect to the in-going one; this makes accurate 
determination of $A_\text{in}^{(\infty)}$ very difficult.  Hence the negative $s$ 
approach is not used in this work.

On the other hand if $s$ is positive, it is easy to accurately extract 
$A_\text{in}^{(\infty)}$ because the out-going solution at infinity is suppressed by $r^{-2s}$.  However, close to the horizon the out-going solution is
exponentially smaller than the in-going one. 
Therefore the solution for $R_s(r)$ can be easily contaminated by 
components of the out-going solution. This problem becomes worse for larger 
$s$ and first becomes significant for $s=1$.
\enlargethispage{-\baselineskip}

For $n=0$, various methods (see e.g.\ \cite{Press:1974}) have been used to 
solve the numerical problems which arise in the gauge boson case.  The 
approach of Bardeen in \cite{Press:1974} is also applicable for $n=1$; some details of this method are briefly outlined below.

The method starts from a mathematically equivalent form of the radial equation in Kerr in-going co-ordinates:\footnote{This was derived, for general $n$, in \cite{Kantiprivate} and for $n=0$ reduces to the result found in \cite{Press:1974}.}
\begin{equation}
\Delta^{-s}\frac{d}{dr} \left(\Delta^{s+1}\frac{dR_s}{dr}\right)-2 i r^2\omega \frac{dR_s}{dr}-\left[2(2s+1)i\omega r-s\Delta''+{}_s\Lambda_{\ell}\right]R_s=0\,,
\end{equation}
where ${}_s\Lambda_{\ell}=\ell(\ell+1)-s(s-1)$.  The asymptotic solutions are now different from those of eqs.~(\ref{near}) and (\ref{far}); they are given by
\begin{equation}
R_s^{(\text{h})}=B_\text{in}^{(\text{h})}+B_\text{out}^{(\text{h})}\Delta^{-s} e^{2i\omega r^{*}},
\end{equation}
and
\begin{equation}
R_s^{(\infty)}=B_\text{in}^{(\infty)}\frac{1}{r^{2s+1}}+B_\text{out}^{(\infty)} \frac{e^{2i\omega r}}{r}\,. 
\end{equation}

We can follow the method in Appendix A of \cite{Press:1974} and apply the co-ordinate transformation
\begin{equation}
x=\frac{r}{r_\text{h}}-1\,.
\end{equation}
This transformation is easily applicable for either $n=0$ or $n=1$ since in both cases $\Delta''=2$. In the case $s=-1$ (which is mathematically equivalent to the $s=1$ discussed earlier) applying this transformation for $n=0$ gives
\begin{equation}
\label{xradial}
x(x+1)\frac{d^2R_{-1}}{dx^2}-2i\omega r_\text{h}\left(x^2+2x+1\right)\frac{dR_{-1}}{dx}-\left[-2i\omega r_\text{h} (x+1)+\lambda\right]R_{-1}=0\,,
\end{equation}
where $\lambda=\ell(\ell+1)$.  The asymptotic solutions can now be written in terms of $x$, with more care being taken over sub-leading terms in the in-going solution at the horizon:
\begin{equation}
R_{-1}^{(\text{h})}=B_\text{in}^{(\text{h})} (1+a_1x+a_2x^2)+B_\text{out}^{(\text{h})} r_\text{h}^2 e^{2i\omega r^*}x(1+b_1x+\ldots)\,.
\end{equation}

It is easy to obtain the coefficients $a_1$ and $a_2$ by substituting the in-going solution into the radial equation and comparing terms in different powers of $x$.  This leads to
\begin{equation}
a_1=1+i\frac{\lambda}{2\omega r_\text{h}}\,,
\label{a1n0}
\end{equation}
and
\begin{equation}
a_2=\frac{1}{2(1-2i\omega r_\text{h})}\left[(\lambda+2i\omega r_\text{h})a_1-2i\omega r_\text{h}\right].
\label{a2n0}
\end{equation}
Instead of solving ${\mathcal L}R=0$, the equation ${\mathcal L}Y=f$ is solved, where $Y=R-(1+a_1x)$. This means that
\begin{equation}
f=2(1-2i\omega r_\text{h})a_2x\,.  
\label{fn0}
\end{equation}
Close to the horizon the leading-order terms in $Y$ and $Y'$ are $a_2x^2$ and $2a_2x$---these are used to apply the horizon boundary condition.  Bardeen found that the equation in this form can now be stably integrated from $x=0$ to $x=\infty$.  The $B_\text{in}$ and $B_\text{out}$ coefficients extracted can be used to obtain the grey-body factors using similar formulae to those discussed later for $A_\text{in}$ and $A_\text{out}$.

For $n=1$, equations~(\ref{xradial}) and (\ref{a1n0})--(\ref{fn0}) are slightly modified and become
\begin{gather}
x(x+2)\frac{d^2R_{-1}}{dx^2}-2i\omega r_\text{h}\left(x^2+2x+1\right)\frac{dR_{-1}}{dx}-\left[-2i\omega r_\text{h} (x+1)+\lambda\right]R_{-1}=0\,;\\
a_1=1+i\frac{\lambda}{2\omega r_\text{h}}\,;
\label{a1n1}\\
a_2=\frac{1}{4(1-i\omega r_\text{h})}\left[(\lambda+2i\omega r_\text{h})a_1-2i\omega r_\text{h}\right];
\label{a2n1}\\
f=4(1-i\omega r_\text{h})a_2x\,.  
\label{fn1}
\end{gather}

Unfortunately no analogous method has been found for $n \geq 2$ because there is then no simple equivalent of eq.~(\ref{xradial}).  Instead, an 
alternative transformation of the radial equation~(\ref{radial}) was used.
Writing $y=r/r_\text{h}$ and $R_1=y\,F(y)\,e^{-i\omega r^*}$, the wave equation becomes
\begin{equation}
(hy^2)\,\frac{d^2F}{dy^2}+2y\,(h-i\omega r_\text{h} y)\,\frac{dF}{dy}-\ell(\ell+1)F=0\,.
\end{equation}
Since on the horizon $y=1$ and $h=0$, the boundary conditions become $F(1)=1$ 
and
\begin{equation}
\left. \frac{dF}{dy} \right|_{y=1}=\frac{i \ell(\ell+1)}{2 \omega r_\text{h}}\,.
\end{equation}
Using both this method and the Bardeen method for the $n=1$ case provided a useful cross-check on the spin-1 numerical results.

For fermions no special transformation was necessary and therefore the radial
equation~(\ref{radial}) was used.  However the application of the 
boundary condition at the horizon is made slightly easier by the 
transformation $P_s=\Delta^s R_s$ so that the asymptotic solution at
the horizon becomes
\begin{equation}
P_s^{(\text{h})}=A_\text{in}^{(\text{h})}\,e^{-i\omega r^{*}}+
A_{out }^{(\text{h})}\,\Delta^s \,e^{i\omega r^{*}}.
\label{nh-new}
\end{equation}
Since we require $A_\text{out}^{(\text{h})}=0$, a suitable boundary condition to apply as $r \rightarrow r_\text{h}$ is that
\begin{equation}
P_s=1\,,
\end{equation}
while using eq.~(\ref{rstarr}) we also obtain
\begin{equation}
\frac{dP_s}{dr}=-i\omega\, \frac{dr^*}{dr}=-\frac{i\omega}{h(r)}\,.
\end{equation}
The above boundary conditions ensure that $|A_\text{in}^{(\text{h})}|^2=1$. 
The asymptotic form for $P_s$ at infinity now looks like
\begin{equation}
P_s^{(\infty)}=A_\text{in}^{(\infty)}\,\frac{e^{-i\omega r}}{r^{1-2s}}+
A_\text{out}^{(\infty)}\, \frac{e^{i\omega r}}{r}\, 
\label{ff-new}
\end{equation}
since $\Delta \rightarrow r^2$ as $r \rightarrow \infty$. 

The numerical work described in this chapter was almost entirely performed using the \texttt{NDSolve} package in \texttt{Mathematica}.  However in some cases Fortran programs calling {\small NAG} routines (specifically \texttt{D02EJF} and its associated routines) were used as a check on the numerical robustness of the results obtained in \texttt{Mathematica}.  The {\small NAG} routines implement the Backward Differentiation Formulae (BDF) whilst in \texttt{Mathematica} either the BDF method or the implicit Adams method is used.

\begin{figure}
\unitlength1cm
\psfrag{x}[][][1.4]{$r_{\infty}$}
\psfrag{y}[][][1.4]{$\hat{\sigma}^{(0)}_\text{abs}/\pi r_\text{h}^2$}
\begin{center}
\begin{minipage}[t]{3.35in}
\scalebox{0.65}{\rotatebox{0}{\includegraphics{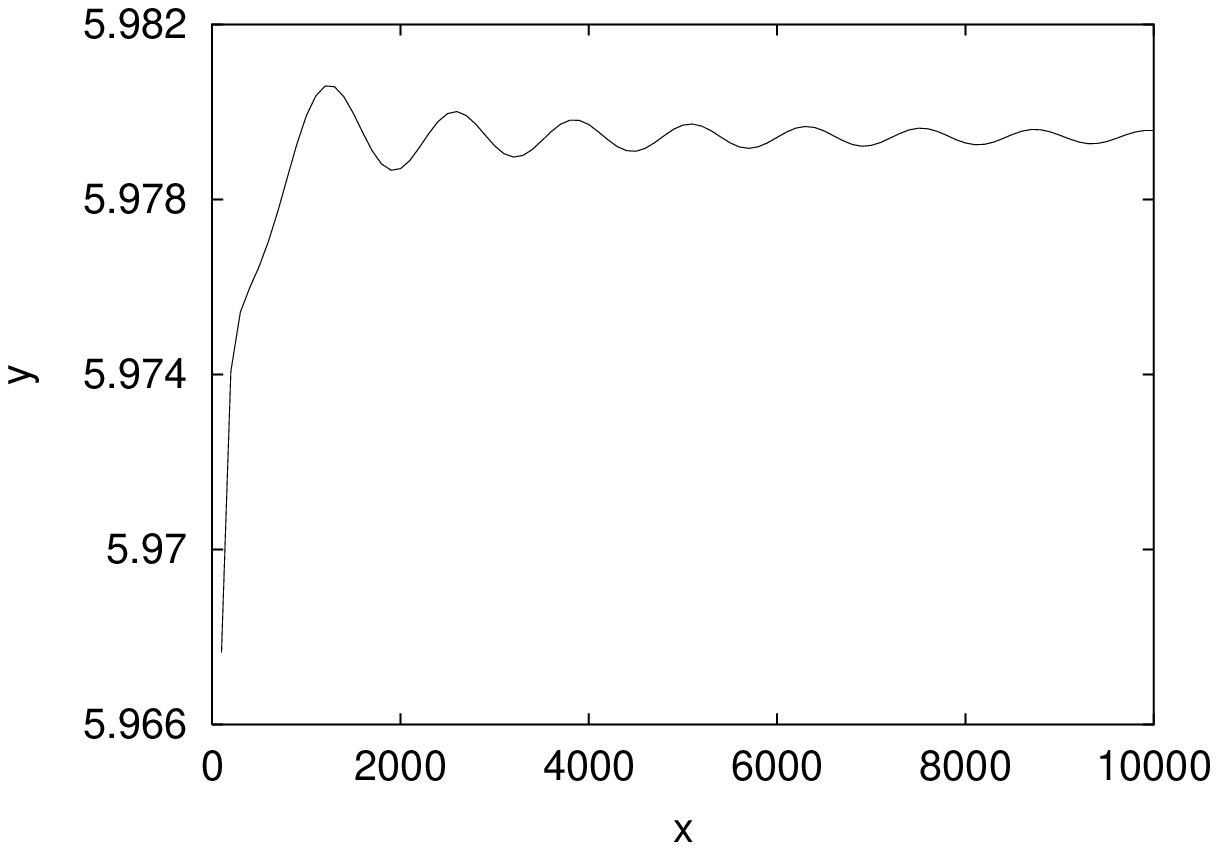}}}
\end{minipage}
\hfill
\begin{minipage}[t]{2.75in}
\scalebox{0.65}{\rotatebox{0}{\includegraphics{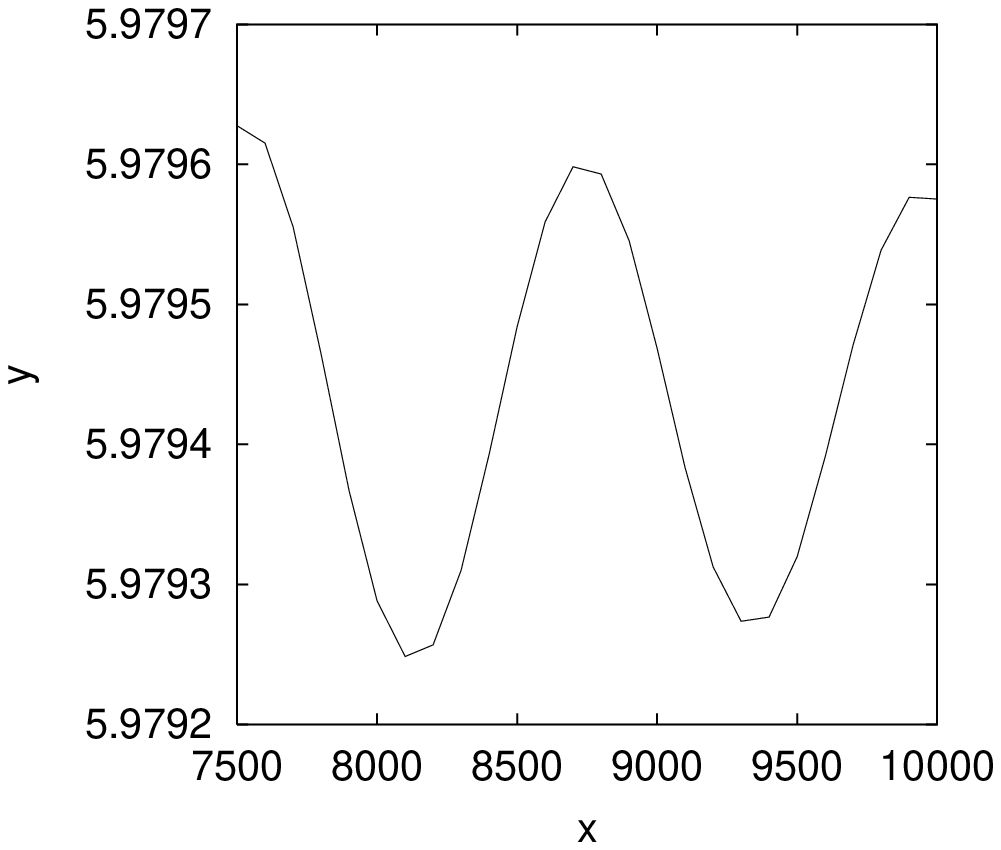}}}
\end{minipage}
\capbox{Stability of grey-body factor for changes in `infinity'}{Stability of the $n=0$, $s=0$, $\omega r_\text{h}=0.5$ grey-body factor as $r_{\infty}$ is changed.  The right plot is a magnified version of part of the left plot to illustrate the accuracy obtained.\label{farstab}}
\end{center}
\end{figure}

There are various considerations which must be taken into account in order to
obtain results to the desired accuracy (at least three significant figures).
Firstly, although the horizon boundary condition cannot be applied exactly at
$r_\text{h}$ (due to singularities in the boundary condition and the differential 
equation) the error introduced by applying the condition at $r=r_\text{n}$ (where $\Delta r_\text{h}=r_\text{n}-r_\text{h} \ll r_\text{h}$) must be small.  This can be investigated by studying changes in the
grey-body factors for order-of-magnitude changes in $\Delta r_\text{h}$.  Similarly it 
must be verified that the value $r_{\infty}$ used as an approximation for `infinity' 
does not introduce errors which will affect the accuracy of the result.  This is illustrated in Figure~\ref{farstab} in which $\hat{\sigma}^{(0)}_\text{abs}=\sum_{\ell}\hat{\sigma}^{(0)}_{\ell}$ is plotted in units of $\pi r_\text{h}^2$.  The oscillations in the numerical result die away as the asymptotic form given in eq.~(\ref{far}) is obtained.  A value of $r_{\infty}=10000$ was used in the work that follows to ensure the grey-body factor was accurate to at least three significant figures.

Care must also be taken that the numerical integration procedure is sufficiently accurate that significant integration errors are avoided even at large values of $r$.  Plots like those in Figure~\ref{farstab} are also useful for investigating this because they can be used to confirm that there is no significant gradient on a straight line drawn through the centre of the oscillations.

Finally, for each energy being considered, enough angular momentum 
modes must be included in the summation so that only modes which do not
contribute significantly are neglected.  For any value of $\ell$ the absorption coefficient will approach unity at high enough energies and the $(2\ell+1)$ factor in eq.~(\ref{brane-loc}) means that higher angular momentum modes will start to dominate the grey-body factor.  It was found that for the highest values of $\omega r_\text{h}$ considered in this work it was necessary to include contributions from in excess of ten angular momentum modes.

\section{Numerical results for brane emission}
\label{numres}

In this section, results are presented for grey-body factors and emission rates for brane-localized scalar, fermion and gauge boson
fields, as obtained by numerically solving the corresponding equations of 
motions.  The relationship between the absorption coefficient ${\cal A}_\ell^{(s)}$
and the $A$ coefficients in equations~(\ref{nh-new}) and (\ref{ff-new}) is different in each case, so each will be considered separately.

\subsection{Spin 0 fields}
\label{numscalar}

The numerical integration of eq.~(\ref{radial}) for $s=0$ yields the solution
for the radial function $R_0(r)$ which smoothly interpolates between the
asymptotic solutions of eqs.~(\ref{nh-new}) and (\ref{ff-new}) in the near-horizon
and far-field regimes respectively. The absorption coefficient is easily
defined in terms of the in-going and out-going energy fluxes at infinity,
or equivalently by the corresponding wave amplitudes ($A_\text{in}^{(\infty)}$ and $A_\text{out}^{(\infty)}$) in the same asymptotic regime.
Therefore eq.~(\ref{absorption}) may be written in the form \cite{Kanti:2002nr}
\begin{equation}
|\hat {\cal A}^{(0)}_\ell|^2=1-|\hat {\cal R}^{(0)}_\ell|^2= 
1-\Biggl|\frac{A_\text{out}^{(\infty)}}{A_\text{in}^{(\infty)}}\Biggr|^2,
\label{scalars}
\end{equation}
where $\hat {\cal R}_\ell$ is the corresponding reflection coefficient. 

\begin{figure}
\begin{center}
\psfrag{x}[][][1.75]{$\omega r_\text{h}$}
\psfrag{y}[][][1.75]{$\hat{\sigma}^{(0)}_\text{abs}(\om)/\pi r_\text{h}^2$}
\scalebox{0.5}
{\rotatebox{-90}{\includegraphics[width=\textwidth]{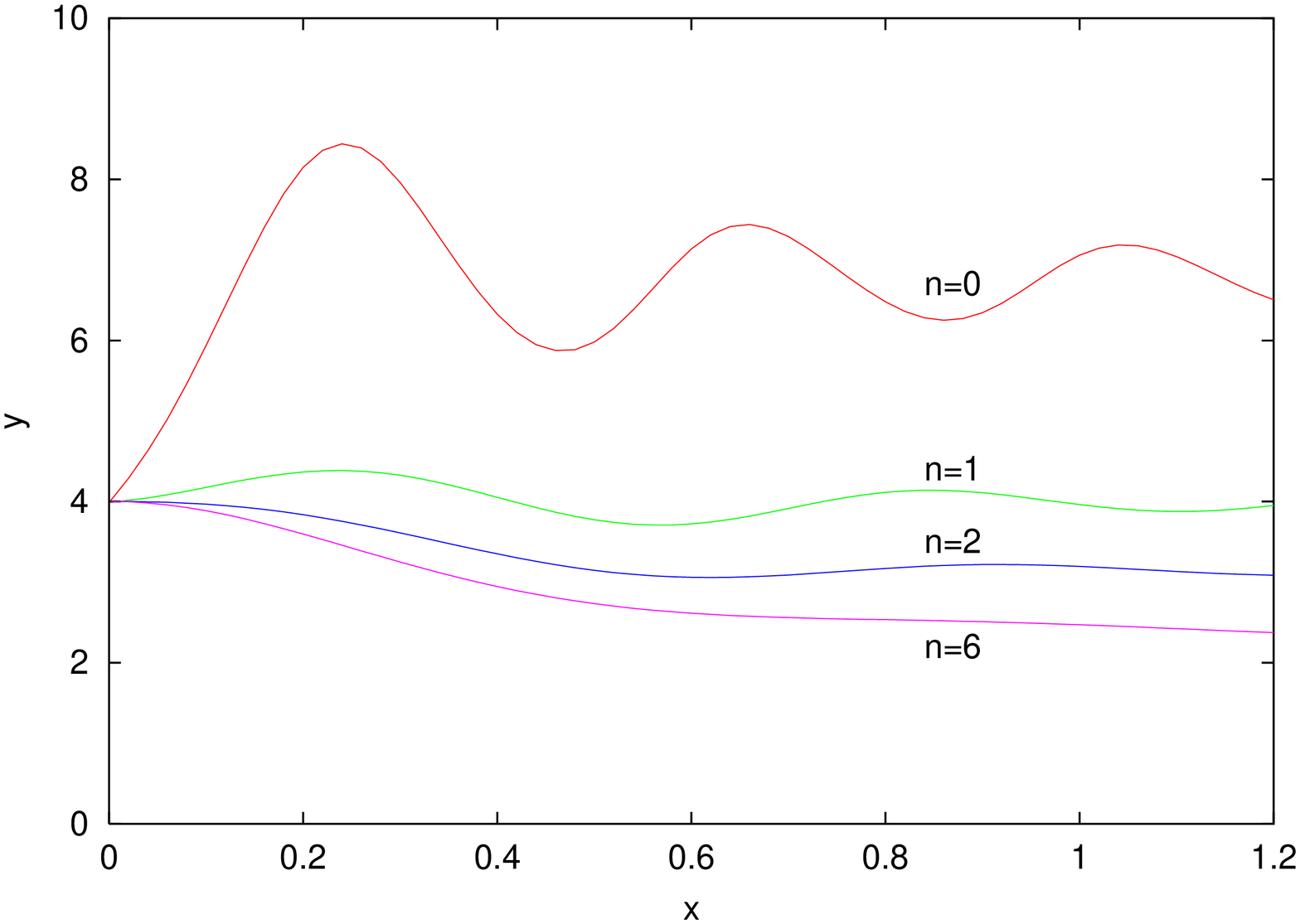}}}
\capbox{Grey-body factors for scalars on the brane}{Grey-body factors for scalar emission on the brane from a $(4+n)$D
black hole.\label{grey0}}
\end{center}
\end{figure}

The plot presented in Figure~\ref{grey0} shows, for several values of $n$,
the grey-body factor for the emission of scalar fields on the brane (for completeness, the plots include values of $n$ ruled out
on astrophysical grounds, i.e.\ $n=1$ and 2).\footnote{Throughout the numerical analyses the horizon radius $r_\text{h}$ is an
arbitrary input parameter which remains fixed.} The grey-body
factor is obtained by using eq.~(\ref{brane-loc}) and summing over the angular momentum number $\ell$.  

For $n=0$ and
$\omega r_\text{h} \rightarrow 0$, the grey-body factor assumes a non-zero value 
which is equal to $4\pi r_\text{h}^2$---that is, the grey-body factor for scalar 
fields with a very low energy is given exactly by the area of the black hole 
horizon. As the energy increases, the factor soon starts oscillating
around the geometrical optics
limit $\hat{\sigma}_\text{g} = 27 \pi r_\text{h}^2/4$ which corresponds to the spectrum of a
black body with an absorbing area of radius $r_\text{c}=3 \sqrt{3}\,r_\text{h}/2$
\cite{Sanchez:1978si,Sanchez:1978vz,misner}.  The $n=0$ result agrees exactly with Page's result presented in Figure 1 of \cite{MacGibbon:1990zk} (this is also found to be the case for the fermion and gauge boson grey-body factors shown later).  
\enlargethispage{-\baselineskip}

In the extra-dimensional case, the grey-body factor has an asymptotic low-energy value which is the same for all values of $n$,
and at high energies it again starts oscillating around a limiting value. This is always lower than the 4-dimensional geometrical optics limit because the effective radius $r_\text{c}$ depends on the dimensionality of the bulk space-time through the metric tensor of the projected space-time in which the particle moves. For arbitrary $n$, it adopts the value \cite{Emparan:2000rs}
\begin{equation}
r_\text{c}=\biggl(\frac{n+3}{2}\biggr)^{1/n+1}\,\sqrt{\frac{n+3}{n+1}}\,\,r_\text{h}\,.
\label{effective}
\end{equation}
The above quantity is a strictly decreasing function of $n$ which causes the
asymptotic grey-body factor, $\hat{\sigma}_\text{g}=\pi r_\text{c}^2$\,, to become more and more
suppressed as the number of extra dimensions projected onto the
brane increases.  The values of $\hat{\sigma}_\text{g}$ are tabulated in Table~\ref{brgolim} for different values of $n$.\footnote{For the larger values of $n$ it was found that these asymptotic values are not approached until relatively high energies, typically $\omega r_\text{h} \sim n$.}

\begin{table}
\def\arraystretch{1.1}
\begin{center}
\begin{tabular}{|c|c|c|c|c|c|c|c|c|}
\hline
$n$ & 0 & 1  & 2  & 3 & 4 & 5 & 6 & 7 \\
\hline
$\hat{\sigma}_\text{g}/\pi r_\text{h}^2$ & 6.75 & 4 & 3.07 & 2.60 & 2.31 & 2.12 & 1.98 & 1.87\\
\hline 
\end{tabular}
\capbox{High-energy limits of grey-body factors for brane emission}{High-energy limits of grey-body factors for brane emission, given in units of $\pi r_\text{h}^2$.\label{brgolim}} 
\end{center}
\def\arraystretch{1.0}
\end{table}

The power series expression of the grey-body factor determined in \cite{Kanti:2002nr} matches the exact solution only in a very limited low-energy
regime. In the limit $\omega r_\text{h} \rightarrow 0$, the asymptotic
value $4\pi r_\text{h}^2$ is recovered as expected; however, as the energy increases
the exact solution rapidly deviates from the behaviour dictated by the
dominant term in the $\omega r_\text{h}$ expansion. This was confirmed in \cite{Kanti:2002ge} where full analytic results for the grey-body
factors were plotted, although still in the low-energy approximation.  The behaviour depicted in Figure 3 of ref. \cite{Kanti:2002ge} is much closer to the exact one, shown here in Figure~\ref{grey0}, and
successfully reproduces some of the qualitative features including the 
suppression of the grey-body factor as the dimensionality of the bulk space-time increases.  Nevertheless, as previously stated, even this result rapidly breaks down as the energy increases; it is particularly unreliable for the $n=0$ case and fails to reproduce the high-energy oscillations for any value of $n$.  This means that the exact numerical solution obtained in this work is the only reliable source of information concerning the form of the extra-dimensional grey-body factor throughout the energy regime.

\begin{figure}
\begin{center}
\psfrag{x}[t][b][1.75]{$\omega r_\text{h}$}
\psfrag{y}[][][1.75]{$r_\text{h} d^2 \hat{E}^{(0)}/dtd\om$}
\scalebox{0.5}{\rotatebox{0}{\includegraphics[width=23cm, height=15cm]{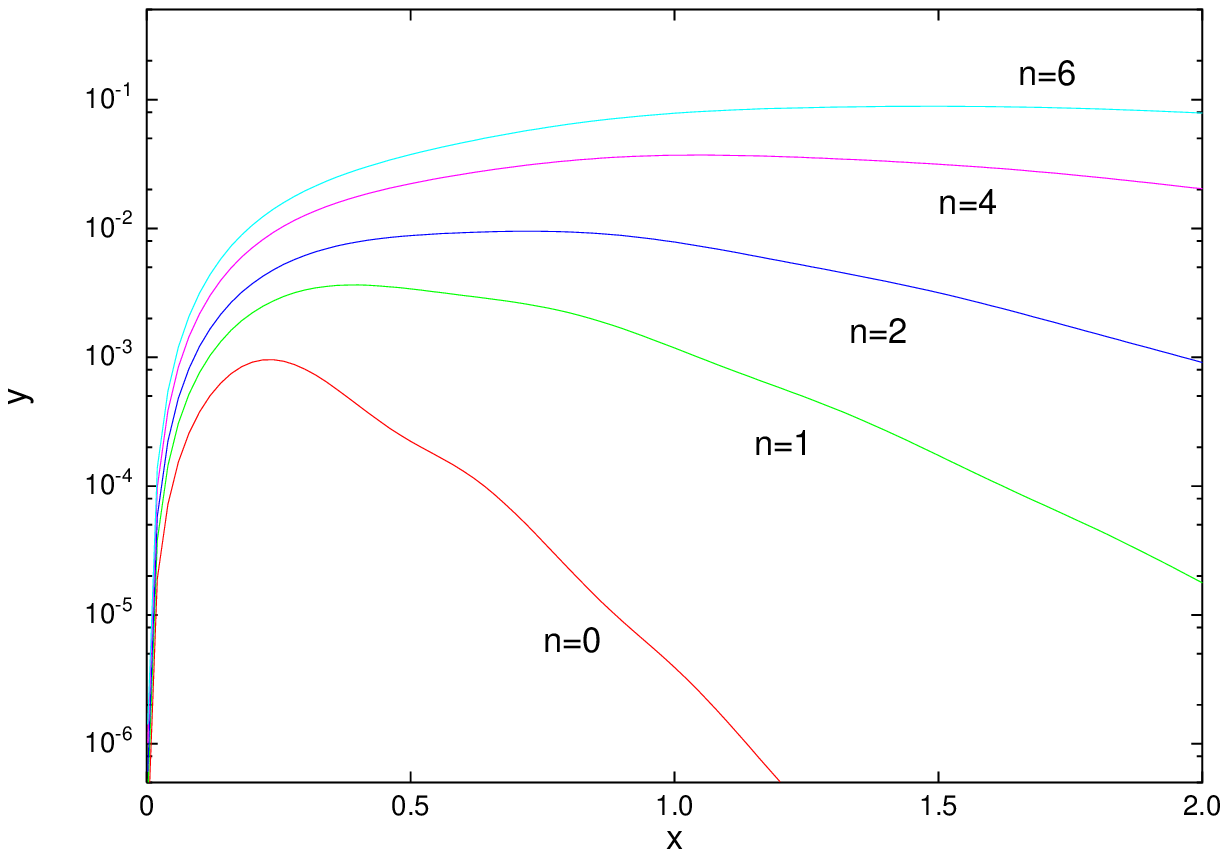}}}
\capbox{Power emission for scalars on the brane}{Energy emission rates for scalar fields on the brane from
a $(4+n)$D black hole.\label{sc-dec}}
\end{center}
\end{figure}

The numerical solution for the grey-body factor allows the computation of the energy emission rate for scalar fields on the brane.  Using eq.~(\ref{pdecay-brane}) it is found that the suppression of the grey-body factor with $n$ does not lead to the suppression of the emission rate itself. The behaviour of the differential energy emission rate is given in Figure~\ref{sc-dec}. As $n$ increases, the increase in the temperature of the black hole, and therefore in its emitting power, overcomes the decrease in the grey-body factor and leads to a substantial enhancement of the energy emission rate.  Figure~\ref{sc-dec} clearly shows that the enhancement of the peak of the emission curve can be up to several orders of magnitude compared to the 4-dimensional case.  In addition the peak in the spectrum is shifted to higher values of the energy parameter $\omega r_\text{h}$ since for fixed $r_\text{h}$ an increase in $n$ corresponds to an increase in the temperature of the radiating body.

In order to quantify the enhancement of the emission rate as the number of extra dimensions projected onto the brane increases, the total flux and power emissivities were computed, for a range of values of $n$, by integrating eqs.~(\ref{4flux}) and (\ref{4power}) with respect to $\om$.  The results obtained are displayed in Table~\ref{sc-tab}.  The relevant emissivities for different values of $n$ have been normalized in terms of those for $n=0$. From the entries of this table, the order-of-magnitude enhancement of both the flux and power radiated by the black hole as $n$ increases is again clear.
\begin{table}
\begin{center}
\begin{tabular}{|l|c|c|}
\hline
$n$& Flux  & Power\\
\hline
0 & 1 & 1\\
1 & 4.75 & 8.94\\
2 & 13.0 & 36.0\\
3 & 27.4 & 99.8\\
4 & 49.3 & 222\\
5 & 79.9 & 429\\
6 & 121 & 749\\
7 & 172 & 1220\\
\hline
\end{tabular}
\capbox{Emissivities for scalar fields on the brane}{Flux and power emissivities for scalar fields on the brane.\label{sc-tab}}
\end{center}
\end{table}


\subsection{Spin 1/2 fields}

Unlike the case of scalar fields, the study of the emission of fields with
non-vanishing spin involves, in principle, fields with more than
one component. Equation~(\ref{radial}) depends on the helicity number $s$ which can be either $+$1/2 or $-$1/2 and this leads to radial equations for the two different components of the field. As mentioned in \secref{numcal}, the upper ($+$1/2) and lower ($-$1/2) components carry mainly the in-going and out-going parts of the field respectively. Although knowledge of both components is necessary to construct the complete solution for the emitted field, the determination of either is sufficient to compute the absorption coefficient
$\hat {\cal A}_j^{(s)}$ ($j$ is the total angular momentum number). For
example, if the in-going wave is known in the case of the emission of
fields with spin $s=$1/2, eq.~(\ref{absorption})
may be directly written as \cite{Kanti:2002ge,Cvetic:1998ap}
\begin{equation}
|\hat {\cal A}^{(1/2)}_j|^2=\Biggl|\frac{A_\text{in}^{(\text{h})}}
{A_\text{in}^{(\infty)}}\Biggr|^2.
\label{fermions}
\end{equation}
The above follows by defining the incoming flux of a fermionic field as the
radial component of the conserved current, 
$J^\mu=\sqrt{2}\,\sigma^\mu_{AB}\,\Psi^A\,\bar \Psi^B$, integrated over a
two-dimensional sphere and evaluated at both the horizon and infinity.

\begin{figure}
\begin{center}
\psfrag{x}[][][1.75]{$\omega r_\text{h}$}
\psfrag{y}[][][1.75]{$\hat{\sigma}^{(1/2)}_\text{abs}(\om)/\pi r_\text{h}^2$}
\scalebox{0.5}{\rotatebox{-90}{\includegraphics[width=\textwidth]{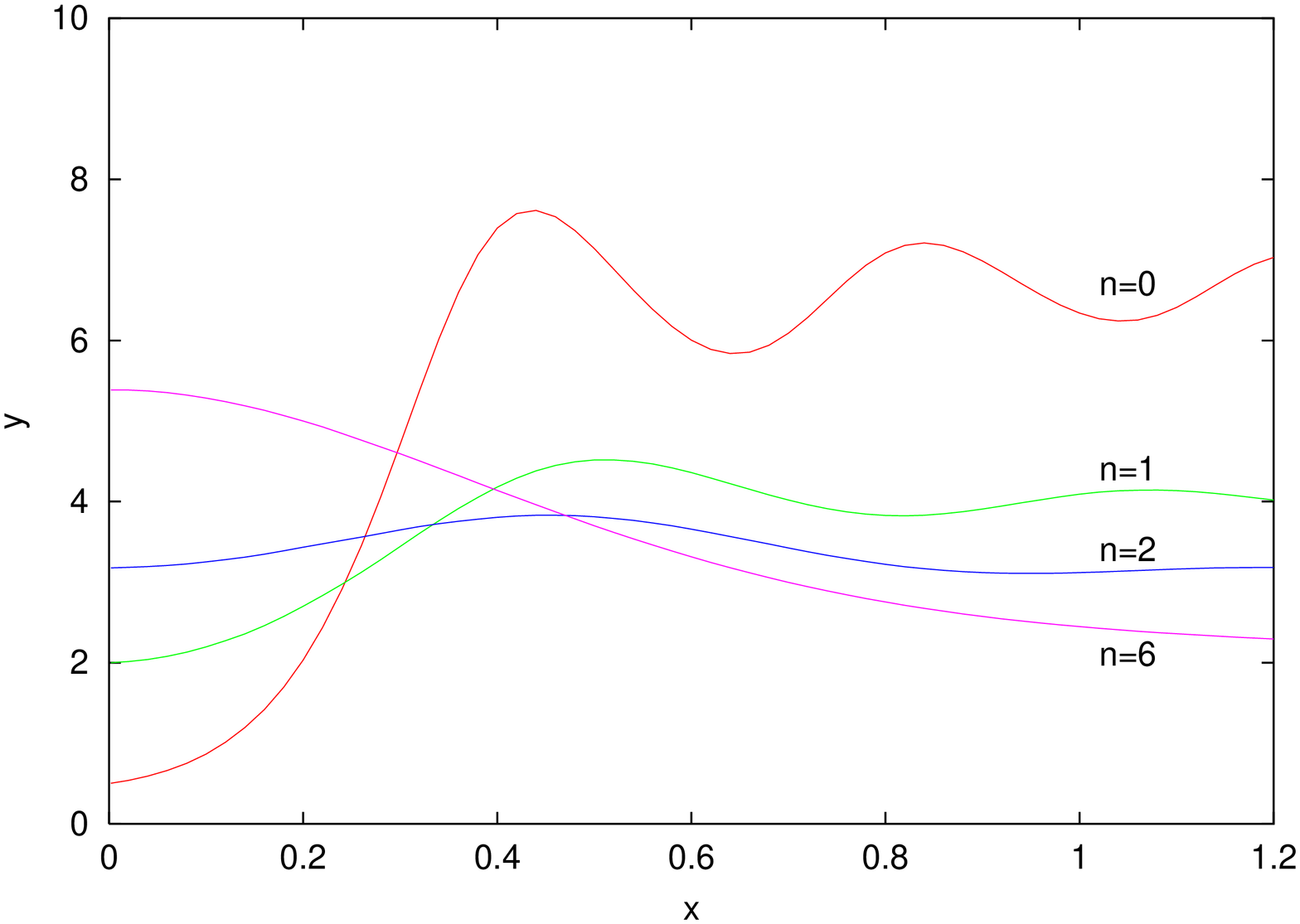}}}
\capbox{Grey-body factors for fermions on the brane}{Grey-body factors for fermion emission on the brane from a $(4+n)$D
black hole.\label{grey05}}
\end{center}
\end{figure}

The grey-body factor is again related to the absorption probability through eq.~(\ref{brane-loc}) with $\ell$ now being replaced by the total angular momentum $j$.  Numerically solving the radial equation~(\ref{radial}) and computing $\hat {\cal A}^{(1/2)}_j$ gives the behaviour of the grey-body factor, in terms of the energy parameter $\omega r_\text{h}$ and the number of extra dimensions $n$.  This is shown in Figure~\ref{grey05} for four different values of $n$. 

As in the scalar case, at low energies the grey-body factor assumes a n-zero asymptotic value; this depends on the dimensionality of
space-time and increases with $n$. The enhancement of 
$\hat{\sigma}^{(1/2)}_\text{abs}(\omega)$ with $n$ in the low-energy regime continues up to intermediate values of $\omega r_\text{h}$ where the situation is reversed: as $n$ becomes larger, the high-energy grey-body factor becomes more and more suppressed as was found in the scalar case.  The full analytic results
derived in \cite{Kanti:2002ge} provide a reasonable description of the low-energy behaviour except in the $n=0$ case; however, as expected, they fail to
give accurate information for the high-energy regime.  As for scalar fields, the high-energy grey-body factors for fermions are shown to oscillate around asymptotic values determined by the effective radius of eq.~(\ref{effective}).

\begin{figure}
\begin{center}
\psfrag{x}[t][b][1.75]{$\omega r_\text{h}$}
\psfrag{y}[][][1.75]{$r_\text{h} d^2 \hat{E}^{(1/2)}/dtd\om$}
\scalebox{0.5}{\rotatebox{0}{\includegraphics[width=23cm, height=15cm]
{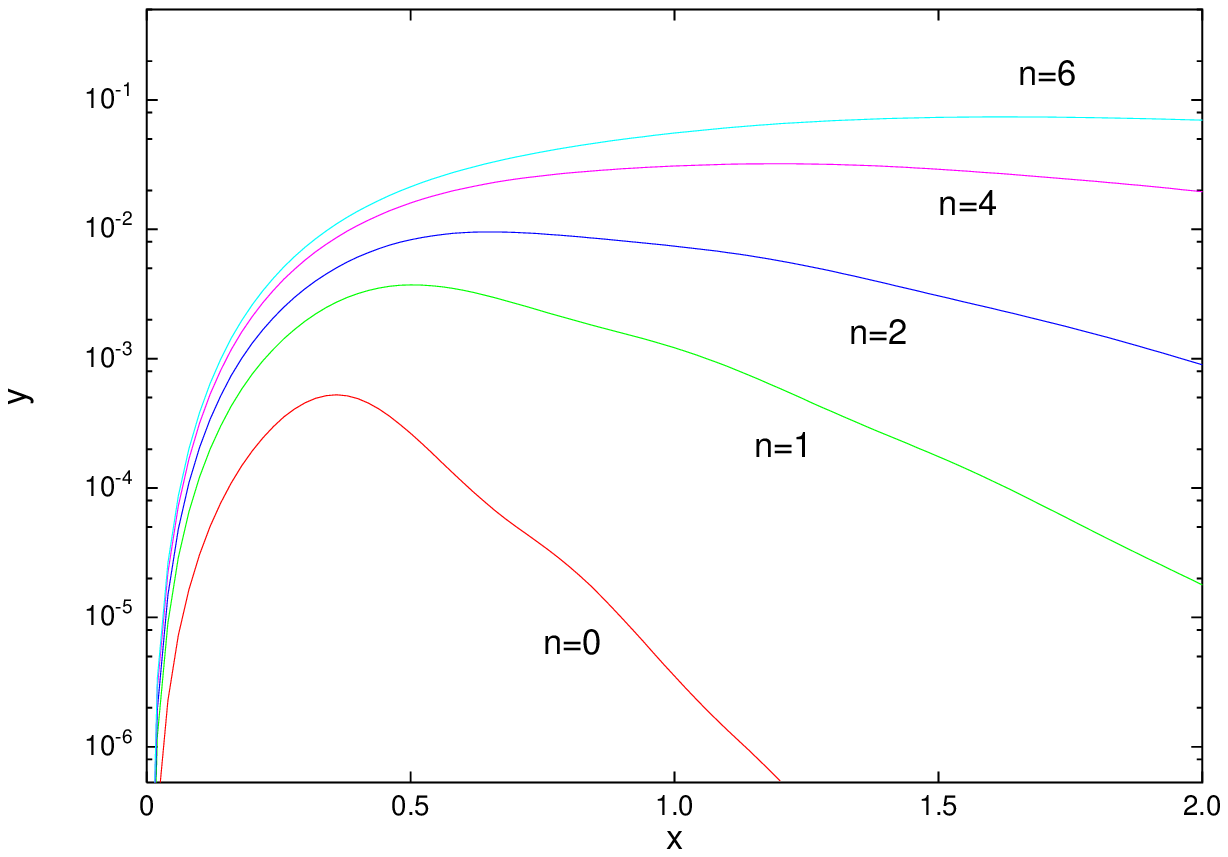}}}
\capbox{Power emission for fermions on the brane}{Energy emission rates for fermions on the brane from a $(4+n)$D black hole.\label{fer-dec}}
\end{center}
\end{figure}
The energy emission rate for fermion fields on the brane is shown in Figure~\ref{fer-dec} for various values of $n$.
As $n$ increases, the power emission is found to be significantly enhanced at both low and high energies. The emission curves show the same features as for the emission of scalar fields, i.e.\ the peak increases in height by several orders of magnitude and shifts towards higher energies (due to the increasing black hole temperature).  Quantitative results regarding the enhancement with $n$ of both the flux and power fermion emission spectra were obtained by integrating the data in Figure~\ref{fer-dec}; they are shown in Table~\ref{fer-tab}. Once again, the enhancement is substantial, and in fact is even more significant than for scalar emission.
\begin{table}[b]
\begin{center}
\begin{tabular}{|l|c|c|}
\hline
$n$& Flux  & Power\\
\hline
0 & 1 & 1\\
1 & 9.05 & 14.2\\
2 & 27.6 & 59.5\\
3 & 58.2 & 162\\
4 & 103 & 352\\
5 & 163 & 664\\
6 & 240 & 1140\\
7 & 335 & 1830\\
\hline
\end{tabular}
\capbox{Emissivities for fermions on the brane}{Flux and power emissivities for fermions on the brane.\label{fer-tab}} 
\end{center}
\end{table}

\subsection{Spin 1 fields}

In the case of the emission of gauge boson fields, the incoming flux can be
computed from the energy-momentum tensor $T^{\mu\nu}=2 \sigma^{\mu}_{AA'} \sigma^{\nu}_{BB'} \Psi^{AB}\,\bar\Psi^{A'B'}$; the time-radial component is
integrated over a two-dimensional sphere and then evaluated at the horizon 
and at infinity. By making use of the solution for the in-going wave, the 
following expression for the absorption probability of eq.~(\ref{absorption}),
is obtained \cite{Kanti:2002ge,Cvetic:1998ap}:
\begin{equation}
|\hat {\cal A}^{(1)}_j|^2=\frac{1}{r_\text{h}^2}\,
\Biggl|\frac{A_\text{in}^{(\text{h})}}{A_\text{in}^{(\infty)}}\Biggr|^2.
\label{bosons}
\end{equation}

\begin{figure}
\begin{center}
\psfrag{x}[][][1.75]{$\omega r_\text{h}$}
\psfrag{y}[][][1.75]{$\hat{\sigma}^{(1)}_\text{abs}(\om)/\pi r_\text{h}^2$}
\scalebox{0.5}{\rotatebox{-90}{\includegraphics{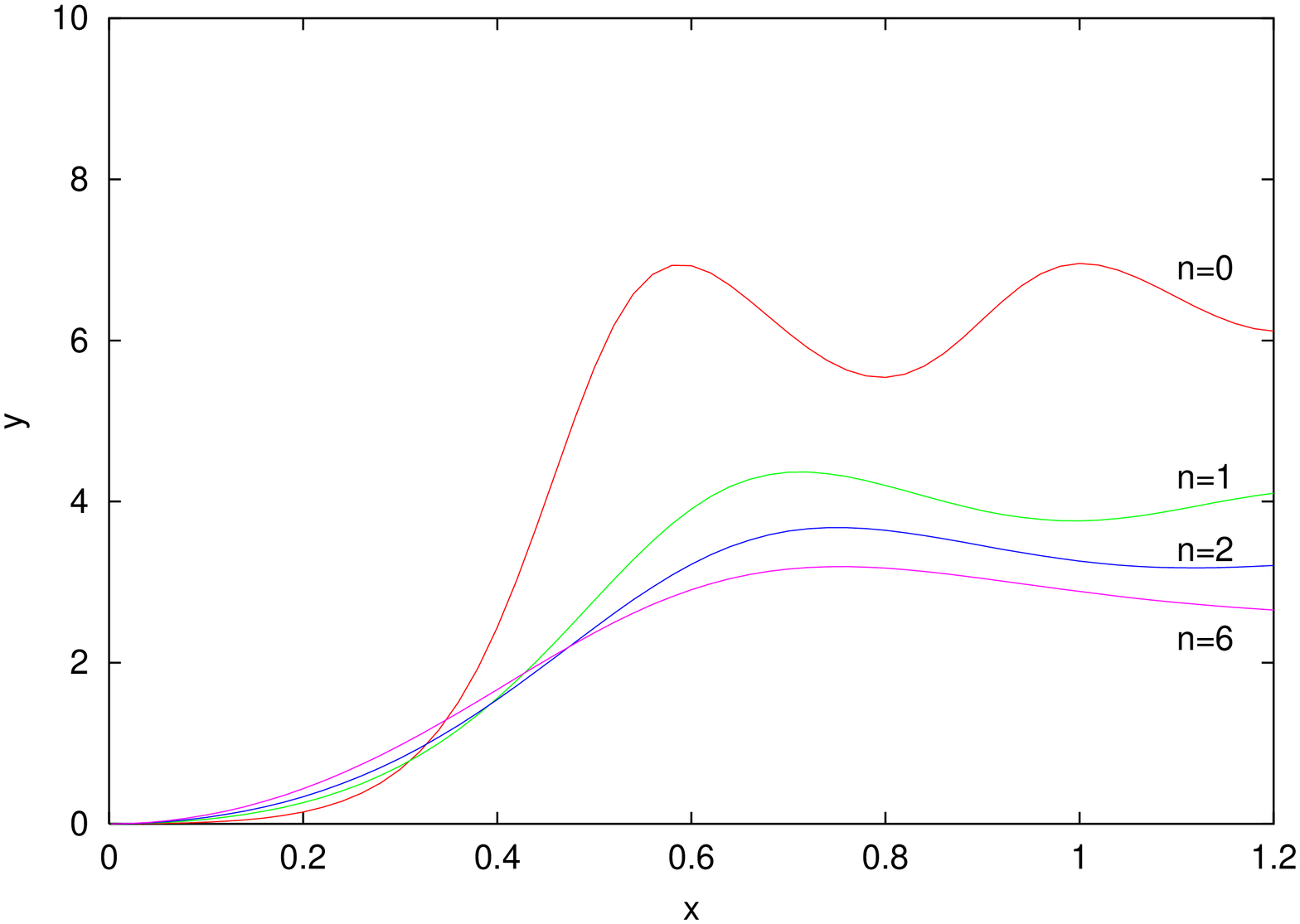}}}
\capbox{Grey-body factors for gauge bosons on the brane}{Grey-body factors for gauge boson emission on the brane from a $(4+n)$D black hole.\label{grey1}}
\end{center}
\end{figure}

\begin{figure}
\begin{center} 
\psfrag{x}[t][b][1.75]{$\omega r_\text{h}$}
\psfrag{y}[][][1.75]{$r_\text{h} d^2 \hat{E}^{(1)}/dtd\om$}
\scalebox{0.5}{\rotatebox{0}{\includegraphics[width=23cm, height=15cm]
{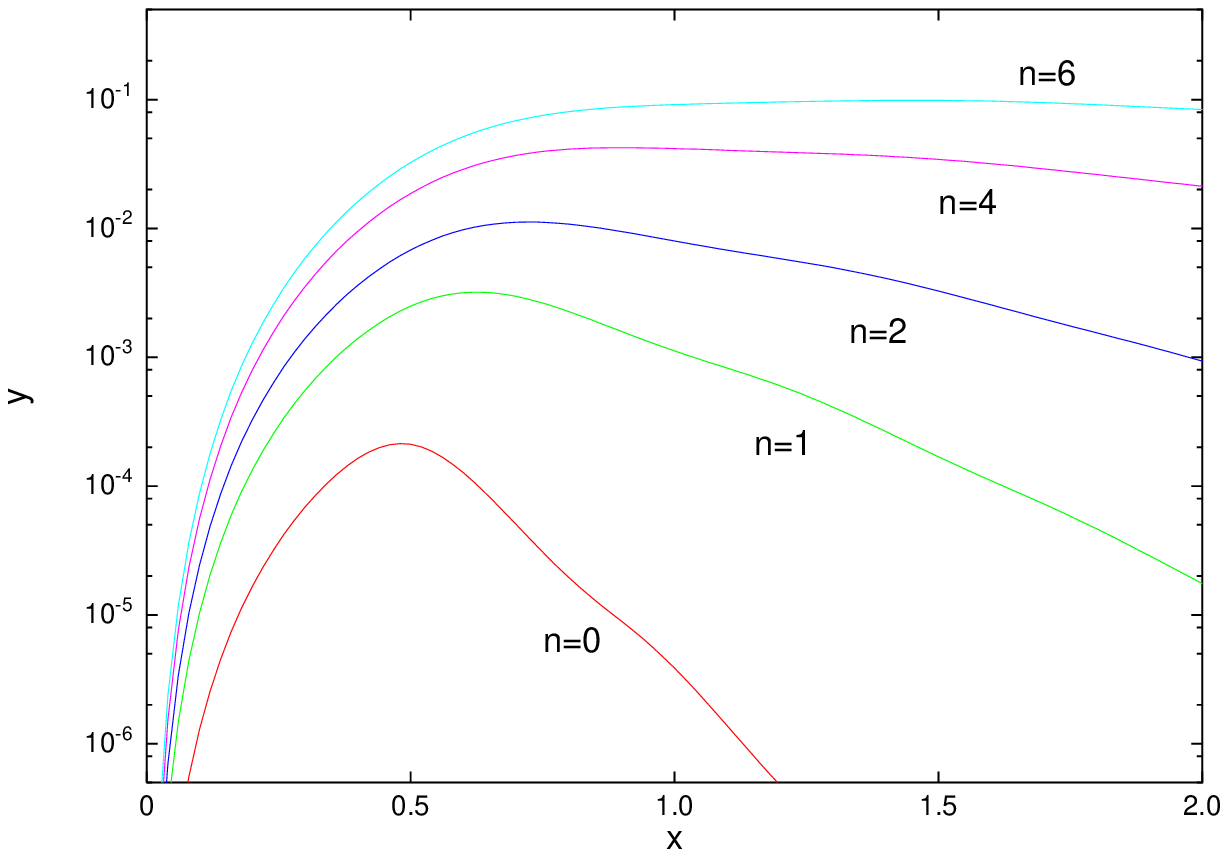}}}
\capbox{Power emission for gauge fields on the brane}{Energy emission rates for gauge fields on the brane from a $(4+n)$D
black hole.\label{gauge-dec}}
\end{center}
\end{figure}

The exact results for the grey-body factors and energy emission rates for 
gauge boson fields are given in Figures~\ref{grey1} and \ref{gauge-dec}
respectively. A distinct feature of the grey-body factor for gauge fields,
previously observed in the 4-dimensional case, is that it vanishes when 
$\omega r_\text{h} \rightarrow 0$. The same behaviour is observed for all values of $n$.  This leads to the suppression of the energy emission rate, in the low-energy regime, compared to those for scalar and fermion fields.  Up to intermediate energies the grey-body factors exhibit enhancement with increasing $n$ as in the case of fermion fields.  An asymptotic behaviour, similar to the previous cases, is observed in the high-energy regime with each grey-body factor assuming, after oscillation, the geometrical optics value (which decreases with increasing $n$). This result establishes the existence of a universal behaviour of all types of particles emitted by black holes at high energies. This behaviour is independent of the particle spin but dependent on the number of extra dimensions projected onto the brane. 
\enlargethispage{-\baselineskip}

Again the total flux and power emissivities for gauge fields on the brane can be obtained for different values of $n$.  The exact results obtained by numerically integrating with respect to energy are given in Table~\ref{gauge-tab}.
As anticipated, the pattern of enhancement with $n$ is also observed 
for the emission of gauge bosons. It is worth noting that 
the relative enhancement observed in this case is the largest among all particle types---this is predominantly due to the large suppression of gauge boson emission in four dimensions.

\begin{table}
\begin{center}
\begin{tabular}{|l|c|c|}
\hline
$n$& Flux  & Power\\
\hline
0 & 1 & 1\\
1 & 19.2 & 27.1\\
2 & 80.6 & 144\\
3 & 204 & 441\\
4 & 403 & 1020\\
5 & 689 & 2000\\
6 & 1070 & 3530\\
7 & 1560 & 5740\\
\hline
\end{tabular}
\capbox{Emissivities for gauge fields on the brane}{Flux and power emissivities for gauge fields on the brane.\label{gauge-tab}} 
\end{center}
\end{table}

\subsection{Relative emissivities for different species}

It is interesting to investigate how the relative numbers of scalars,
fermions and gauge bosons emitted on the brane change
as the number of extra dimensions projected onto the brane varies. In other
words, this means finding out what type of particles the black hole prefers
to emit, for different values of $n$, and what fraction of the total energy
each particular type of particle carries away during emission. 

\begin{figure}
\unitlength1cm
\begin{center}
\psfrag{x}[][][1.25]{$\omega r_\text{h}$}
\psfrag{y}[][][1.25]{$r_\text{h} d^2 \hat{E}^{(0)}/dtd\om$}
\begin{minipage}[t]{3.05in}
\scalebox{0.75}{\rotatebox{0} 
{\includegraphics{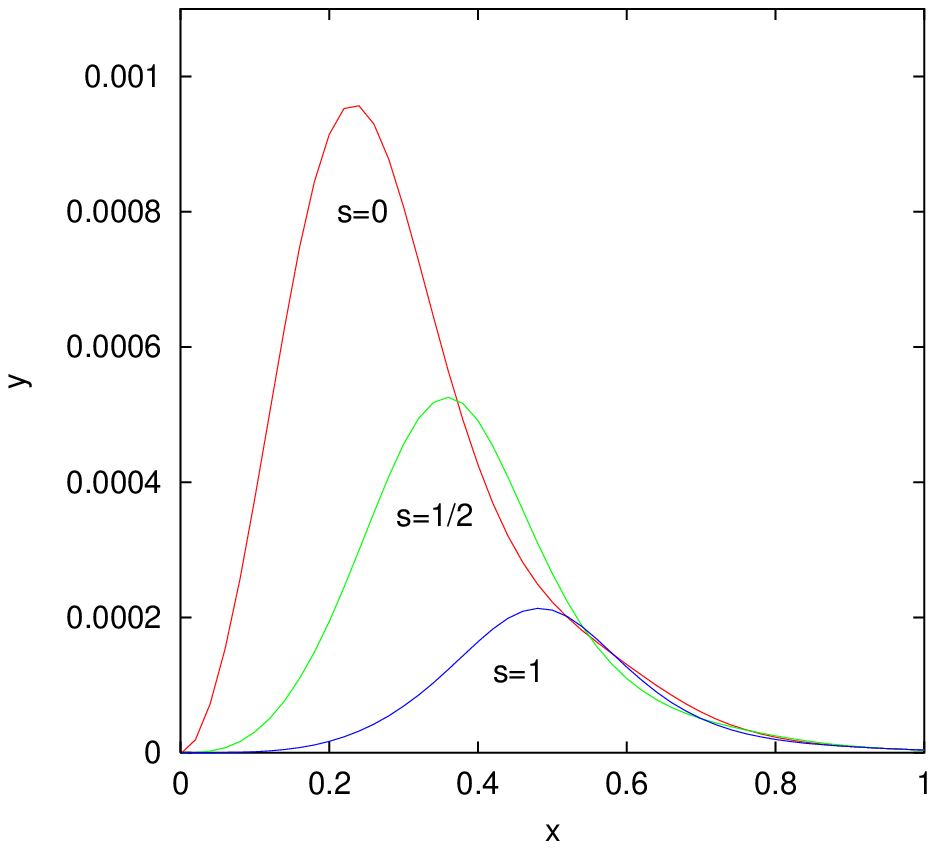}}}
\end{minipage}
\hfill
\begin{minipage}[t]{3.05in}
\scalebox{0.75}{\rotatebox{0}
{\includegraphics{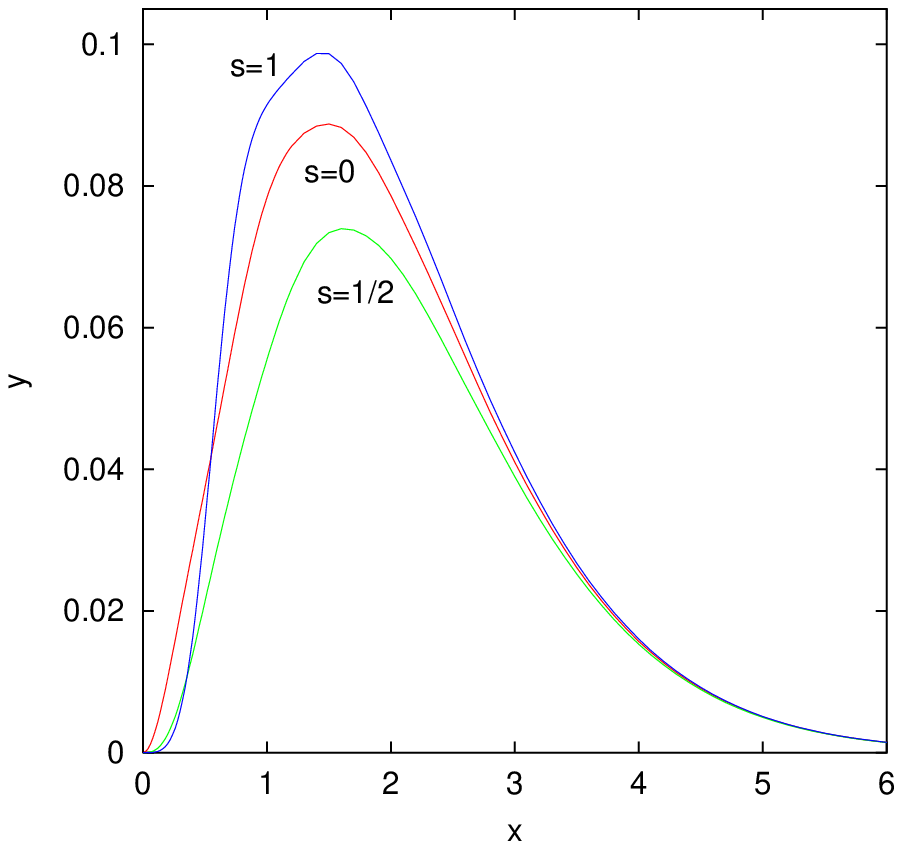}}}
\end{minipage}
\capbox{Power emission on the brane for $n=0$ and $n=6$}{Energy emission rates for the emission of scalars, fermions and gauge bosons on the brane with {\bf(a)} $n=0$ and {\bf(b)} $n=6$.\label{relan6}}
\end{center}
\end{figure}

Comparing the energy emission rates for different types of particles and for 
fixed $n$, gives the qualitative behaviour.   Figures~\ref{relan6}(a) and \ref{relan6}(b) show (with a linear scale) the a power spectra for $n=0$ and $n=6$ respectively; these two figures demonstrate very clearly the effects already discussed, namely the orders-of-magnitude enhancement of the emission rates and the displacement of the peak to higher energies, as $n$ increases. Figure~\ref{relan6}(a) reveals that, in the absence of any extra dimensions, most of the energy of the black hole emitted on the brane is in the form of scalar particles;
the next most important are the fermion fields, and less significant are the 
gauge bosons. As $n$ increases, the emission rates for all species are enhanced but at different rates.  Figure~\ref{relan6}(b) clearly shows that, for a 
large number of extra dimensions, the most effective `channel' during
the emission of brane-localized modes is that of gauge bosons;
the scalar and fermion fields follow second and third respectively.
The changes in the flux spectra are similar as $n$ increases. 

\begin{table}[b]
\begin{minipage}[t]{3.05in}
\begin{center}
\begin{tabular}{|c|c|c|c|}
\hline {\rule[-3mm]{0mm}{8mm} }
$\!\!n$ & $s=0$ & $s=\frac{1}{2}$ & $s=1$ \\
\hline
0&1&0.37&0.11\\
1&1&0.70&0.45\\
2&1&0.77&0.69\\
3&1&0.78&0.83\\
4&1&0.76&0.91\\
5&1&0.74&0.96\\
6&1&0.73&0.99\\
7&1&0.71&1.01\\
\hline 
Black body&1&0.75&1\\
\hline
\end{tabular}
\capbox{Flux emission ratios}{Flux emission ratios.\label{fratios}}
\end{center}
\end{minipage}
\hfill
\begin{minipage}[t]{3.05in}
\begin{center}
\begin{tabular}{|c|c|c|c|}
\hline {\rule[-3mm]{0mm}{8mm} }
$\!\!\!n$ & $s=0$ & $s=\frac{1}{2}$ & $s=1$ \\
\hline
0&1&0.55&0.23\\
1&1&0.87&0.69\\
2&1&0.91&0.91\\
3&1&0.89&1.00\\
4&1&0.87&1.04\\
5&1&0.85&1.06\\
6&1&0.84&1.06\\
7&1&0.82&1.07\\
\hline 
Black body&1&0.87&1\\
\hline
\end{tabular}
\capbox{Power emission ratios}{Power emission ratios.\label{pratios}}
\end{center}
\end{minipage}
\end{table}

The above behaviour can be more helpfully quantified by computing the relative emissivities for scalars, fermions and gauge bosons emitted on the brane by integrating the flux and power emission spectra. The relative emissivities obtained in this way are shown in Tables~\ref{fratios} and \ref{pratios} (they are normalized to the scalar values).  The ratios for $n\geq 1$ are available for
the first time in the literature as a result of this numerical work, while
the $n=0$ results would appear to be the most accurate ones available.  From \cite{MacGibbon:1990zk}, flux and power ratios are found to be 1\,:\,0.36\,:\,0.11 and 1\,:\,0.56\,:\,0.23 respectively, in good agreement with those ratios shown in Tables~\ref{fratios} and \ref{pratios}.  In \cite{Giddings:2001bu} the power ratio for $n=0$ is given as 40\,:\,19\,:\,7.9 i.e.\ 1\,:\,0.48\,:\,0.20 which is in less good agreement.  More careful examination shows that the relative power emitted by the $s=1/2$ and $s=1$ degrees of freedom agrees exactly with Table~\ref{pratios} and so the disagreement seems to come from the scalar value.  This is unsurprising since the scalar value quoted in \cite{Giddings:2001bu} was only estimated from a plot in \cite{Sanchez:1978si}.
\enlargethispage{-\baselineskip}

The entries in these tables reflect the qualitative behaviour
illustrated above for some extreme values of the number of extra dimensions.
For $n=0$, the scalar fields are the type of particle which are
most commonly produced and the ones which carry away most of the energy of
the black hole emitted on the brane; the fermion and gauge fields carry
approximately 1/2 and 1/4, respectively, of the energy emitted in scalar fields, and their fluxes are only 1/3 and 1/10 of the scalar 
flux. For intermediate values of $n$, the fermion and
gauge boson emissivities are considerably enhanced compared to
the scalar one and have become of approximately the same magnitude---e.g.\
for $n=2$ the amount of energy spent by the black hole in the emission
of fermions and gauge bosons is exactly the same, although the net number
of gauge bosons is still subdominant. For large values of $n$, the situation
is reversed: the gauge bosons dominate both flux and power spectra, with
the emission of fermions being the least effective channel both in
terms of number of particles produced and energy emitted.  The reader is reminded that the above results refer to the emission of individual scalar,
fermionic or bosonic degrees of freedom and not to the elementary particles.  However it is easy to combine the ratios in Table~\ref{pratios} with the relevant numbers of degrees of freedom (see Table~\ref{pprobs}) to obtain relative emissivities for different particle types. 

These numerical results confirm that the relative emissivities of different particles produced by small, higher-dimensional black holes depend on the number of extra dimensions projected onto the brane.   Therefore, if Hawking radiation from such objects is detected, this could provide a way of determining the number of extra dimensions existing in nature.

\subsection{Total flux and power emitted}
\label{totfandp}

Whilst it is often only necessary to know the relative emissivities of the fields of different spins, sometimes (for example, when estimating the black hole lifetime) the absolute values are required.  For the case of brane emission from a non-rotating black hole, the flux emitted in a field of spin $s$ can be written as
\begin{equation}
F^{(s)}=\int_0^{\infty} \frac{\Gamma^{(s)} (\omega r_\text{h})}{\exp\left(\omega/T_\text{H}\right)\mp1}\frac{d\omega}{2\pi}=\int_0^{\infty} \frac{\Gamma^{(s)} (\omega r_\text{h})}{\exp\left(\frac{4\pi\omega r_\text{h}}{n+1}\right)\mp1}\frac{d\omega}{2\pi}\,.
\label{totflux}
\end{equation}
This expression is obtained by integrating equation~(\ref{fdecay-brane}) using the definition
\begin{equation}
\Gamma^{(s)}(\omega r_\text{h})=\sum_{\ell} (2\ell +1) |\hat {\cal A}^{(s)}_\ell|^2.
\end{equation}
Since the integrand in eq.~(\ref{totflux}) is entirely a function of $\omega r_\text{h}$, the substitution $x=\omega r_\text{h}$ can be used to express $F^{(s)}$ as
\begin{equation}
F^{(s)}=\frac{1}{r_\text{h}}\int_0^{\infty}\frac{\Gamma^{(s)}(x)}{\exp\left(\frac{4\pi x}{n+1}\right)\mp1}\frac{dx}{2\pi}\,,
\end{equation}
where the only dependence on the radius (or equivalently the mass) of the black hole is contained in the factor outside the integral.

A similar procedure, starting from equation~(\ref{4power}), shows the power emitted in a field of spin $s$ to be
\begin{equation}
P^{(s)}=\frac{1}{r_\text{h}^2}\int_0^{\infty}\frac{x\Gamma^{(s)}(x)}{\exp\left(\frac{4\pi x}{n+1}\right)\mp1}\frac{dx}{2\pi}\,.
\end{equation}

\enlargethispage{-\baselineskip}
These results of these integrations are given in Tables~\ref{fvalues} and \ref{pvalues}.  The values in these tables can be combined with the total numbers of degrees of freedom for each spin.  These are given in the last row
of Table~\ref{pprobs} and are 4, 90 and 24 for $s=0$, 1/2 and 1 respectively.
Therefore the total flux $F$ and power $P$ emitted by black holes of different numbers of dimensions can be calculated---the results are shown in Table~\ref{totfp}.

\begin{table}
\def\arraystretch{1.1}
\begin{center}
\begin{tabular}{|l|c|c|c|}
\hline
$n$& $r_\text{h} F^{(0)}$  & $r_\text{h} F^{(1/2)}$ & $r_\text{h} F^{(1)}$ \\
\hline
0 & 0.00133 & 0.000486 & 0.000148\\
1 & 0.00631 & 0.00439 & 0.00283\\ 
2 & 0.0173 & 0.0134 & 0.0119\\
3 & 0.0364 & 0.0283 & 0.0301\\ 
4 & 0.0655 & 0.0499 & 0.0596\\
5 & 0.106 & 0.0789 & 0.102\\
6 & 0.160 & 0.116 & 0.159\\
7 & 0.229 & 0.163 & 0.231\\
\hline
\end{tabular}
\capbox{Flux for different spin fields}{Total flux emitted in fields of different spins.\label{fvalues}} 
\end{center}
\def\arraystretch{1.0}
\end{table}

\begin{table}
\def\arraystretch{1.1}
\begin{center}
\begin{tabular}{|l|c|c|c|}
\hline
$n$& $r_\text{h}^2 P^{(0)}$  & $r_\text{h}^2 P^{(1/2)}$ & $r_\text{h}^2 P^{(1)}$ \\
\hline
0 & 0.000298 & 0.000164 & 6.72$\times 10^{-5}$\\
1 & 0.00266 & 0.00232 & 0.00182\\
2 & 0.0107 & 0.00973 & 0.00971\\
3 & 0.0297 & 0.0265 & 0.0296\\
4 & 0.0661 & 0.0575 & 0.0686\\
5 & 0.128 & 0.109 & 0.135\\
6 & 0.223 & 0.187 & 0.237\\
7 & 0.362 & 0.299 & 0.386\\
\hline
\end{tabular}
\capbox{Power for different spin fields}{Total power emitted in fields of different spins.\label{pvalues}} 
\end{center}
\def\arraystretch{1.0}
\end{table}

\begin{table}
\def\arraystretch{1.1}
\begin{center}
\begin{tabular}{|l|c|c|}
\hline
$n$& $r_\text{h} F$  & $r_\text{h}^2 P$\\
\hline
0 & 0.0526 & 0.0175\\
1 & 0.489 & 0.263\\
2 & 1.56 & 1.15\\
3 & 3.41 & 3.21\\
4 & 6.18 & 7.09\\
5 & 9.98 & 13.5\\
6 & 14.9 & 23.4\\
7 & 21.1 & 37.6\\
\hline
\end{tabular}
\capbox{Total flux and power emitted}{Total flux and power emitted from black holes of different numbers of dimensions.\label{totfp}} 
\end{center}
\def\arraystretch{1.0}
\end{table}

It is possible to use these values for the total flux emitted from extra-dimensional black holes to comment on the usual assumption that black hole decays can
be considered as quasi-stationary.  The simplest way of trying to verify this is to compare the typical time between emissions (given by $F^{-1}$) with $r_\text{h}$ (the time for light to cross the black hole radius in the natural units being used).  To be sure that the quasi-stationary approach is valid, we would require $F^{-1}\gg r_\text{h}$ or equivalently $r_\text{h} F \ll 1$.  From Table~\ref{totfp}, this can be seen to be true only for the cases $n=0$ and $n=1$.

\enlargethispage{-\baselineskip}
Another requirement for the validity of a semi-classical description of black hole production and decay is that the lifetime $\tau \gg 1/M_\text{BH}$ so that the black hole is a well-defined resonance.  This assumption can also be tested here by using the values of $P$ in Table~\ref{totfp}.  The relationship
\begin{equation}
P=-\frac{dM_\text{BH}}{dt}=\frac{p}{r_\text{h}^2}\,,
\end{equation}
can be integrated as follows:
\begin{equation}
p\int_0^\tau dt = - \int_{M_\text{BH}}^0 r_\text{h}^2 dM_\text{BH} = \frac{1}{\pi M_{\text{P}(4+n)}^2}\left(\frac{8 \Gamma\left(\frac{n+3}{2}\right)}{n+2}\right)^{\frac{2}{n+1}} \int_0^{M_\text{BH}} \left(\frac{M_\text{BH}}{M_{\text{P}(4+n)}}\right)^{\frac{2}{n+1}} dM_\text{BH}\,. 
\end{equation}
In the above, the final Planck phase and any complications caused by kinematically forbidden emissions have been ignored.  Proceeding with the integration gives
\begin{equation}
\tau M_\text{BH} = \frac{1}{\pi p}\left(\frac{8 \Gamma\left(\frac{n+3}{2}\right)}{n+2}\right)^{\frac{2}{n+1}} \frac{n+1}{n+3} \left(\frac{M_\text{BH}}{M_{\text{P}(4+n)}}\right)^{\frac{2(n+2)}{n+1}}.
\end{equation}
Values of $\tau M_\text{BH}$ for different values of $n$ are shown in Table~\ref{taum} both in the case $M_\text{BH}=5M_{\text{P}(4+n)}$ and in the case $M_\text{BH}=10M_{\text{P}(4+n)}$. It should be noted that switching to convention `a' for the definition of $M_{\text{P}(4+n)}$ the high-$n$ values in the table would be significantly altered (for example, the $n=7$ values would be multiplied by a factor of 13.2).  The `long'-lifetime requirement is clearly different to the quasi-stationary issue discussed above since it depends on the initial mass of the black hole.  However given the limits on $M_{\text{P}(4+n)}$ and the energy available at the LHC, it would again seem that this requirement may not be satisfied for the higher values of $n$.

\begin{table}
\def\arraystretch{1.1}
\begin{center}
\begin{tabular}{|c|c|c|}
\hline
& \multicolumn{2}{c|}{$\tau M_\text{BH}$} \\
\cline{2-3}
$n$& $M_\text{BH}=5M_{\text{P}(4+n)}$  & $M_\text{BH}=10M_{\text{P}(4+n)}$\\
\hline
0 & 47500 & 761000\\
1 & 202 & 1610\\
2 & 23.3 & 148\\
3 & 6.60 & 37.4\\
4 & 2.77 & 14.6\\
5 & 1.43 & 7.23\\
6 & 0.846 & 4.12\\
7 & 0.544 & 2.59\\
\hline
\end{tabular}
\capbox{Values of $\tau M_\text{BH}$}{Values of $\tau M_\text{BH}$ for different values of $n$.\label{taum}} 
\end{center}
\def\arraystretch{1.0}
\end{table}

\enlargethispage{\baselineskip}
The results presented in Table~\ref{taum} differ significantly from equivalent results which can be obtained from an expression given in \cite{Dimopoulos:2001en}.  Partly this is due to inclusion of the grey-body factors here, but this does not account for order-of-magnitude differences.  It would appear that the lifetime is incorrect in \cite{Dimopoulos:2001en} because of the failure to take all the degrees of freedom of the emitted particles into account.

The numerical results presented in Table~\ref{totfp} can also be used to calculate the average particle multiplicity, again completely ignoring the complications of the Planck phase and kinematic constraints on the decay.  Defining $r_\text{h} F=f$, the average particle multiplicity can be expressed as
\begin{equation}
\langle N\rangle\,=\,\frac{f}{p}\int_0^{M_\text{BH}} r_\text{h} \, dM_\text{BH} =  \frac{1}{\sqrt{\pi} M_{\text{P}(4+n)}}\left(\frac{8 \Gamma\left(\frac{n+3}{2}\right)}{n+2}\right)^{\frac{1}{n+1}}\!\!\int_0^{M_\text{BH}} \left(\frac{M_\text{BH}}{M_{\text{P}(4+n)}}\right)^{\frac{1}{n+1}} dM_\text{BH}\,.
\end{equation}
Performing the integration gives
\begin{equation}
\langle N\rangle\,=\,\frac{f}{p}\frac{1}{\sqrt{\pi}}\left(\frac{8 \Gamma\left(\frac{n+3}{2}\right)}{n+2}\right)^{\frac{1}{n+1}} \frac{n+1}{n+2} \left(\frac{M_\text{BH}}{M_{\text{P}(4+n)}}\right)^{\frac{n+2}{n+1}}=\frac{f}{p}\frac{n+1}{4\pi}S_\text{BH}\,,
\label{avn}
\end{equation}
where the final identification has been made by referring back to equation~(\ref{bhent}).  Table~\ref{avntab} shows the average multiplicities obtained for different values of $n$ along with the (initial) entropy $S_\text{BH}$.  The average multiplicity values given improve on those in \cite{Dimopoulos:2001hw} (where $\langle N\rangle\sim M_\text{BH}/2\,T_\text{H}$ was assumed based on the constant temperature approximation) and \cite{Cavaglia:2003hg} (where a more careful approach was taken but still without the full numerical grey-body factors).

\begin{table}
\def\arraystretch{1.1}
\begin{center}
\begin{tabular}{|c|c|c|c|c|}
\hline
& \multicolumn{2}{c|}{$\langle N\rangle$} & \multicolumn{2}{c|}{$S_\text{BH}$}\\
\cline{2-5}
$n$& $M_\text{BH}=5M_{\text{P}(4+n)}$  & $M_\text{BH}=10M_{\text{P}(4+n)}$ & $M_\text{BH}=5M_{\text{P}(4+n)}$ & $M_\text{BH}=10M_{\text{P}(4+n)}$\\
\hline
0 & 75.3 & 301  & 314 & 1260 \\
1 & 12.8 & 36.1 & 43.1 & 122\\
2 & 6.82 & 17.2 & 21.0 & 52.9\\
3 & 4.78 & 11.4 & 14.2 & 33.7\\
4 & 3.80 & 8.75 & 11.0 & 25.2\\
5 & 3.22 & 7.23 & 9.13 & 20.5\\
6 & 2.80 & 6.20 & 7.92 & 17.5\\
7 & 2.52 & 5.50 & 7.06 & 15.4\\
\hline
\end{tabular}
\capbox{Values of $\langle N\rangle$ and $S_\text{BH}$}{Values of $\langle N\rangle$ and $S_\text{BH}$ (calculated using eqs.~(\ref{avn}) and (\ref{bhent}) respectively) for different values of $n$.\label{avntab}} 
\end{center}
\def\arraystretch{1.0}
\end{table}

The average multiplicities in the first column clearly do not satisfy $\langle N\rangle\gg 1$ for $n>3$ which means that a significant amount of the decay will be affected by kinematic and Planck phase considerations.  However using the alternative definition of $M_{\text{P}(4+n)}$ these results would again be modified (for $n=7$ they would be multiplied by a factor of 3.64).  There is, of course, no fundamental difference between these conventions---setting $M_{\text{P}(4+n)}=1$~TeV in convention `a' simply corresponds to a lower value of the Planck mass in convention `d' and hence for a fixed black hole mass the Hawking temperature is lower and the average multiplicity higher.  

The tabulated entropies allow comparison with the requirement that $1/\sqrt{S_\text{BH}}$ should be much less than unity if statistical fluctuations of the number of micro-canonical degrees of freedom are to be small (as is necessary for a semi-classical description of the black hole decay to be valid).


\section{Emission in the bulk}
\label{embulk}

An important question regarding the emission of particles by
higher-dimensional black holes is how much of this energy is radiated
onto the brane and how much is lost in the bulk. In the former case,
the emitted particles are zero-mode gravitons and scalar fields as 
well as Standard Model fermions and gauge bosons, while in the latter
case all emitted energy is in the form of massive Kaluza-Klein gravitons
and, as discussed in \secref{gbfintro}, possibly also scalar fields. In \cite{Emparan:2000rs}, it was shown that the whole
tower of KK excitations of a given particle carries approximately the
same amount of energy as a massless particle emitted on the
brane. Combining this result with the observation that many more types of particles
live on the brane than in the bulk, it was concluded that most of the
energy of the black hole goes into brane modes. The results obtained
in \cite{Emparan:2000rs} were only approximate since the dependence of the grey-body factor on the energy of the emitted particle was ignored and the
geometric expression for the area of the horizon (valid in the high-energy regime) was used instead.  

In order to provide an accurate answer to the question of how much energy
is emitted into the bulk compared to on the brane, it is necessary to take into account the dependence of the grey-body factor on both energy and number of extra dimensions.   There is a similar discussion in \cite{Cavaglia:2003hg} but it is incomplete because full numerical results for the grey-body factor in all energy regimes were not available.  In this section the emission of scalar modes in the bulk is thoroughly investigated.  Exact numerical results are produced for the grey-body factors and energy emission rates in terms of the energy and number of extra dimensions.  This allows the questions mentioned above to be addressed by calculating the total bulk-to-brane relative emissivities for different values of $n$.

\subsection{Grey-body factors and emission rates}

The analysis in this section is relevant for gravitons and possibly scalar fields, and requires knowledge of the solutions of the
corresponding equations of motion in the bulk.  Only the case of scalar fields is considered here since the bulk equation of motion is known---the emission of bulk scalar modes was previously studied analytically, in the low-energy regime, in \cite{Kanti:2002nr}.\footnote{After this work was completed, some plots of grey-body factors for bulk scalars were discovered in Figure 1 of \cite{Frolov:2002as}.  It is unclear how these were calculated and only the $\ell=0$ contributions for $n=$1--3 are shown; however, they provided a useful cross-check on some of the results presented here.}

A scalar field propagating in the background of a higher-dimensional,
non-rotating (Schwarzschild-like) black hole, with line-element given by eq.~(\ref{metric-D}), satisfies the following equation of motion \cite{Kanti:2002nr}
\begin{equation}
\frac{h(r)}{r^{n+2}}\,\frac{d \,}{dr}\,
\biggl[\,h(r)\,r^{n+2}\,\frac{d R}{dr}\,\biggr] +
\biggl[\,\om^2 - \frac{h(r)}{r^2}\,\ell\,(\ell+n+1)\,\biggr] R =0 \, .
\label{scalareqn}
\end{equation}
This radial equation was obtained by using a separable solution similar to equation~(\ref{sepsoln}) but with ${}_sS^m_{\ell}(\theta)$ replaced by $S^m_{\ell}(\theta_1,\theta_2,\ldots,\theta_n)$. The functions $e^{im\varphi}\,S^m_{\ell}(\theta_1,\theta_2,\ldots,\theta_n)$ can be written as $\tilde{Y}^m_{\ell}(\Omega_{2+n})$\,; these are generalizations of the usual spherical harmonic functions to the case of $(3+n)$ spatial dimensions \cite{muller} and account for the $\ell(\ell+n+1)$ term in eq.~(\ref{scalareqn}).  

As for the emission of particles on the brane, the determination of the grey-body factor for bulk emission requires the above equation to be solved over the whole radial domain. The exact solution for the radial function must interpolate between the near-horizon and far-field asymptotic solutions, given by
\begin{equation}
\label{near-bulk}
R^{(\text{h})}=A_\text{in}^{(\text{h})}\,e^{-i\omega r^{*}}+ A_{out }^{(\text{h})}\,e^{i\omega r^{*}},
\end{equation}
and
\begin{equation}
\label{far-bulk}
R^{(\infty)}=A_\text{in}^{(\infty)}\,\frac{e^{-i\omega r}}{\sqrt{r^{n+2}}}+
A_\text{out}^{(\infty)}\,\frac{e^{i\omega r}}{\sqrt{r^{n+2}}}\,,
\end{equation}
respectively.  Again the boundary condition that no out-going solution should exist near the horizon of the black hole is imposed which means that $A_\text{out}^{(\text{h})}=0$. The solution at infinity comprises, as usual, both in-going and out-going modes. 

The absorption probability $|\tilde {\cal A}_\ell|^2$ may
then be calculated either by using eq.~(\ref{scalars}) or from the ratio $|A_\text{in}^{(\text{h})}/A_\text{in}^{(\infty)}|^2$ (this latter approach is found to be more numerically stable for low energies and/or larger values of $n$).  A tilde will now denote bulk quantities, as opposed to brane quantities which carry a hat.  The grey-body factor $\tilde \si_\ell(\om)$ may be determined by using eq.~(\ref{greydef}). The dimensionality of the grey-body factor changes as the number of extra dimensions varies; therefore, in order to be able to compare its values for different $n$, it is normalized to the horizon area of a $(4+n)$-dimensional black hole.  The discussion of the normalization procedure and the subsequent comments on the geometrical optics limit are due to Kanti \cite{Kantiprivate,Harris:2003eg}.

First equation~(\ref{greydef}) is re-written in the form
\begin{equation}
\tilde \si_\ell(\om) = \frac{2^{n}}{\pi}\,
\Ga\Bigl[\frac{n+3}{2}\Bigr]^2\,
\frac{\tilde A_\text{h}}{(\om r_\text{h})^{n+2}}\, \tilde N_\ell\,
|\tilde {\cal A}_\ell|^2,
\label{greyb}
\end{equation}
where $\tilde N_\ell$ is the multiplicity of states corresponding to the
same partial wave $\ell$.  For a $(4+n)$-dimensional space-time it is given by
\begin{equation}
\tilde N_\ell= \frac{(2\ell+n+1)\,(\ell+n)!}{\ell! \,(n+1)!}\,,
\label{bulk-mult}
\end{equation}
and the horizon area in the bulk is defined as
\begin{equation}
\begin{split}
\tilde A_\text{h} =&\, 
r_\text{h}^{n+2}\,\int_0^{2 \pi} \,d \varphi \,\prod_{k=1}^{n+1}\,
\int_0^\pi\,\sin^k\theta_{n+1}\,
\,d(\sin\theta_{n+1}) \\[3mm]
=&\, r_\text{h}^{n+2}\,(2\pi)\,\prod_{k=1}^{n+1}\,\sqrt{\pi}\,\,\frac{\Ga[(k+1)/2]}
{\Ga[(k+2)/2]}\\[3mm]
=& 
\,r_\text{h}^{n+2}\,(2\pi)\,\pi^{(n+1)/2}\,\Ga\Bigl[\frac{n+3}{2}\Bigr]^{-1}.
\end{split}
\end{equation}

Equation~(\ref{greyb}) allows the computation of the grey-body factor's low-energy limit once the corresponding expression for the absorption
coefficient is determined. Analytic results for $\tilde {\cal A}_\ell$
were derived in \cite{Kanti:2002nr} by solving eq.~(\ref{scalareqn}) in the two
asymptotic regimes (near-horizon and far-field) and matching them in an intermediate zone. It was found that the low-energy expression of the $\ell=0$ absorption coefficient has the form
\begin{equation}
|\tilde {\cal A}_0|^2 = \biggl(\frac{\omega r_\text{h}}{2}\biggl)^{n+2}
\,\frac{4 \pi}{\Gamma[(n+3)/2]^2} + \ldots \,,
\end{equation}
where the ellipsis denotes higher-order terms in the $\omega r_\text{h}$ power series. These terms, as well as the corresponding expressions for higher partial waves, vanish quickly in the limit $\omega r_\text{h} \rightarrow 0$, leaving the above term as the dominant one. Substituting
into eq.~(\ref{greyb}) it is clear that, in the low-energy regime, the grey-body factor is given by the area $\tilde A_\text{h}$ of the black hole horizon. This behaviour is similar to the 4-dimensional case except that the area of the horizon now changes with $n$. 

In the high-energy regime, it is anticipated that an equivalent of the geometrical optics limit discussed in \secref{numscalar} will be recovered. In four dimensions, the low- and high-energy asymptotic limits are $4\pi r_\text{h}^2$ and $\pi r_\text{c}^2$ respectively.  This has led to the na\"{\i}ve generalization that, in an arbitrary number of dimensions, the high-energy expression for the grey-body factor will be approximately $\Omega_{n+2}\,r_\text{c}^{n+2}/4$, where $\Omega_{n+2}$ is defined as in eq.~(\ref{omegapdef}). It will shortly be shown that this is in
fact an over-estimate of the high-energy limit. 

As in four dimensions it is assumed, for high energy particles, that the grey-body factor
becomes equal to the area of an absorptive body of radius $r_\text{c}$, projected on a plane parallel to the orbit of the moving particle
\cite{misner}. According to ref. \cite{Emparan:2000rs}, the value of the effective radius $r_\text{c}$ remains the same for both bulk and brane particles and
is given by eq.~(\ref{effective}). The area of the absorptive body depends
strongly on the dimensionality of space-time and its calculation requires
one of the azimuthal angles to be set to $\pi/2$. A careful calculation
reveals that the `projected' area is given by
\begin{equation}
\tilde A_\text{p} = \frac{2 \pi}{(n+2)}\,\frac{\pi^{n/2}}{\Gamma[(n+2)/2]}\,
r_\text{c}^{n+2} = \frac{1}{n+2}\,\Omega_{n+1}\,r_\text{c}^{n+2}\,.
\end{equation}

The above relation reduces to the usual 4-dimensional result
($\tilde A_\text{p} = \pi r_\text{c}^2$) for $n=0$ but, compared to the \,na\"{\i}ve $\Omega_{n+2}\,r_\text{c}^{n+2}/4$, leads to values reduced by 50\% for higher values of $n$. Assuming that the grey-body factor at high
energies becomes equal to the absorptive area $\tilde A_\text{p}$, it can be explicitly written as
\begin{equation}
\begin{split}
\tilde \si_\text{g} = & \,\frac{1}{n+2}\,\frac{\Omega_{n+1}}{\Omega_{n+2}}\,
\biggl(\frac{r_\text{c}}{r_\text{h}}\biggr)^{n+2}\,\tilde A_\text{h} \\[3mm]
 = &\,\frac{1}{\sqrt{\pi}\,(n+2)}\,\frac{\Gamma[(n+3)/2]}{\Gamma[(n+2)/2]}\,
\biggl(\frac{n+3}{2}\biggr)^{(n+2)/(n+1)}\,
\biggl(\frac{n+3}{n+1}\biggr)^{(n+2)/2}\,\tilde A_\text{h}\,.
\label{high}
\end{split}
\end{equation}
In the above the same normalization is used as in the low-energy regime---that is, the normalization is in terms of $\tilde{A}_\text{h}$, the area of the $(4+n)$-dimensional horizon.  The values predicted by eq.~(\ref{high}) are tabulated in Table~\ref{golim}, along with the more na\"{\i}ve $\Omega_{n+2}\,r_\text{c}^{n+2}/4$ prediction.

\begin{table}
\def\arraystretch{1.1}
\begin{center}
\begin{tabular}{|l|c|c|}
\hline
$n$ & $\phantom{\Bigl.}\tilde \si_\text{g}/\tilde{A}_\text{h}\phantom{\Bigr.}$ & $\Omega_{n+2}\,r_\text{c}^{n+2}/4\tilde{A}_\text{h}$\\
\hline
0 & 1.69 & 1.69\\
1 & 1.70 & 2\\
2 & 1.77 & 2.36\\
3 & 1.85 & 2.72\\
4 & 1.93 & 3.09\\
5 & 2.01 & 3.45\\
6 & 2.08 & 3.81\\
7 & 2.16 & 4.17\\
\hline
\end{tabular}
\capbox{High-energy limits of grey-body factors for bulk emission}{High-energy limits of grey-body factors for bulk emission, given in units of the $(4+n)$-dimensional area $\tilde{A}_\text{h}$.\label{golim}} 
\end{center}
\def\arraystretch{1.0}
\end{table}

\begin{figure}
\begin{center}
\psfrag{x}[t][b][1.75]{$\omega r_\text{h}$}
\psfrag{y}[][][1.75]{$\tilde{\sigma}^{(0)}_\text{abs}(\om)/\tilde{A}_\text{h}$}
\scalebox{0.5}{\rotatebox{0}{\includegraphics[width=22cm, height=15.4cm]
{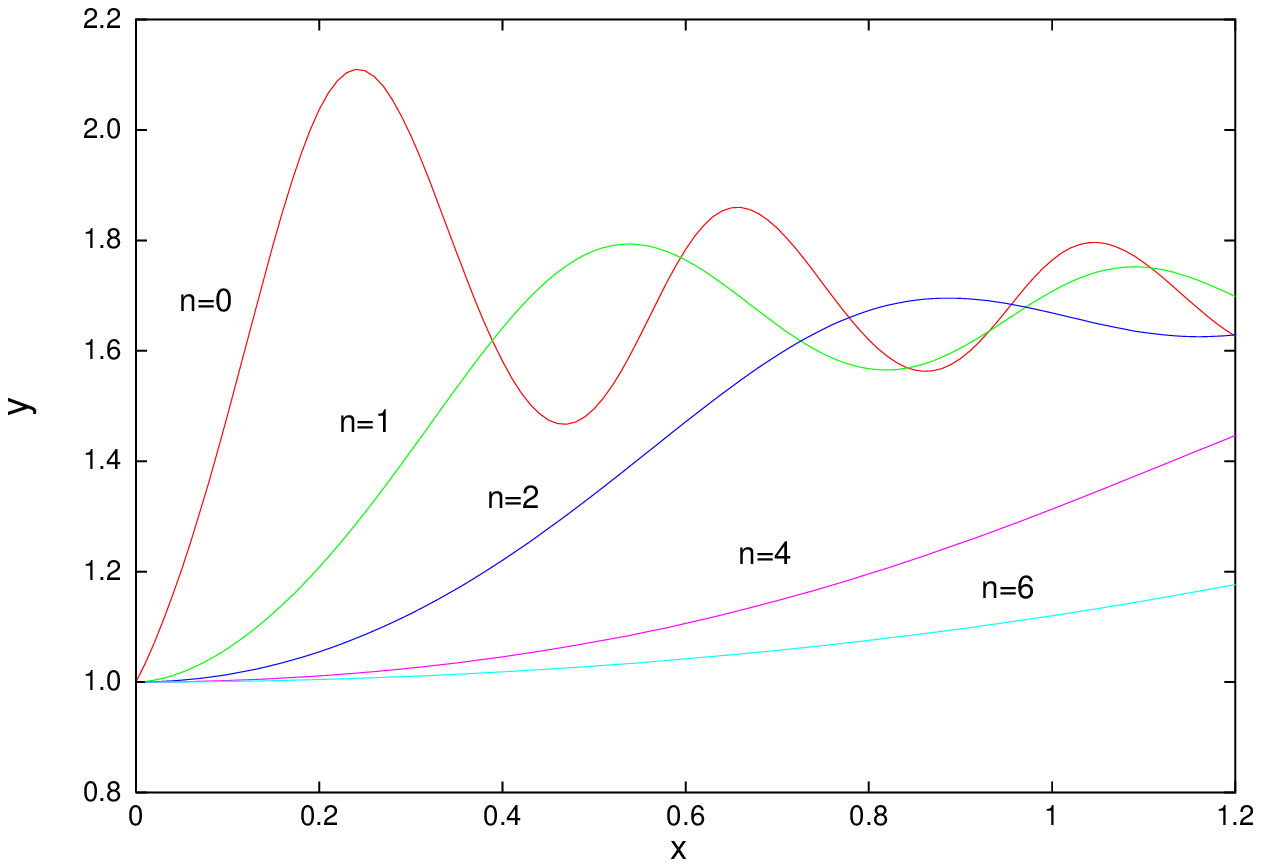}}}
\capbox{Grey-body factors for scalars in the bulk}{Grey-body factors for scalar emission in the bulk from a $(4+n)$D
black hole.\label{grey0-bulk}}
\end{center}
\end{figure}

Turning now to the numerical analysis, the grey-body factors can be found by using eq.~(\ref{greyb}) and the exact numerical results for the absorption
coefficients.  Their behaviour is shown in Figure~\ref{grey0-bulk}. As
it was anticipated after the above discussion, the normalized grey-body 
factors, in the low-energy regime, tend to unity for all values of $n$
as each one adopts the value of the black hole horizon area
to which it has been normalized. As for the emission of 
scalar fields on the brane, the grey-body factors are suppressed with increasing $n$ in the low-energy regime, start oscillating at intermediate energies and then tend to their asymptotic high-energy limits. A simple numerical analysis shows that the na\"{\i}ve expression
$\Omega_{n+2}\,r_\text{c}^{n+2}/4$ fails to describe the high-energy asymptotic
limits for all values of $n$ larger than zero.  In contrast, eq.~(\ref{high}) gives asymptotic values which are verified by the numerical results.  As in the brane emission case, it is found that relatively large values of $\omega r_\text{h}$ are required before the asymptotic values are approached for the higher values of $n$ (this can be seen in Figure~\ref{grey0-bulk} in which the $n=4$ and 6 curves clearly do not attain the asymptotic values given in Table~\ref{golim}).

\begin{figure}
\begin{center}
\psfrag{x}[t][b][1.75]{$\omega r_\text{h}$}
\psfrag{y}[][][1.75]{$r_\text{h} d^2\tilde{E}^{(0)}/dtd\om$}
\scalebox{0.5}{\rotatebox{0}{\includegraphics[width=23cm, height=15cm]
{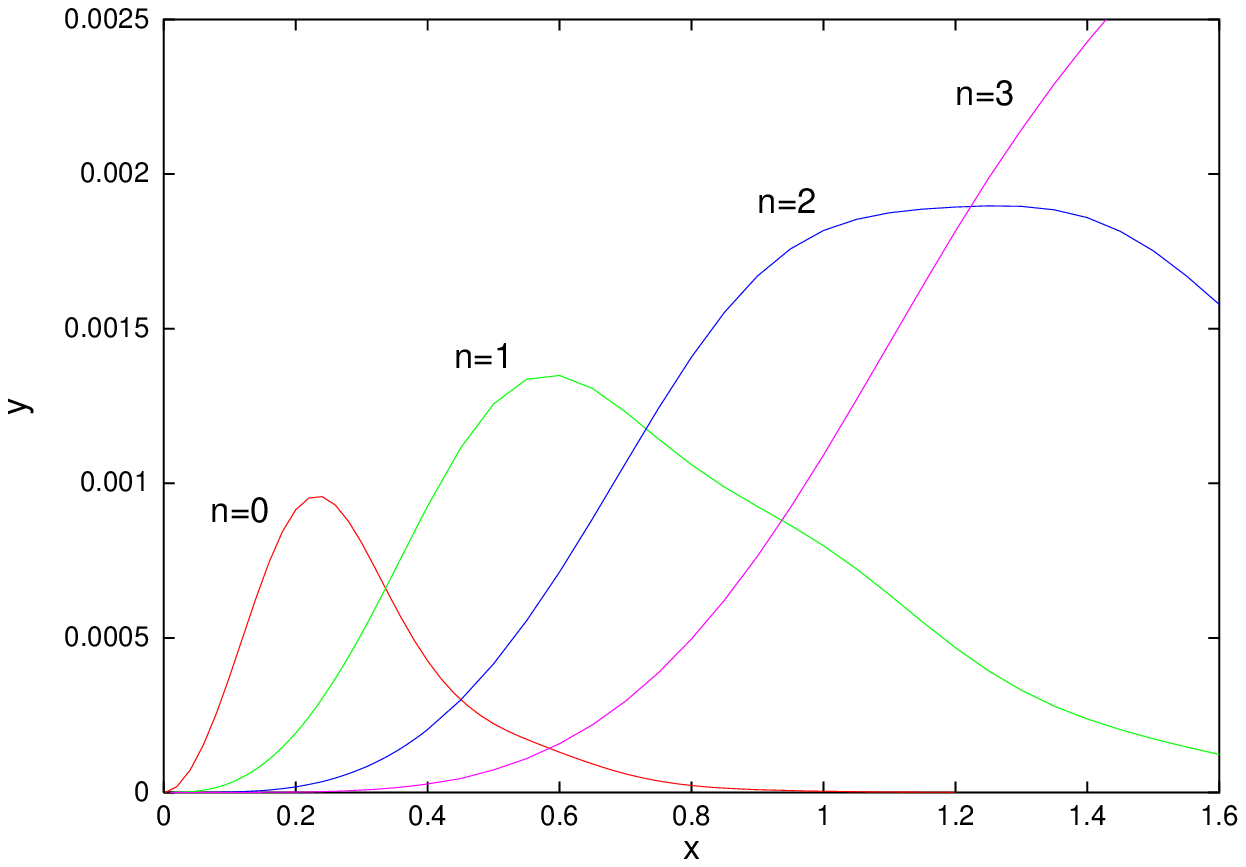}}}
\capbox{Power emission for scalars in the bulk}{Energy emission rates for scalar fields in the bulk from a $(4+n)$D
black hole.\label{rate-bulk}}
\end{center}
\end{figure}

In general, the suppression of the grey-body factor for bulk emission at low
energies is milder than the one for brane emission. However this does not
lead to higher emission rates for bulk modes compared to those for brane
modes: the integration over the phase-space in eq.~(\ref{4power})
involves powers of $\om r_\text{h}$ which cause an increasingly
suppressive effect in the low-energy regime as $n$ increases. Nevertheless,
the increase in the temperature of the black hole (still given by $T_\text{H}=(n+1)/4 \pi r_\text{h}$) eventually overcomes the decrease in the grey-body factor and causes the enhancement of the emission rate with $n$ at
high energies (as well as shifting the spectrum peak to higher energies).  The behaviour of the differential energy emission rates as a function of the energy parameter $\om r_\text{h}$ is shown in Figure~\ref{rate-bulk} for some indicative values of $n$.  Comparing these numerical results with the analytic results of ref. \cite{Kanti:2002nr} shows that the earlier results were reasonably successful in describing the low-energy behaviour of both the grey-body factors and the energy emission rates for scalar fields in the bulk.

\subsection{Bulk-to-brane relative emissivities}

\enlargethispage{-3\baselineskip}
The aim of this section is to perform an analysis which provides an answer
to the question of the relative bulk-to-brane emissivity.  This requires the differential energy emission rates in the bulk and on the brane to be evaluated; the two quantities are then compared for different numbers of extra dimensions. 

Equation~(\ref{4power}) for energy emission in the bulk may alternatively be
written, in terms of the absorption coefficient, as
\begin{equation}
\frac{d \tilde E(\om)}{dt} = 
\sum_{\ell} \tilde N_\ell\, |\tilde {\cal A}_\ell|^2\,
\frac{\om}{\exp\left(\om/T_\text{H}\right) - 1}\,\,\frac{d \om}{2\pi}\,.
\label{alter-bulk}
\end{equation}
The above can be compared with the corresponding expression for the emission
of brane-localized modes given, as in eq.~(\ref{pdecay-brane}), by
\begin{eqnarray}
\frac{d \hat E (\om)}{dt} =  \sum_{\ell} \hat N_\ell\, |\hat {\cal  A}_\ell|^2\,
\frac{\om}{\exp\left(\om/T_\text{H}\right) - 1}\,\,\frac{d \om}{2\pi}\,,
\label{emission-br}
\end{eqnarray}
where $\hat N_\ell =2\ell +1$.
Since the temperature is the same for both bulk and brane modes, the bulk-to-brane ratio of the two energy emission rates will be simply given by the expression 
\begin{figure}
\begin{center}
\psfrag{x}[t][b][1.75]{$\omega r_\text{h}$}
\scalebox{0.5}{\rotatebox{0}{\includegraphics[width=23cm, height=15cm]
{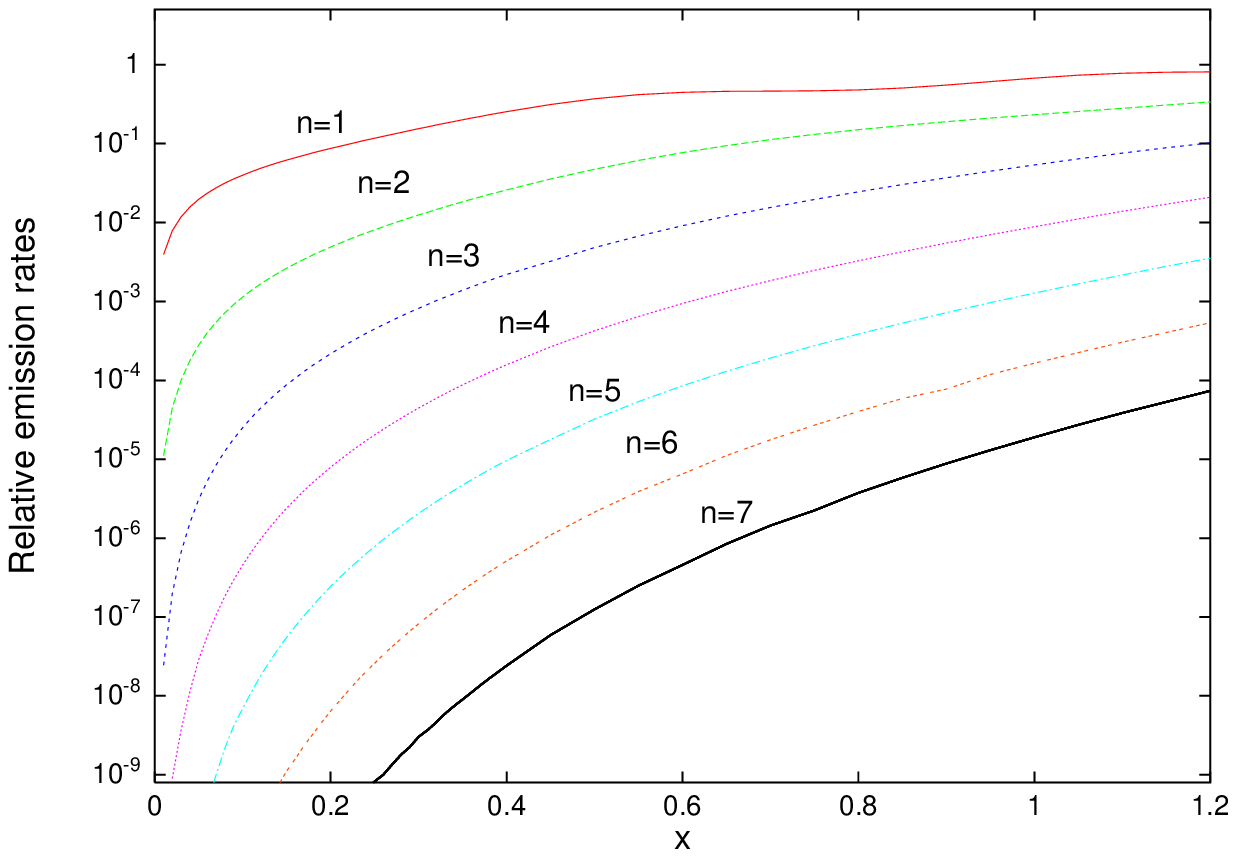}}}
\capbox{Bulk-to-brane relative emission rates for scalars}{Bulk-to-brane energy emission rates for scalar fields from a
$(4+n)$D black hole.\label{bb-ratio}}
\end{center}
\end{figure}
\begin{equation}
\frac{d \tilde E/dt}{d \hat E/dt} = \frac{\sum_{\ell} \tilde N_\ell\,
|\tilde {\cal A}_\ell|^2}{\sum_{\ell} \hat N_\ell\,
|\hat {\cal A}_\ell|^2}\,,
\label{ratio}
\end{equation}
\enlargethispage{-\baselineskip}and will depend on the scaling of the multiplicities of states and the
absorption coefficients with $n$.\footnote{The absorption coefficients are related to the grey-body factors through eq.~(\ref{greydef}). This includes a multiplicative coefficient which depends on both $\omega r_\text{h}$ and $n$, and so the coefficients might have a completely different behaviour from the grey-body factors themselves.}  

As is clear from eq.~(\ref{bulk-mult}), the
multiplicity of bulk modes $\tilde N_\ell$ increases quickly for increasing
$n$, while $\hat N_\ell$ remains the same. However, it turns out that the
enhancement with $n$ of the absorption probability $|\hat {\cal A}_\ell|^2$ for
brane emission is considerably greater than
the one for bulk emission. This leads to the dominance of the emission of
brane-localized modes over bulk modes, particularly for intermediate values of $n$. The behaviour of this ratio is shown in Figure~\ref{bb-ratio}.  This figure shows that, in the low-energy regime, the ratio for large $n$ is suppressed by many orders of magnitude, compared to the value of unity for $n=0$. In the high-energy regime on the other hand, the suppression becomes smaller and the ratio seems to approach unity.  A more careful examination reveals that bulk modes in fact dominate over brane modes in a limited high-energy regime which becomes broader as $n$ increases.\footnote{For the very high energies for which this is true, it is not clear that the expressions for higher-dimensional Hawking radiation will remain valid.}

A definite conclusion regarding the relative amount of energy emitted in the two `channels'---bulk and brane---can only be drawn if
the corresponding total energy emissivities are computed. By integrating
the areas under the bulk and brane energy emission rate curves, the relative energy emission rates are determined.  The results obtained, for values of $n$ from 1 to 7, are given in Table~\ref{bb-ratios}.

\begin{table}
\def\arraystretch{1.1}
\begin{center}
\begin{tabular}{|c|c|c|c|c|c|c|c|c|}
\hline
$n$ & 0 & 1  & 2  & 3 & 4 & 5 & 6 & 7 \\
\hline
Bulk/Brane & 1 & 0.40 & 0.24 & 0.22 & 0.24 & 0.33 & 0.52 & 0.93\\
\hline 
\end{tabular}
\capbox{Relative bulk-to-brane emission rates for scalars}{Relative bulk-to-brane energy emission rates for scalar fields.
\label{bb-ratios}}
\end{center}
\def\arraystretch{1.0}
\end{table}

From the entries of the Table~\ref{bb-ratios}, it becomes clear that the emission of
brane-localized scalar modes is dominant, in terms of the energy emitted, for all the values of $n$ considered. As $n$ increases, the
ratio of bulk-to-brane emission gradually becomes smaller, and is
particularly suppressed for intermediate values, i.e.\ $n=2$--5; in these cases, the total energy emitted in a bulk mode varies between approximately 1/3 and 1/4 of that emitted in
a brane mode. As $n$ increases further, the high-energy dominance of the bulk
modes mentioned above gives a boost to the value of the bulk-to-brane
ratio---nevertheless, the energy ratio never exceeds unity.

The above analysis provides exact results for the energy emission rates of brane and bulk scalar modes and gives considerable support to
earlier, more heuristic, arguments \cite{Emparan:2000rs}, according to which a
$(4+n)$-dimensional black hole emits mainly brane modes. A complete confirmation would mean performing a similar analysis for the emission of gravitons, but there are still theoretical and numerical issues which prevent this. 

\section{Rotating black holes}
\label{rotbh}

We have already seen that a black hole created in a hadron collider is expected to have some angular momentum $J$ about an axis perpendicular to the plane of parton collision.  
Using the same na\"{\i}ve argument as in \secref{bhprod}, we assume $J=bM_\text{BH}/2$ (where $b$ is the impact parameter of the colliding partons) and that $b<2r_\text{h}$ if a black hole is to form.  The maximum possible value of the rotation parameter defined in eq.~(\ref{astardef}) is then found to be
\begin{equation}
\label{astarmax}
a_*^\text{max}=\frac{n+2}{2}\,.
\end{equation}
The equivalent of eqs.~(\ref{fdecay-brane}) and (\ref{pdecay-brane}) are now
\begin{equation}
\frac{d \hat N^{(s)}(\om)}{dt} = \sum_{\ell,m} |\hat {\cal A}^{(s)}_{\ell,m}|^2\,
\frac{1}{\exp\left[(\om-m\Omega)/T_\text{H}\right] \mp 1}\,\frac{d\om}{2\pi}\,,
\label{rflux}
\end{equation}
and
\begin{equation}
\frac{d \hat N^{(s)}(\om)}{dt} = \sum_{\ell,m} |\hat {\cal A}^{(s)}_{\ell,m}|^2\,
\frac{\om}{\exp\left[(\om-m\Omega)/T_\text{H}\right] \mp 1}\,\frac{d\om}{2\pi}\,,
\label{rpower}
\end{equation}
where the Hawking temperature is now given by
\begin{equation}
T_\text{H}=\frac{(n+1)+(n-1)a_*^2}{4\pi(1+a_*^2)r_\text{h}}\,,
\end{equation}
and $\Omega$ is defined by
\begin{equation}
\Omega=\frac{a_*}{(1+a_*^2)r_\text{h}}\,.
\end{equation}

For the rotating case, the metric takes on a more complicated form than previously.  In general $(4+n)$-dimensional objects are described by $(n+3)/2$ angular momentum parameters; however, it is assumed that here there is only one non-zero parameter (about an axis in the brane).  This is reasonable because the partons which collide to produce the black hole are themselves on the brane.  Hence the metric reduces to
\begin{equation}
\begin{split}
ds^2=\left(1-\frac{\mu}{\Sigma r^{n-1}}\right)dt^2&+\frac{2 a\mu\sin^2\theta}{\Sigma r^{n-1}}dtd\varphi-\frac{\Sigma}{\Delta}dr^2 \\[3mm]
&-\Sigma d\theta^2-\left(r^2+a^2+\frac{a^2\mu\sin^2\theta}{\Sigma r^{n-1}}\right)\sin^2\theta d\varphi^2, 
\end{split}
\end{equation}
where
\begin{equation}
\Delta=r^2+a^2-\frac{\mu}{r^{n-1}} \quad\mbox{ and } \quad\Sigma=r^2+a^2\cos^2\theta,
\end{equation}
with $a=a_*r_\text{h}$ and $\mu=r_\text{S}^{n+1}$ ($r_\text{S}$ is the Schwarzschild radius of a non-rotating black hole of the same mass).

Again the field equation is separable using a solution of the same form as in the non-rotating case, but now the radial equation is
\begin{equation}
\Delta^{-s} \frac{d}{dr}\left(\Delta^{s+1}\frac{R_s}{dr}\right)+\left(\frac{K^2-isK\Delta'}{\Delta}+4is\omega r+s\Delta''-{}_s\Lambda^m_{\ell} \right)R_s=0\,,
\end{equation}
where
\begin{equation}
K=(r^2+a^2)\omega-am \quad\mbox{ and } \quad{}_s\Lambda^m_{\ell}={}_sE^m_{\ell}+2s+a^2\omega^2-2am\omega\,.
\end{equation}
The angular equation is now
\begin{multline}
\label{spinang}
\frac{1}{\sin\theta} \frac{d}{d\theta}\left(\sin\theta\frac{d \:{}_sT^m_{\ell}(\theta)}{d\theta}\right)+ \\[3mm]
\left(-\frac{2ms\cot\theta}{\sin\theta}-\frac{m^2}{\sin^2\theta}+a^2\omega^2\cos^2\theta-2a\omega s\cos\theta+s-s^2\cot^2\theta+{}_sE^m_{\ell}\right){}_sT^m_{\ell}(\theta)=0\,,
\end{multline}
where $e^{im\varphi}{}_sT^m_{\ell}(\theta)$ are known as spin-weighted spheroidal harmonics.

\begin{figure}
\unitlength1cm
\begin{center}
\begin{minipage}[t]{2.0in}
\scalebox{0.55}{\rotatebox{0}{\includegraphics{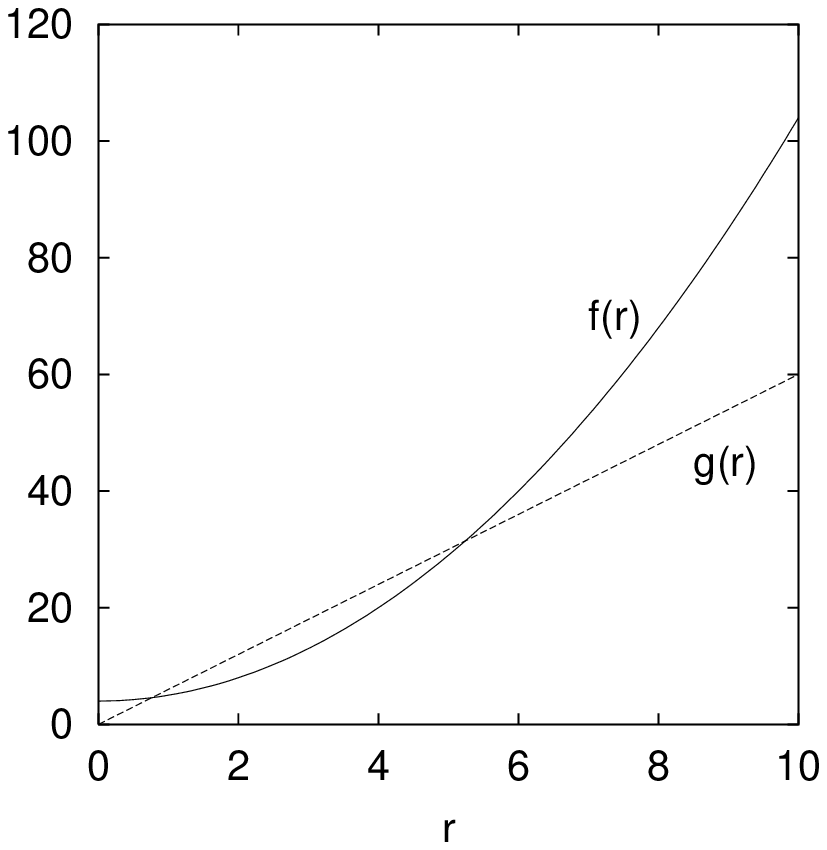}}}
\end{minipage}
\hfill
\begin{minipage}[t]{2.0in}
\scalebox{0.55}{\rotatebox{0}{\includegraphics{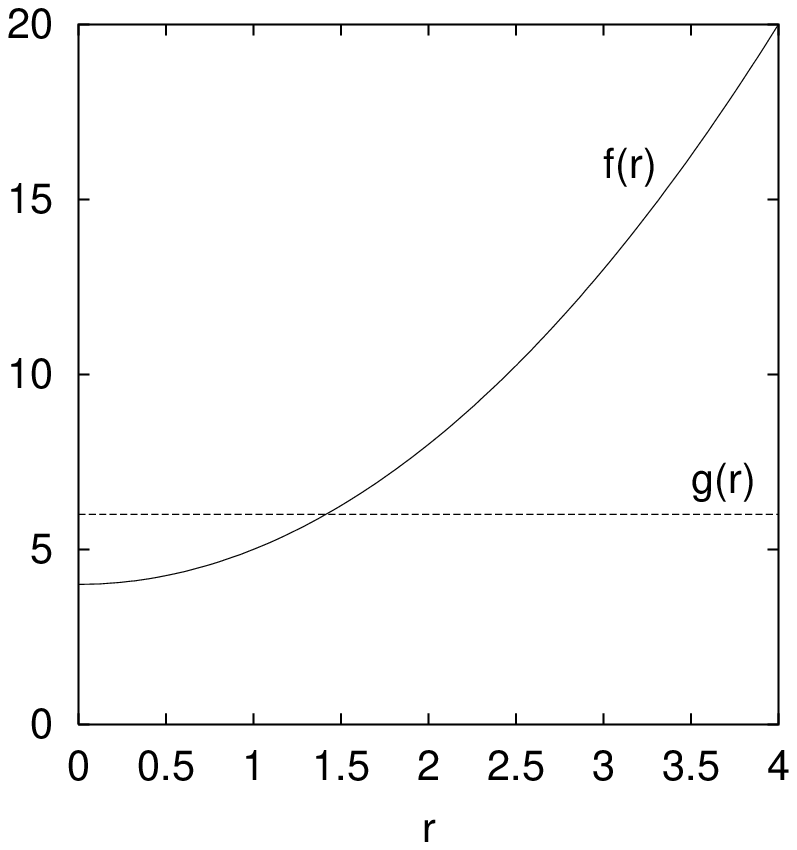}}}
\end{minipage}
\hfill
\begin{minipage}[t]{2.0in}
\scalebox{0.55}{\rotatebox{0}{\includegraphics{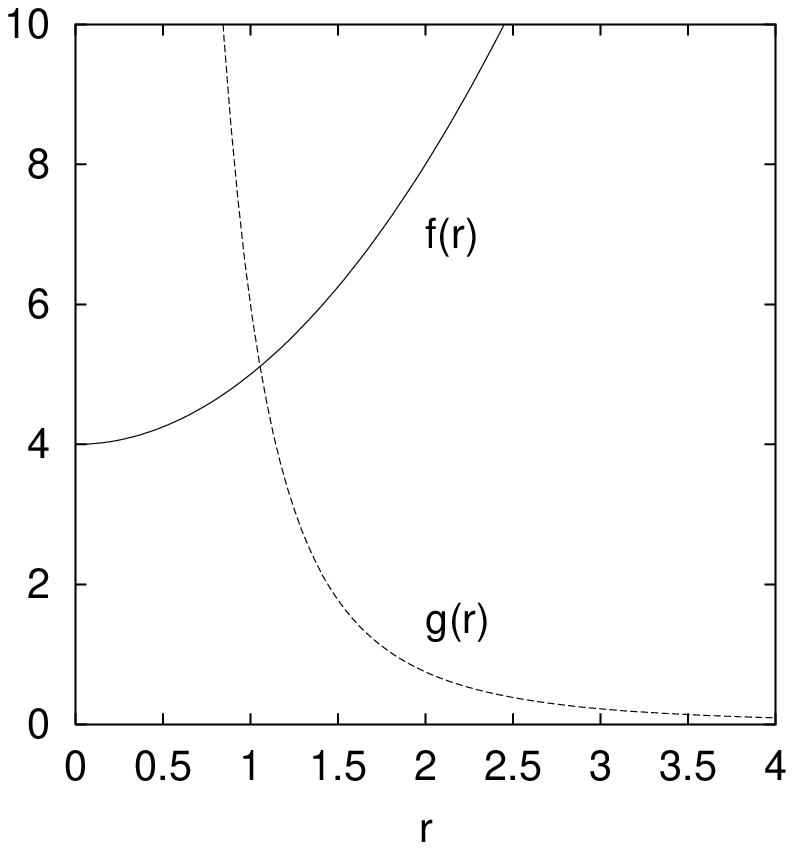}}}
\end{minipage}
\capbox{Number of solutions of $\Delta(r)=0$ for different $n$}{Number of solutions of $\Delta(r)=0$ for $n=0,1 \mbox{ and } 4$.  On all three plots, the two functions shown are $f(r)=r^2+a^2$ and $g(r)=\mu r^{1-n}$ for particular values of $a$ and $\mu$.\label{deltacomp}}
\end{center}
\end{figure}

The black hole horizon is given by solving $\Delta(r)=0$.  Unlike the 4D Kerr black hole for which where are inner and outer solutions for $r_\text{h}$, for $n\geq1$ there is only one solution of this equation (as can easily be seen graphically---Figure~\ref{deltacomp}).  In the $n=0$ case the maximum possible value of $a_*$ is 1 otherwise there are no solutions of $\Delta=0$ (this is forbidden as it would mean there is no horizon and hence there is a naked ring singularity at $r=0$).  For $n>1$ there is no fundamental upper bound on $a_*$ but only the bound given in eq.~(\ref{astarmax}) which was argued geometrically.  For general $n$, the horizon radius $r_\text{h}$ is found to be
\begin{equation}
r_\text{h}=\frac{r_\text{S}}{(1+a_*^2)^{\frac{1}{n+1}}}\,.
\end{equation}

The solution of the radial equation to obtain the grey-body factor is carried out in exactly the same same way as for the non-rotating black hole.  There is a slight difference in the asymptotic solution close to the horizon in that the $\omega$ in eq.~(\ref{nh-new}) is replaced by $k$ where
\begin{equation}
k=w-\frac{ma}{r_\text{h}^2+a^2}\,,
\end{equation}
and that the equivalent of eq.~(\ref{rstarr}) is now
\begin{equation}
\frac{dr^*}{dr}=\frac{r^2+a^2}{\Delta(r)}\,.
\end{equation}

\subsection{Calculation of the eigenvalue}

The only other complication is to determine ${}_sE^m_{\ell}$ from the angular equation.  The $\theta$ wave functions ${}_sT^m_{\ell}$ are no longer like those of the spin-weighted spherical harmonics and so ${}_sE^m_{\ell} \neq {}_s\lambda_{\ell}$\,.  The eigenvalues of the spin-weighted spheroidal harmonics are functions of $a\omega$ and can be obtained by a continuation method \cite{Wasserstrom:1972}.  This is a generalization of perturbation theory which can be applied for arbitrarily large changes in the initial Hamiltonian for which the eigenvalues are known.  Making the $a\omega$-dependence of both $T$ and $E$ clear, eq.~(\ref{spinang}) can be written as
\begin{equation}
\label{opang}
({\mathcal H}_0+{\mathcal H}_1)\, {}_sT^m_{\ell}(\theta,a\omega) =-{}_sE^m_{\ell} (a\omega)\, {}_sT^m_{\ell}(\theta,a\omega)\,, 
\end{equation}
where 
\begin{equation}
{\mathcal H}_0=\frac{1}{\sin\theta}\frac{d}{d\theta}\left(\sin\theta\frac{d}{d\theta}\right)+\left(-\frac{2ms\cot\theta}{\sin\theta}-\frac{m^2}{\sin^2\theta}+s-s^2\cot^2\theta\right),
\end{equation}
and
\begin{equation}
{\mathcal H}_1=(a^2\omega^2\cos^2\theta-2a\omega s\cos\theta)\,.
\end{equation}
From these equations it is evident that 
\begin{equation}
{}_sT^m_{\ell}(\theta,0)= {}_sS^m_{\ell}(\theta)\,, 
\label{init1}
\end{equation}
and that
\begin{equation}
{}_sE^m_{\ell}(0)={}_s\lambda_{\ell}=\ell(\ell+1)-s(s+1)\,.
\label{init2}
\end{equation}

The continuation method was outlined in \cite{Press:1973} although this paper unfortunately seems to contain a number of errors.  First consider the case in which $a\omega$ is small and normal perturbation theory can be used.  This means we have
\begin{equation}
\label{pte}
{}_sE^m_{\ell}(a\omega)={}_s\lambda_{\ell}-\langle s\ell m|{\mathcal H}_1|s\ell m\rangle+\ldots\,,
\end{equation}
and
\begin{equation}
\label{pts}
{}_sT^m_{\ell}(a\omega)={}_sS^m_{\ell}-\sum_{\ell'\neq l} \frac{\langle s\ell'm|{\mathcal H}_1|s\ell m\rangle}{\ell(\ell+1)-\ell'(\ell'+1)} {}_sS^m_{\ell'}+\ldots\,,
\end{equation}
where
\begin{equation}
\label{prod}
\langle s\ell'm|{\mathcal H}_1|s\ell m\rangle \equiv \int d\Omega\,({}_sS_{\ell'}^m)^* {}_sS_{\ell}^m \,{\mathcal H}_1\,. 
\end{equation} 
Care must be taken about the signs of the terms in eqs.~(\ref{pte}) and (\ref{pts}) due to the sign of the eigenvalue ${}_sE^m_{\ell}$ in eq.~(\ref{opang}).  This seems to account for some differing results in \cite{Press:1973}.

The product in eq.~(\ref{prod}) can be split into terms for which the following standard formulae are useful:
\begin{equation}
\langle s\ell'm|\cos^2\theta|s\ell m\rangle=\frac{1}{3}\delta_{\ell\ell'}+\frac{2}{3}\left(\frac{2\ell +1}{2\ell'+1}\right)^{\frac{1}{2}}\langle\ell2m0|\ell'm\rangle\langle\ell2(\!-\!s)0|\ell'(\!-\!s)\rangle,
\end{equation}
and
\begin{equation}
\langle s\ell'm|\cos\theta|s\ell m\rangle=\left(\frac{2\ell +1}{2\ell'+1}\right)^{\frac{1}{2}}\langle\ell1m0|\ell'm\rangle\langle\ell1(\!-\!s)0|\ell'(\!-\!s)\rangle,
\end{equation}
where $\langle\ell_1\ell_2m_1m_2|LM\rangle$ are Clebsch-Gordan coefficients.  Using the above we obtain
\begin{equation}
{}_sE^m_{\ell}= \left\{ \begin{array}{ll}
		{}_s\lambda_{\ell}-2a\omega \frac{s^2m}{\ell(\ell+1)}+{\mathcal O}[(a\omega)^2] & \mbox{if $s \neq 0$}\,;\\
\\
		{}_s\lambda_{\ell}+a^2\omega^2\left[\frac{2m^2+1-2\ell(\ell+1)}{(2\ell-1)(2\ell-3)}\right]+{\mathcal O}[(a\omega)^4] & \mbox{if $s=0$}\,.\end{array} \right. 
\end{equation}
Again this differs (in the $s=0$ case) from the result given in \cite{Press:1973}.  However it agrees with \cite{Seidel:1989ue} where several discrepancies in the literature are helpfully clarified.

To employ the continuation method we write the ${}_sT^m_{\ell}(\theta,a\omega)$ functions in the basis of the $\theta$-parts of the spin-weighted spherical harmonics:
\begin{equation}
{}_sT^m_{\ell}(\theta,a\omega)=\sum_{\ell'}{}_sB_{\ell\ell'}^m(a\omega)S_{s\ell'}^m(\theta)\,.
\end{equation}
By differentiating eq.~(\ref{opang}) and applying the same techniques as in perturbation theory it is possible to obtain the results
\begin{equation}
\label{conte}
\frac{d \, {}_sE_{\ell}^m}{d(a\omega)}=-\sum_{\alpha,\beta}{}_sB_{\ell\alpha}^m \,{}_sB_{\ell\beta}^m\langle\alpha|\beta\rangle,
\end{equation}
and
\begin{equation}
\label{contb}
\frac{d \, {}_sB_{\ell\ell'}}{d(a\omega)}=-\sum_{\alpha,\beta,\gamma \neq l}\frac{{}_sB_{\gamma\alpha} \, {}_sB_{\ell\beta}}{{}_sE^m_{\ell}-{}_sE_{\gamma}^m}\langle\alpha|\beta\rangle{}_sB_{\gamma \ell'},
\end{equation}
where $\langle\alpha|\beta\rangle \equiv \langle s\alpha m|d{\mathcal H}_1/d(a\omega)|s\beta m\rangle$.  The initial conditions are obtained from eqs. (\ref{init1}) and (\ref{init2}) which imply that ${}_sB_{\ell\ell'}^m(0)=\delta_{\ell\ell'}$.  By integrating eqs.~(\ref{conte}) and (\ref{contb}) it is possible to obtain the eigenvalues of the spin-weighted spheroidal functions for any $s,\ell$ and $m$ and for arbitrarily large values of $a\omega$.  This integration was performed by using the \texttt{NDSolve} package in \texttt{Mathematica} to implement a Runge-Kutta method.  The computational effort involved increases significantly for larger values of $\ell$ as it is necessary to include a larger number of terms in the summations, increasing the number of differential equations to be solved.

\begin{figure}
\begin{center}
\psfrag{x}[][][1.5]{$a\omega$}
\psfrag{y}[][][1.5]{${}_sE^m_{\ell}(a\omega)$}
\scalebox{0.7}{\rotatebox{0}{
\includegraphics[width=\textwidth]{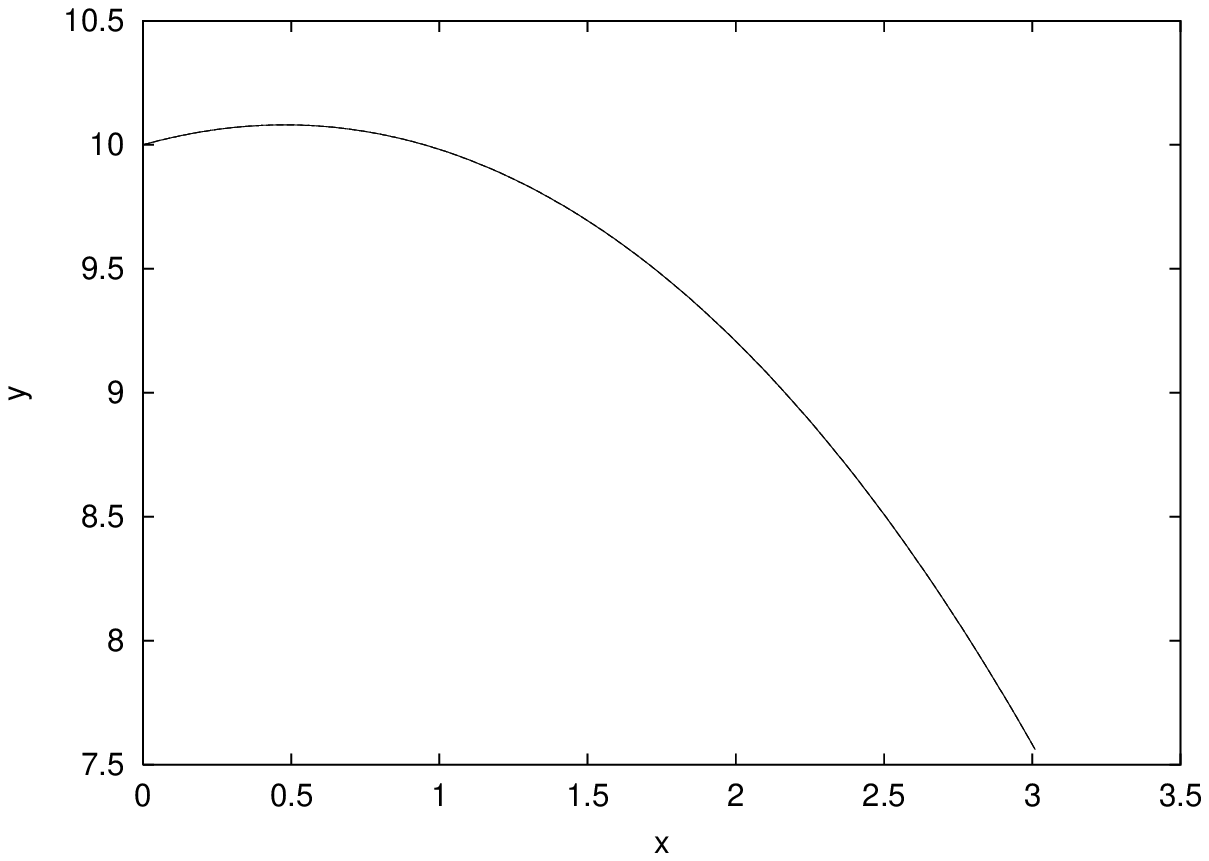}}}
\capbox{Comparison of eigenvalue with previous result}{Comparison of the numerical result obtained in this work for the $s=1,\ell=3,m=-2$ eigenvalue (solid line) with the polynomial approximation in \cite{Press:1974} (dashed line).  The lines are exactly on top of each other (the maximum difference between them is $\sim 10^{-5}$).\label{eigen}}
\end{center}
\end{figure}

Although only the $s=0$ eigenvalues were required here, as a check the continuation method was also used to calculate the $s=1$ eigenvalues as a function of $a\omega$.  This allowed comparison with the polynomial approximations given in \cite{Press:1974}.  Excellent agreement was found---see Figure~\ref{eigen} for an example.

\subsection{Calculation of the grey-body factor}

In order to calculate the grey-body factors in the rotating case, it is necessary to know the equivalent of eqs.~(\ref{scalars}), (\ref{fermions}) and (\ref{bosons}).  Unfortunately, for fermions and bosons these expressions are not available in the literature.  However for the scalar case, since eq.~(\ref{scalars}) only involves coefficients at infinity, exactly the same formula holds even when the black hole is rotating.  This means that for scalars it has been possible to numerically calculate the grey-body factors for rotating black holes.

Since the denominator in eqs.~(\ref{rflux}) and (\ref{rpower}) has $m$-dependence, it is less useful to plot an equivalent of the $\hat{\sigma}$ used in the non-rotating case,\footnote{The absorption cross section $\hat{\sigma}_\text{abs}$ summed over all angular momentum modes \emph{can} still be calculated; the high-energy asymptotic value appears to increase with $a_*$, apparently in contradiction with \cite{Page:1976df} which states that roughly the same value is expected as in the non-rotating case.} so only the power spectra are shown.  They are presented as a function of $\omega r_\text{h}$ for different values of $a_*$ and hence cannot be used to directly compare the spectra from two black holes of the same mass (but different $J$) since this is itself a function of $M_\text{BH}$. 

\begin{figure}
\begin{center}
\psfrag{x}[][][1.2]{$\omega r_\text{h}$}
\psfrag{y}[][][1.2]{$r_\text{h} d^2 \hat{E}^{(0)}/dtd\om$}
\scalebox{0.8}{\rotatebox{0}{\includegraphics[width=\textwidth]{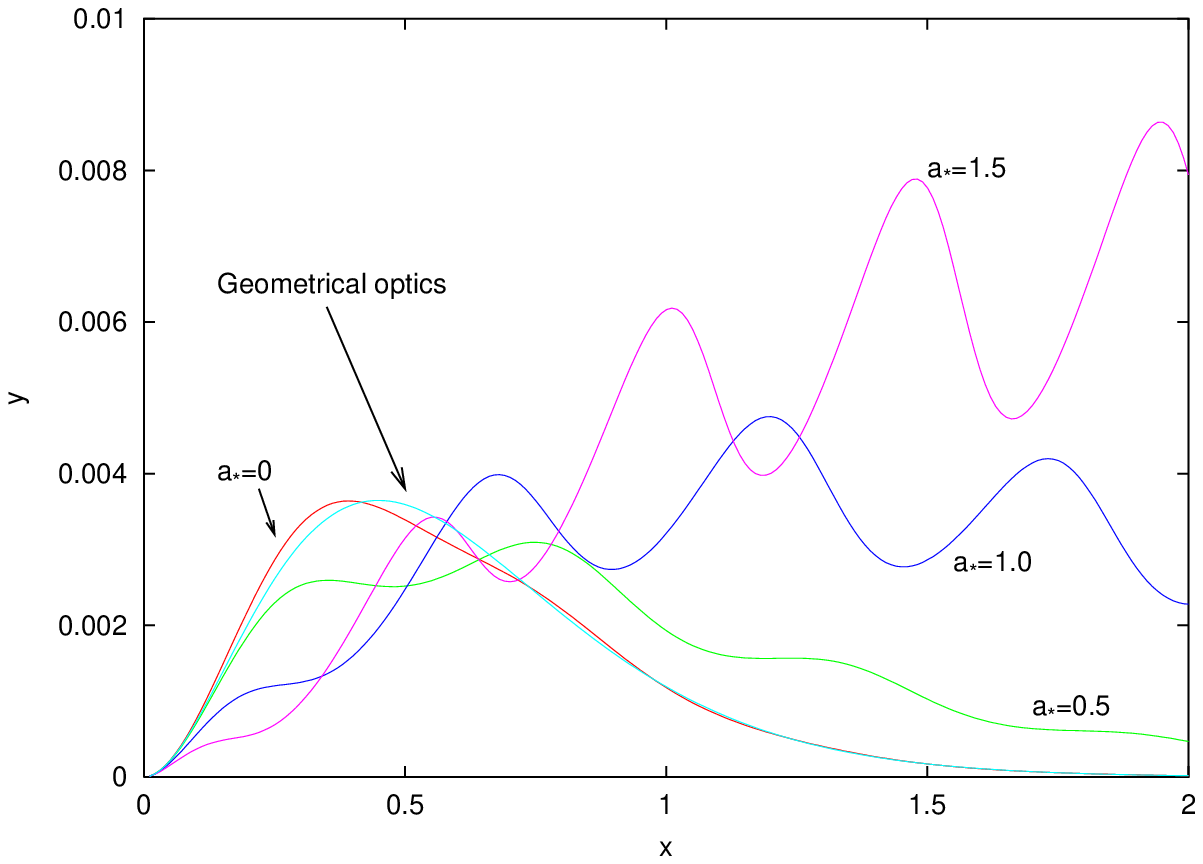}}}
\capbox{Power spectra for scalar emission from rotating black holes ($n=1$)}{Power spectra for scalar emission from rotating black holes ($n=1$).\label{s0a}}
\end{center}
\end{figure}

Figure~\ref{s0a} can be directly compared to Figure 2 in \cite{Ida:2002ez} and shows markedly different behaviour for values of $\omega r_\text{h}$ larger than $\sim 0.2$.  Almost all of this effect is due to the low-energy expansion used by those authors to obtain the grey-body factors.  Their approximation that ${}_sE^m_{\ell}={}_s\lambda_{\ell}=\ell(\ell+1)-s(s+1)$ is much less significant except at the very highest values of $\omega r_\text{h}$.

These numerical results for the grey-body factors also allowed the statement in \cite{Giddings:2001bu} that emission is dominated by modes with $\ell=m$ to be tested.  It was found that a plot like Figure~\ref{s0a} looks the same at the $\sim$~90\% level if only the the $\ell=m$ modes are included in the sum of eq.~(\ref{rpower})
 
\subsection{Super-radiance}

The numerical results obtained allow the possibility of super-radiance (i.e.\ negative grey-body factors) to be investigated in an extra-dimensional context.  Previously an analytic approach \cite{Frolov:2002xf} was used to confirm super-radiance for bulk scalars incident on 5D black holes, but there do not seem to be any other results in the literature for higher-dimensional black holes.  To study this phenomenon it is useful to consider angular momentum modes individually and, for comparison with the 4D scalar results presented in \cite{Press:1972}, the quantity on the horizontal axis of the plots is
\begin{equation}
\frac{\omega}{m \Omega}=\frac{(1+a_*^2)(\omega r_\text{h})}{m a_*}\,,
\end{equation}
rather than the $\omega r_\text{h}$ used in previous plots.  The vertical axis is the absorption probability (in fact $-|\hat {\cal A}_{\ell,m}|^2$) expressed as a percentage so that it gives the percentage energy amplification of an incident wave.
\begin{figure}
\begin{center}
\psfrag{x}[][][1.5]{$\omega/m\Omega$}
\psfrag{a}[][][1.5][18]{\textsf{l=m=1}}
\psfrag{b}[][][1.5][30]{\textsf{l=m=2}}
\psfrag{c}[][][1.5][37]{\textsf{l=m=3}}
\psfrag{d}[][][1.5][40]{\textsf{l=m=4}}
\psfrag{e}[][][1.5][45]{\textsf{l=m=5}}
\psfrag{f}[][][1.5]{\textsf{l=2, m=1}}
\scalebox{0.5}{\rotatebox{0}{\includegraphics[width=\textwidth]{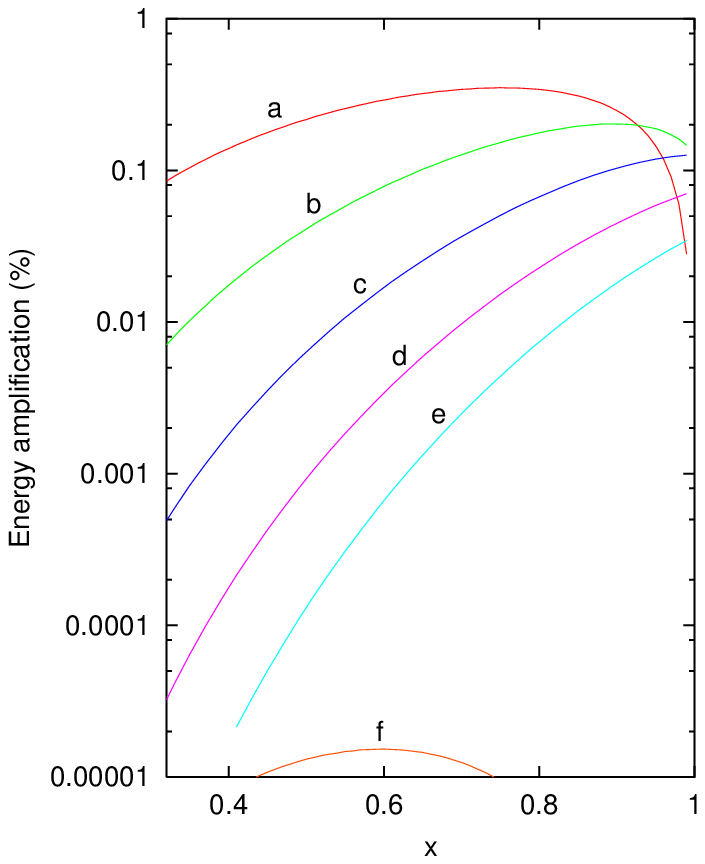}}}
\capbox{Super-radiant scattering by 4D black holes with $a_*=1$}{Super-radiant scattering of scalars by maximally rotating ($a_*=1$) 4-dimensional black holes.\label{super1}}
\end{center}
\end{figure}
Figure~\ref{super1} shows excellent agreement with the results in Figure 1 of \cite{Press:1972} which are for $n=0$ and $a_*=1$ (equivalent to $a=M_\text{BH}$ in the 4D case).  Figure~\ref{super2} is for $n=2$ and 6 but still with the same value of the rotation parameter.  It is clear that the peak amplification is more significant than previously and becomes comparable to that for gauge bosons in the 4D case.  It is still found that the $\ell=m=1$ mode provides the greatest amplification.  However Figure~\ref{super3} shows that for larger values of $a_*$ the maximum amplification can occur in modes with larger values of $\ell$.  This trend continues as $n$ and $a_*$ increase further: for $n=6$, equation~(\ref{astarmax}) shows that the geometric maximal rotation is $a_*^\text{max}=4.0$ and for this value of the rotation parameter the maximum energy amplification is found to be just under 9\% (and occurs in the $\ell=m=7$ mode).

\begin{figure}
\unitlength1cm
\begin{center}
\psfrag{x}[][][0.8]{$\omega/m\Omega$}
\psfrag{a}[][][0.8][18]{\textsf{l=m=1}}
\psfrag{b}[][][0.8][28]{\textsf{l=m=2}}
\psfrag{c}[][][0.8][35]{\textsf{l=m=3}}
\psfrag{d}[][][0.8][40]{\textsf{l=m=4}}
\psfrag{e}[][][0.8][43]{\textsf{l=m=5}}
\psfrag{f}[][][0.8]{\textsf{l=2, m=1}}
\psfrag{g}[][][0.8][20]{\textsf{l=m=1}}
\psfrag{h}[][][0.8][30]{\textsf{l=m=2}}
\psfrag{i}[][][0.8][37]{\textsf{l=m=3}}
\psfrag{j}[][][0.8][42]{\textsf{l=m=4}}
\psfrag{k}[][][0.8][45]{\textsf{l=m=5}}
\psfrag{l}[][][0.8]{\textsf{l=2, m=1}}
\begin{minipage}[t]{3.05in}
{\rotatebox{0}{\includegraphics{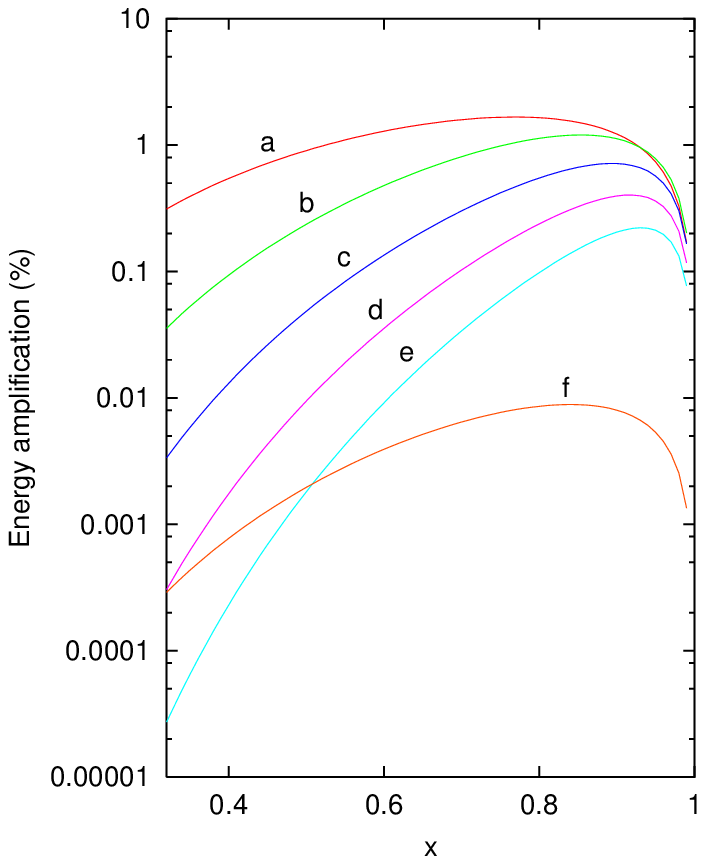}}}
\end{minipage}
\hfill
\begin{minipage}[t]{3.05in}
{\rotatebox{0}{\includegraphics{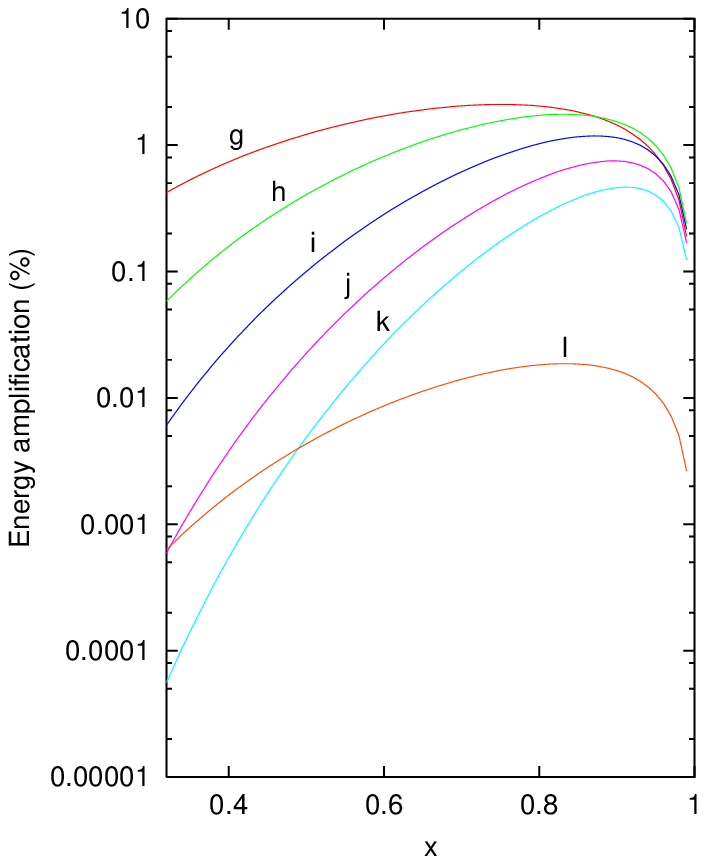}}}
\end{minipage}
\capbox{Super-radiant scattering by $(4+n)$D black holes with $a_*=1$}{Super-radiant scattering of scalars by $(4+n)$-dimensional black holes with $a_*=1$: {\bf(a)} $n=2$ and {\bf(b)} $n=6$.\label{super2}}
\end{center}
\end{figure}

\begin{figure}
\unitlength1cm
\begin{center}
\psfrag{x}[][][1]{$\omega/m\Omega$}
\psfrag{a}[r][][1]{\textsf{l=m=1}}
\psfrag{b}[r][][1]{\textsf{l=m=2}}
\psfrag{c}[][][1][60]{\textsf{l=m=3}}
\psfrag{d}[][][1][60]{\textsf{l=m=4}}
\psfrag{e}[l][][1]{\textsf{l=m=5}}
\psfrag{f}[r][][1]{\textsf{l=m=1}}
\psfrag{g}[r][][1]{\textsf{l=m=2}}
\psfrag{h}[][][1]{\textsf{l=m=3}}
\psfrag{i}[][][1][64]{\textsf{l=m=4}}
\psfrag{j}[l][][1]{\textsf{l=m=5}}
\begin{minipage}[t]{3.05in}
\scalebox{0.8}
{\rotatebox{0}{\includegraphics{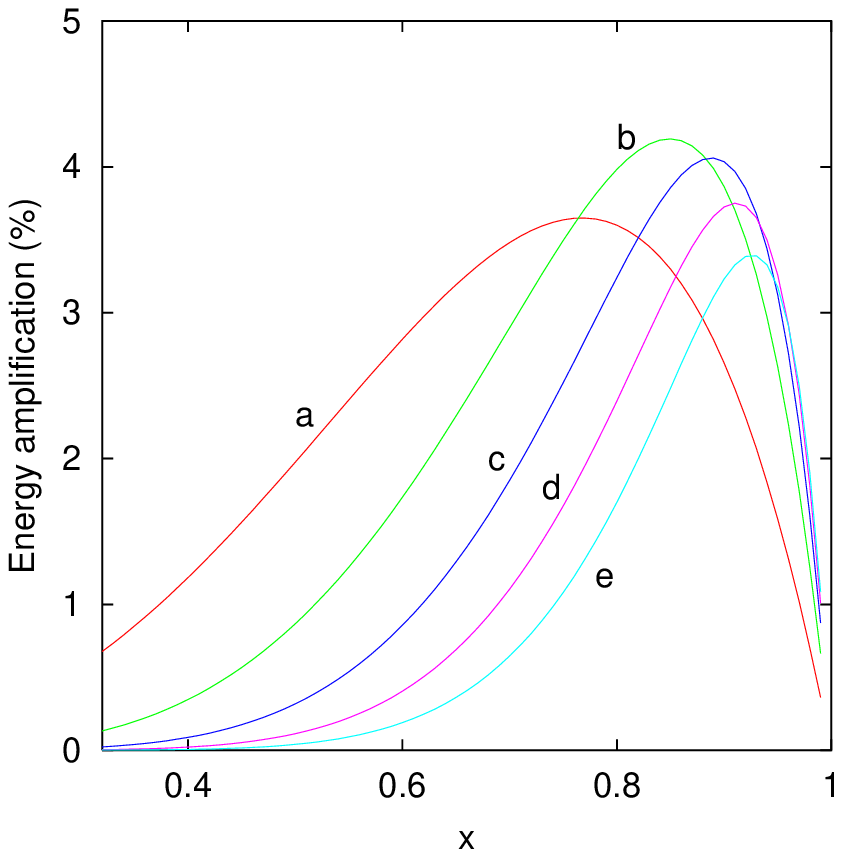}}}
\end{minipage}
\hfill
\begin{minipage}[t]{3.05in}
\scalebox{0.8}
{\rotatebox{0}{\includegraphics{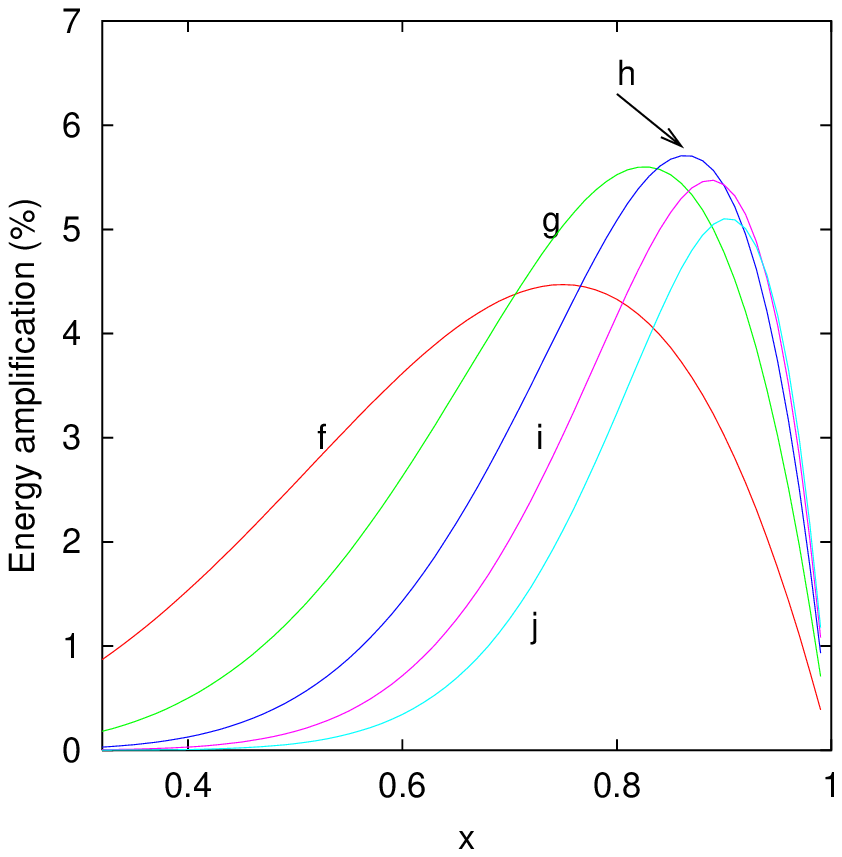}}}
\end{minipage}
\capbox{Super-radiant scattering by $(4+n)$D black holes with $a_*>1$}{Super-radiant scattering of scalars by $(4+n)$-dimensional black holes: {\bf(a)} $n=2$, $a_*=1.5$ and {\bf(b)} $n=6$, $a_*=2$.\label{super3}}
\end{center}
\end{figure}

As in the 4-dimensional case, the region in which the grey-body factors are negative exactly corresponds to the region in which the thermal factor in the denominator of equations like (\ref{rflux}) and (\ref{rpower}) is negative; hence the quantum emission rate always remains positive.  For all angular momentum modes the grey-body factor is positive for $\omega > m \Omega$ and the change from negative to positive at $\omega = m \Omega$ occurs such that the flux and energy spectra are smooth positive functions.  Therefore although super-radiance can in principle be used to extract energy from rotating black holes, its significance when considering the Hawking radiation spectra is limited.

\section{Conclusions}
\label{gbfconc}

The revival of the idea of extra space-like dimensions
in nature has led to the formulation of theories which allow gravity to become strong at a significantly lower energy scale. This has opened the way for the proposal of the creation of miniature higher-dimensional black holes during collisions of energetic particles in the earth's atmosphere or at ground-based particle colliders.  The Hawking radiation emitted by such black holes is described by the relevant grey-body factors; a wide variety of these factors have been calculated numerically in the work described in this chapter and, at the same time, some open questions from previous analyses in the literature have been addressed. 

The exact numerical results obtained in \secref{numres} for the grey-body factors and emission rates of scalars, fermions and gauge bosons confirm that previous analytic results \cite{Kanti:2002nr,Kanti:2002ge} were only valid in limited low-energy regimes.  The grey-body factors depend on both the dimensionality of space-time and the spin.  They adopt the same high-energy asymptotic values (which decrease as $n$ increases) for all particle species, confirming the geometrical optics limit of \cite{Emparan:2000rs}.  In the low-energy limit the grey-body factors are found to be enhanced as $n$ increases for $s=1$ and 1/2 and suppressed for $s=0$. 

Using these grey-body factors it was possible to accurately calculate the flux and power emission spectra for the black holes.  As the number of extra dimensions projected on the brane increases there is a substantial (orders of magnitude) enhancement in the energy emission rate, mainly as a result of the increase in black hole temperature with $n$.  However the effect of including the grey-body factors is that the increase in the rate depends on the spin of the particle studied.  The computed relative emissivities show that scalar fields, which are the dominant degree of freedom emitted by $n=0$ black holes, are just outnumbered by gauge bosons for large values of $n$, with the fermions becoming the least effective emission channel.  Therefore the emission spectra and the relative particle emissivities may possibly both lead to the determination of the number of extra dimensions.

The emission spectra for modes of different spins were combined with the total number of degrees of freedom for all the particle types which could be emitted from an extra-dimensional black hole.  This made it possible to calculate the total flux and power emission and hence also the expected time for, and average multiplicity of, the black hole decay.  These calculations suggest that some of the assumptions usually made when modelling black hole decay by sequential emission of Hawking radiation may not be valid for the larger values of $n$.
\enlargethispage{-\baselineskip}

In section \ref{embulk}, the details of the emission of bulk scalar modes were investigated with the aim being to provide an accurate estimate for the amount of energy lost into the bulk.  Comparing the total bulk and brane emission rates for scalar modes, integrated over the whole energy regime, the conclusion is that more of the black hole energy is emitted in brane modes than bulk modes, for all values of $n$ considered. The total emissivity in the bulk is less than 1/4 of that on the brane for $n=2$--4, while it becomes substantial for the extreme value of $n=7$ without, however, exceeding the brane value.  The accurate results presented in this chapter provide firm support to the heuristic arguments made in ref. \cite{Emparan:2000rs}.

No results are presented in this work concerning the emission of gravitons either on the brane or in the bulk. The derivation of a consistent equation, which can describe the motion of gravitons in the induced black hole background on the brane, is still under investigation.  Nevertheless it is expected that, as for the emission of fields with spin $s=0$, 1/2 and 1 the graviton emission becomes enhanced as the dimensionality of space-time increases, but remains subdominant compared to the other species at least for small values of $n$. 

The work on rotating black holes in \secref{rotbh} suggests that, at least for the emission of scalars, the power spectra can be significantly modified compared to those in the Schwarzschild-like case.  The spectra are dominated by the angular momentum modes with $\ell=m$ and as the rotation parameter $a_*$ increases, the peak in the spectrum shifts to much larger values of $\omega r_\text{h}$.  The phenomenon of super-radiance was also observed for higher-dimensional rotating black holes.  The peak energy amplification was found to be significantly larger than in the four-dimensional case and occurred in higher angular momentum modes for larger values of the rotation parameter.
\enlargethispage{\baselineskip}

A final comment is appropriate here concerning the validity of the results obtained in this work. As pointed out in the text, the horizon of the black hole is an input parameter of the analysis.  The results are applicable for all values of $r_\text{h}$ which are smaller than the size of the extra dimensions $R$, no matter how small or large $R$ is (as long as it remains considerably larger than $\ell_\text{P}$ to avoid quantum corrections).  The analysis therefore remains valid for all theories postulating the existence of flat extra dimensions and a fundamental scale of gravity even a few orders of magnitude lower than $M_\text{P}$. If the emission of Hawking radiation from these black holes is successfully detected, either at next-generation colliders or in the more distant future, the distinctive features discussed here may help in the determination of the number of extra dimensions existing in nature.  Chapter~\ref{generator} describes a black hole event generator which incorporates these grey-body effects and would be a useful tool in such an analysis.
\clearpage{\pagestyle{empty}\cleardoublepage}
\chapter{CHARYBDIS: A Black Hole Event Generator}
\chaptermark{Event Generation}
\label{generator}
\section{Introduction}

The work on black holes in this thesis was motivated by extra dimension models which allow the fundamental Planck scale to be of order a TeV\@.  These models are attractive because they solve the hierarchy problem, but they are also exciting because they mean that gravity is strong at these energy scales and so black holes can be investigated at the LHC and other next-generation machines.  If these models are realized in nature, we would expect particle accelerators at TeV-scale energies to be able to produce miniature black holes.  These would then decay rapidly by Hawking evaporation, giving rise to characteristic high-multiplicity final states.

There has already been much discussion of this issue in the literature (for example, in many of the works cited in Chapters~\ref{bhintro} and \ref{greybody}) but 
little work has been done in trying to realistically simulate black holes at
the LHC\@.  This chapter describes the implementation of a simple model of black hole production and decay which can be interfaced to existing Monte Carlo programs using the Les Houches accord \cite{Boos:2001cv}.  The major new theoretical input to the generator is the inclusion of the results from Chapter~\ref{greybody}---the grey-body factors for black holes in extra dimensions \cite{Harris:2003eg}.  The recoil and change of temperature of the black hole during decay is also taken into account, and various models are provided for the termination of the decay process.

Section~\ref{proddec} reviews important aspects of the theory of extra-dimensional black holes.  Then, in \secref{bhgen}, there is a description of the {\small CHARYBDIS}\footnote{In Greek mythology, Charybdis was a nymph daughter of Poseidon and Gaia who was turned into a monster by Zeus.  She lived in a cave at one side of the Strait of Messina (opposite the monster Scylla) and sucked water in and out three times each day.} event generator itself including a discussion of the theoretical assumptions involved, details of the Fortran code and instructions on how to use the program.  P.~Richardson was involved in the development of some of this code, specifically in ensuring baryon number and charge conservation, defining the colour flow required by general-purpose event generators, and making {\small CHARYBDIS} compatible with the Les Houches accord.   An event display and some sample production cross sections, particle spectra and particle emissivities produced using the event generator are presented in \secref{output}.  Details of some preliminary experimental studies are given in \secref{expstud} and there is a concluding discussion in \secref{discuss}.

\section{Black hole production and decay}
\label{proddec}

Most of the details of the production and decay of black holes in extra dimension models have already been discussed in Chapter~\ref{bhintro}.  Here some of the key features of the theory are re-iterated and some comments made on those which are particularly relevant for the task of event generation.
\enlargethispage{\baselineskip}

A fundamental Planck scale as low as $\sim$ TeV means that it is possible for 
tiny black holes to be produced at the LHC when two partons pass within the 
horizon radius set by their centre-of-mass energy, $\sqrt{\hat{s}}$.  Using the results of \cite{Myers:1986un}, the horizon radius for a non-spinning black hole is found to be
\begin{equation}
r_\text{S}=\frac{1}{\sqrt{\pi}M_{\text{P}(4+n)}}\left(\frac{M_\text{BH}}{M_{\text{P}(4+n)}}\right)^{\frac{1}{n+1}}\left(\frac{8\Gamma\left(\frac{n+3}{2}\right)}{n+2}\right)^{\frac{1}{n+1}},
\end{equation}
where $M_\text{BH}$ is the mass of the black hole and, as throughout the rest of this chapter, $M_{\text{P}(4+n)}$ is the fundamental $(4+n)$-dimensional Planck scale in convention `d' of Table~\ref{consum}.  Numerical calculations, discussed in \secref{bhprod}, show that $M_\text{BH} \le \sqrt{\hat{s}}$.  Provided there is enough energy to be in the trans-Planckian regime, the parton-level cross section is $F_n \pi r_\text{S}^2$ where $F_n \sim 1$.

Black holes can be produced with any gauge and spin quantum numbers so to determine the p-p or $\bar{\text{p}}$-p production cross section it is necessary to sum over all possible quark and gluon pairings.  Although the parton-level cross sections grow with black hole mass, the parton distribution functions (PDFs) fall rapidly at high energies and so the cross section also falls off quickly.

These miniature black holes are expected to decay instantaneously on LHC detector time scales (typical lifetimes are $\sim 10^{-26}$ s).  The decay is made up of four phases (discussed in more detail in Chapter~\ref{bhintro}): the balding phase, the spin-down phase, the Schwarzschild phase (accounting for the greatest proportion of the mass loss in four dimensions \cite{Page:1976df}) and the Planck phase.

A black hole of a particular mass is characterized by a Hawking temperature and as the decay progresses the black hole mass falls and the temperature rises.  For an uncharged, non-rotating black hole the decay spectrum is described, as in eq.~(\ref{fdecay-brane}), by
\begin{equation}
\label{spec}
\frac{d \hat N^{(s)}(\om)}{dt} = \sum_{\ell} (2l+1) |\hat {\cal A}^{(s)}_\ell|^2\,
\frac{1}{\exp\left(\om/T_\text{H}\right) \mp 1}\,\frac{d\om}{2\pi}\,,
\end{equation}
where the Hawking temperature is given by
\begin{equation}
T_\text{H}=\frac{n+1}{4\pi r_\text{S}}\,,
\end{equation}
and $|\hat {\cal A}^{(s)}_\ell|^2$ is the absorption probability from which the grey-body factor can be obtained (see eq.~(\ref{greydef}) in Chapter~\ref{greybody}). Equation~(\ref{spec}) can be used to determine the decay spectrum for
a particular particle, by considering the number of degrees of freedom (Table~\ref{pprobs} shows the relevant numbers for all the particle types which the black hole can emit).  It is relevant for predicting both the spectra and the relative emission probabilities of the different particle types.

The grey-body factors modify the spectra of emitted particles from that of a
perfect thermal black body\cite{Hawking:1975sw}; their spin dependence means that they are also necessary to determine the relative emissivities of different particle types from a black hole.  Until the work in Chapter~\ref{greybody}, the full numerical grey-body factors were only available in the literature for the 4D case \cite{Page:1976df,Sanchez:1978si}.  

It is usually assumed that a quasi-stationary approach to the black hole decay is valid---that is, after each particle emission the black hole has time to come into equilibrium at the new temperature before the next particle is emitted.\footnote{Note that the results of Chapter~\ref{greybody} have cast doubt on this assumption for cases with $n \ge 3$.}

It was argued in \cite{Emparan:2000rs} that the majority of energy in Hawking 
radiation is emitted into modes on the brane (i.e.\ as Standard Model 
particles) but that a small amount is also emitted into modes in the bulk 
(i.e.\ as gravitons).  The work described in Chapter~\ref{greybody} provides support for this statement based on the relative emissivities of scalars in the bulk and on the brane.

\section{Event generator} 
\label{bhgen}

\subsection{Features and assumptions of the event generator}
\label{egassump}

There are a number of features of the {\small CHARYBDIS} generator which, within the uncertainties of much of the theory, allow reliable simulation of black hole events.  Most notable is that, unlike other generators (e.g.\ \cite{Dimopoulos:2001en}), the grey-body effects are fully included.  The dependence on both spin and the dimensionality of space-time means that grey-body factors must be taken into account in any attempt to determine the number of extra dimensions by studying the energy spectra and relative emission probabilities of particles from a black hole.  When studying black hole decay, other experimental variables may also be sensitive to these grey-body effects.

The generator also has an option to allow the black hole temperature to vary as the decay progresses and is designed for simulations with either p-p or $\bar{\text{p}}$-p.

{\small CHARYBDIS} only attempts to model in detail the Hawking evaporation phase which is expected to account for the majority of the mass loss.  To provide a further simplification only non-spinning black holes are modelled.  This is perhaps a less good approximation but comparison with the 4D situation suggests that most of the angular momentum will be lost in a relatively short spin-down phase \cite{Page:1976ki}.  The balding phase is difficult to model and is neglected---this is equivalent to the assumption that $M_\text{BH}=\sqrt{\hat{s}}$ in spite of the evidence that this will not necessarily be the case.  A related assumption is that the cross section calculation assumes $F_n$ of equation~(\ref{fndef}) is equal to unity (i.e.\ the parton-level cross section is assumed to be $\pi r_\text{S}^2$).

\enlargethispage{\baselineskip}
It is hoped that the way in which the black hole decay is terminated will provide a reasonable approximation to the Planck phase of the decay.  The generator also assumes that energy lost from the black hole in graviton emission can be neglected, based on the work in \cite{Emparan:2000rs} and in Chapter~\ref{greybody}.

As discussed in \secref{protonbh}, it is possible that black hole decay does not conserve baryon number.  However the treatment of processes which do not conserve baryon number is complicated, in both the QCD evolution and hadronization, and has only been studied for a few specific processes \cite{Gibbs:1995cw,Gibbs:1995bt,Dreiner:1999qz,Sjostrand:2002ip}.  At the same time, the violation of baryon number is extremely difficult to detect experimentally and so the effect of including baryon number violation is not expected to be experimentally observable.  Therefore {\small CHARYBDIS} conserves baryon number in black hole production and decay; lepton number, however, is not conserved.

\subsection{General description}
\label{gendes}

The black hole event generator developed attempts to model the theory as 
outlined in \secref{proddec}, within the assumptions of \secref{egassump}.  There are several related parameters and switches which can be set in the first part of the Les Houches subroutine \texttt{UPINIT} \cite{Boos:2001cv}.  No other part of the charybdis1000.F code should be modified.

Firstly the properties of the beam particles must be specified.  \texttt{IDBMUP(1)} and \texttt{IDBMUP(2)} are their Particle Data Group (PDG) codes (only protons and anti-protons are allowed) and the corresponding energies are \texttt{EBMUP(1)} and \texttt{EBMUP(2)}.

As already discussed, the geometric parton-level cross section ($\sigma=\pi r_\text{S}^2$) is used but the parameters \texttt{MINMSS} and \texttt{MAXMSS} allow the mass range for the black holes produced to be specified.  This means that it is possible to adjust the lower mass limit at which this cross section for the trans-Planckian production of black holes is thought to become valid.  The details of the Monte Carlo (MC) cross~section calculation are given in Appendix~\ref{appc}.

Three other parameters which must be set before using the event generator are
\texttt{TOTDIM}, \texttt{MPLNCK} and \texttt{MSSDEF}.  The total number of dimensions in the model being used is given by \texttt{TOTDIM} (this must be set between \texttt{6} and \texttt{11}).  There are a number of different definitions of the Planck mass (set using \texttt{MPLNCK}) but the parameter \texttt{MSSDEF} can be set to three different values to allow easy interchange between the three conventions outlined in Appendix A of \cite{Giddings:2001bu}.  The conversions between these conventions are summarized in Table~\ref{massdefs} (in fact options \texttt{1},\texttt{2} and \texttt{3} correspond to \texttt{MPLNCK} being set in the conventions a, d and b of Table~\ref{consum} respectively).

\begin{table}
\def\arraystretch{1.2}
\begin{center}
\begin{tabular}{|c|l|c|}
\hline
\texttt{MSSDEF} & Conversion & Convention\\
\hline
1 & $\mbox{\texttt{MPLNCK}}=(2^{n-2}\pi^{n-1})^{\frac{1}{n+2}}M_{\text{P}(4+n)}$ & a\\
2 & $\mbox{\texttt{MPLNCK}}=M_{\text{P}(4+n)}$ & d\\
3 & $\mbox{\texttt{MPLNCK}}=(2^{n-3}\pi^{n-1})^{\frac{1}{n+2}}M_{\text{P}(4+n)} $ & b\\
\hline
\end{tabular}
\capbox{Definitions of the Planck mass}{\label{massdefs}Definitions of the Planck mass.}
\end{center}
\def\arraystretch{1.0}
\end{table}

It has been suggested that since black hole formation is a non-perturbative 
process, the momentum scale for evaluating the PDFs should be the inverse 
Schwarzschild radius rather than the black hole mass.  The switch 
\texttt{GTSCA} should be set to \texttt{.TRUE.}\ for the first of these options
and \texttt{.FALSE.}\ for the second.\footnote{As confirmed in \cite{Rizzoprivate}, the cross sections quoted in reference \cite{Giddings:2001bu} were actually calculated with the latter PDF scale.}  The PDFs to be used are set using the Les Houches parameters \texttt{PDFGUP} and \texttt{PDFSUF}\@.

As discussed in \secref{proddec}, the Hawking temperature of the black 
hole will increase as the decay progresses so that later emissions will 
typically be of higher energy.  However to allow comparison with other work 
which has ignored this effect, the \texttt{TIMVAR} switch can be 
used to set the time-variation of the Hawking temperature as on 
(\texttt{.TRUE.}) or off (\texttt{.FALSE.}).  The indication in \secref{numres} that a quasi-stationary approach to the black hole decay might not be valid for $n \ge 3$ means that the \texttt{.FALSE.}\ option might in fact give a better description of the decay for higher values of $n$.

The emission probabilities of different types of particles are set according to the new theoretical results of Chapter~\ref{greybody} (see particularly Table~\ref{fratios}).  Heavy particle production is allowed and can be controlled by setting the value of the \texttt{MSSDEC} parameter to \texttt{2} for top quark, W and Z production, or \texttt{3} to include the Higgs boson as well (\texttt{MSSDEC=1} gives only light particles).  The production spectra for heavy particles may be unreliable if the initial Hawking temperature is below the rest mass of the particle in question.

If \texttt{GRYBDY} is set as \texttt{.TRUE.}\ the particle types and energies are chosen according to the grey-body modified emission probabilities and spectra.  If instead the \texttt{.FALSE.}\ option is selected, the black-body emission probabilities and spectra are used.  The choice of energy is made in the rest frame of the black hole prior to the emission, which is isotropic in this frame.  As overall charge must be conserved, when a charged particle is to be emitted the particle or anti-particle is chosen such that the magnitude of the black hole charge decreases.  This reproduces some of the features of the charge-dependent emission spectra in \cite{Page:1977um} whilst at the same time making it easier for the event generator to ensure that charge is conserved for the full decay.

Although the Planck phase at the end of decay cannot be well modelled as it is
not well understood, the Monte Carlo event generator must have some way of
terminating the decay.  There are two different possibilities for this, each with a range of options for the terminal multiplicity.  

If \texttt{KINCUT=.TRUE.}\ termination occurs when the chosen energy for an emitted particle is ruled out by the kinematics of a two-body decay.  At this point an isotropic \texttt{NBODY} decay is performed on the black hole remnant where \texttt{NBODY} can be set between \texttt{2} and \texttt{5}.  The \texttt{NBODY} particles are chosen according to the same probabilities used for the first part of the decay. The selection is then accepted if charge and baryon number are conserved, otherwise a new set of particles is picked for the \texttt{NBODY} decay. If this does not succeed in conserving charge and baryon number after \texttt{NHTRY} attempts the whole decay is rejected and a new one generated.  If the whole decay process fails for \texttt{MHTRY} attempts then the initial black hole state is rejected and a new one generated. 

In the alternative termination of the decay (\texttt{KINCUT=.FALSE.}) particles are emitted according to their energy  spectra until $M_\text{BH}$ falls below \texttt{MPLNCK}; then an \texttt{NBODY} decay as described above is performed.  Any chosen energies which are kinematically forbidden are simply discarded. 

In order to perform the parton evolution and hadronization the general-purpose event generators require a colour flow to be defined.  This colour flow is defined in the large number of colours ($N_\text{c}$) limit in which a quark can be considered as a colour line, an anti-quark as an anti-colour line and a gluon both a colour and anti-colour line.  A simple algorithm is used to connect all the lines into a consistent colour flow.  This algorithm starts with a colour line (from either a quark or a gluon) and then randomly connects this line with one of the unconnected anti-colour lines (from either a gluon or an anti-quark).  If the selected partner is a gluon the procedure is repeated to find the partner for its colour line; if it is an anti-quark one of the other unconnected quark colour lines is selected.  If the starting particle was a gluon the colour line of the last parton is connected to the anti-colour line of the gluon.  Whilst there is no deep physical motivation for this algorithm it at least ensures that all the particles are colour-connected and the showering generator can proceed to evolve and hadronize the event.

After the black hole decay, parton-level information is written into the Les Houches common block \texttt{HEPEUP} to enable a general-purpose event generator to fragment all emitted coloured particles into hadron jets, and perform the decays of any unstable particles.

\subsection{Control switches, constants and options}

Those parameters discussed in \secref{gendes} which are designed to be set by the user are summarized in Table~\ref{parameters}.

\begin{table}[t]
\def\arraystretch{1.1}
\begin{center}
\begin{tabular}{|c|l|c|c|}
\hline
Name & Description & Values & Default\\
\hline
\texttt{IDBMUP(2)} & PDG codes of beam particles & $\pm \mbox{\texttt{2212}}$ & \texttt{2212}\\
\texttt{EBMUP(2)} & Energies of beam particles (GeV) & & \texttt{7000.0}\\
\texttt{PDFGUP(2)} & {\small PDFLIB} codes for PDF author group & & $-\mbox{\texttt{1}}$\\
\texttt{PDFSUP(2)} & {\small PDFLIB} codes for PDF set & & $-\mbox{\texttt{1}}$\\  
\hline
\texttt{MINMSS} & Minimum mass of black holes (GeV) & $<\mbox{\texttt{MAXMSS}}$ & \texttt{5000.0}\\
\texttt{MAXMSS} & Maximum mass of black holes (GeV) & $\leq\mbox{c.m. energy}$ & c.m. energy\\
\texttt{MPLNCK} & Planck mass (GeV) & $\leq\mbox{\texttt{MINMSS}}$ & \texttt{1000.0}\\
\texttt{MSSDEF} & Convention for \texttt{MPLNCK} (see Table~\ref{massdefs})& \texttt{1}--\texttt{3} & \texttt{2}\\
\texttt{TOTDIM} & Total number of dimensions ($4+n$) & \texttt{6}--\texttt{11} & \texttt{6}\\
\texttt{GTSCA} & Use $r_\text{S}^{-1}$as the PDF momentum scale & \texttt{LOGICAL} & \texttt{.FALSE.}\\
& rather than the black hole mass & &\\
\texttt{TIMVAR} & Allow $T_\text{H}$ to change with time & \texttt{LOGICAL}& \texttt{.TRUE.}\\
\texttt{MSSDEC} & Choice of decay products & \texttt{1}--\texttt{3} & \texttt{3}\\
\texttt{GRYBDY} & Include grey-body effects & \texttt{LOGICAL} & \texttt{.TRUE.}\\
\texttt{KINCUT} & Use a kinematic cut-off on the decay & \texttt{LOGICAL} & \texttt{.FALSE.}\\
\texttt{NBODY} & Number of particles in remnant decay & \texttt{2}--\texttt{5} & \texttt{2}\\
\hline
\end{tabular}
\capbox{List of {\small CHARYBDIS} parameters}{\label{parameters}List of parameters with brief descriptions, allowed values and default settings.}
\end{center}
\def\arraystretch{1.0}
\end{table}

\subsection{Using charybdis1000.F}
\enlargethispage{-\baselineskip}

The generator itself only performs the production and parton-level decay of the black hole. It is interfaced, via the Les Houches accord, to either {\small HERWIG} \cite{Marchesini:1992ch,Corcella:2000bw} or {\small PYTHIA} \cite{Sjostrand:2000wi} to perform the parton shower evolution, hadronization and particle decays.  This means that it is also necessary to have a Les Houches accord compliant version of either {\small HERWIG} or {\small PYTHIA} with both the dummy Les Houches routines (\texttt{UPINIT} and \texttt{UPEVNT}) and the dummy {\small PDFLIB} subroutines (\texttt{PDFSET} and \texttt{STRUCTM}) deleted.  For {\small HERWIG}, the first Les Houches compliant version is {\small HERWIG6.500} \cite{Corcella:2002jc}; for {\small PYTHIA}, version 6.220 \cite{Sjostrand:2001yu} or above is required.\footnote{Versions of {\small PYTHIA} above 6.200 support the Les Houches accord but cannot handle more than 7 out-going particles, which is necessary in black hole decays.}

The black hole code is available\footnote{\url{ http://www.ippp.dur.ac.uk/montecarlo/leshouches/generators/charybdis/}} as a gzipped tar file which includes

\begin{itemize}
\item{charybdis1000.F (code for the black hole generator),} 
\item{dummy.F (dummy routines needed if not using PDFLIB),} 
\item{mainpythia.f (example main program for {\small PYTHIA}),} 
\item{mainherwig.f (example main program for {\small HERWIG}),} 
\item{charybdis1000.inc (include file for the black hole generator).} 
\end{itemize}

The Makefile must specify which general-purpose event generator is to be used (i.e.\ \texttt{GENERATOR=HERWIG} or \texttt{GENERATOR=PYTHIA}) and also whether {\small PDFLIB} is to be used (\texttt{PDFLIB=PDFLIB} if required, otherwise \texttt{PDFLIB=~}).  The name of the {\small HERWIG} or {\small PYTHIA} source and the location of the {\small PDFLIB} library must also be included. 
\enlargethispage{\baselineskip}

If the code is extracted to be run separately then the following should be taken into account:

\begin{itemize}
\item{charybdis1000.F will produce the {\small HERWIG} version by default when compiled; the flag \texttt{-DPYTHIA} should be added if the {\small PYTHIA} version is required.} 
\item{dummy.F will by default produce the version for use without {\small PDFLIB}; the flag \texttt{-DPDFLIB} should be added if {\small PDFLIB} is being used.} 
\end{itemize}

\subsection{List of subroutines}

Table~\ref{subs} contains a list of all the subroutines of the generator along with their functions.  Those labelled by {\small HW/PY} are 
{\small HERWIG/PYTHIA} dependent and are pre-processed according to the \texttt{GENERATOR} flag in the Makefile.  Many of the utility routines are identical to routines which appear in the {\small HERWIG} program.

\begin{table}[p]
\def\arraystretch{1.1}
\begin{center}
\begin{tabular}{|c|l|}
\hline
Name & Description\\
\hline
& Les Houches routines\\
\hline
\texttt{UPINIT} & Initialization routine\\
\texttt{UPEVNT} & Event routine\\
\hline
& Particle decays\\
\hline
\texttt{CHDFIV} & Generates a five-body decay\\
\texttt{CHDFOR} & Generates a four-body decay\\
\texttt{CHDTHR} & Generates a three-body decay\\
\texttt{CHDTWO} & Generates a two-body decay\\
\hline
& Hard sub-process and related routines\\
\hline
\texttt{CHEVNT} & Main routine for black hole hard sub-process\\
\texttt{CHFCHG} & Returns charge of a SM particle\\
\texttt{CHFMAS} & Returns mass of a SM particle ({\small HW/PY})\\
\texttt{CHHBH1} & Chooses next particle type if \texttt{MSSDEC=1}\\
\texttt{CHHBH2} & Chooses next particle type if \texttt{MSSDEC=2}\\
\texttt{CHHBH3} & Chooses next particle type if \texttt{MSSDEC=3}\\
\texttt{CHPDF} & Calculates the PDFs ({\small HW/PY})\\
\hline
& Random number generators\\
\hline
\texttt{CHRAZM} & Randomly rotates a 2-vector\\
\texttt{CHRGEN} & Random number generator ({\small HW/PY})\\
\texttt{CHRLOG} & Random logical\\
\texttt{CHRUNI} & Random number: uniform\\
\hline
& Miscellaneous utilities\\
\hline
\texttt{CHUBHS} & Chooses particle energy from spectrum\\
\texttt{CHULB4} & Boost: rest frame to lab, no masses assumed\\
\texttt{CHULOB} & Lorentz transformation: rest frame $\rightarrow$ lab\\
\texttt{CHUMAS} & Puts mass in 5th component of vector\\
\texttt{CHUPCM} & Centre-of-mass momentum\\
\texttt{CHUROB} & Rotation by inverse of matrix {\bf R}\\
\texttt{CHUROT} & Rotation by matrix {\bf R}\\
\texttt{CHUSQR} & Square root with sign retention\\
\texttt{CHUTAB} & Interpolates in a table\\
\hline
& Vector manipulation\\
\hline
\texttt{CHVDIF} & Vector difference\\
\texttt{CHVEQU} & Vector equality\\
\texttt{CHVSUM} & Vector sum\\
\hline
\end{tabular}
\capbox{List of {\small CHARYBDIS} subroutines}{\label{subs}List of {\small CHARYBDIS} subroutines with brief descriptions.}
\end{center}
\def\arraystretch{1.0}
\end{table}

\section{Event generator output}
\label{output}

\subsection{Sample cross sections}

In this section some indicative black hole production cross sections are shown for various different settings of the {\small CHARYBDIS} parameters.  These are chosen so as to allow direct comparison with cross sections quoted in \cite{Giddings:2001bu} and \cite{Dimopoulos:2001hw}.  The results in Table~\ref{crosstab} are found to be in good agreement with the previously published values.

\begin{table}

\begin{center}
\def\arraystretch{1.1}
\begin{tabular}{|c|c|c|c|}
\hline
\texttt{MPLNCK} & \texttt{MINMSS} & \texttt{TOTDIM} & Cross section (fb)\\
\hline
1000.0 & 5000.0 & 8 & $0.155\times10^6$\\
1000.0 & 7000.0 & 8 & $6.0\times10^3$\\
1000.0 & 10000.0 & 8 & 6.9\\
1000.0 & 5000.0 & 10 & $0.233\times10^6$\\
1000.0 & 7000.0 & 10 & $8.7\times10^3$\\
1000.0 & 10000.0 & 10 & 9.7\\
\hline
2000.0 & 2000.0 & 11 & $0.511\times10^6$\\
6000.0 & 6000.0 & 7 & 0.12\\
\hline
\end{tabular}
\capbox{Black hole production cross sections}{\label{crosstab} Black hole production cross sections for various different parameters.  Those in the upper part of the table can be directly compared with values quoted in \cite{Giddings:2001bu} (they are calculated using \texttt{MSSDEF}=1, \texttt{PDFGUP(I)}=4 and \texttt{PDFSUP(I)}=48) whereas those in the lower part can be compared with \cite{Dimopoulos:2001hw} (they are calculated using \texttt{MSSDEF}=2, \texttt{PDFGUP(I)}=3 and \texttt{PDFSUP(I)}=34).  In all cases \texttt{MAXMSS}=14000.0.}
\end{center}
\def\arraystretch{1.0}
\end{table}

\subsection{Sample event display}

\begin{figure}[p]
\begin{center}
\epsfig{file=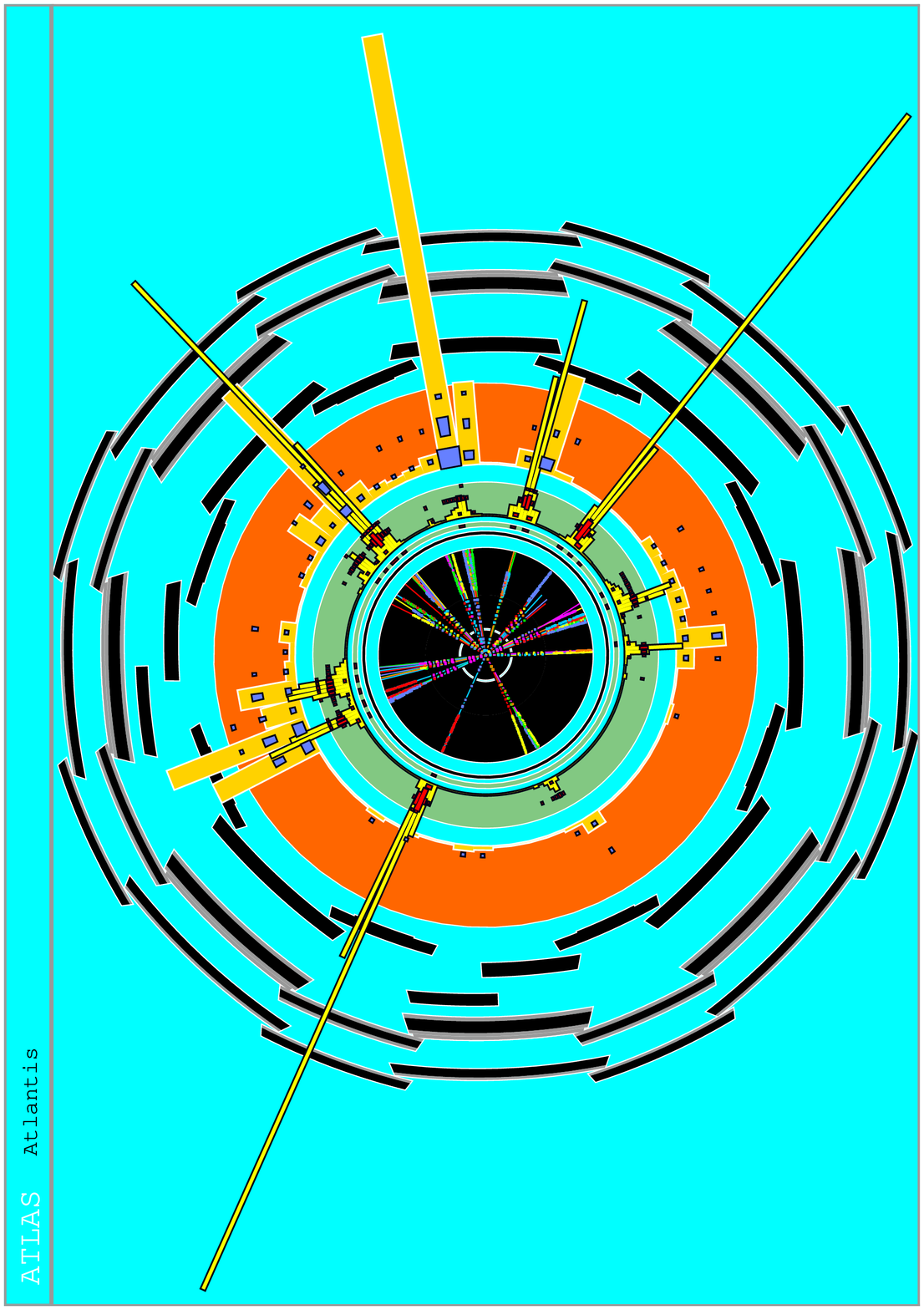, angle=0, width=0.95\textwidth}
\capbox{Sample event display}{Sample event display (details in text).\label{display}}
\end{center}
\end{figure} 

Figure~\ref{display} is an example black hole event display produced from the {\small CHARYBDIS} output using the {\small ATLANTIS} program\footnote{\url{http://atlantis.web.cern.ch/atlantis/}} which is being developed for the ATLAS experiment.  The black hole event used was generated with an earlier version of {\small CHARYBDIS} which did not fully take into account the grey-body effects; however it can still be considered as a `typical' black hole event which displays many of the expected features.  The event shown is for a black hole mass of $\sim$ 8~TeV in a scenario with \texttt{TOTDIM=10} and \texttt{MPLNCK=1000.0} (\texttt{MSSDEF=1}).  The event includes 9 quarks, a gluon, an electron, a positron and three neutrinos.

\subsection{Sample particle spectra and emissivities}

\begin{figure}
\begin{center}
\epsfig{file=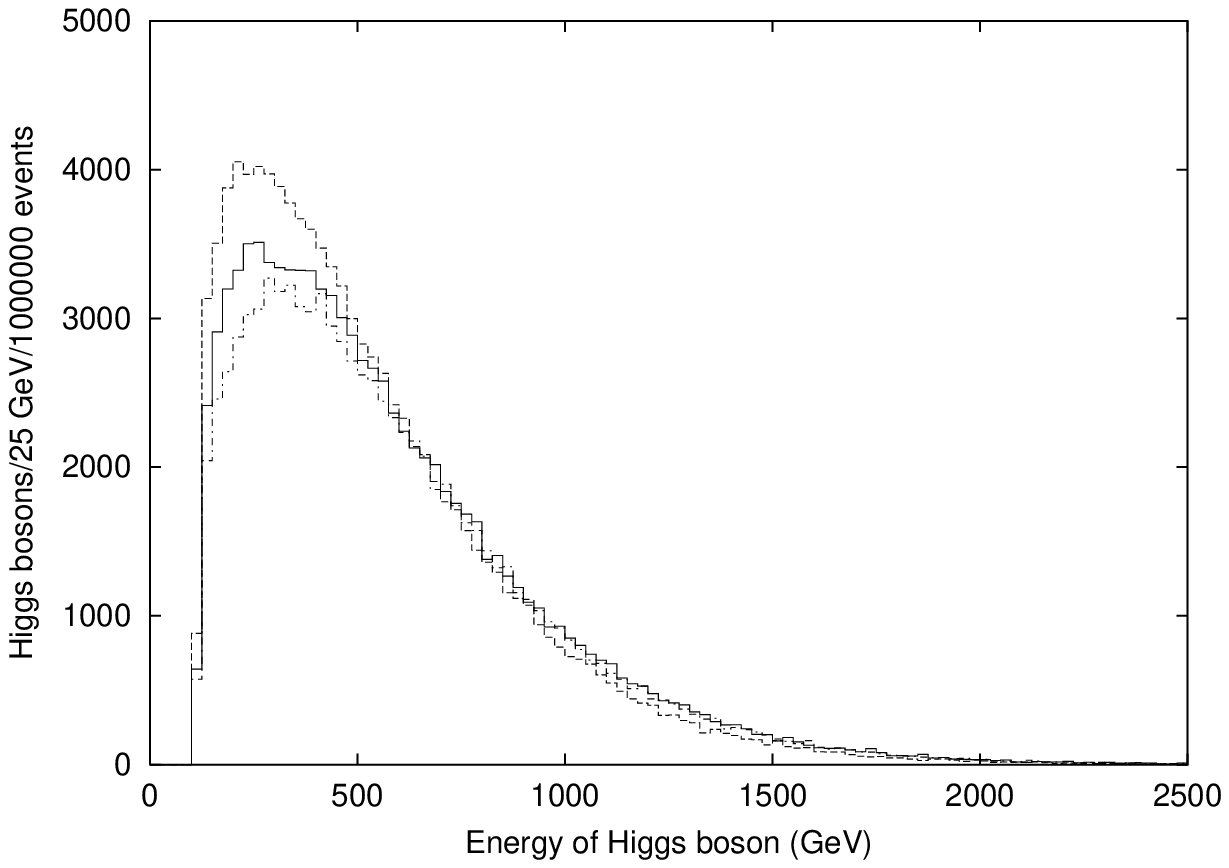, angle=0, width=.78\textwidth}
\capbox{Parton-level energy spectra of Higgs bosons, $m_\text{H} = 115$~GeV}{Parton-level energy spectra of Higgs bosons, $m_\text{H} = 115$~GeV. Solid: predicted energy spectrum of Higgs bosons from decay of black holes with initial masses 5.0--5.5 TeV.  Dashed: neglecting time-variation of temperature.  Dot-dashed: neglecting grey-body factors.\label{higgs}}
\end{center}
\end{figure}

\begin{figure}
\begin{center}
\epsfig{file=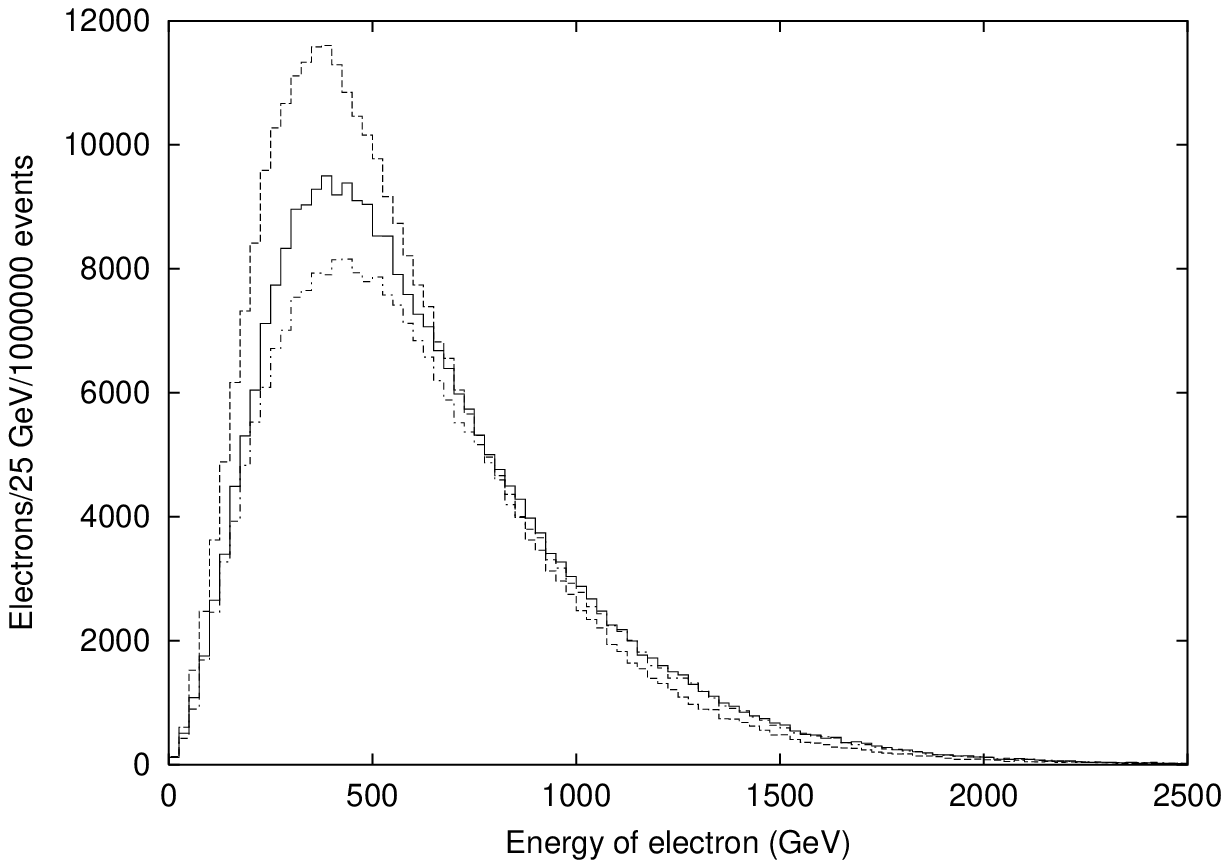, angle=0, width=.78\textwidth}
\capbox{Parton-level energy spectra of electrons and positrons}{Parton-level energy spectra of electrons and positrons.  As Figure~\ref{higgs} but for electron and positron spectra.\label{electron}}
\end{center}
\end{figure}

\begin{figure}
\begin{center}
\epsfig{file=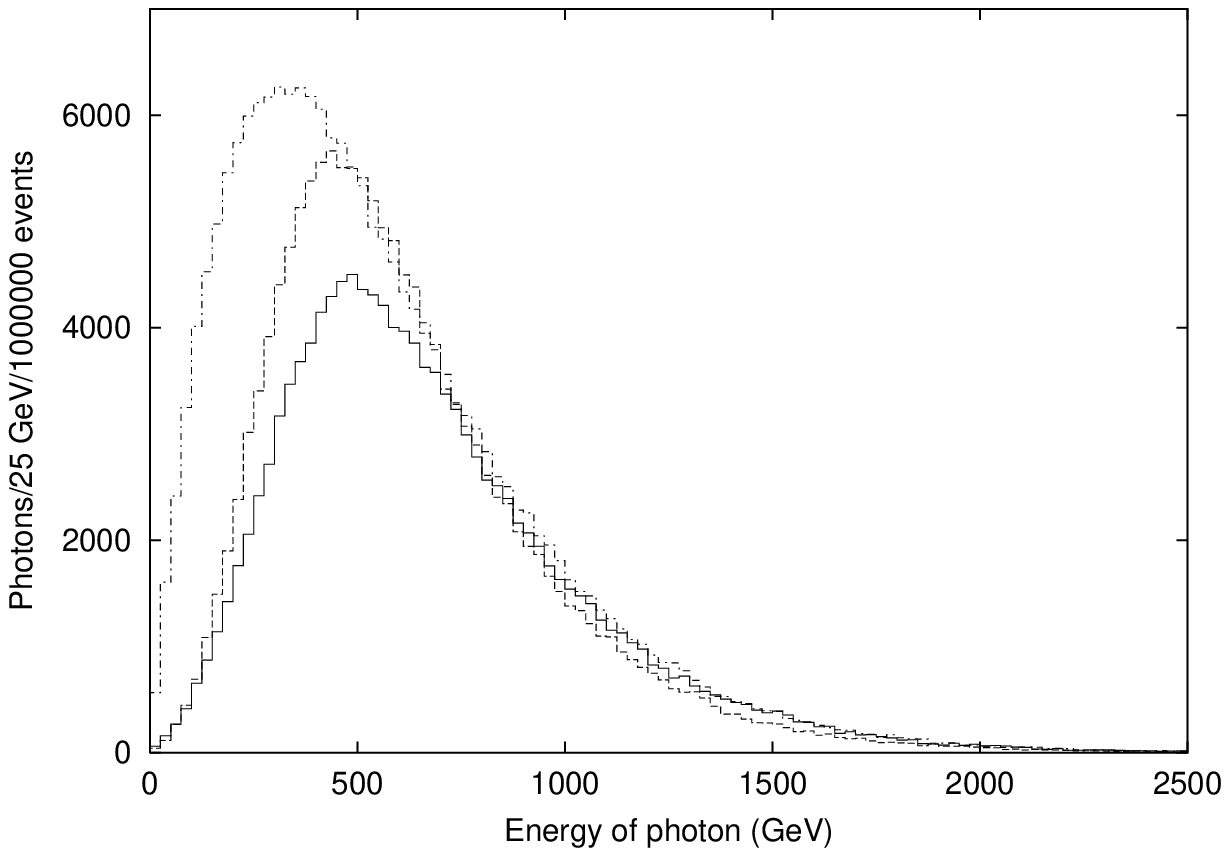, angle=0, width=.78\textwidth}
\capbox{Parton-level energy spectra of photons}{Parton-level energy spectra of photons.  As Figure~\ref{higgs} but for photon spectra.\label{photon}}
\end{center}
\end{figure}

Figures~\ref{higgs}--\ref{photon} show the results, at parton level, of neglecting the time-variation of the black hole temperature (\texttt{TIMVAR=.FALSE.}, dashed line) or the grey-body factors (\texttt{GRYBDY=.FALSE.}, dot-dashed line) for initial black hole masses in the range from \texttt{MINMSS=5000.0} to \texttt{MAXMSS=5500.0} (with the default values for the other parameters).  The solid line is for simulations with the default parameter settings (but with the same reduced range of initial black hole masses used in the other two cases).

The effect of time-variation is to harden the spectra of all particle species.
However, the effect of the grey-body factors depends on the spin, in this case slightly softening the spectra of scalars and fermions but hardening the spectrum of gauge bosons.

Relative numbers of primary emitted particles were also calculated and are shown as percentages in Table~\ref{probtab}.  The same parameters were used as for the spectra, except that in both cases the time-variation of the temperature was included (\texttt{TIMVAR=.TRUE.}).  For comparison, the theoretical values used by the event generator (and easily calculable using Tables~\ref{pprobs} and \ref{fratios}) are also shown.  The differences are due to any kinematic constraints on the massive particles and also the imposed conservation of charge and baryon number; however since the average particle multiplicity per event is relatively high in these examples, the decay of the black hole is not significantly constrained.

\begin{table}
\begin{center}
\begin{tabular}{|l|c|c|c|c|}
\hline
& \multicolumn{4}{c|}{Particle emissivity (\%)} \\
\cline{2-5}
& \multicolumn{2}{c|}{\texttt{GRYBDY=.TRUE.}} & \multicolumn{2}{c|}{\texttt{GRYBDY=.FALSE.}}\\
\cline{2-5}
Particle type & Generator & Theory & Generator & Theory\\
\hline
Quarks & 63.9 & 61.8 & 58.2 & 56.5\\
Gluons & 11.7 & 12.2 & 16.9 & 16.8\\
Charged leptons  & 9.4 & 10.3 & 8.4 & 9.4\\
Neutrinos & 5.1 & 5.2 & 4.6 & 4.7\\
Photon & 1.5 & 1.5 & 2.1 & 2.1\\
Z$^0$ & 2.6 & 2.6 & 3.1 & 3.1\\
W$^+$ and W$^-$ & 4.7 & 5.3 & 5.7 & 6.3\\
Higgs boson & 1.1 & 1.1 & 1.0 & 1.1\\
\hline
\end{tabular}
\capbox{Relative emissivities of primary partons}{Comparison of relative emissivities of primary partons with their theoretical values.\label{probtab}} 
\end{center}
\end{table}

\section{Experimental studies}
\label{expstud}

Results of a full experimental study of black hole production and decay at the LHC will be presented in \cite{ali}.  However some of the experimental issues will be highlighted here with reference to the theoretical work of Chapter~\ref{greybody}.

If extra dimensions do exist, then a priority will be to try to determine how many there are.  The black hole generator described in this chapter was constructed with the aim of being able to more realistically examine ways of doing this experimentally.  The grey-body factors of the previous chapter were incorporated for the same reason.

One method for determining $n$ is described in \cite{Dimopoulos:2001hw}: Black hole events are isolated using the experimental cuts mentioned in \secref{expsig}, events are put into 500~GeV bins based on the estimated black hole mass and then, for each mass bin, photon or electron energy spectra are produced.  If Hawking temperatures can be determined by fitting the spectra, then the correlation between $T_\text{H}$ and black hole mass can be used to try to determine the number of extra dimensions (and potentially also the value of $M_{\text{P}(4+n)}$).  

\enlargethispage{-\baselineskip}
It was argued in \cite{Dimopoulos:2001hw} that this method is quite successful but a number of unjustified approximations were made.  These include a failure to take account of the temperature increase as the decay progresses, the grey-body effects described in Chapter~\ref{greybody}, and the secondary photons and electrons.  Isolation cuts can be used to reduce the problem of secondary particles, but using the {\small CHARYBDIS} generator (with {\small GETJET} providing a very simple detector simulation) it was found that the time-dependence of the Hawking temperature significantly reduces the viability of such a method.  The effect of the black hole not always being at rest is also found to have a significant effect on the energy spectra (although this is itself a matter of some uncertainty since the way in which a black hole recoils against high-energy emissions may be complicated \cite{Page:1980tc,Frolov:2002as,Frolov:2002gf}).

Attempts were made to perform a more sophisticated version of this method by using a more realistic spectrum, obtained by numerically integrating the Hawking spectrum, for the fitting procedure.  This approach was significantly more successful for $n=2$ but less promising for larger values of $n$.  This is because, as the emitted particle multiplicity decreases, it is difficult to know what cut-off to use when calculating the time-integrated spectrum, particularly without making assumptions about the value of $M_{\text{P}(4+n)}$.  This highlights a problem which was found repeatedly when dealing with the larger values of $n$: the expected particle multiplicity at the LHC is not particularly high (see calculations in \secref{totfandp}) and so it is difficult to extract information in a way which is not affected by the kinematic constraints on the decay, and the way in which the Planck phase is modelled.

Experimentalists from the ATLAS collaboration have subsequently spent a significant amount of time trying to find event shape variables and other distributions which can be used to extract the number of extra dimensions from black hole events; however results are as yet inconclusive.  It has hoped that by using the different event generator options for modelling the Planck phase it will be possibly to confirm that any proposed method is relatively insensitive to this final phase of the decay.  However most of methods investigated have been limited by the problem discussed above---that is, the low multiplicity at the LHC for large numbers of extra dimensions (and with $M_{\text{P}(4+n)}$ assumed to be $\sim$~TeV).  A higher energy collider in the more distant future would alleviate many of these problems.  The rate of fractional change of $n$ decreases with $n$ which also makes it progressively more difficult to distinguish $n$ from $n+1$ extra dimensions.  It is unfortunate that these problems are most significant for large values of $n$ since these have the best theoretical motivation and are the least constrained astrophysically.

\section{Discussion}
\label{discuss}

The {\small CHARYBDIS} program described in this chapter would appear to be the most sophisticated black hole event generator available.  Figures~\ref{higgs}--\ref{photon} and Table~\ref{probtab} confirm that the inclusion of the grey-body factors can have a significant effect on the particle energy spectra and emissivities.

However it will have become clear to the reader that there are many assumptions and approximations made in the way the production and decay of black holes is modelled.  Various improvements could be made, but most of them would be impossible without further theoretical work.

A major advance would be to correctly model stages of the black hole decay other than the Schwarzschild phase.  The balding phase could be taken into account if the amount of energy trapped by the black hole horizon was more accurately known.  It was recently pointed out \cite{Anchordoqui:2003ug} that this might have a significant effect on the black hole discovery potential of the LHC\@.  Further numerical calculations to estimate the typical amount of trapped energy could also provide information about the expected distribution of initial angular momenta.  This information, together with more theoretical work on the grey-body factors for rotating black holes, would allow the spin-down phase of the decay to be simulated in the event generator.  If attempting to model the spin-down phase, the assumption of isotropic particle emission from the black hole would no longer be a good one.   Grey-body factors are calculated for individual values of $\ell$ and $m$ and so it might be possible for the angular distribution of emitted particles to be modelled correctly (although to do this fully would also require the spin-weighted spheroidal harmonics of equation~(\ref{spinang}) to be calculated).  

The assumption that the emission of energy into the bulk as gravitons (and possibly scalars) can be neglected is not ideal.  The emission of gravitons from the black hole could be taken into account but a full implementation of this would again require more theoretical work on the grey-body factors.

Other possible improvements would be to more carefully account for the modifications in the energy spectra of massive gauge bosons and other heavy particles, or to allow the black holes to emit supersymmetric particles (since extra dimension models motivated from string theory are expected to be supersymmetric).

One relatively simple modification to {\small CHARYBDIS} would be to try to simulate black holes in a separated fermions model \cite{Arkani-Hamed:1999za,Arkani-Hamed:1999dc} which would avoid some of the proton decay issues (see discussion in \secref{protonbh}).  A model like this would be implemented by altering the number of degrees of freedom available for decays as described in \cite{Han:2002yy}.  However, since the values of $M_{\text{P}(4+n)}$ required in such models make black hole production experimentally inaccessible at the next generation of colliders, there is little motivation for such a modification.

It would be desirable to have a better understanding of both the validity of the quasi-stationary approach to the decay of higher-dimensional black holes, and of the Planck phase of the decay.  In the latter case, it seems unlikely that this will be possible without a full theory of quantum gravity and in the meantime the best approach would seem to be to try to extract information which is independent of the way in which the Planck phase is modelled.

\clearpage{\pagestyle{empty}\cleardoublepage}
\chapter{Conclusions}

The individual chapters of this thesis contain their own conclusions, so here there is only a brief overview, and a few comments about the future of the field.  Two different possibilities for new physics beyond the Standard Model have been examined: supersymmetry and extra dimensions.  Although the phenomenology of both models is diverse the particular aspects addressed were the detection of heavy leptons in intermediate scale supersymmetric models, and the production and decay of black holes in extra dimension scenarios.  In both cases it was shown that the next generation of particle colliders will be able to investigate these models.

The study of the detection of charged heavy leptons at the LHC using a time-of-flight technique showed that, in the particular model considered, the discovery mass range extends up to 950~GeV\@.  In addition it was found that it would be difficult to use their differing angular distributions to distinguish the heavy leptons from scalar leptons if the lepton mass was greater than 580~GeV\@.

The majority of the thesis addressed various aspects of the production and decay of miniature black holes.  The grey-body factors which were calculated numerically in Chapter~\ref{greybody} are useful in that they allow a more realistic Monte Carlo event generator to be constructed, as described in Chapter~\ref{generator}.  Whilst the experimental signatures produced by black hole decays are expected to make events unmissable, it is important to model grey-body effects correctly if information (like the number of extra dimensions) is to be accurately extracted.  It was found that the grey-body factors modify the particle energy spectra and emissivities and so might be important in such a task.  The work on grey-body factors also allowed some of the usual assumptions and approximations made in black hole decay to be examined more carefully.  The assumption that the relative emission of particles into the bulk is small was shown to be correct, but some doubt was cast on the usual approximation that a higher-dimensional black hole decay can be treated in a quasi-stationary manner.  

One conclusion from this work was that if the number of extra dimensions is more than three or four, the centre-of-mass energy available at the LHC is barely sufficient for the semi-classical approximations used to be legitimate.  Therefore it might need a higher energy collider to investigate black hole production in detail.  The LHC would still be expected to provide the first indications of extra dimensions through graviton production, Planckian effects and the first signs of black hole production.

The question of how likely it is that any particular model for new physics is realized in nature is inevitably impossible to quantify.  Every physicist has their own prejudices and in general supersymmetric models seem to be the favoured option.  Extra dimension models are `wacky' enough to attract the attention of the popular science media and are quickly dismissed by some physicists.  However, as outlined in Chapter~\ref{introch}, they can be well motivated theoretically.

Almost all high-energy physicists are convinced that the LHC will discover new physics of some kind.  Leaving aside all the theoretical arguments, it would be unprecedented in the history of the field if the order-of-magnitude increase in available energy did not reveal something new.  It may well be that nature surprises us and reveals physics that nobody has so far suggested.  Even if extra dimensions do exist the compactification scheme may make the physics much more complicated than the simple cases discussed here.  

With half a decade still to wait before CERN's Large Hadron Collider has produced a substantial amount of data, there is good reason for theorists to continue investigating these models, and refining and extending their phenomenological studies.

\appendix
\renewcommand{\chaptermark}[1]{\markboth{\bf Appendix \thechapter:  {#1}}{}}
\renewcommand{\sectionmark}[1]{\markright{\bf \thesection\ #1}}
\clearpage{\pagestyle{empty}\cleardoublepage}
\chapter{Cross Section for Heavy Lepton Pair Production}
\chaptermark{Heavy Lepton Pair Production Cross Section}
\label{appa}

\section{$2\rightarrow2$ body scattering in centre-of-mass frame}

The differential cross section for $2\rightarrow2$ scattering in the centre-of-mass frame is given by
\begin{equation}
\frac{d\sigma}{d\Omega^*} = \frac{1}{64\pi^2 s} \frac{p_\text{f}^*}{p_\text{i}^*} 
\langle\,|M_\text{fi}|^2\,\rangle.
\label{diffxs}
\end{equation}
In the above expression, the asterisk indicates the centre-of-mass 
frame, subscripts i and f refer to initial- and final-state particles, and $s$ is the square of the total centre-of-mass energy.

Therefore the major work in calculating the cross section is to determine the 
spin averaged square of the transition matrix element $\langle\,|M_\text{fi}|^2\,\rangle$, evaluated in the centre-of-mass frame.

\section{Definitions and conventions}
\label{defs}

The 4-momenta of the particles involved are defined in this section.  Subscripts 1, 2, 3 and 4 refer to the quark, anti-quark, lepton and anti-lepton
respectively, and the angle $\theta^*$ is between the out-going lepton and the 
incoming quark in the centre-of-mass frame.  Hence the $(E,p)$ 4-momenta are 
as follows (note that the incoming quarks are treated as massless):
\begin{equation}
\begin{split}
p_1 &= (E,0,0,E)\,;\\[2mm]
p_2 &= (E,0,0,-E)\,;\\[2mm]
p_3 &= (E_3,p_3 \sin \theta^*,0,p_3 \cos \theta^*)\,;\\[2mm]
p_4 &= (E_4,-p_3 \sin \theta^*,0,-p_3 \cos \theta^*)\,.
\end{split}
\end{equation}

Later we will require the Lorentz dot products of several pairs of 
these 4-vectors.  The useful ones are listed below:
\begin{equation}
\begin{split}
p_1.p_2 &= 2E^2\,;\\[2mm]
p_1.p_3 &= EE_3-Ep_3\cos\theta^*\,; \\[2mm]
p_2.p_4 &= EE_4-Ep_3\cos\theta^*\,; \\[2mm]
p_1.p_4 &= EE_4+Ep_3\cos\theta^*\,; \\[2mm]
p_2.p_3 &= EE_3+Ep_3\cos\theta^*.
\end{split}
\end{equation}

\section{Matrix element calculation}

Using the Feynman rules, the matrix element for Z boson exchange is
\begin{multline}
-iM_\text{fi}^\text{Z} = \left[\bar{v}(p_2).-ig_\text{Z}\gamma^{\mu}\frac{1}{2}\left(c_\text{V}^\text{i}-c_\text{A}^\text{i}\gamma^5\right)
u(p_1)\right]\frac{-ig_{\mu\nu}}{s-m_\text{Z}^2+im_\text{Z}\Gamma_\text{Z}}\\[3mm]
\times\left[\bar{u}(p_3).-ig_\text{Z}\gamma^{\nu} \frac{1}{2}\left(c_\text{V}^\text{f}-c_\text{A}^\text{f}\gamma^5\right)v(p_4)\right], 
\end{multline}
which gives
\begin{multline}
M_\text{fi}^\text{Z} = -\frac{g_\text{Z}^2}{s-m_\text{Z}^2+im_\text{Z}\Gamma_\text{Z}}g_{\mu\nu}
\left[\bar{v}(p_2)\gamma^{\mu}\frac{1}{2}\left(c_\text{V}^\text{i}-c_\text{A}^\text{i}\gamma^5\right)u(p_1)\right]\\[3mm]
\times\left[\bar{u}(p_3)\gamma^{\nu} \frac{1}{2}\left(c_\text{V}^\text{f}-c_\text{A}^\text{f}\gamma^5\right)v(p_4)\right].
\end{multline}
Note that the matrix element for $\gamma$-exchange is easily 
obtained from this by setting $m_\text{Z}$ and $c_\text{A}$ to zero, $c_\text{V}$ to 2, and 
$g_\text{Z}^2$ to $q^\text{i}q^\text{f}e^2$.

Taking into account Z- and $\gamma$-exchange the full matrix element may be 
written as $M_\text{fi}=M_\text{fi}^\text{Z}+M_\text{fi}^{\gamma}$.  Hence
\begin{equation}
\begin{split}
|M_\text{fi}|^2&=|M_\text{fi}^\text{Z}|^2+|M_\text{fi}^{\gamma}|^2+(M_\text{fi}^\text{Z})^*M_\text{fi}^{\gamma}
+M_\text{fi}^\text{Z}(M_\text{fi}^{\gamma})^*\\[2mm]
&=|M_\text{fi}^\text{Z}|^2+|M_\text{fi}^{\gamma}|^2+2 \Re\{(M_\text{fi}^\text{Z})^*M_\text{fi}^{\gamma}\}\,.
\end{split}
\end{equation}
We start by calculating the first term in this expression (from which the 
second term is easily obtained using the substitutions mentioned above).  
Finally the third term which contains the Z-$\gamma$ interference information 
is calculated.

To obtain the spin-averaged square of the transition matrix element which 
appears in equation~(\ref{diffxs}), we must average over the spins of the 
incoming quarks and sum over the spins of out-going leptons such that
\begin{equation}
\langle\,|M_\text{fi}|^2\,\rangle = \frac{1}{4} \sum_\text{spins} |M_\text{fi}|^2.
\end{equation}

\subsection{Z-exchange}

Firstly we want to calculate $\sum_\text{spins} |M_\text{fi}^\text{Z}|^2$.  This is given by
\begin{equation}
\begin{split}
\sum_\text{spins} |M_\text{fi}^\text{Z}|^2 = \sum_{r,r',s,s'}&
\frac{g_\text{Z}^4}{(s-m_\text{Z}^2)^2+m_\text{Z}^2\Gamma_\text{Z}^2} \\[3mm]
&\times\bar{v}_{r'}(p_4)\left(c_\text{V}^\text{f}+c_\text{A}^\text{f}\gamma^5\right)\frac{\gamma^{\mu}}{2}u_{s'}(p_3)
\left[\bar{u}_s(p_1)\left(c_\text{V}^\text{i}+c_\text{A}^\text{i}\gamma^5\right)\frac{\gamma_{\mu}}{2}v_r(p_2) \right.\\[3mm]
&\times\left.\bar{v}_r(p_2)\frac{\gamma_{\nu}}{2}\left(c_\text{V}^\text{i}-c_\text{A}^\text{i}\gamma^5\right)u_s(p_1)\right]
\bar{u}_{s'}(p_3)\frac{\gamma^{\nu}}{2}\left(c_\text{V}^\text{f}-c_\text{A}^\text{f}\gamma^5\right)v_{r'}(p_4).
\end{split}
\label{zex1}
\end{equation}
To obtain the above expression the following have been used:
\begin{equation}
\begin{split}
(\bar{u}\gamma^{\nu}v)^*&=(u^{\dagger}\gamma^0\gamma^{\nu}v)^\dagger
=v^{\dagger}(\gamma^{\nu})^{\dagger}(\gamma^0)^{\dagger}u
=v^{\dagger}(\gamma^0\gamma^{\nu}\gamma^0)\gamma^0u
=\bar{v}\gamma^{\nu}u\,;\\[2mm]
(\bar{u}\gamma^{\nu}\gamma^5v)^*&=(u^{\dagger}\gamma^0\gamma^{\nu}
\gamma^5v)^\dagger
=v^{\dagger}(\gamma^5)^{\dagger}(\gamma^{\nu})^{\dagger}(\gamma^0)^{\dagger}u
=v^{\dagger}\gamma^5(\gamma^0\gamma^{\nu}\gamma^0)\gamma^0u\\[2mm]
&=v^{\dagger}\gamma^5\gamma^0\gamma^{\nu}u
=-v^{\dagger}\gamma^0\gamma^5\gamma^{\nu}u
=-\bar{v}\gamma^5\gamma^{\nu}u\,.
\end{split}
\end{equation}

We now make use of familiar trace identities ($\bar{u}_{\alpha}
\Gamma_{\alpha\beta}u_{\beta}= \mbox{Tr}\,(u\bar{u}\Gamma)$ etc. with 
$\alpha,\beta$ Dirac matrix indices) as well as the following spinor identities:
\begin{equation}
\begin{split}
\sum_\text{spins}u(p)\bar{u}(p)=\gamma.p+m\,; \\[2mm]
\sum_\text{spins}v(p)\bar{v}(p)=\gamma.p-m\,.
\end{split}
\end{equation}
Using these identities (and temporarily ignoring the propagators) eq.~(\ref{zex1}) becomes
\begin{equation}
\begin{split}
&\sum_{r,r',s,s'}
\bar{v}_{r'}(p_4)\left(c_\text{V}^\text{f}+c_\text{A}^\text{f}\gamma^5\right)\frac{\gamma^{\mu}}{2}u_{s'}(p_3)
\left[\bar{u}_s(p_1)\left(c_\text{V}^\text{i}+c_\text{A}^\text{i}\gamma^5\right)\frac {\gamma_{\mu}}{2}v_r(p_2)\right.\\[3mm]
&\quad\:\times\left.\bar{v}_r(p_2)\frac{\gamma_{\nu}}{2}\left(c_\text{V}^\text{i}-c_\text{A}^\text{i}\gamma^5\right)u_s(p_1)\right]
\bar{u}_{s'}(p_3)\frac{\gamma^{\nu}}{2}\left(c_\text{V}^\text{f}-c_\text{A}^\text{f}\gamma^5\right)v_{r'}(p_4)\\[3mm]
&=\mbox{Tr}\biggl[\gamma.p_1\left(c_\text{V}^\text{i}+c_\text{A}^\text{i}\gamma^5\right)\frac{\gamma_{\mu}}{2}\gamma.p_2
\frac{\gamma_{\nu}}{2}\left(c_\text{V}^\text{i}-c_\text{A}^\text{i}\gamma^5\right)\biggr]\\[3mm]
&\quad\:\times\mbox{Tr}\left[(\gamma.p_4-m_4)\left(c_\text{V}^\text{f}+c_\text{A}^\text{f}\gamma^5\right)\frac{\gamma^{\mu}}{2}(\gamma.p_3
+m_3)\frac{\gamma^{\nu}}{2}\left(c_\text{V}^\text{f}-c_\text{A}^\text{f}\gamma^5\right)\right].
\end{split}
\end{equation}
\enlargethispage{-\baselineskip}

We now evaluate the traces using the properties that the trace of 
the product of an odd number of gamma matrices is zero, 
$\mbox{Tr}\,(\gamma^{\mu}\gamma^{\nu})=4g^{\mu\nu}$, $\mbox{Tr}\,(\gamma^{\mu}\gamma^{\nu}
\gamma^{\lambda}\gamma^{\sigma})=4(g^{\mu\nu}g^{\lambda\sigma}-g^
{\mu\lambda}g^{\nu\sigma}+g^{\mu\sigma}g^{\nu\lambda})$, 
$\mbox{Tr}\,(\gamma^5\gamma^{\mu}\gamma^{\nu})=0$, and
$\mbox{Tr}\,(\gamma^5\gamma^{\mu}\gamma^{\nu}\gamma^{\lambda}\gamma^{\sigma})
=4i\epsilon^{\mu\nu\lambda\sigma}$.  Therefore the second trace becomes
\begin{equation}
\begin{split}
&\frac{1}{4}\left[\left(c_\text{V}^\text{f}\right)^2 \mbox{Tr}\,(\gamma.p_4\gamma^{\mu}\gamma.p_3\gamma^{\nu}
-m_3m_4\gamma^{\mu}\gamma^{\nu})\right.\\[3mm]
&\quad\left.-\left(c_\text{A}^\text{f}\right)^2 \mbox{Tr}\,(\gamma.p_4\gamma^5\gamma^{\mu}
\gamma.p_3\gamma^5 -m_3m_4\gamma^5\gamma^{\mu}\gamma^{\nu}\gamma^5)
+2c_\text{V}^\text{f}c_\text{A}^\text{f} \mbox{Tr}\,(\gamma.p_4\gamma^5\gamma^{\mu}\gamma.p_3\gamma^{\nu})\right]\\[3mm]
&=\left[\left(c_\text{V}^\text{f}\right)^2+\left(c_\text{A}^\text{f}\right)^2\right](p_4^{\mu}p_3^{\nu}+p_4^{\nu}p_3^{\mu}-
p_3.p_4 g^{\mu\nu})-\left[\left(c_\text{V}^\text{f}\right)^2-\left(c_\text{A}^\text{f}\right)^2\right]m_3m_4g^{\mu\nu}\\[3mm]
&\quad\,-2c_\text{V}^\text{f}c_\text{A}^\text{f}i\epsilon^{\alpha\mu\beta\nu}(p_4)_{\alpha}(p_3)_{\beta}\,,
\end{split}
\end{equation}
where the properties of $\gamma^5$ (it squares to 1 and anti-commutes with the other $\gamma$-matrices) have also been used.  The first trace is similar, but simpler as we are treating the incoming quarks
as massless.  It becomes
\begin{equation}
\left[\left(c_\text{V}^\text{i}\right)^2+\left(c_\text{A}^\text{i}\right)^2\right]\left[(p_1)_{\mu}(p_2)_{\nu}+(p_1)_{\nu}(p_2)_{\mu}-
p_1.p_2 g_{\mu\nu}\right]
- 2c_\text{V}^\text{i}c_\text{A}^\text{i}i\epsilon_{\gamma\mu\delta\nu}p_1^{\gamma}p_2^{\delta}\,.
\end{equation}

Finally we must multiply these two terms together, noting which 
terms do not contribute because of the anti-symmetry/symmetry of the indices.
We obtain
\begin{equation}
\begin{split}
&\left[\left(c_\text{V}^\text{f}\right)^2+\left(c_\text{A}^\text{f}\right)^2\right]\left[\left(c_\text{V}^\text{i}\right)^2+\left(c_\text{A}^\text{i}\right)^2\right]2\left[\,(p_1.p_4)(p_3.p_2)+(p_1.p_3)(p_2.p_4)\,\right]\\[3mm]
&\;+\left[\left(c_\text{V}^\text{f}\right)^2-\left(c_\text{A}^\text{f}\right)^2\right]\left[\left(c_\text{V}^\text{i}\right)^2+\left(c_\text{A}^\text{i}\right)^2\right]2m_3m_4p_1.p_2
-4c_\text{V}^\text{f}c_\text{A}^\text{f}c_\text{V}^\text{i}c_\text{A}^\text{i}\epsilon_{\gamma\mu\delta\nu}\epsilon_{\alpha\mu\beta
\nu}p_1^{\gamma}p_2^{\delta}(p_4)_{\alpha}(p_3)_{\beta}\,.
\end{split}
\end{equation}
To complete the calculation we use the identity 
$\epsilon_{\mu\nu\alpha\beta}\epsilon_{\mu\nu\gamma\delta}=
2(g_{\alpha}^{\delta}g_{\beta}^{\gamma}
-g_{\alpha}^{\gamma}g_{\beta}^{\delta})$, which means that
\begin{equation}
\begin{split}
\epsilon_{\gamma\mu\delta\nu}\epsilon_{\alpha\mu\beta\nu}
p_1^{\gamma}p_2^{\delta}(p_4)_{\alpha}(p_3)_{\beta}
&=\epsilon_{\mu\nu\alpha\beta}\epsilon_{\mu\nu\gamma\delta}
p_1^{\gamma}p_2^{\delta}(p_4)_{\alpha}(p_3)_{\beta}\\[2mm]
&=2\left(g_{\alpha}^{\delta}g_{\beta}^{\gamma}-g_{\alpha}^{\gamma}
g_{\beta}^{\delta}\right)p_1^{\gamma}p_2^{\delta}(p_4)_{\alpha}(p_3)_{\beta}\\[2mm]
&=2\left[\,(p_1.p_3)(p_2.p_4)-(p_1.p_4)(p_2.p_3)\,\right].
\end{split}
\end{equation}
\enlargethispage{\baselineskip}Putting all of the above together gives us that
\begin{equation}
\begin{split}
\langle\,|M_\text{fi}^\text{Z}|^2\,\rangle=&\frac{1}{4}\frac{g_\text{Z}^4}{(s-m_\text{Z}^2)^2+m_\text{Z}^2\Gamma_\text{Z}^2}\\[3mm]
&\times\Bigl(\left[\left(c_\text{V}^\text{f}\right)^2+\left(c_\text{A}^\text{f}\right)^2\right]\left[\left(c_\text{V}^\text{i}\right)^2+\left(c_\text{A}^\text{i}\right)^2\right]2\left[\,(p_1.p_4)(p_3.p_2)+(p_1.p_3)
(p_2.p_4)\,\right]\\[3mm]
&\quad\quad+\left[\left(c_\text{V}^\text{f}\right)^2-\left(c_\text{A}^\text{f}\right)^2\right]\left[\left(c_\text{V}^\text{i}\right)^2+\left(c_\text{A}^\text{i}\right)^2\right]2m_3m_4p_1.p_2\\[3mm]
&\quad\quad+8c_\text{V}^\text{f}c_\text{A}^\text{f}c_\text{V}^\text{i}c_\text{A}^\text{i}\left[\,(p_1.p_4)(p_2.p_3)-(p_1.p_3)(p_2.p_4)\,\right]\Bigr).
\end{split}
\end{equation}

\subsection{$\gamma$-exchange}

Using the replacements outlined above, it is immediately obvious that
\begin{equation}
\langle\,|M_\text{fi}^{\gamma}|^2\,\rangle=\frac{1}{4}\frac{e^4(q^\text{f})^2(q^\text{i})^2}{s^2}32\left[\,(p_1.p_4)(p_3.p_2)+(p_1.p_3)(p_2.p_4)+m_3m_4p_1.p_2)\,\right].
\end{equation}

\subsection{Interference term}

The final part of the squared matrix element to be calculated is 
$\Re\{(M_\text{fi}^\text{Z})^*M_\text{fi}^{\gamma}\}$.  Neglecting the propagators for now, the 
spin sum is
\begin{equation}
\begin{split}
&\sum_{r,r',s,s'}
\bar{v_{r'}}(p_4)\left(c_\text{V}^\text{f}+c_\text{A}^\text{f}\gamma^5\right)\frac{\gamma^{\mu}}{2}u_{s'}(p_3)
\left[\bar{u_s}(p_1)\left(c_\text{V}^\text{i}+c_\text{A}^\text{i}\gamma^5\right)\frac{\gamma_{\mu}}{2}v_r(p_2)\right.\\[2mm]
&\quad\:\times\Bigl.\bar{v}_r(p_2)\gamma_{\nu}u(p_1)\Bigr]\bar{u}_{s'}(p_3)\gamma^{\nu}v_{r'}(p_4) \\[3mm]
&=\mbox{Tr}\biggl[\gamma.p_1\left(c_\text{V}^\text{i}+c_\text{A}^\text{i}\gamma^5\right)\frac{\gamma_{\mu}}{2}\gamma.p_2
\gamma_{\nu}\biggr]\\[3mm]
&\quad\:\times\mbox{Tr}\left[(\gamma.p_4-m_4)\left(c_\text{V}^\text{f}+c_\text{A}^\text{f}\gamma^5\right)\frac{\gamma^{\mu}}{2}(\gamma.p_3+m_3)\gamma^{\nu}\right].
\end{split}
\end{equation}
Now the second trace is
\begin{multline}
\frac{1}{2}\left[c_\text{V}^\text{f} \mbox{Tr}\left(\gamma.p_4\gamma^{\mu}\gamma.p_2\gamma^{\nu}
-m_3m_4\gamma^{\mu}\gamma^{\nu}\right)
+c_\text{A}^\text{f} \mbox{Tr}\left(\gamma.p_4\gamma^5\gamma^{\mu}\gamma.p_3\gamma^{\nu}\right)\right]\\[2mm]
=2c_\text{V}^\text{f}\left(p_4^{\mu}p_3^{\nu}+p_4^{\nu}p_3^{\mu}-p_3.p_4 g^{\mu\nu}
-m_3m_4g^{\mu\nu}\right)-2c_\text{A}^\text{f}i\epsilon^{\alpha\mu
\beta\nu}(p_4)_{\alpha}(p_3)_{\beta}\,,
\end{multline}
and the first
\begin{equation}
2c_\text{V}^\text{i}\left[\,(p_1)_{\mu}(p_2)_{\nu}+(p_1)_{\nu}(p_2)_{\mu}-p_1.p_2 g_{\mu\nu}\right]
- 2c_\text{A}^\text{i}i\epsilon_{\gamma\mu\delta\nu}p_1^{\gamma}p_2^{\delta}\,.
\end{equation}
Multiplying these together and using the same identity as before for
the $\epsilon$-tensor we obtain
\begin{equation}
\begin{split}
\langle(M_\text{fi}^\text{Z})^*M_\text{fi}^{\gamma}\rangle=&\frac{1}{4}\frac{g_\text{Z}^2e^2q^\text{f}q^\text{i}}
{s(s-m_\text{Z}^2-im_\text{Z}\Gamma_\text{Z})}\\[3mm]
&\times8\Bigl(c_\text{V}^\text{f}c_\text{V}^\text{i}\left[\,(p_1.p_4)(p_3.p_2)+(p_1.p_3)(p_2.p_4)+m_3m_4p_1.p_2\right]\Bigr.\\[3mm]
&\quad\quad\;\Bigl.+c_\text{A}^\text{f}c_\text{A}^\text{i}\left[\,(p_1.p_4)(p_2.p_3)-(p_1.p_3)(p_2.p_4)\right]\Bigr).
\end{split}
\end{equation}

\subsection{Complete matrix element squared}

Making the change of variables 
\begin{equation}
d_\text{V} = \frac {c_\text{V} g_\text{Z}} {2e},
\end{equation}
and remembering that the real part of the interference term 
contributes twice, we obtain
\begin{multline}
\langle\,|M_\text{fi}|^2\,\rangle=\frac{8e^4}{s^2}\Bigl(C_1\left[\,(p_1.p_4)(p_3.p_2)+(p_1.p_3)(p_2.p_4)\,\right]+C_2m_3m_4p_1.p_2\Bigr.\\[2mm]
\Bigl.+C_3\left[\,(p_1.p_4)(p_2.p_3)-(p_1.p_3)(p_2.p_4)\,\right]\Bigr),
\end{multline}
where
\begin{gather}
C_1 = \frac {\left[\left(d_\text{V}^\text{f}\right)^2 + \left(d_\text{A}^\text{f}\right)^2 
\right] \left[ \left(d_\text{V}^\text{i}\right)^2 + \left(d_\text{A}^\text{i}\right)^2 \right]s^2 } 
{\left(s-m_\text{Z}^2\right)^2 + m_\text{Z}^2\Gamma_\text{Z}^2} + \left(q^\text{f}\right)^2
\left(q^\text{i}\right)^2 + \frac {2 q^\text{f} q^\text{i} d_\text{V}^\text{f} d_\text{V}^\text{i}s
\left(s-m_\text{Z}^2\right)}{\left(s-m_\text{Z}^2\right)^2 + m_\text{Z}^2\Gamma_\text{Z}^2}\,,
\\[3mm]
C_2 = \frac {\left[\left(d_\text{V}^\text{f}\right)^2 - \left(d_\text{A}^\text{f}\right)^2 
\right] \left[ \left(d_\text{V}^\text{i}\right)^2 + \left(d_\text{A}^\text{i}\right)^2 \right]s^2 } 
{\left(s-m_\text{Z}^2\right)^2 + m_\text{Z}^2\Gamma_\text{Z}^2} + \left(q^\text{f}\right)^2
\left(q^\text{i}\right)^2 + \frac {2 q^\text{f} q^\text{i} d_\text{V}^\text{f} d_\text{V}^\text{i}s 
\left(s-m_\text{Z}^2\right)}{\left(s-m_\text{Z}^2\right)^2 + m_\text{Z}^2\Gamma_\text{Z}^2}\,,
\\[3mm]
C_3 = 2 \left( \frac {2 d_\text{V}^\text{f} d_\text{A}^\text{f} d_\text{V}^\text{i} d_\text{A}^\text{i}s^2} 
{\left(s-m_\text{Z}^2 \right)^2 + m_\text{Z}^2\Gamma_\text{Z}^2} + 
\frac {q^\text{f} q^\text{i} d_\text{A}^\text{f} d_\text{A}^\text{i} s \left(s-m_\text{Z}^2\right)} 
{\left(s-m_\text{Z}^2\right)^2 + m_\text{Z}^2\Gamma_\text{Z}^2} \right).
\end{gather}

\section{Differential cross section}

Substituting the squared matrix element into equation~(\ref{diffxs}) we have
\begin{multline}
\frac{d\sigma}{d\Omega^*} = \frac{e^4}{8\pi^2}\frac{1}{s^3} \frac{p_3}{E} 
\Bigl(C_1\left[\,(p_1.p_4)(p_3.p_2)+(p_1.p_3)(p_2.p_4)\,\right]
+C_2m_3m_4p_1.p_2\Bigr.\\[2mm]
\Bigl.+C_3\left[\,(p_1.p_4)(p_2.p_3)-(p_1.p_3)(p_2.p_4)\,\right]\Bigr).
\end{multline}
From \secref{defs} we see that
\begin{equation}
\begin{split}
(p_1.p_4)(p_2.p_3)&=E^2[E_3E_4+(E_3+E_4)p_3\cos\theta^*+p_3^2\cos^2\theta^*]\,,\\[2mm]
(p_1.p_3)(p_2.p_4)&=E^2[E_3E_4-(E_3+E_4)p_3\cos\theta^*+p_3^2\cos^2\theta^*]\,,
\end{split}
\end{equation}
and so, noting that $s=4E^2$ and $E_3+E_4=2E$,
\begin{equation}
\frac{d\sigma}{d\Omega^*} = \frac{e^4}{16\pi^2}\frac{1}{s^2}\frac{p_3}{E} 
\left[C_1(E_3E_4+p_3^2\cos^2\theta^*)+C_2m_3m_4+C_3(2Ep_3\cos\theta^*)\,\right].
\end{equation}

The final result for the parton-level cross section (indicated by the `hat') 
includes a factor of 1/3 since the quarks must both be of the same colour if they are to annihilate.  So we have
\begin{equation}
\frac {d\hat{\sigma}}{d\Omega^*} (\text{q}\bar{\text{q}}\rightarrow \text{L}^-\text{L}^+) 
= \frac{e^4}{48\pi^2} \frac {1} {\hat{s}^2} \frac {p_3} {E}
\left[ C_1 \left( E_3E_4 + p_3^2\cos^2\theta^*\right) + C_2 m_3 m_4 +
2 C_3 E p_3 \cos\theta^* \right].
\end{equation}
Although the notation is different, this expression is found to agree with the equivalent result in \cite{Azuelos:1994qu}.

\section{Full cross section}

To integrate this we use $d\Omega^*=2{\pi}d(\cos\theta^*)$, and so obtain
\begin{equation}
\hat{\sigma} (\text{q}\bar{\text{q}}\rightarrow \text{L}^-\text{L}^+)= 
\frac {e^4} {12\pi} \frac {1} {\hat{s}^2} \frac {p_3} {E} 
\left[ C_1 \left(E_3E_4 + \frac {p_3^2} {3}\right) + C_2 m_3 m_4 \right].
\end{equation}

\section{Slepton Drell-Yan cross section}

The slepton calculation is in many ways similar to the lepton one above, 
although it is made slightly simpler by the scalar coupling of the sleptons
to Z and $\gamma$.  Left-handed and right-handed scalar states can mix but we
just consider the cross section for production of one or the other (L below).

For Z the vertex factor is then $i\frac{g_\text{Z}}{2}g_\text{L}(p_3-p_4)^{\mu}$, and hence
\begin{equation}
-iM_\text{fi}^\text{Z} = \left[\bar{v}(p_2).-ig_\text{Z}\gamma^{\mu}\frac{1}{2}\left(c_\text{V}^\text{i}-c_\text{A}^\text{i}\gamma^5\right)
u(p_1)\right] \frac{-ig_{\mu\nu}}{s-m_\text{Z}^2+im_\text{Z}\Gamma_\text{Z}}
\biggl[ig_\text{Z}\frac{g_\text{L}}{2}(p_3-p_4)^{\nu}\biggr],
\end{equation}
which gives
\begin{equation}
M_\text{fi}^\text{Z} = \frac{g_\text{Z}^2}{s-m_\text{Z}^2+im_\text{Z}\Gamma_\text{Z}}g_{\mu\nu}
\left[\bar{v}(p_2)\gamma^{\mu}\frac{1}{2}\left(c_\text{V}^\text{i}-c_\text{A}^\text{i}\gamma^5\right)u(p_1)\right]
\biggl[\frac{g_\text{L}}{2}(p_3-p_4)^{\nu}\biggr].
\end{equation}
Note that the first square bracket is exactly as previously, so 
contributes the same factor when we evaluate $\sum_\text{spins} |M_\text{fi}^\text{Z}|^2$.  
The second square bracket contributes
\begin{equation}
\frac{g_\text{L}^2}{4}(p_3-p_4)^{\mu}(p_3-p_4)^{\nu}.
\end{equation}
Multiplying these together we obtain
\begin{equation}
\begin{split}
\langle\,|M_\text{fi}^\text{Z}|^2\,\rangle&=\frac{1}{4}\frac{g_\text{Z}^4}{(s-m_\text{Z}^2)^2+m_\text{Z}^2\Gamma_\text{Z}^2}\\[3mm]
&\times\Bigl(\left[\left(c_\text{V}^\text{i}\right)^2+\left(c_\text{A}^\text{i}\right)^2\right]\frac{g_\text{L}^2}{4}
\left[\,2p_1.(p_3-p_4)p_2.(p_3-p_4)-p_1.p_2(p_3-p_4).(p_3-p_4)\,\right]\Bigr).
\end{split}
\end{equation}
\enlargethispage{-2\baselineskip}

This time to obtain the $\gamma$-exchange version it is necessary to make the
additional substitution of $g_\text{L}=2$.  Therefore
\begin{equation}
\langle\,|M_\text{fi}^{\gamma}|^2\,\rangle=\frac{1}{4}\frac{e^4(q^\text{f})^2(q^\text{i})^2}{s^2} 4\left[\,2p_1.(p_3-p_4)p_2.(p_3-p_4)-p_1.p_2(p_3-p_4).(p_3-p_4)\,\right].
\end{equation}

The final term to be calculated is once again the interference term.  
Neglecting the propagators, the spin sum is
\begin{multline}
\sum_{r,s}
\frac{g_\text{L}}{2}(p_3-p_4)^{\mu}
\left[\bar{u_s}(p_1)\left(c_\text{V}^\text{i}+c_\text{A}^\text{i}\gamma^5\right)\frac {\gamma_{\mu}}{2}.v_r(p_2)
\bar{v_r}(p_2)\gamma_{\nu}.u(p_1)\right]
(p_3-p_4)^{\nu} \\[3mm]
=\mbox{Tr}\left[\gamma.p_1\left(c_\text{V}^\text{i}+c_\text{A}^\text{i}\gamma^5\right)\frac{\gamma_{\mu}}{2}\gamma.p_2
\gamma_{\nu}\right]\frac{g_\text{L}}{2}(p_3-p_4)^{\mu}(p_3-p_4)^{\nu}.
\end{multline}

Multiplying these terms (once again noting that the trace is the same as 
appeared in the lepton cross section) we obtain
\begin{multline}
\langle(M_\text{fi}^\text{Z})^*M_\text{fi}^{\gamma}\rangle=\frac{1}{4}\frac{g_\text{Z}^2e^2q^\text{f}q^\text{i}}
{s(s-m_\text{Z}^2-im_\text{Z}\Gamma_\text{Z})}\\[3mm]
\times c_\text{V}^\text{i}g_\text{L}\left[\,2p_1.(p_3-p_4)p_2.(p_3-p_4)-p_1.p_2(p_3-p_4).(p_3-p_4)\,\right].
\end{multline}

Making the additional change of variables $h_\text{L}=g_\text{L}g_\text{Z}/2e$, we arrive at the total matrix element squared:
\begin{equation}
\langle\,|M_\text{fi}|^2\,\rangle=\frac{e^4}{s^2}
D\left[\,2p_1.(p_3-p_4)p_2.(p_3-p_4)-p_1.p_2(p_3-p_4).(p_3-p_4)\,\right],
\end{equation}
where
\begin{equation}
D = \frac {h_\text{L}^2 \left[\left(d_\text{V}^\text{i}\right)^2 +\left(d_\text{A}^\text{i}\right)^2 \right] 
s^2} {\left(s-m_\text{Z}^2\right)^2 + m_\text{Z}^2\Gamma_\text{Z}^2} + 
\left(q^\text{f}\right)^2 \left(q^\text{i}\right)^2 +
\frac {2 q^\text{f} q^\text{i} h_\text{L} d_\text{V}^\text{i}s\left(s-m_\text{Z}^2\right)}
{\left(s-m_\text{Z}^2\right)^2 + m_\text{Z}^2\Gamma_\text{Z}^2}\,.
\end{equation}

Therefore the differential cross section is
\begin{equation}
\frac{d\sigma}{d\Omega^*} = \frac{e^4}{64\pi^2}\frac{1}{s^3} \frac{p_3}{E} 
D\left[\,2p_1.(p_3-p_4)p_2.(p_3-p_4)-p_1.p_2(p_3-p_4).(p_3-p_4)\,\right].
\end{equation}
The 4-momenta in \secref{defs} give us
\begin{equation}
\begin{split}
p_1.(p_3-p_4)p_2.(p_3-p_4)=&E^2\left[\,(E_3-E_4)^2-4p_3^2\cos^2\theta^*\right],\\[2mm]
p_1.p_2(p_3-p_4).(p_3-p_4)=&2E^2\left[\,(E_3-E_4)^2-4p_3^2\right].
\end{split}
\end{equation}
Therefore
\begin{equation}
2p_1.(p_3-p_4)p_2.(p_3-p_4)-p_1.p_2(p_3-p_4).(p_3-p_4)
=8E^2p_3^2\sin^2\theta^*\,,
\end{equation}
and the differential cross~section is
\begin{equation}
\frac{d\sigma}{d\Omega^*} = \frac{e^4}{32\pi^2}D\frac{1}{s^2}\frac{p_3}{E} 
p_3^2\sin^2\theta^*.
\end{equation}
The final result for either left- or right-handed sleptons, including the 
1/3 colour factor, is therefore
\begin{equation}
\frac {d\hat{\sigma}}{d\Omega^*}(\text{q}\bar{\text{q}}\rightarrow\tilde{\text{l}}_\text{L/R}
\tilde{\text{l}}^*_\text{L/R})=\frac {e^4} {96\pi^2} D \frac {1}{\hat{s}^2} \frac {p_3} 
{E} p_3^2 \sin^2\theta^*.
\end{equation}
\clearpage{\pagestyle{empty}\cleardoublepage}
\chapter{Black Hole Master Equation}
\chaptermark{Black Hole Master Equation}
\label{appb}

In this appendix, Kanti's derivation of a master equation is reproduced from \cite{Harris:2003eg}.  The Newman-Penrose formalism is used to obtain this equation describing the motion of a particle with spin $s$ in the background of a higher-dimensional,  non-rotating, neutral black
hole projected onto a 3-brane. The corresponding 4-dimensional metric tensor is
given in eq.~(\ref{non-rot}). 

We first need to choose a tetrad basis of null
vectors $(\ell^\mu, n^\mu, m^\mu, \bar m^\mu)$, where $\ell$ and $n$ are real
vectors and $m$ and $\bar m$ are a pair of complex conjugate vectors. They
satisfy the relations ${\bf l} \cdot {\bf n}=1$ and ${\bf m}\cdot {\bf \bar m}=-1$,
with all other products being zero. Such a tetrad basis is given by
\begin{eqnarray}
\ell^\mu=\Bigl(\frac{1}{h},\,1, \,0,\,0\Bigr)\,, &\quad &
n^\mu=\Bigl(\frac{1}{2}, \,-\frac{h}{2}, \,0, \,0\Bigr)\,, \nonumber \\[3mm]
m^\mu=\Bigl(0, \,0, \,1, \,\frac{i}{\sin\theta}\Bigr)\,
\frac{1}{\sqrt{2} r}\,, &\quad&
\bar m^\mu=\Bigl(0, \,0, \,1, \,\frac{-i}{\sin\theta}\Bigr)\,
\frac{1}{\sqrt{2} r}\,. 
\end{eqnarray}

The $\lambda_{abc}$ coefficients, which are used to construct the spin coefficients, are defined by
$\lambda_{abc}=(e_b)_{i,j}\Bigl[(e_a)^i (e_c)^j-(e_a)^j (e_c)^i \Bigr]$,
where $e_a$ stands for each one of the null vectors and $(i,j)$ denote the
components of each vector. Their non-vanishing components are found to be
\begin{equation}
\lambda_{122}=-\frac{h'}{2}\,, \quad \lambda_{134}=\frac{1}{r}\,,
\quad \lambda_{234}=-\frac{h}{2r}\,, \quad
\lambda_{334}=\frac{\cos\theta}{\sqrt{2} r \sin\theta}\,.
\end{equation}
The above components must be supplemented by those that follow from the
symmetry $\lambda_{abc}=-\lambda_{cba}$ and the complex conjugates obtained
by replacing an index 3 by 4 (or vice versa) or by interchanging 3 and 4.

We may now compute the spin coefficients defined by
$\gamma_{abc}=(\lambda_{abc}+\lambda_{cab} -\lambda_{bca})/2$. Particular
components, or combinations, of the spin coefficients can be directly used
in the field equations \cite{Newman:1962qr,chandra}. They are found to have the
following values:
\begin{eqnarray}
\kappa=\sigma=\lambda=\nu=\tau=\pi=\epsilon=0\,; \nonumber
\end{eqnarray}
\begin{equation}
\rho=-\frac{1}{r}\,; \quad \mu=-\frac{h}{2r}\,; \quad \gamma=\frac{h'}{4}\,; 
\quad\alpha=-\beta=-\frac{\cot\theta}{2\sqrt{2} r}\,.
\end{equation}

In what follows, we will also employ the Newman-Penrose operators
\begin{equation}
\hat D=\frac{1}{h}\,\frac{\partial \,}{\partial t} +
\frac{\partial \,}{\partial r}\,, \qquad
\hat \Delta = \frac{1}{2}\,\frac{\partial \,}{\partial t} -
\frac{h}{2}\,\frac{\partial \,}{\partial r}\,, \qquad
\hat \delta = \frac{1}{\sqrt{2} r}\,\Bigl(\frac{\partial \,}{\partial \theta} +
\frac{i}{\sin\theta}\,\frac{\partial \,}{\partial \varphi}\Bigr)\,,
\end{equation}
and make use of the field factorization
\begin{equation}
\Psi_s(t,r,\theta,\varphi)= e^{-i\om t}\,e^{i m \varphi}\,R_{s}(r)
\,{}_sS^{m}_{\ell}(\theta)\,,
\label{facto}
\end{equation}
where ${}_sY^m_{\ell}(\theta,\varphi)=e^{i m \varphi}\, {}_sS^{m}_{\ell}(\theta)$ are the spin-weighted
spherical harmonics \cite{Goldberg:1967uu}. We will now consider each type of field
separately.

\section{Gauge bosons ($s=1$)}

In the Newman-Penrose formalism, there are only three `degrees of freedom'
for a gauge field (namely $\Phi_0=F_{13}$, $\Phi_1=(F_{12}+ F_{43})/2$ and
$\Phi_2=F_{42}$) in terms of which the different components of the Yang-Mills 
equation for a massless gauge field are written as
\begin{align}
(\hat D-2 \rho)\,\Phi_1 - (\hat \delta^*-2 \alpha) \,\Phi_0 =& 0\,, 
\label{B1}\\
\hat \delta\,\Phi_1 - (\hat \Delta + \mu-2 \gamma) \,\Phi_0 =& 0\,, 
\label{B2}\\
(\hat D-\rho)\,\Phi_2 - \hat \delta^*\,\Phi_1 =& 0\,, \label{B3}\\
(\hat \delta+2 \beta)\,\Phi_2 - (\hat \Delta +2 \mu) \,\Phi_1 =& 0\,, 
\label{B4}
\end{align}
where $\hat \delta^*$ stands for the complex conjugate of $\hat \delta$.
Rearranging eqs. (\ref{B1}) and (\ref{B2}), we can see that $\Phi_1$ decouples
leaving behind an equation involving only $\Phi_0$. Using the explicit forms
of the operators and spin coefficients, as well as the factorized ansatz
of eq.~(\ref{facto}), this can be separated into an angular equation,
\begin{equation}
\frac{1}{\sin\theta}\,\frac{d \,}{d \theta}\,
\biggl(\sin\theta\,\frac{d\, {}_1S^m_{\ell}}{d \theta}\biggr)
+ \biggl[ -\frac{2 m \cot\theta}{\sin\theta} -\frac{m^2}{\sin^2\theta}
+ 1 - \cot^2\theta + {}_1\lambda_{\ell} \biggl] {}_1S^m_{\ell}(\theta)=0\,,
\end{equation}
with eigenvalue ${}_s\lambda_{\ell}=\ell\,(\ell +1) - s\,(s+1)$, and a radial
equation,
\begin{equation}
\frac{1}{\Delta}\,\frac{d \,}{d r}\biggl(
\Delta^2\,\frac{d R_1}{d r}\biggr) + \biggl[
\frac{\omega^2 r^2}{h} + 2 i \omega r -\frac{i \omega r^2 h'}{h}
+ (\Delta'' -2) -{}_1\lambda_{\ell}\biggr]\,R_1(r)=0\,,
\end{equation}
where $\Delta=h r^2$. 

\section{Fermion fields ($s=1/2$)}

For a massless two-component spinor field, the Dirac equation can be
written as
\begin{align}
(\hat \delta^*-\alpha)\,\chi_0 =& (\hat D-\rho)\,\chi_1\,, \\
(\hat \Delta + \mu -\gamma)\,\chi_0 =& (\hat \delta + \beta)\,\chi_1\,.
\end{align}
Performing a similar rearrangement as in the case of bosons, we find that
$\chi_1$ is decoupled and that the equation for $\chi_0$ reduces to the
following set of angular,
\begin{equation}
\frac{1}{\sin\theta}\,\frac{d \,}{d \theta}\,
\biggl(\sin\theta\,\frac{d\, {}_{1/2}S^m_{\ell}}{d \theta}\biggr)
+ \biggl[ -\frac{m \cot\theta}{\sin\theta} -\frac{m^2}{\sin^2\theta}
+ \frac{1}{2} - \frac{1}{4}\,\cot^2\theta + {}_{1/2}\lambda_{\ell} \biggl] 
{}_{1/2}S^m_{\ell}(\theta)=0\,,
\end{equation}
and radial,
\begin{equation}
\frac{1}{\sqrt{\Delta}}\,\frac{d \,}{d r}\biggl(
\Delta^{3/2}\,\frac{d R_{1/2}}{d r}\biggr) + \biggl[
\frac{\omega^2 r^2}{h} + i \omega r -\frac{i \omega r^2 h'}{2h}
+ \frac{1}{2}\,(\Delta'' -2) -{}_{1/2}\lambda_{\ell}\biggr]\,R_{1/2}(r)=0\,,
\end{equation}
equations, with the same definitions for $\Delta$ and ${}_s\lambda_{\ell}$
as before.  

\section{Scalar fields ($s=0$)}

For completeness the equation of motion for a scalar field
propagating in the same background is included here. This equation can be determined quite
easily by evaluating the double covariant derivative $g^{\mu\nu} D_\mu D_\nu$
acting on the scalar field. It finally leads to this pair of equations:
\begin{equation}
\frac{1}{\sin\theta}\,\frac{d \,}{d \theta}\,\biggl(\sin\theta\,
\frac{d\, {}_0S^m_{\ell}}{d \theta}\,\biggr) + \biggl[-\frac{m^2}{\sin^2\theta}
+ {}_0\lambda_{\ell} \biggr]\,{}_0S^m_{\ell}=0\,;
\end{equation}
\begin{equation}
\frac{d \,}{dr}\,\biggl(\Delta\,\frac{d R_0}{dr}\biggr) +
\Bigl(\frac{\om^2 r^2}{h} - {}_0\lambda_{\ell}\Bigr) R_0(r) =0\,.
\label{scalar}
\end{equation}
Here, ${}_0Y^m_{\ell}(\theta, \varphi)=e^{i m \varphi}\,{}_0S^{m}_{\ell}(\theta)$ are the usual
spherical harmonics $Y^m_\ell(\theta, \varphi)$ and
${}_0\lambda_{\ell}=\ell (\ell+1)$.
The above equations were used in \cite{Kanti:2002nr} for the analytic determination
of grey-body factors for the brane emission of scalar particles by higher-dimensional black holes.

\section{Master equation for a field with arbitrary spin}

Combining the equations derived above for bosons, fermions and scalar fields, we may now rewrite them in the form of a master equation, valid for field types. The radial equation then takes
the form
\begin{equation}
\Delta^{-s}\,\frac{d \,}{dr}\,\biggl(\Delta^{s+1}\,\frac{d R_s}{dr}\,\biggr) +
\biggl(\frac{\om^2 r^2}{h} + 2i s\,\om\,r -\frac{i s \om\,r^2 h'}{h}
+s\,(\Delta''-2) - {}_s\lambda_{\ell} \biggr)\,R_s(r)=0\,,
\label{master1}
\end{equation}
while the angular equation reads
\begin{equation}
\frac{1}{\sin\theta}\,\frac{d \,}{d \theta}\,\biggl(\sin\theta\,
\frac{d S^m_{s,\ell}}{d \theta}\,\biggr) + \biggl[-\frac{2 m s \cot\theta}
{\sin\theta} - \frac{m^2}{\sin^2\theta} + s - s^2 \cot^2\theta 
+ {}_s\lambda_{\ell} \biggr]\,S^m_{s,\ell}=0\,.
\label{master2}
\end{equation}
\enlargethispage{-0.5\baselineskip}

The latter equation is identical to the one derived by Teukolsky \cite{Teukolsky:1973ha}
in the case of a non-rotating, spherically-symmetric black hole. However the radial one
differs by the extra factor $s\,(\Delta''-2)$ because for the metric
tensor here this combination is not zero (unlike for the
4-dimensional Schwarzschild and Kerr metrics). The $\Delta''$-term can be removed
if we make the redefinition $R_s=\Delta^{-s}\, P_s$.  Then, we obtain
\begin{equation}
\Delta^{s}\,\frac{d \,}{dr}\,\biggl(\Delta^{1-s}\,\frac{d P_s}{dr}\,\biggr) +
\biggl(\frac{\om^2 r^2}{h} + 2i s\,\om\,r -\frac{i s \om\,r^2 h'}{h}
- {}_s\Lambda_\ell \biggr)\,P_s(r)=0\,,
\label{master3}
\end{equation}
where now ${}_s\Lambda_\ell={}_s\lambda_{\ell} +2 s =\ell\,(\ell+1)-s\,(s-1)$. The above
form of the radial equation was used in \cite{Kanti:2002ge} to determine analytically the grey-body factors and
emission rates for fermions and gauge bosons on the brane.
\clearpage{\pagestyle{empty}\cleardoublepage}
\chapter{Black Hole Monte Carlo Method}
\label{appc}

\section{Monte Carlo calculation of the cross section}

The main features of the Monte Carlo (MC) method can be described by the following
mathematical relationship:
\begin{equation}
I=\int_{x_1}^{x_2} f(x)\, dx = (x_2-x_1) \left<f(x)\right> \approx (x_2-x_1) \frac{1}{N} \sum_{i=1}^{N} f(x_i),
\label{mcdef}
\end{equation}
where the $f(x_i)$ are values of $f(x)$ at the $N$ randomly chosen values of $x \in[x_1,x_2]$.

In this case the integral we require is for the black hole production cross section $\sigma$:
\begin{equation}
\sigma = \int_{E_\text{min}}^{E_\text{max}} \hat{\sigma}(E) p(E)\, dE, 
\end{equation}
where $E$ is the centre-of-mass energy of the partons involved and $\hat{\sigma}$ is the parton-level cross section.  Using the substitution $y(E)=dY/dE$ we obtain
\begin{equation}
\sigma = \int_{Y_1=Y(E_\text{min})}^{Y_2=Y(E_\text{max})} \frac{\hat{\sigma}(E) p(E)}{y(E)}\, dY.
\end{equation}

The MC procedure will be most efficient if the integrand is as flat as
possible as a function of $E$.  The parton-level geometrical cross section for black hole production means that $\hat{\sigma} (E) \sim E^{\beta}$ where $\beta = 2/(n+1)$.  In addition we expect that $p(E) \sim E^{-8}$ because of the behaviour of the parton distribution functions (PDFs).  Hence a sensible choice for $y(E)$ is
\begin{equation}
y(E)=\alpha E^{\alpha -1} \Rightarrow Y(E)=E^{\alpha},
\end{equation}
where $\alpha = \beta -7$.  Returning to eq.~(\ref{mcdef}) we have
\begin{equation}
\sigma = (Y(E_\text{max})-Y(E_\text{min})) \left<\frac{\hat{\sigma} (E) p(E)}{y(E)} \right> = \frac{E_\text{max}^{\alpha}-E_\text{min}^{\alpha}}{\alpha} \left< \frac{\hat{\sigma} (E) p(E)}{E^{\alpha-1}}\right>,
\end{equation}
where $Y(E)$ is chosen at random from a uniform distribution between the maximum and minimum values.
\enlargethispage{-2\baselineskip}

$p(E)$ is calculated from the PDFs and is more commonly written as a function of $\tau=\hat{s}/s$ (hence $\tau \propto E^2$).  Therefore $p(E)$ can be re-written as
\begin{equation}
p(E)=\frac{2\tau}{E} p(\tau)=\frac{2\tau}{E} \int_{\tau}^{1} \frac{dx}{x} f_1(x) f_2\left(\frac{\tau}{x}\right).
\end{equation}
This integral can also be evaluated by a MC procedure, and once again
a change of variable is useful.  We may conveniently write $p(\tau)$ as
\begin{equation}
p(\tau)=\frac{1}{\tau}\int_{\tau}^{1}\frac{dx}{x} x f_1(x) \frac{\tau}{x} f_2\left(\frac{\tau}{x}\right)=
\frac{1}{\tau}\int_{\tau}^{1} \frac{dx}{x} h(x,\tau)\,,
\end{equation}
and the substitution $z(x)=dZ/dx$ gives us that
\begin{equation}
p(\tau)=\frac{1}{\tau}\int_{Z(\tau)}^{Z(1)} \frac{h(x,\tau)}{x z(x)} \, dZ\,.
\end{equation}
The assumption that $h(x,\tau)$ is fairly smooth (since $f(x)$ is usually of the approximate form $1/x$) means that the MC procedure should be efficient if 
\begin{equation}
z(x)=\frac{1}{x} \Rightarrow Z(x)=\ln x\,.
\end{equation}
So, using eq.~(\ref{mcdef}), we conclude that
\begin{equation}
p(\tau)=\frac{(Z(1)-Z(\tau))}{\tau}\left<h(x,\tau)\right>= -\frac{\ln\tau}{\tau} \left<h(x,\tau)\right>,
\end{equation}
where $Z(x)$ is chosen at random from a uniform distribution between the maximum and minimum values.

Combining these two MC procedures, a good approximation to the total cross section $\sigma$ is given by
\begin{equation}
\frac{E_\text{max}^{\alpha}-E_\text{min}^{\alpha}}{\alpha}\left<\frac{\hat{\sigma}}{E^{\alpha-1}}\frac{2}{E}(-\ln\tau) h(x,\tau)\right>,
\end{equation}
where firstly $Y(E)$ is chosen at random ($\rightarrow E \mbox{ and hence } \tau$) and then, given this value of $\tau$, $Z(x)$ is chosen at random ($\rightarrow x$).

\section{Monte Carlo generation of unweighted events}

As well as estimating the total cross section, the MC program must generate events drawn from this distribution.  This can be done using essentially the same procedure---$E$ and then $x$ are chosen and the cross section value is computed.  The calculated weight is then accepted or rejected against the maximum weight found by an initial search.  A weight $w_i$ is accepted with probability $w_i/w_\text{max}$.

\clearpage{\pagestyle{empty}\cleardoublepage}
\begin{spacing}{1.5}
\fancyhead[LO,RE]{\leftmark}
\addcontentsline{toc}{chapter}{Bibliography}
\bibliography{thesis}
\bibliographystyle{cmhJHEP}
\end{spacing}

\end{document}